\documentclass[atlasdraft=false, cernpreprint, UKenglish, texmf, orcidlogo]{atlasdoc}
\usepackage{atlaspackage}
\usepackage{atlasbiblatex}

\usepackage[BSM=true,jetetmiss=true]{atlasphysics}
\graphicspath{{logos/}{figures/}}

\usepackage{ANA-EXOT-2023-14-PAPER-defs}

\AtlasTitle{Exploration at the high-energy frontier: ATLAS Run~2 searches investigating the exotic jungle beyond the Standard Model}

\AtlasAbstract{%
This report presents a comprehensive collection of searches for new physics performed by the ATLAS Collaboration during the Run~2 period of data taking at the Large Hadron Collider, from 2015 to 2018, corresponding to about 140~fb$^{-1}$ of $\sqrt{s}=13$~TeV proton--proton collision data. These searches cover a variety of beyond-the-standard model topics such as dark matter candidates, new vector bosons, hidden-sector particles, leptoquarks, or vector-like quarks, among others. Searches for supersymmetric particles or extended Higgs sectors are explicitly excluded as these are the subject of separate reports by the Collaboration. For each topic, the most relevant searches are described, focusing on their importance and sensitivity and, when appropriate, highlighting the experimental techniques employed. In addition to the description of each analysis, complementary searches are compared, and the overall sensitivity of the ATLAS experiment to each type of new physics is discussed. Summary plots and statistical combinations of multiple searches are included whenever possible.
}

\AtlasRefCode{EXOT-2023-14}

\PreprintIdNumber{CERN-EP-2024-075}

\AtlasJournal{Physics Reports}
\AtlasJournalRef{Phys. Rep. 1116 (2025) 301-385}
\AtlasDOI{10.1016/j.physrep.2024.10.001}

\hypersetup{pdftitle={ATLAS document},pdfauthor={The ATLAS Collaboration}}

\begin{document}

\maketitle

\tableofcontents

\section{Introduction}

While describing a vast amount of data with excellent precision, the Standard Model (SM) of particle physics leaves many questions open, many of which are at the core of the actual particle physics search strategy, as also discussed in~Refs.~\cite{Bose:2022obr,EuropeanStrategyforParticlePhysicsPreparatoryGroup:2019qin}.
The Large Hadron Collider (LHC)~\cite{Evans:2008zzb} at CERN stands at the forefront of particle physics investigation, providing a wealth of collision data and unprecedented centre-of-mass energy. During \RunTwo, from 2015 to 2018, 140~\ifb of $\sqrt{s}=13$~\TeV proton--proton collision data were collected by the ATLAS experiment. These data were thoroughly scrutinized by the ATLAS Collaboration to search for hints of physics beyond the SM (BSM), as summarized in this report. The diagram in Figure~\ref{fig:diagram} provides a graphical representation of how the different search topics briefly introduced below and covered in this report fit with some of the open questions of the SM and within its structure.

The fermionic sector of the SM is rather well known, the last piece of the puzzle falling into place in 2000 with the observation of the tau neutrino by the DONUT Collaboration~\cite{DONUT:2000fbd}. There are however remaining questions concerning the known fermions. Their three-generation structure is not explained in the SM; it is thus tantalizing to question whether these fermions are really fundamental or if they could instead be composed of even smaller constituents whose rules of assembly could explain the lepton and quark families. This compositeness could be evinced by searches looking for excited fermionic states. The SM is also silent on the neutrino masses, assuming they are identically zero. Are the tiny neutrino masses simply due to seemingly unnaturally small Yukawa couplings, or are they explained by mechanisms involving additional heavy neutral leptons? Multiple final states can be probed in searching for these heavy neutrino partners and their eventual associated BSM particles.

But besides these questions, is the structure evoked above even complete? Additional lepton- or quark-like states may exist; while adding additional chiral ones is heavily disfavoured, vector-like fermions can be more easily accommodated by existing constraints as their masses is not linked to the Higgs mechanism. An extensive search program targeting their production, either alone or in pairs, has been carried out. Particles carrying both baryon and lepton numbers, the leptoquarks, are also postulated by some BSM theories trying to explain the similarity between the lepton and quark sectors of the SM. Like the vector-like fermions, they can be singly or pair produced and their searches cover a broad set of final states.

Similar considerations also apply to the gauge sector of the SM: while it describes well the known interactions, the set of gauge groups could be enlarged, predicting new vector bosons, like the \Zprime and the \Wprime, interacting with the SM particles. Depending on their masses and decays, resonant and non-resonant searches for such states can be conducted. The $Z$ boson decays have also been investigated to search for signs of charged-lepton flavour violation and searches for a \Zprime in the same final states have also been made.

New gauge interactions, or interactions through the Higgs boson could also connect the SM particles to an entire sector of BSM that remains \enquote{dark}. These new particles would be hidden from the interactions of the SM and interact with it only through a heavy or feebly coupled mediator. This could lead to peculiar signatures of long-lived neutral particles. Such dark sector theories and others predicting weakly interacting massive stable particles can provide candidates to explain the nature of dark matter. The possibility of producing dark matter candidates at the LHC, through the mediators that would connect them to SM particles, has led to the development of a vibrant research activity covering many final states. Long-lived particles  could also be electrically multi-charged, or highly ionizing particles such as magnetic monopoles could exist, leading to very exotic signatures that can be probed in ATLAS.

Finally, various theories try to explain the smallness of the gravitational coupling compared to the electroweak scale as an artefact of extra dimensions. They can predict gravitons and quantum black holes which are also the focus of dedicated searches.

Besides the searches for these exotic phenomena, searches for supersymmetry or additional Higgs bosons, as well as precision measurements of SM particles and their interactions, have also been pursued and are reviewed in other reports~\cite{SUSY-2023-10,HDBS-2023-15,HIGG-2023-11,STDM-2023-20,TOPQ-2023-19}. A similarly broad search and measurement program has also been pursued with the \RunTwo proton--proton collision data by the CMS Collaboration and is documented in other reports~\cite{CMS:2024bni,CMS:2024zqs,CMS-B2G-23-002,CMS-EXO-23-007,CMS-TOP-23-003,CMS:2024gzs}.

This report is organized as follows. After Section~\ref{sec:tools}, which introduces the common methods and tools used in the analyses, each of the topics discussed above is covered in a dedicated section, showing how the ATLAS Collaboration analysed the \RunTwo data in its quest for answers. The compositeness hypothesis has been probed for quarks and leptons and is addressed in Section~\ref{sec:compositness}, while searches for heavy neutral leptons are discussed in Section~\ref{sec:leptons}.  Sections~\ref{sec:veclike} and~\ref{sec:lq} summarize the searches for vector-like quarks and leptons, and for leptoquarks, respectively. Section~\ref{sec:gauge} reviews the resonant and non-resonant searches for \Zprime and \Wprime gauge bosons. The searches for charged-lepton flavour violation in $Z$ or \Zprime decays are discussed in Section~\ref{sec:lfv}.  Sections~\ref{sec:hidden} and~\ref{sec:dm} review the searches for long-lived neutral particles coming from hidden sectors and for dark matter candidates, respectively. Analyses looking for multi-charged and highly ionizing particles have also been conducted, as discussed in Section~\ref{sec:mcp}. Finally, Section~\ref{sec:gravity} reviews the searches for gravitons and quantum black holes, and a summary is given in Section~\ref{sec:conclusion}.

\begin{figure}[tb]
\begin{center}
\includegraphics[width=0.9\textwidth]{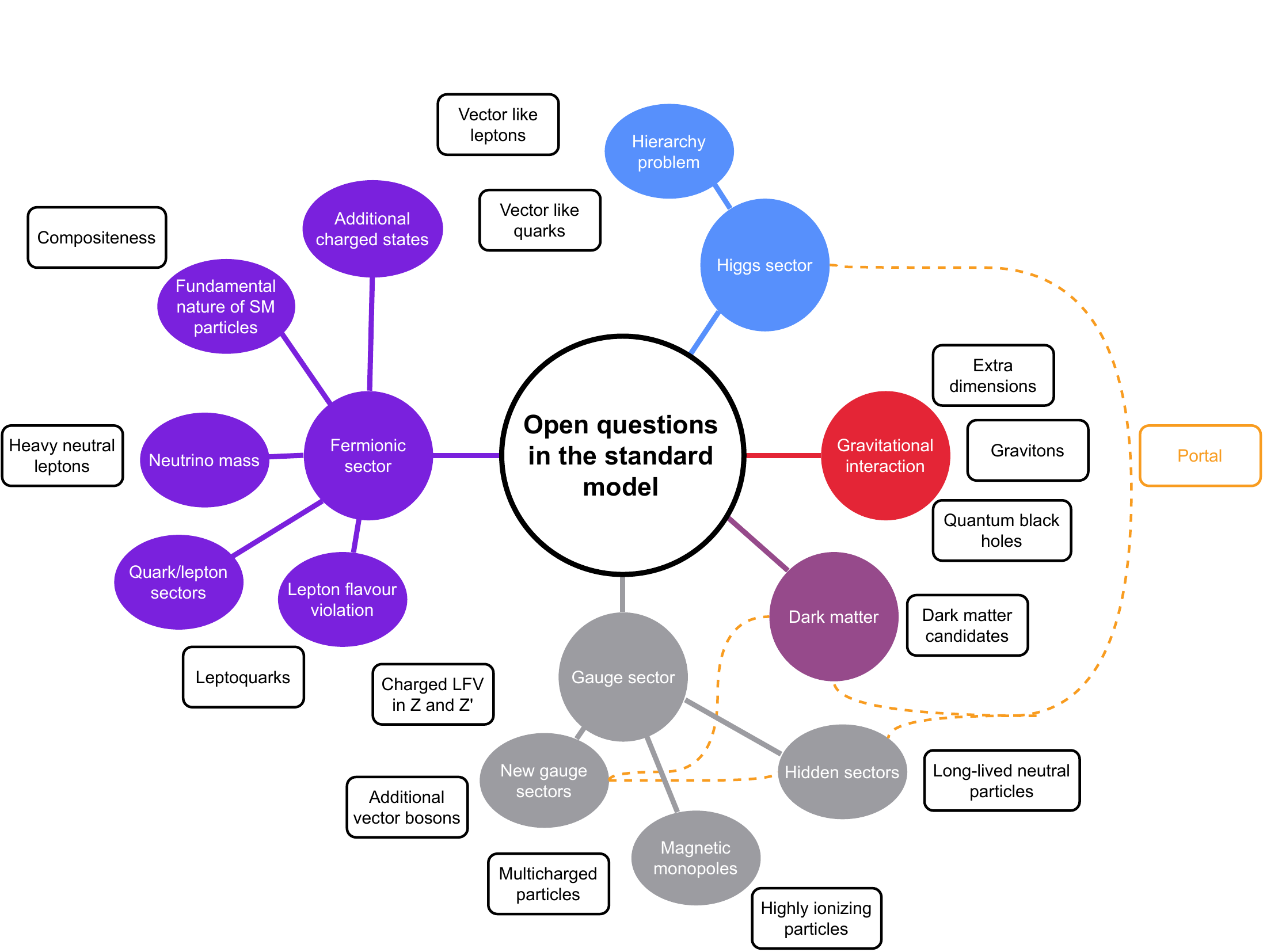}
\end{center}
\caption{Graphical representation of the different search topics covered by this report and their relationship with some of the open questions in the SM.}
\label{fig:diagram}
\end{figure}

\label{sec:intro}
%


\section{Tools and methods}
\label{sec:tools}

%
\newcommand{\AtlasCoordFootnote}{%
ATLAS uses a right-handed coordinate system with its origin at the nominal interaction point (IP)
in the center of the detector and the \(z\)-axis along the beam pipe.
The \(x\)-axis points from the IP to the center of the LHC ring,
and the \(y\)-axis points upwards.
Polar coordinates \((r,\phi)\) are used in the transverse plane,
\(\phi\) being the azimuthal angle around the \(z\)-axis.
The pseudorapidity is defined in terms of the polar angle \(\theta\) as \(\eta = -\ln \tan(\theta/2)\).
Angular distance is measured in units of \(\Delta R \equiv \sqrt{(\Delta\eta)^{2} + (\Delta\phi)^{2}}\).}

\subsection{The ATLAS detector}
\label{sec:detector}

The ATLAS detector~\cite{PERF-2007-01} at the LHC covers nearly the entire solid angle around the collision point.\footnote{\AtlasCoordFootnote}
It consists of an inner tracking detector surrounded by a thin superconducting solenoid, electromagnetic and hadron calorimeters,
and a muon spectrometer incorporating three large superconducting air-core toroidal magnets.

The inner-detector system (ID) is immersed in a \qty{2}{\tesla} axial magnetic field
and provides charged-particle tracking in the range \(|\eta| < 2.5\).
The high-granularity silicon pixel detector covers the vertex region and typically provides four measurements per track,
the first hit normally being in the insertable B-layer (IBL) installed before Run~2~\cite{ATLAS-TDR-19,PIX-2018-001}.
It is followed by the silicon microstrip tracker (SCT), which usually provides eight measurements per track.
These silicon detectors are complemented by the transition radiation tracker (TRT) which is made of straws filled with a Xenon- or Argon-gas-based mixture
and which enables radially extended track reconstruction up to \(|\eta| = 2.0\).
The TRT also provides electron identification information
based on the fraction of hits (typically 30 in total) above a higher energy-deposit threshold corresponding to transition radiation.

The calorimeter system covers the pseudorapidity range \(|\eta| < 4.9\).
Within the region \(|\eta|< 3.2\), electromagnetic calorimetry is provided by barrel and
endcap high-granularity lead/liquid-argon (LAr) calorimeters,
with an additional thin LAr presampler covering \(|\eta| < 1.8\)
to correct for energy loss in material upstream of the calorimeters.
Hadron calorimetry is provided by the steel/scintillator-tile calorimeter,
segmented into three barrel structures within \(|\eta| < 1.7\), and two copper/LAr hadron endcap calorimeters.
The solid angle coverage is completed with forward copper/LAr and tungsten/LAr calorimeter modules
optimized for electromagnetic and hadronic energy measurements respectively.

Surrounding the calorimeter system is the muon spectrometer (MS), which comprises separate trigger and
high-precision tracking chambers measuring the deflection of muons in a magnetic field generated by the superconducting air-core toroidal magnets.
The field integral of the toroids ranges between \num{2.0} and \qty{6.0}{\tesla\metre}
across most of the detector.
Three layers of precision chambers, each consisting of layers of monitored drift tubes (MDT), cover the region \(|\eta| < 2.7\),
complemented by cathode-strip chambers (CST) in the forward region, where the background is highest.
The muon trigger system covers the range \(|\eta| < 2.4\) with resistive-plate chambers (RPC) in the barrel, and thin-gap chambers (TGC) in the endcap regions.

Interesting events are selected by the first-level trigger system implemented in custom hardware,
followed by selections made by algorithms implemented in software in the high-level trigger~\cite{TRIG-2016-01}.
The first-level trigger accepts events from the \qty{40}{\MHz} bunch crossings at a rate below \qty{100}{\kHz},
which the high-level trigger further reduces in order to record events to disk at about \qty{1}{\kHz}.

An extensive software suite~\cite{ATL-SOFT-PUB-2021-001} is used in data simulation, in the reconstruction
and analysis of real and simulated data, in detector operations, and in the trigger and data acquisition
systems of the experiment.


\subsection{Data and Monte Carlo samples}
\label{sec:dataMC}
The analyses described in this report, when not specified otherwise, are based on the full \RunTwo dataset collected by ATLAS between 2015 and 2018. The data sample is selected by requiring good conditions for the beams and the ATLAS detector and corresponds to an integrated luminosity of $140.1\pm1.2$~\ifb~\cite{DAPR-2021-01}, with some analyses included in this report using a preliminary value of $139.0\pm2.4$~\ifb~\cite{ATLAS-CONF-2019-021}. Various triggers are used to select the data, depending on the analysis. For example, the lowest unprescaled-trigger thresholds during \RunTwo were at a \pT of 420~\GeV for a single small-$R$ jet~\cite{TRIG-2012-01}, 140~\GeV for a single photon~\cite{TRIG-2018-05}, and varied between 20 and 26~\GeV for a single electron or muon~\cite{TRIG-2018-01}, while the lowest unprescaled-trigger threshold for missing transverse momentum varied between 70 and 110~\GeV~\cite{TRIG-2019-01}.

Monte Carlo (MC) samples, which are used to model signal and background events, were produced using a variety of generators, depending on the process of interest. For the signal samples, unless specified, the generator used was either \PYTHIA[8]~\cite{Sjostrand:2007gs,Sjostrand:2014zea}, \MGNLO~\cite{Alwall:2014hca} or \POWHEGBOX[v2]~\cite{Nason:2004rx,Frixione:2007vw,Alioli:2010xd}, with \PYTHIA[8] being used in all cases to model the parton shower, hadronization and underlying event.  The decays of $b$- and $c$-hadrons were performed consistently with the \EVTGEN[1.2.0] decay package~\cite{Lange:2001uf}, except for processes modelled using \Sherpa~\cite{Bothmann:2019yzt}. To account for additional proton--proton interactions (\pileup) in the same and neighbouring bunch crossings, a number of inelastic $pp$ interactions, generated with \Pythia[8.186] using the \NNPDF[2.3lo] PDF set~\cite{Martin:2009iq} and the ATLAS A3 set of tuned parameters~\cite{ATL-PHYS-PUB-2016-017}, were superimposed on the hard-scattering events. The simulated events were weighted so that the distributions of the average number of collisions per bunch crossing in simulation match those in data. The generated MC samples were passed through either a full ATLAS detector simulation~\cite{SOFT-2010-01} using \GEANT~\cite{Agostinelli:2002hh} or a fast simulation which relies on a parameterization of the calorimeter response~\cite{ATL-PHYS-PUB-2010-013}. The simulated events were weighted so that the distributions of the average number of collisions per bunch crossing in simulation match those in data and are processed with the same reconstruction algorithms as data.


\subsection{Object reconstruction and identification}
\label{sec:reco}

Various algorithms are used to reconstruct the objects in the events. They were improved during \RunTwo, bringing better identification efficiency, background rejection, measurement accuracy or precision. Object reconstruction algorithms used in most analyses discussed in this report are reviewed briefly here.

\textbf{Tracks}\\
Track reconstruction~\cite{ATL-PHYS-PUB-2015-051,PERF-2015-08} in the inner detector is seeded from combinations of hits found in the pixel and SCT detectors. These seeds are used to initiate a search for additional hits along the track candidate's trajectory in an iterative procedure using a Kalman filter. Since hits can be shared by multiple tracks, an ambiguity-resolving step then follows, and the remaining tracks are then extrapolated to the TRT detector to form an \enquote{inside-out} track collection. An \enquote{outside-in} reconstruction is also available, which instead extrapolates TRT track segments back into the silicon detectors, using silicon hits that were not selected by the inside-out algorithm. Tracks are required to have a transverse momentum $\pt > 0.5$~\GeV, $|\eta|<2.5$, and an origin compatible with the interaction region, ensured by imposing constraints on their transverse ($d_0$) and longitudinal ($z_0$) impact parameters. They must also satisfy good-quality criteria related to their number of silicon hits (or missing silicon hits, when some hits are expected but not seen along their trajectory). This algorithm is not efficient in reconstructing tracks from the highly displaced vertices predicted in some models of exotic long-lived particles. A dedicated large-radius tracking (LRT) algorithm is employed in these cases: it uses as inputs the hits omitted by standard tracking algorithms and imposes looser requirements, notably on $d_0$ and $z_0$~\cite{ATL-PHYS-PUB-2017-014}.

\textbf{Vertices}\\
From the collision vertices, whose iterative-reconstruction seed positions are based on the beam spot position and which are reconstructed from at least two tracks, the primary vertex~\cite{ATL-PHYS-PUB-2015-026} is selected as the one with the highest $\Sigma \pt^2$ of the associated tracks. Analysed events are required to possess a primary vertex (PV).

\textbf{Jets} \\
Jets, made of collimated showers of hadrons, are reconstructed in the detector by using algorithms that seek to identify them by clustering together different types of inputs. Input constituents considered in jet reconstruction can be inner-detector tracks, calorimeter energy deposits, or a combination of these. The calorimeter energy-deposit inputs are called topological clusters; they are formed as groups of contiguous calorimeter cells with an energy deposit significantly above the noise level~\cite{PERF-2014-07}. The algorithm most commonly used by ATLAS in \RunTwo is the particle-flow algorithm, which exploits both types of constituents~\cite{PERF-2015-09}: to improve the accuracy of the charged-hadron measurement, this algorithm computes their momenta from inner-tracker information only, and retains the calorimeter energy-deposit measurements for the determination of neutral-particle energies. These constituents are in turn grouped into jets using the \antikt algorithm~\cite{Cacciari:2008gp,Fastjet} with either a small ($R = 0.4$) or large ($R = 1.0$) radius parameter value. While the small-$R$ jets are commonly used by analyses concerned with quark- and gluon-initiated jets, large-$R$ jets are usually considered for so-called boosted topologies, where the decay products of a $W/Z/H$ boson or a top quark are collimated. The jets' energies and directions are then corrected by a calibration procedure~\cite{PERF-2016-04}. In special cases, a variable-$R$ algorithm~\cite{Krohn:2009zg} is also employed, in which the radius parameter's value depends on the \pt of the constituents being clustered. A different type of jet is used in some analyses to take advantage of boosted topologies without having to explicitly reconstruct large-$R$ jets from low-level inputs. They are called reclustered jets and are constructed by the anti-\kt algorithm, using small-$R$ jets as inputs.

Jets compatible with noise bursts, beam-induced background or cosmic rays are usually discarded~\cite{ATLAS-CONF-2015-029}, and to reduce the effect of \pileup interactions, a jet-vertex-tagger (JVT) algorithm is usually applied to track information to select jets originating from the PV~\cite{PERF-2014-03}.

\textbf{\btagged jets}\\
Jets originating from \bquarks can be tagged by exploiting the long lifetime of $b$-hadrons. A multivariate algorithm~\cite{FTAG-2019-07,FTAG-2018-01,ATL-PHYS-PUB-2017-013} is used, based on the impact parameters of displaced tracks and the properties of vertices in the jets. Operating points are defined for different targeted efficiencies as measured in simulated $t\bar{t}$ samples.

\textbf{Large-$R$ jet tagging}\\
Jets originating from heavy resonances (top, $W$, $Z$ or $H$) decaying into hadrons can be identified as such by using a variety of techniques based on substructure information. In the case of top-tagging, a deep neural network (DNN)~\cite{JETM-2018-03} that uses a large number of jet-substructure variables as input is often employed, with operating points defined similarly to those for \btagged jets.

\textbf{Photons}\\
Photons can reach the calorimeters directly or convert into $e^{+}e^{-}$ in the material composing the detector. They are reconstructed~\cite{EGAM-2018-01}  from clusters of energy deposits in the electromagnetic calorimeter, along with information from inner-detector tracks which is used to classify them either as converted, if the cluster is matched to a reconstructed conversion vertex, or as unconverted, if there is no match between the cluster and a conversion vertex or an electron track. Based on observables which reflect the shape of the electromagnetic shower in the calorimeter, loose or tight identification criteria are used to separate the photon candidates from background. Isolation criteria based on calorimeter and/or inner-detector information are also applied to avoid contamination such as from neutral-hadron decays into almost-collinear photon pairs.

\textbf{Electrons}\\
Electrons are reconstructed by matching a cluster of energy deposits in the electromagnetic calorimeter to an inner-detector track which is identified as originating from the primary vertex via impact parameter selections~\cite{PERF-2017-01}. Loose, medium or tight identification criteria based, for example, on the cluster's shape are applied depending on the desired levels of purity and background rejection~\cite{EGAM-2018-01}.

\textbf{Muons}\\
Muon reconstruction is usually based on the presence of a PV-compatible inner-detector track that is matched to a track reconstructed in the muon spectrometer. The two tracks are combined to better determine the muon kinematics~\cite{MUON-2018-03}. Identification criteria are applied, with their tightness depending on the targeted signal-to-background ratio.

For both electrons and muons, a veto is usually imposed on the nearby presence of additional significant calorimeter energy deposits and/or tracking activity: these isolation selections aim to remove non-prompt leptons coming, for example, from heavy-flavour decays.

\textbf{Tau leptons}\\
Hadronically decaying $\tau$-leptons (\tauhad) are reconstructed from energy clusters in the calorimeters that are matched to inner-detector tracks~\cite{ATLAS-CONF-2017-029,ATL-PHYS-PUB-2015-045,PERF-2013-06}, according to any of five categories: one matched track with zero, one or more neutral particles; and three matched tracks with zero or more neutral particles. Selections are then made on the reconstructed electric charge of the $\tau$-lepton, which must have an absolute value of one, and on the output score of an identification algorithm based on a recurrent neural network~\cite{ATL-PHYS-PUB-2019-033}. Leptonically decaying $\tau$-leptons are usually identified as muons or electrons.

\textbf{Missing transverse momentum}\\
The missing transverse momentum \met is defined as the magnitude of \ptmiss, which is the negative vectorial sum of the transverse momenta of the objects identified in the event, including a so-called \emph{soft term} which takes into account the remaining soft particle tracks from the PV which are not matched to any object~\cite{JETM-2020-03}. When only a subset of objects is used and no soft term is added, the \met is denoted by $H_\mathrm{T}^\mathrm{miss}$.

The missing transverse momentum's significance, $S_{\met}$, is a variable which can be used to discriminate between the events in which the reconstructed \met originates from weakly interacting particles and the events in which the \met is consistent with resolution effects and inefficiencies in particle measurements. The value of $S_{\met}$ can be approximated by $\met/\sqrt{\HT}$ (event-level significance), where $\HT$ is the scalar sum of the transverse momenta from all the reconstructed objects, or can be computed in a more detailed way (object-level significance) by considering the expected resolution and mismeasurement likelihood of all the objects that enter the \met reconstruction, as detailed in Ref.~\cite{ATLAS-CONF-2018-038}.

\subsection{Useful kinematic variables}

Kinematic variables are used to increase the signal-to-background ratio or to preferentially select some types of backgrounds in order to study them in more detail. Some of them are used by multiple analyses discussed in this report and are therefore briefly described here. For a more detailed review, see for example Ref.~\cite{Franceschini:2022vck}.

\textbf{Invariant masses}\\
The invariant mass of a two-body system, $m_{XY}$ (where $X$ and $Y$ can be jets ($j$), \btagged jets ($b$), top-quark candidates ($t$), light charged leptons ($\ell=e,\mu$) or photons ($\gamma$), for example), is especially useful when resonances are expected. Since $\tau$ decays involve at least one neutrino, the collinear approximation (which is well realized at large momenta) is normally used: the neutrinos produced in the $\tau$ decays are assumed to travel in the same direction as the visible $\tau$-decay products and to be the only source of \met, providing information to estimate the \pT of the $\tau$-lepton and making it possible to compute a collinear invariant mass, $m_{X\tau}^\text{col}$. When reconstructing the mass of a resonance decaying into a pair of $\tau$-leptons, a technique called the missing mass calculator (MMC)~\cite{Elagin:2010aw} is sometimes used. This technique is based on the maximization of a likelihood that replaces the assumptions of the collinear approximation with a requirement that the decay products are consistent with the mass and decay kinematics of a $\tau$-lepton. This accurately reconstructs the mass of the original resonance without the limitations of the collinear approximation.

\textbf{Transverse masses}\\
When part of the decay consists of \met, computing the transverse mass of the visible object $X$ and the \met can be helpful:  $\mT(X,\met)=\sqrt{2\pT^{X}\met(1-\cos{\phi_{X\met}})}$.  The $\mT(\ell,\met)$ variable can be particularly helpful in discriminating against the presence of a leptonically decaying $W$ boson from $W$+jets or $\ttbar$ background processes. The leptonic stransverse mass, $m_\mathrm{T2}$~\cite{Lester:1999tx,Barr:2003rg}, is instead based on two visible leptons and \met. It assumes that the leptons come from the two-body decay of a pair of identical particles into a visible one and an invisible one. For dileptonic \ttbar events, the $m_\mathrm{T2}$ distribution has an endpoint at the $W$ boson mass, while higher values are expected for the signal.  The leptonic asymmetric stransverse mass~\cite{Konar:2009qr,Lester:2014yga}, $am_\mathrm{T2}$, is a variation on $m_\mathrm{T2}$, and is based on one lepton only and can be used to reduce the background from dileptonic \ttbar events in which one of the leptons is missed. The contransverse mass ($m_\text{CT}$)~\cite{Tovey:2008ui} can also be used in the search for heavy pair-produced particles decaying semi-visibly and to reduce of the \ttbar background.

\textbf{Scalar sums of transverse momenta}\\
The scalar sum of the transverse momenta of some objects (denoted by $S_\text{T}^{\mathrm{objects}}$ or $H_\text{T}^\mathrm{objects}$ below, with the objects considered specified) can also be useful in discriminating between more energetic signal events and the background. When this sum runs over all selected objects and also includes the \met value, it is called the effective mass, $m_\mathrm{eff}$,  because it would be correlated with the mass scale of the initially produced heavy \enquote{beyond the SM} (BSM) particles whose decay would have led to the measured objects.


\subsection{Analysis strategy}

For each analysis, one or more signal regions (SRs) are defined by a set or sets of event selections that usually optimize the statistical significance of the signal relative to the expected background. A SR can be defined as a list of requirements on reconstructed objects or kinematic variables, or be based on the output of more sophisticated machine-learning algorithms taking these as inputs. Although machine-learning techniques have been used for quite some time in particle physics, their use has increased recently, helping to improve  object tagging and signal event selection, as reviewed in Ref.~\cite{Karagiorgi:2021ngt}. Once a SR is defined, a reliable way to estimate the expected background along with its associated systematic uncertainties must be devised. Only when this full strategy is in place is the data in the SR unblinded to perform the actual search. There are several ways to estimate the background in a given SR, the choice of method depending on the process of interest and the expected associated uncertainties: using MC simulations only, which is often chosen for sources of background which are expected to contribute only a small proportion of the events in the SR; using MC simulations of events whose yields or distribution shapes are corrected to data in control regions (CRs), a selection of events orthogonal to the SR where a given type of background is enhanced relative to the others and the signal contribution is expected to be small; or in a fully data-driven way, especially useful for events whose simulation might be difficult or unreliable. Validation regions (VRs), which are selected to be close but orthogonal to both the SRs and CRs and should have small expected signal contributions, are also defined sometimes in order to validate the background estimation strategy prior to the unblinding. Some common data-driven background estimation methods used by multiple analyses presented in this report are described in more detail below.

If the signal is expected to peak over a smooth SM background distribution, such as in searches for narrow resonances in an invariant-mass distribution, the background in the SR can often be estimated directly, by fitting a parametric function to the data itself~\cite{ATL-PHYS-PUB-2020-028}. The functional form chosen for the fit of the background should minimize the number of free parameters while being flexible enough to minimize a \emph{spurious signal} (the number of signal events extracted from a signal-plus-background fit performed on a background-only distribution) and allowing a signal to be extracted if present.

Another data-driven technique is known as the ABCD (or 2D-sideband) method. In the most straightforward implementation of the ABCD method, the background in the targeted signal region, A, is estimated from the amounts of data in three contiguous control regions, B, C, and D. Regions B and D are each built by inverting one of the selection criteria used to define the signal region. In contrast, region C is built by inverting both of these criteria simultaneously. The method relies on the assumptions that the two criteria used in constructing the regions are uncorrelated for background events and that the signal contamination in the control regions is minor. If this is the case, then the background in the SR, A, can be predicted from the background-enriched regions via $N_\text{A}=N_\text{D}\times N_\text{B}/N_\text{C}$.

The contribution from \emph{fake} objects in the SR, such as a jet being falsely reconstructed as a lepton, a $\tau$-lepton or a photon, can be evaluated in a fully data-driven way via two different methods: the matrix method and the fake-factor method~\cite{EGAM-2019-01}. In the matrix method, the numbers of targeted objects (e.g.\ leptons) in the SR passing the nominal identification or a loosened identification are counted separately. They are linked to the numbers of true or fake targeted objects by a matrix of the efficiencies for a true or fake object to satisfy the loosened or nominal criteria, which can be measured in dedicated data CRs. Inversion of this matrix can thus provide an estimate of the number of fake objects in the SR. In the fake-factor method, the probability of a fake object to pass the SR identification requirements, derived in a fake-enriched CR, is applied to events found in a region similar to the SR but in which the object identification requirements are relaxed.


\subsection{Statistical interpretation}
\label{sec:stat}

In order to test for the presence of new physics, templates for both the background and signal, obtained through MC simulations or data-driven methods, are compared with the data. Unless indicated otherwise, a binned maximum-likelihood fit is used, while the variable (or variables) being fitted depends on the specific analysis. Several regions may be fitted simultaneously, including the signal regions and control regions. Systematic uncertainties are included as nuisance parameters~\cite{Conway:2011in} with either log-normal or Gaussian constraints.

When no significant deviation from the expected background is observed in a search, constraints on various signal models that would produce an excess are  extracted at 95\% confidence level (CL) using the CL$_\text{s}$ method~\cite{Read:2002hq} with a profile likelihood ratio as the test statistic. The asymptotic approximation to the test statistic's distribution is usually assumed~\cite{Cowan:2010js} when extracting expected limits, except in cases where it is seen to fail (e.g.\ due to low statistical precision), in which case pseudo-experiments are generated.



%
\section{Compositeness}
\label{sec:compositness}

In pursuit of the elementary constituents of matter, physicists since the late 1800s have delved into matter at progressively smaller scales, proving that what were at first deemed elementary, in turn atoms, nuclei and nucleons, were in fact composed of even smaller constituents. The SM currently considers, in line with the experimental results so far, that the leptons and quarks are indeed the elementary particles, but could their three generations be explained by yet smaller constituents? -- i.e.\ could they too be composite? With the unprecedented centre-of-mass energy of the LHC, it is indeed possible to probe even smaller scales in order to see whether they have an internal structure. A tell-tale signature of compositeness would be the existence of excited states at higher masses~\cite{Baur:1987ga}.
These excited states could be produced at the LHC through known gauge interactions (GI), for example the production of an excited quark $q^*$ through a gluonic excitation $qg\rightarrow q^*$, or through a four-fermion contact interaction (CI), for example the production of an excited lepton $\ell^*$ though $q\bar{q}\to \bar{\ell}\ell^*$. The decay of these excited states can proceed in similar ways: through a GI (e.g.\ $q^*\rightarrow gq$, $q^*\to \gamma q$ or  $\ell^*\to W\nu$), or through a CI (e.g.\ $\ell^*\to q \bar{q}\ell$ or $q^*\to q \bar{q} q$).  The limits obtained are summarized in Table~\ref{tab:compositness}.

\subsection{Excited quarks}
\label{sec:qstar}
The existence of a $q^*$ in the GI model could be revealed by observing a resonance of the type $q^*\rightarrow gq$ in the dijet invariant mass distribution, i.e.\ a peak above the smoothly falling QCD multijet background. A search for such excited quarks was performed~\cite{EXOT-2019-03} by scanning the \mjj distribution from 1.1 to 8~\TeV. The lower bound on \mjj is motivated by the fact that the events must pass the trigger selection, which requires a single, high-\pT small-$R$ jet, and that lower values of $m_{q^*}$ were  excluded in \RunOne~\cite{EXOT-2010-01} or by previous experiments~\cite{CDF:2008ieg}.  Events are further selected by requiring two small-$R$ jets with \pT$>150$~\GeV, separated by an azimuthal angle of at least $\Delta\phi(j_1,j_2)=1.0$ and a rapidity difference of at most $y^{*} = (y_{j_1} - y_{j_2})/2 = 0.6$. This selection was used to search inclusively for excited quarks, and a specific search for excited \bquarks was also performed. The latter also requires the presence of at least one $b$-tagged jet and imposes a $|\eta|<2.0$ requirement on each of the two jets, but allows $y^*$ values up to 0.8.

The signal model assumes  %
that the excited quarks are spin-1/2 particles, have the same couplings as SM quarks, and that the compositeness scale is $m_{q^*}$. Two scenarios are considered: $q^*$ comprising both $u^*$ and $d^*$ or being only a $b^*$. The SM background is dominated by QCD multijet events and is estimated in each bin by fitting a parametric function of the type $f (x) = p_1(1-x)^{p_2} x^{p_3+p_4 \ln{x}}$ in a sliding window over \mjj.
The uncertainty in the fitted parameters is taken as a systematic uncertainty of the background estimation, along with an uncertainty concerning the choice of fitting function, and another related to the amount of spurious signal. The \mjj distribution is shown for the inclusive search in Figure~\ref{fig:dijet_mjj} for the data and two examples of an excited quark signal.

\begin{figure}[tb]
\begin{center}
\includegraphics[width=0.4\textwidth]{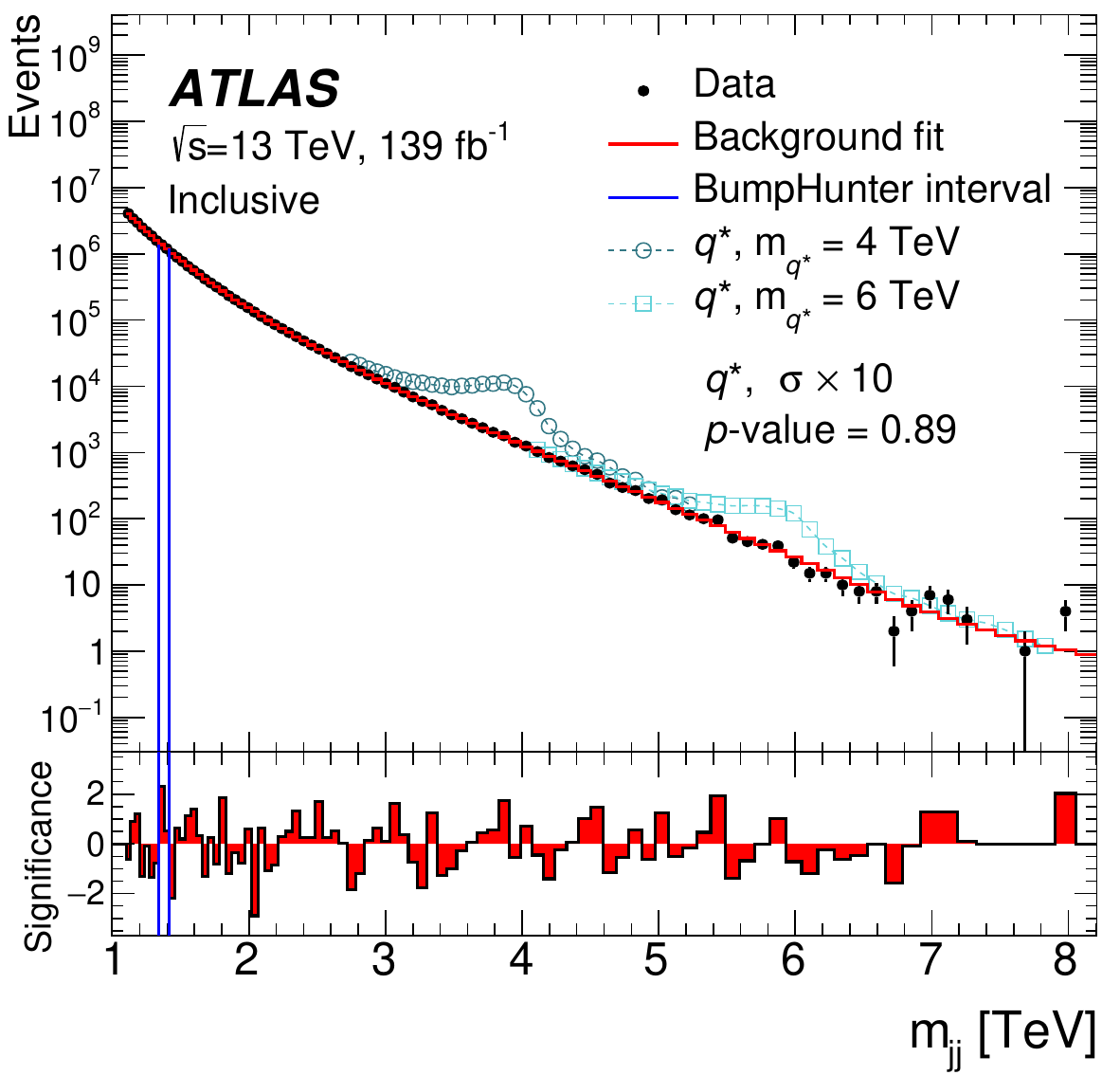}
\end{center}
\caption{Distribution of \mjj in the inclusive search for excited quarks~\cite{EXOT-2019-03}. The vertical lines indicate the most discrepant \mjj interval identified, for which the $p$-value is stated in the figure. Two example $q^*$ resonances are also shown, with an enhanced cross section for visibility.}
\label{fig:dijet_mjj}
\end{figure}

Since no significant excess is seen, a lower limit on the excited quark's mass is set at $m_{q^*} = 6.7$~\TeV in the inclusive search, while for the specific $b^*$ search, the limit is set at 3.2~\TeV. These correspond to compositeness length scales below $3.0\times10^{-5}$~fm and $6.2\times10^{-5}$~fm, respectively.

\subsection{Excited leptons}
\label{sec:taustar}

Searches for excited electrons and muons have been performed by ATLAS, using a partial \RunTwo dataset and the \RunOne dataset, respectively~\cite{EXOT-2017-22,EXOT-2015-01}. A more recent addition is a search for excited $\tau$-leptons, $\tau^*$, assuming that they are produced and decay through a CI, as shown in Figure~\ref{fig:taustar}(a), leading to a final state comprising two $\tau$-leptons and two jets (the signal model includes all the quark flavours that are kinematically allowed). These four objects are used to compute an $S_\mathrm{T}$ value which is exploited in a search which focuses on hadronic $\tau$ decays~\cite{EXOT-2020-18}.

The analysis is based on a di-$\tau$ trigger. The SR requires the events to have exactly two \tauhad of opposite electric charge
which are well separated in $\DeltaR$.
There must also be at least two small-$R$ jets with high \pT
and no reconstructed electron or muon.
The main background, \Ztautaujets, is suppressed by requiring $m_{\tau\tau}^\text{col}>110$~\GeV. For each \tauhad, the ratio of the visible-decay \pT to the total \pT is usually positive for signal, while it can become negative for fake \tauhad because now the \MET points in a random direction; requiring this ratio to exceed a minimum value reduces the fake-\tauhad background. Finally, the scalar sum of the \pT of the two \tauhad, denoted by $L_\mathrm{T}$, is required to be larger than 140~\GeV.
The major remaining backgrounds are \Ztautaujets, \ttbar and single-top events, and events with jets falsely reconstructed as \tauhad.

\begin{figure}[tb]
\begin{center}
\subfloat[]{\raisebox{0.5\height}{\includegraphics[width=0.3\textwidth]{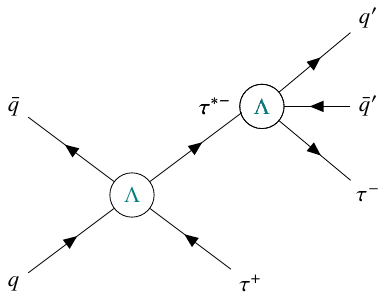}}}
\qquad
\subfloat[]{\includegraphics[width=0.4\textwidth]{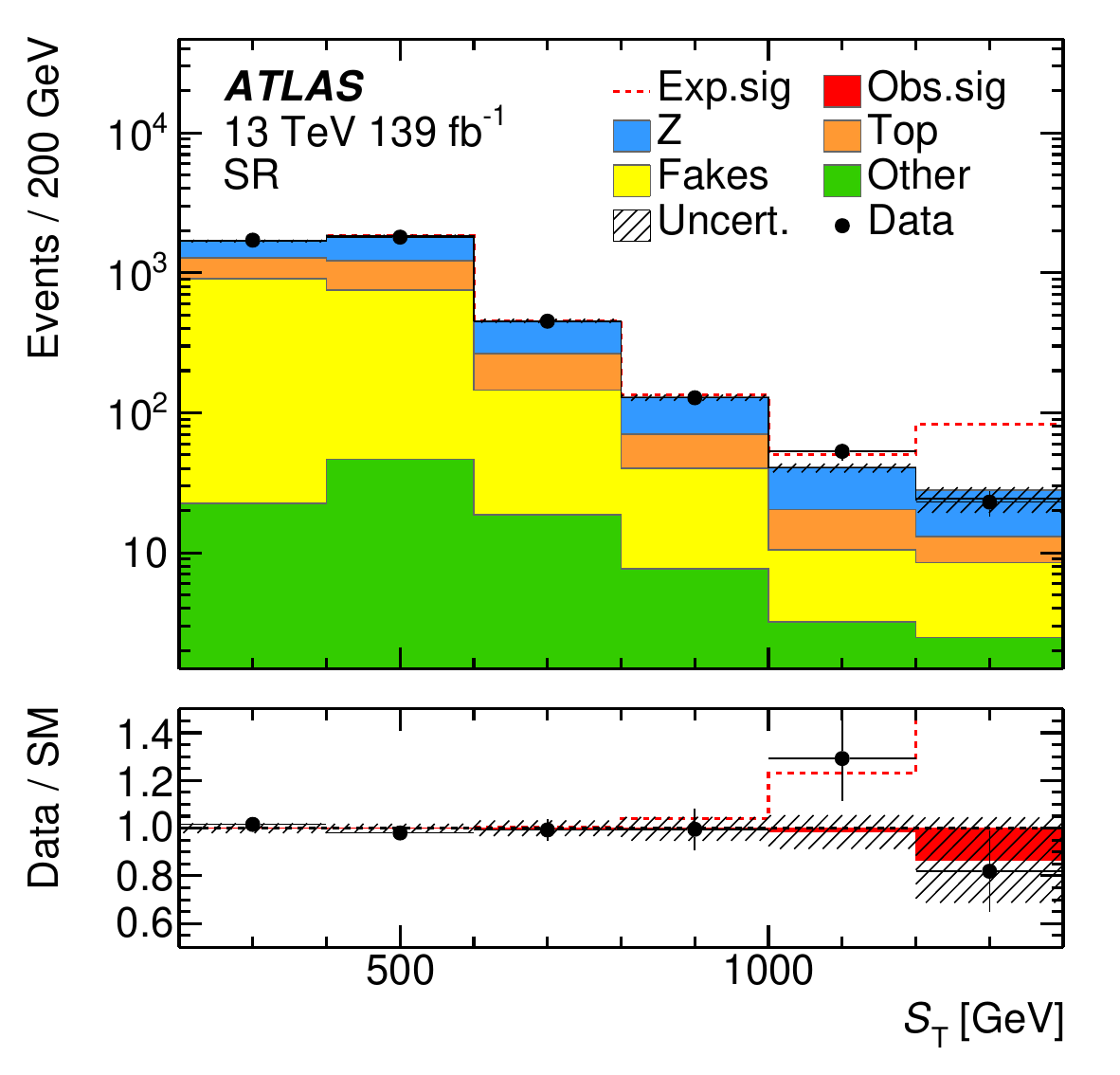}}
\end{center}
\caption{(a) Production and decay of an excited $\tau$-lepton via a contact interaction with a compositeness scale $\Lambda$ and (b) $S_\mathrm{T}$ distribution in the signal region for data compared to the estimated background contribution with its uncertainty and an example signal ($m_{\tau^*}=1.5$~\TeV, $\Lambda=10$~\TeV) with a nominal signal strength (Exp.) and with the signal strength extracted from the fit to the data (Obs.)~\cite{EXOT-2020-18}.}
\label{fig:taustar}
\end{figure}

The \Ztautaujets and top-related background yields are extracted using dedicated CRs. For the \Ztautaujets background, the CR is built by requiring a $Z$-compatible $m_{\tau\tau}^\text{col}$ and inverting the selection on $L_\mathrm{T}$. For the top-related events, the CR requires one \tauhad with high \pT up to 450~\GeV, along with at least four jets amongst which two must be $b$-tagged, and a large \met to avoid fakes. Finally, the fake-factor method is used to estimate the fake background.
The main sources of systematic uncertainty in the final background estimation are the top-related background modelling and the $\tau$-lepton energy scale.

As shown in  Figure~\ref{fig:taustar}(b), good agreement between the data and the estimated background is observed in the SR. A lower limit is hence set on the excited $\tau$-lepton's compositeness scale. For $\Lambda=m_{\tau^*}$, the limit obtained is at 4.6~\TeV.

\begin{table*}[htp!]
\begin{center}
\caption{95\% CL lower mass limits obtained by various analyses on different excited fermion ($f^*$) models produced either by gauge interaction (GI) or contact interaction (CI) for $\Lambda=m_{f^*}$ (see the text for more details).  }
\begin{tabular}{l c c c}
\hline
\hline
Model & Production mode & Observed lower limit on $m_{f^*}$ [$\text{T\electronvolt}$] & Section \\\hline
$q^*$ ($q=u,d$) &  GI & 6.7 ~\cite{EXOT-2019-03} & \ref{sec:qstar} \\\hline
$b^*$        &  GI & 3.2 ~\cite{EXOT-2019-03} & \ref{sec:qstar} \\\hline
$e^*$  (36.1 \ifb) & CI &  4.8 ~\cite{EXOT-2017-22} & \ref{sec:taustar} \\\hline
$\mu^*$ (\RunOne) & CI & 2.8~\cite{EXOT-2015-01} & \ref{sec:taustar} \\\hline
$\tau^*$ & CI & 4.6~\cite{EXOT-2020-18} & \ref{sec:taustar} \\\hline
\hline
\end{tabular}
\label{tab:compositness}
\end{center}
\end{table*}


%
\section{Heavy neutral leptons}
\label{sec:leptons}

The origin of the non-zero neutrino masses, brought to light by the observation of neutrino oscillations~\cite{Super-Kamiokande:2005mbp,SNO:2002tuh}, is not known. While Yukawa couplings can be invoked, their extremely small values compared to the other fermionic masses appear unnatural. One solution is the seesaw mechanism, which introduces a neutrino mass matrix containing Majorana mass terms and tiny neutrino mass eigenvalues which are due to the existence of heavier BSM particles. Several such seesaw theories exist depending on their field contents~\cite{Ma:1998dn}, classified into Type-I with right-handed neutrinos, Type-II with a scalar triplet, and Type-III with fermion triplets. Searches for heavy leptons in the context of these theories have been performed in multiple final states~\cite{EXOT-2020-06,EXOT-2019-29,EXOT-2019-39,EXOT-2018-33,EXOT-2020-02,EXOT-2019-36}, as reviewed in this section.

\subsection{Type-I seesaw}
\label{sec:typeI}

In the Type-I seesaw mechanism, the heavy Majorana neutrinos $N$ couple to SM particles through their mixing with the light left-handed neutrinos of flavour $\alpha=\{e,\mu,\tau\}$, the size of the mixing being  determined by the dimensionless coefficient $V_{\alpha N}$.

Due to the Majorana nature of the heavy neutrino, a same-sign (SS) dilepton final state can be created, for example in the scattering of $W$ bosons, as shown in Figure~\ref{fig:HNL}(a). This signature was exploited to search for these new particles for non-zero values of $V_{\mu N}$~\cite{EXOT-2020-06}. Besides requiring the presence of exactly two SS muons, the vector-boson scattering (VBS) production signature is exploited by requiring two jets with a large rapidity separation and a high \mjj. To reject background events, $b$-tagged jets are vetoed and the \met significance is required to be small. The SR is binned in subleading-muon \pT, and the shape of this distribution is exploited to enhance the sensitivity. The two main background processes, VBS $W^\pm W^\pm jj$ and $WZ$ production, are estimated via CRs which reverse the \met significance or third-lepton veto requirement, respectively. The fake-muon background is evaluated with the fake-factor method. Since the data  agree with the background expectations within statistically dominated uncertainties, upper limits are set on $|V_{\mu N}|^2$ as a function of the heavy neutrino mass $m_N$, as shown in Figure~\ref{fig:HNL}(c).

A complementary search is also conducted for lower values of $m_N$ and $|V_{\alpha N}|^2$~\cite{EXOT-2019-29}, focusing on production via $W\to N\ell_\alpha$ where the heavy neutrino subsequently decays as $N\to\ell_\beta\ell_\gamma\nu_\gamma$ (via a $W^*$) or as $N\to\nu_\beta\ell_\gamma\ell_\gamma$ (via a $Z^*$), where $\alpha,\beta,\gamma=e$ or $\mu$ (see Figure~\ref{fig:HNL}(b)). For the low values of $m_N$ and $|V_{\alpha N}|^2$ probed in this search, the heavy neutrino is long-lived, leading to a displaced vertex (DV) of two charged leptons ($\ell_\beta\ell_\gamma$ or $\ell_\gamma\ell_\gamma$).
The secondary vertexing is run on standard and LRT tracks using a procedure drawing inspiration from Ref.~\cite{ATL-PHYS-PUB-2019-013} and requiring the selected DVs to have exactly two leptons of opposite sign, and no additional tracks. The SR events are selected by requiring the presence of such a DV and one additional prompt lepton $\ell$, with a combined invariant mass $40<m_{\mathrm{DV}+\ell}<90$~\GeV, and a reconstructed $m_N$ below 20~\GeV, where the $m_N$ calculation uses the known $W$ mass, assumes massless charged leptons, and takes the $N$ flight direction to be the vector connecting the $pp$ collision PV to the DV. Further selections are applied to suppress the cosmic-ray muon background (by removing DVs with back-to-back muons), the $ee$ background from interactions with material in the inner detector (by removing DVs found in regions with detector material), the background from $J/\Psi$ or other heavy-flavour decays (by setting a lower bound on $m_\mathrm{DV}$), the random-lepton-crossing background (by requiring the prompt lepton and displaced lepton with the same flavour to have opposite charge), and the $Z\to\ell\ell$ background (by removing events in which the prompt lepton and the displaced lepton with the same flavour have a mass compatible with a $Z$ boson). After these selections, the random-crossing background dominates; it is estimated in a data-driven way by using a CR at higher $m_N$ and utilizing the fact that opposite-sign leptons and same-sign leptons should be equally probable in the DVs for this source of background. In all the lepton flavour combinations probed, between 0.2 and 2.8 background events are predicted in the SR, with 0 to 2 events observed in the data. Figure~\ref{fig:HNL}(d) shows the excluded parameter space in the same plane as used in the prompt heavier-neutrino search discussed above, illustrating the complementary coverage of the two analyses.

\begin{figure}[tb]
\begin{center}
\qquad
\subfloat[]{\includegraphics[width=0.35\textwidth]{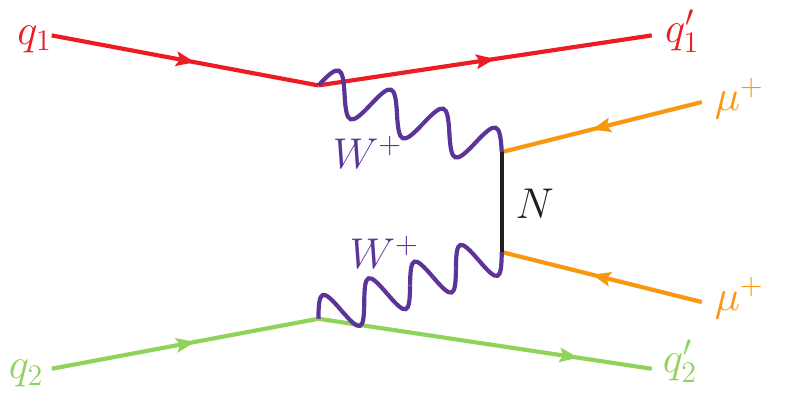}}
\qquad
\subfloat[]{\includegraphics[width=0.35\textwidth]{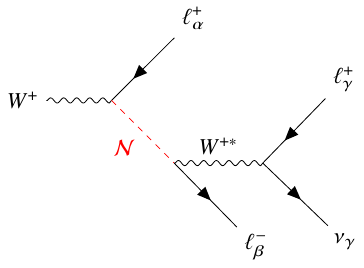}}
\qquad
\subfloat[]{\includegraphics[width=0.45\textwidth]{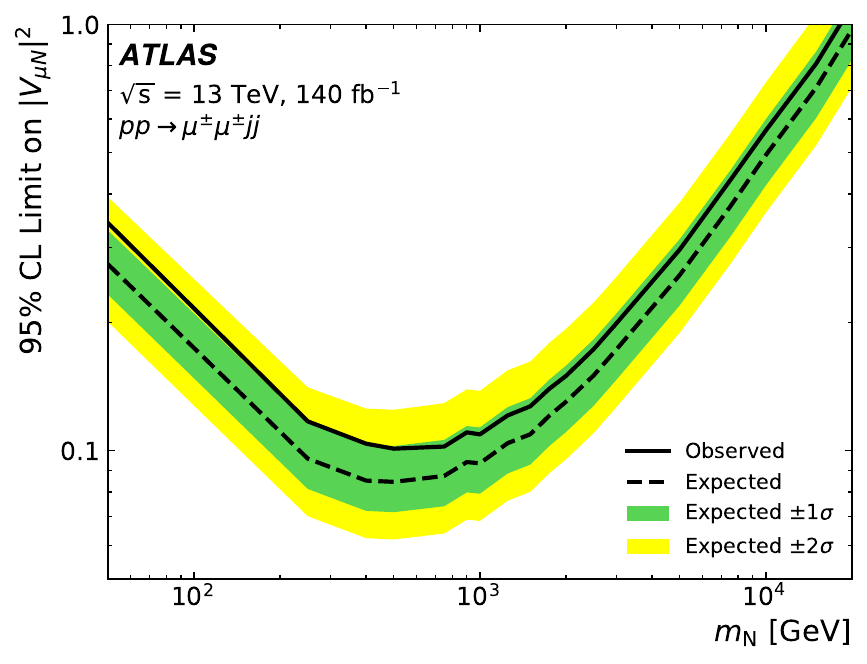}}
\qquad
\subfloat[]{\includegraphics[width=0.45\textwidth]{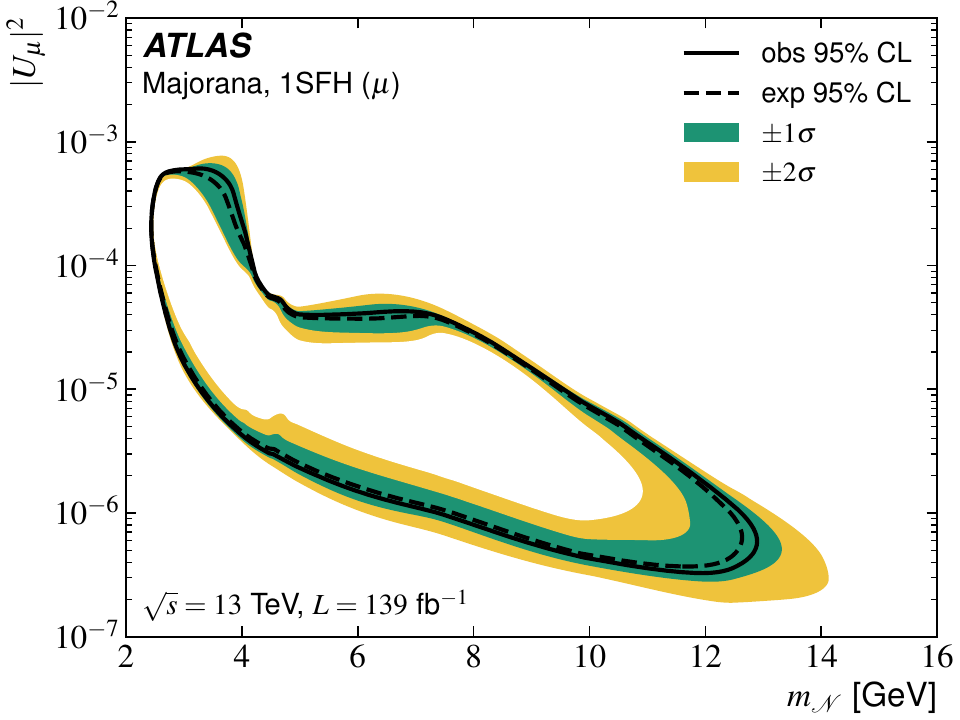}}
\end{center}
\caption{Examples (a,b) of the production and decay of the heavy neutrinos and (c,d) limits set on the heavy Majorana neutrino mixing element $|V_{\mu N}|^2$ (also shown here as $|U_{\mu}|^2$) as a function of $m_N$ in a Type-I seesaw mechanism by (a,c) the same-sign $WW$ scattering search~\cite{EXOT-2020-06} and (b,d) the displaced dilepton vertex search~\cite{EXOT-2019-29}. }
\label{fig:HNL}
\end{figure}

\subsection{Left--right symmetric model}

Type-I and Type-II models can also be included in larger left--right symmetric models (LRSM)~\cite{Pati:1974yy,PhysRevD.11.2558,PhysRevD.12.1502} whose aim is to explain the broken parity symmetry of the SM weak interaction. In these models, right-handed counterparts to the $W$ and $Z$ bosons can be introduced along with the right-handed heavy neutrinos $N_\mathrm{R}$. A search for such Majorana or Dirac neutrinos and right-handed $W$ gauge bosons ($W_\mathrm{R}$) has been performed~\cite{EXOT-2019-39}, looking for the Keung--Senjanović process~\cite{PhysRevD.12.1502} with a semileptonic final state $W_\mathrm{R}\to\ell N_\mathrm{R}\to\ell\ell q\bar{q}'$ ($\ell=\{e,\mu\}$), as shown in Figure~\ref{fig:LRSM}(a).
Majorana and Dirac heavy neutrinos will differ in that the Majorana neutrinos will produce same-sign leptons half of the time. For $m(W_\mathrm{R})\gg m(N_\mathrm{R})$, the $N_\mathrm{R}$ decay products will be merged, due to a Lorentz boost, into a large-$R$ jet. Two sets of SRs are therefore defined: one for the resolved decay of the $N_\mathrm{R}$, and one for the boosted regime.

In the resolved regime, different SRs are defined for opposite-sign (OS) or same-sign (SS) electron and muon pairs. Since a high-mass $W_\mathrm{R}$ is targeted, high values are required for the \pT of the two leading jets, their \mjj, and the $H_\mathrm{T}$ computed from these jets and leptons. A high value of \mll is also required, to suppress the $Z$+jets background, and, for SS leptons, the dileptonic angular separation must not be too large, to avoid a mismodelled region of the phase space for the diboson background. The SRs are binned in either $m_{\ell\ell jj}$ (OS case with $m(W_\mathrm{R})>m(N_\mathrm{R})$), \mjj (OS case with $m(W_\mathrm{R})<m(N_\mathrm{R})$),  or $H_\mathrm{T}$ (SS case), to increase the sensitivity. CRs to constrain the main leptonic background processes ($Z$+jets, \ttbar and diboson) are built either at lower \mll values or with opposite-flavour leptons, with the non-prompt-lepton background estimated with a fake-factor method. In the boosted channel, one of the leptons must be well separated from the large-$R$ jet in azimuth and one is likely to be inside the large-$R$ jet -- while such a muon can still be readily identified, this may not be the case for an electron whose associated calorimeter energy clusters are embedded in the jet. Therefore, besides a $2\mu$ and a $2e$ SR with a high \mll, a $1e$ SR is also defined. In the electron SRs, the $W$+jets background is suppressed by imposing an upper bound on \met and additionally, in the $1e$ SR, by imposing a lower bound on the polar angle  of the electron from an assumed $W$ boson decay in this boson's rest frame (with respect to the boson's flight direction in the laboratory frame). Furthermore, in the $1e$ SR, the $\gamma$+jets and dijet contributions are reduced by selecting events in which the $\eta$ difference between the jet and the electron is not too large. Finally,  in order to suppress the \ttbar background, $b$-tagged jets are vetoed in all SRs and, in the $2\mu$ channel, the \pT of the dimuon system is required to be large. The SRs are binned in the reconstructed $W_\mathrm{R}$ mass, with $m(W_\mathrm{R})>3$~\TeV being the SR, and lower values being used as CRs to estimate the background. Other CRs are used in the $1e$ SR, at lower polar angle, to estimate $W$+jets background, or with a looser electron identification, to estimate fakes.

In all SRs, the data agree with the background expectations within uncertainties, the background MC sample size being one of the dominant systematic uncertainties in the SS SRs, while modelling uncertainties become important for the other SRs. Limits in the plane of $m(N_\mathrm{R})$ versus $m(W_\mathrm{R})$ are shown for the Dirac scenario in the electron channel in Figure~\ref{fig:LRSM}(b), which illustrates the complementarity of the resolved and boosted SRs. Similar limits are obtained in the muon channel and for a Majorana scenario.

\begin{figure}[tb]
\begin{center}
\subfloat[]{\raisebox{0.5\height}{\includegraphics[width=0.45\textwidth]{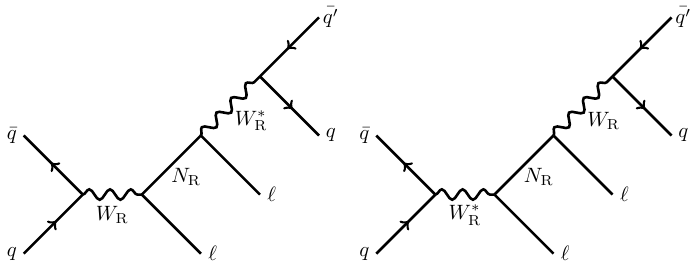}}}
\qquad
\subfloat[]{\includegraphics[width=0.45\textwidth]{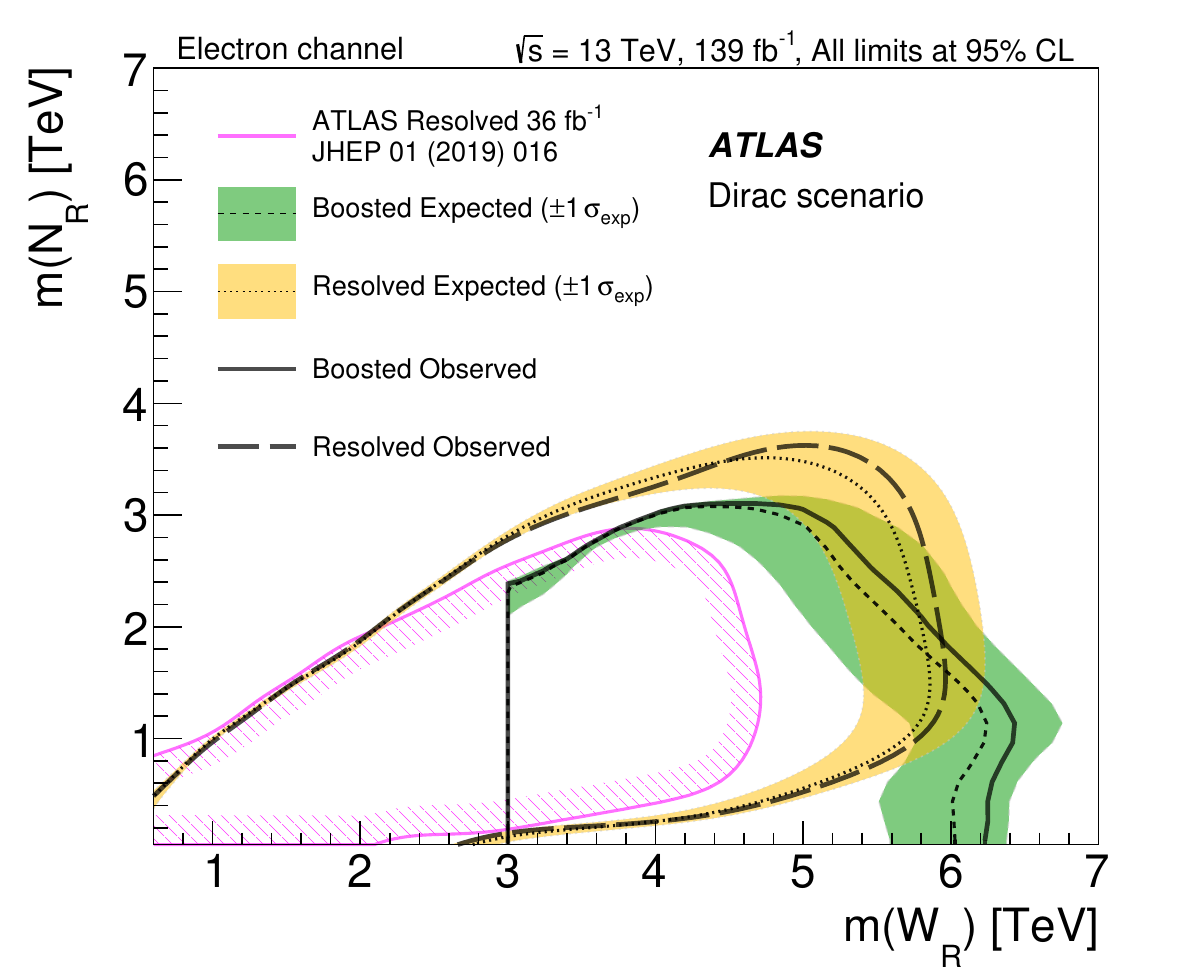}}
\end{center}
\caption{(a) Diagrams of the Keung--Senjanović process
with different mass orderings and an off-shell $W_\mathrm{R}^*$. (b) Limits obtained in the resolved and boosted channels of the analysis searching for left--right symmetric model $W_\mathrm{R}$ and $N_\mathrm{R}$~\cite{EXOT-2019-39} in the case of a Dirac neutrino in the electron channel.}
\label{fig:LRSM}
\end{figure}

\subsection{Type-III seesaw}
\label{sec:typeiii}

The minimal Type-III seesaw model described in Ref.~\cite{Biggio:2011ja} is the target of searches in multilepton final states~\cite{EXOT-2018-33,EXOT-2020-02}. This model introduces a single fermionic triplet containing one neutral Majorana lepton $N$ and two oppositely charged leptons $L^\pm$, which are degenerate in mass\footnote{There could be a small mass-splitting due to radiative corrections which does not affect the phenomenology.} and decay into a SM lepton (with which they mix with coupling $V_\alpha$ where $\alpha=\{e,\mu,\tau\}$) and a $W$, $Z$ or Higgs boson. The dominant production mechanism is shown with an example decay in Figure~\ref{fig:typeIII}(a).

The two-lepton search~\cite{EXOT-2018-33} is based on two SRs, with either SS or OS leptons ($e$ or $\mu$) having either the same or different flavours and a large \mll, and with at least two jets with an invariant mass compatible with a $W$ boson. Large values are also required for the \pT of the dijet system and of the dilepton system, for the significance of the \met and for $H_\mathrm{T}+\met$, where $H_\mathrm{T}$ is based on the leptons. Finally, to suppress the \ttbar background, no $b$-tagged jet is allowed to be present in the final state and, in the OS SRs, the \met must point away from the leptons. The remaining dominant backgrounds, \ttbar in the OS SRs and dibosons in the SS SRs, are estimated from CRs built mainly by reversing the $b$-tagged veto and \mjj requirements, respectively.
The fake-lepton background contribution is estimated via a fake-factor method. The SRs are binned in $H_\mathrm{T}+\met$ to improve the sensitivity.

The three- and four-lepton searches~\cite{EXOT-2020-02} are also divided into various SRs. In the three-lepton channel, three SRs are defined and all require a large \met significance. The first one targets a dileptonic $Z$ boson in the decay chain, and assumes that the SM boson coming from the other heavy lepton decays hadronically. It thus requires an opposite-sign same-flavour (OSSF) lepton pair compatible with the $Z$ boson mass, along with a large tri-lepton mass and large $m_\mathrm{T}(\ell,\met)$ for the two leading leptons, whose angular separation is also constrained. A complementary SR vetoes the presence of a $Z$ boson candidate, requiring large values of $H_\mathrm{T}$, \met and the scalar sum of the \pT of the SS leptons, with an upper bound placed on the dijet invariant mass to reduce the diboson contribution. Finally, the last three-lepton SR targets leptonically decaying EW bosons, thus requiring a low jet multiplicity, and a lower bound on the summed scalar \pT of the leptons and on the OSSF lepton pair's invariant mass. Requirements are also placed on the transverse mass and angular separation of the two leading leptons, as in the $Z$ SR. Since diboson production is the main background in the three-lepton channel, a CR is defined by selecting events with two jets and a low value of $m_\mathrm{T}(\ell_{2},\met)$, where $\ell_2$ is the subleading lepton.

The four-lepton channel is divided into two SRs, according to the sum of the electric charges of the leptons ($q=0$ or $|q|=2$), both requiring a large value of the four-lepton invariant mass and of $H_\mathrm{T}+\met$. In the $q=0$ SR, which has more background to suppress, further selections are made: a $b$-tagged-jet veto, a $Z$ boson veto and the requirement of a high \met significance. Two CRs are built to estimate the diboson and rare top backgrounds for the $q=0$ SR, the first at lower four-lepton mass and the second by inverting the $b$-jet veto. The background due to charge misidentification in the $|q|=2$ channel is estimated by correcting the simulation with factors derived in a $Z\to ee$ CR. The fake-lepton background contributions are obtained via a fake-factor method. The data are found to agree well with the expected background's post-fit distributions of three-lepton transverse mass and $H_\mathrm{T}+\met$, which are the discriminating variables used in the likelihood fit in the three- and four-lepton channels respectively.

\begin{figure}[tb]
\begin{center}
\subfloat[]{\raisebox{0.3\height}{\includegraphics[width=0.3\textwidth]{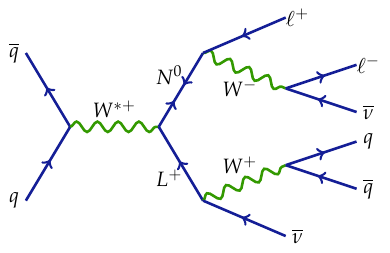}}}
\qquad
\subfloat[]{\includegraphics[width=0.5\textwidth]{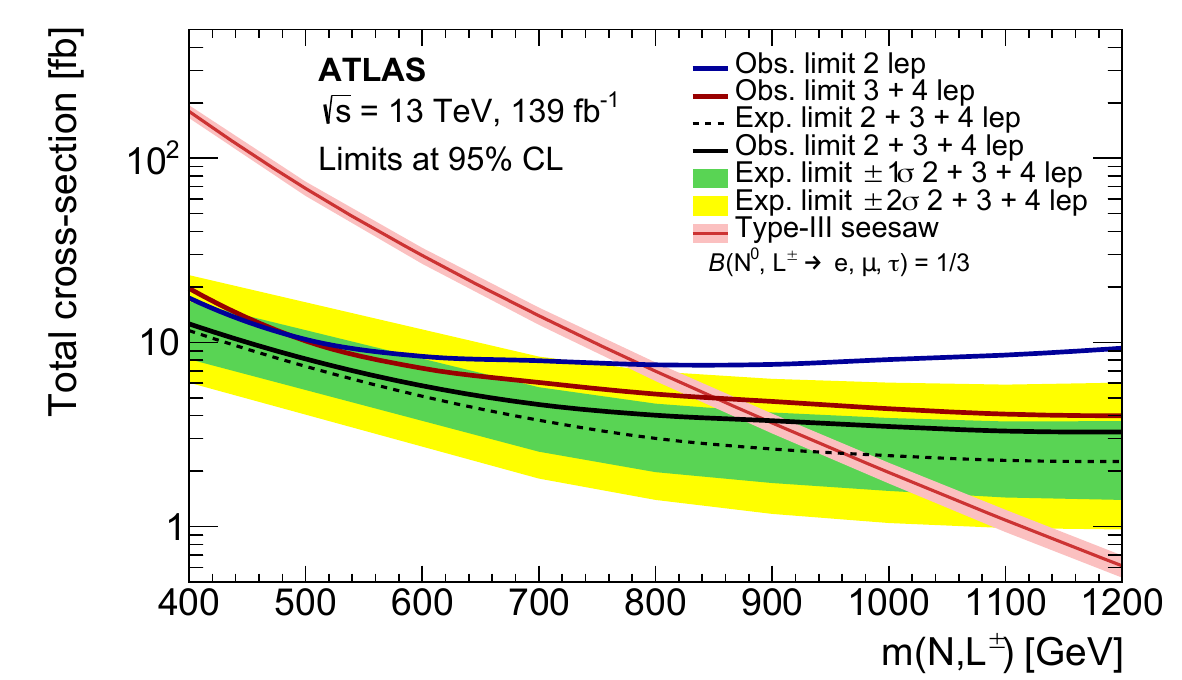}}
\end{center}
\caption{(a) The dominant production mechanism for Type-III seesaw heavy leptons with an example of their decay and (b) the limits placed on their mass in the various channels and their combination~\cite{EXOT-2018-33,EXOT-2020-02}.}
\label{fig:typeIII}
\end{figure}

The data in all channels are found to agree with the estimated background within uncertainties, which are dominated by the number of data events. The two-, three- and four-lepton channels are used, individually and in combination, to derive lower limits on the mass of the heavy leptons, as shown in Figure~\ref{fig:typeIII}(b). The exclusion region reaches a mass of 910~\GeV. A general, model-independent search for excesses in three- or four-lepton final states has also been performed~\cite{EXOT-2019-36} and is interpreted in terms of this model as well. However, due to the more general nature of this search in comparison with the targeted SRs described above, its sensitivity to this particular model is lower.

\FloatBarrier


%
\section{Vector-like lepton and quarks}
\label{sec:veclike}

One of the best-known open questions in the SM concerns the origin of the huge difference between the electroweak scale, ${\sim}250$~\GeV, (as well as the Higgs boson's mass itself) and the Planck scale, ${\sim}10^{19}$~\GeV. Many possible solutions have been proposed for defining a new mechanism that cancels out radiative corrections to the Higgs boson mass without fine-tuning~\cite{PhysRevD.20.2619}. Although a chiral fourth generation of fermions is strongly disfavoured by Higgs boson measurements and electroweak precision data~\cite{Djouadi:2012ae,Eberhardt:2012gv,Lenz:2013iha}, the presence of vector-like leptons and quarks is much less constrained, since these can have electroweak-singlet masses that dominate the mass contributions from their Yukawa couplings to the Higgs boson~\cite{Frampton:1999xi}. Discovery of vector-like partners of the SM fermions could shed light on their mass and mixing patterns.

Vector-like quarks (VLQ) are colour-triplet fermions and are common in models that use a new strongly interacting sector to resolve the fine-tuning problem. In these models, the Higgs boson would be a pseudo Nambu--Goldstone boson~\cite{Hill:2002ap} of a spontaneously broken global symmetry. The composite Higgs model~\cite{Dimopoulos:1981xc,Kaplan:1983fs,Kaplan:1983sm} is a particular realization of this idea.

Vector-like leptons (VLL) are colour-singlet fermions. They often appear in phenomenological models motivated by string theory~\cite{Hewett:1988xc,Gursey:1975ki} or large extra dimensions~\cite{Asaka:2003iy}. They also appear in weak-scale supersymmetry~\cite{Moroi:1991mg,Martin:1997ns,Endo:2011mc}, where the mass of the lightest Higgs scalar boson can be raised by introducing new vector-like heavy chiral supermultiplets with large Yukawa couplings.

This section focuses on the searches for these two types of particles in ATLAS. It discusses one VLL result and a large part of the extensive search programme for VLQs.

\subsection{Vector-like leptons}
\label{sec:vll}

In a scenario with a pure VLL doublet $L' = (\nu_{\tau}',\tau')$, this doublet comprises two fermions of approximately equal mass that couple only to the third generation of SM leptons~\cite{Kumar:2015tna}. In such a scenario the VLL production cross section is dominated by the $pp\rightarrow \nu_{\tau}'\tau'$ process, with a rate approximately four times larger than for other possible combinations, such as $pp\rightarrow \nu_{\tau}'\bar{\nu}_{\tau}'$ or  $pp\rightarrow \tau'^{+}\tau'^{-}$. The $\nu_{\tau}'$ decays exclusively into $W^{+}\tau^{-}$, while the $\tau'$ can decay into $Z\tau$ or $H\tau$, with the former dominating for low masses. To cover this wide range of possibilities in the most efficient way, VLLs are searched for~\cite{EXOT-2020-07} in final states containing at least two charged light leptons, $e^{\pm}$ or $\mu^{\pm}$, zero or more $\tau$-leptons decaying hadronically, and a significant amount of \met. To maximize the sensitivity to the different final states, seven independent boosted decision trees~\cite{Freund:1997xna} are trained in seven different categories of events based on the numbers of reconstructed light leptons and $\tau$-leptons. Thirty-four kinematic and topological variables for each event are used to train the BDTs, achieving good separation between signal and background samples. Seven SRs are then defined using the corresponding BDT's output score in each category. In addition to the SRs, three CRs are used to normalize the dominant background processes ($\ttbar+Z$, $WZ$, and $ZZ$), and a final CR is used to assess the fake-$\tau$ background originating from gluon-initiated jets and \pileup.

The statistical analysis uses the BDT score in the signal regions, and the event yield in the CRs. Templates for signal and background processes are obtained from MC simulations. The fake contribution is estimated using the fake-factor method. The result of a background-only fit in all of the SRs is shown in Figure~\ref{fig:vll_postfit} together with a fit to one of the SRs, corresponding to a selection of two same-sign same-flavour light leptons and one $\tau$-lepton (2L SSSF, 1$\tau$). Since no significant excess is found, upper limits are set on the combined production cross section of the three VLL processes considered. Masses below 900~\GeV (970~\GeV) are observed (expected) to be excluded for the doublet model used as a benchmark.

\begin{figure}[tb]
\begin{center}
\subfloat[]{\includegraphics[width=0.4\textwidth]{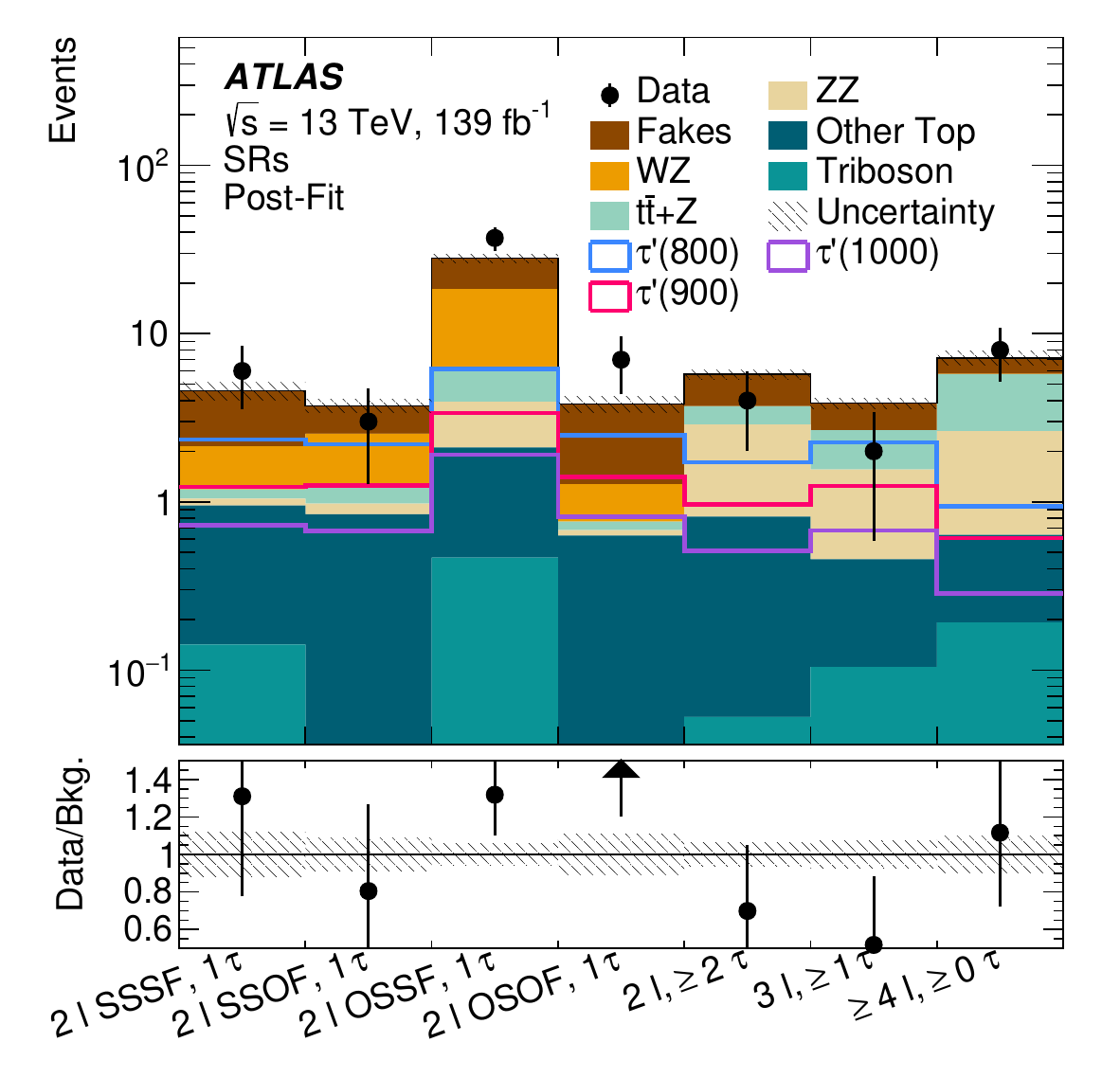}}
\qquad
\subfloat[]{\includegraphics[width=0.4\textwidth]{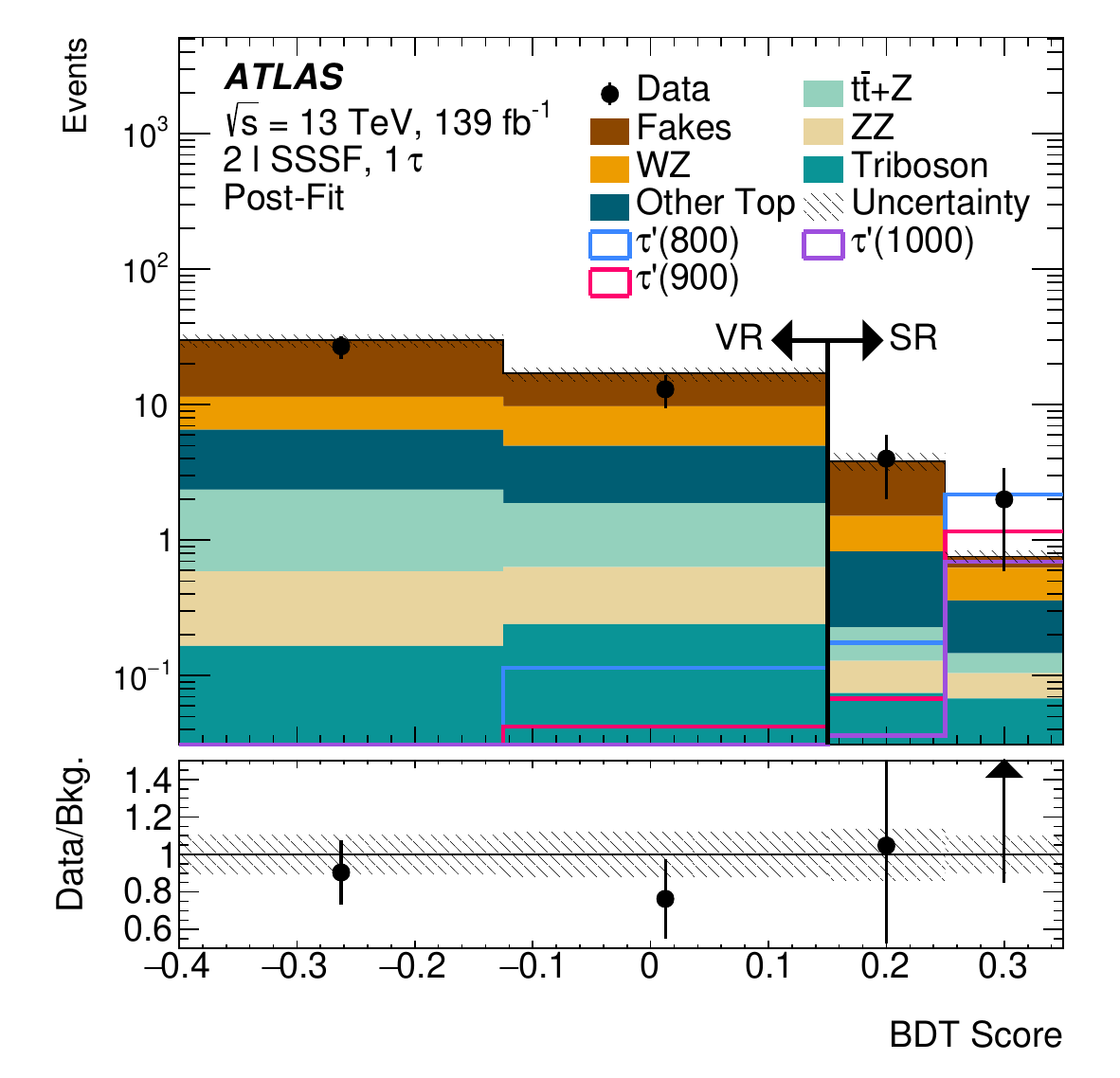}}
\end{center}
\caption{(a) The post-fit yield for all of the signal regions and (b) BDT score distribution for one of the signal regions used in the search for vector-like $\tau$-leptons~\cite{EXOT-2020-07}. Distributions of data, background, and pre-fit simulated signal are shown.}
\label{fig:vll_postfit}
\end{figure}

\subsection{Vector-like quarks}
\label{sec:vlq}

Vector-like quarks may appear in one of seven possible multiplets: singlets ($T^{2/3}$) or ($B^{1/3}$); doublets ($X^{5/3}$, $T^{2/3}$), ($T^{2/3}$, $B^{-1/3}$) or ($B^{-1/3}$, $Y^{-4/3}$); or triplets ($X^{5/3}$, $T^{2/3}$, $B^{-1/3}$) or ($T^{2/3}$, $B^{-1/3}$, $Y^{-4/3}$) where the superscript represents the charge of the new particle. The $T$ and $B$ quarks have the same charge as the SM top and bottom quarks but different masses, while the up-type $X$ quarks and down-type $Y$ quarks have non-standard isospins. Most analyses described in this report consider only the singlet and doublet scenarios containing only $T$ and $B$.

Light VLQs, with masses below ${\sim}1$~\TeV, would be produced mostly in pairs at the LHC via the strong interaction, illustrated in Figure~\ref{fig:vlq_proc} for a $T$ quark. However, this process is kinematically suppressed for higher masses, and electroweak single production (also illustrated in Figure~\ref{fig:vlq_proc}) becomes important and can dominate depending on the VLQ coupling strength. Searching for pair-produced VLQs is a relatively model-independent approach, as the pair-production cross section depends only on the assumed mass of the VLQ. At the same time, single production has the advantage of providing direct constraints on the coupling of VLQs to SM quarks. In most models, VLQs couple preferentially to third-generation quarks and can decay through many different channels involving $W$, $Z$, or $H$ bosons. The relative couplings, and therefore branching fractions, for the different decay modes depend on the considered multiplet representation.

This wealth of final states and production modes allows a variety of searches for pair-produced and singly produced VLQs. In many cases, they complement one another and aim to cover the full range of multiplet configurations, masses, and types of VLQs. ATLAS has searched for VLQs in final states with a hadronically decaying Higgs boson appearing together with top and bottom quarks~\cite{EXOT-2019-04, EXOT-2019-07}, in events with multiple leptons~\cite{EXOT-2020-01,EXOT-2018-58}, in events with large \met~\cite{EXOT-2019-08}, and in events with a lepton and a large number of jets~\cite{EXOT-2018-52}. The production and decay modes explored by each search discussed in this section are summarized in Table~\ref{tab:vlqStates}.

\begin{figure}[tb]
\begin{center}
\subfloat[]{\includegraphics[width=0.3\textwidth]{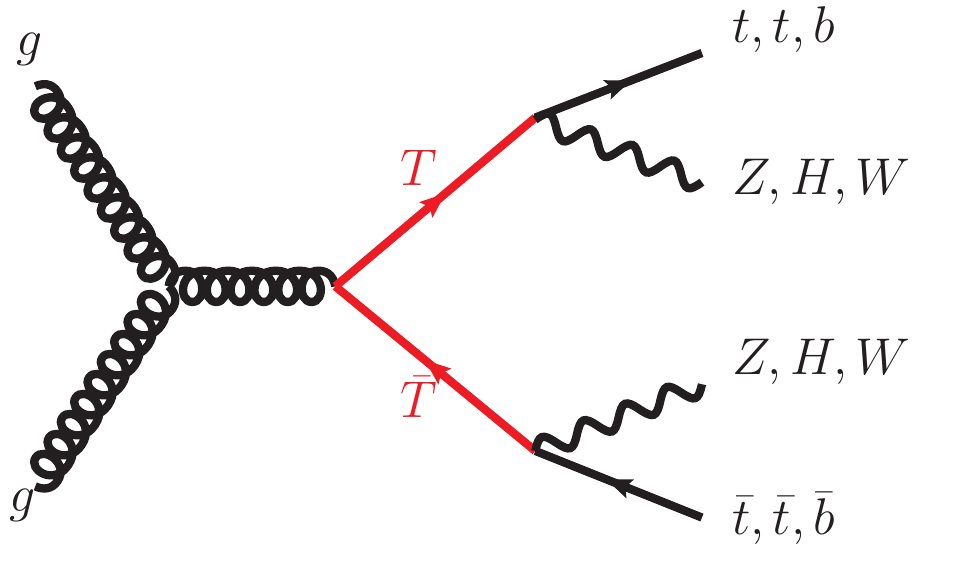}}
\qquad
\subfloat[]{\includegraphics[width=0.3\textwidth]{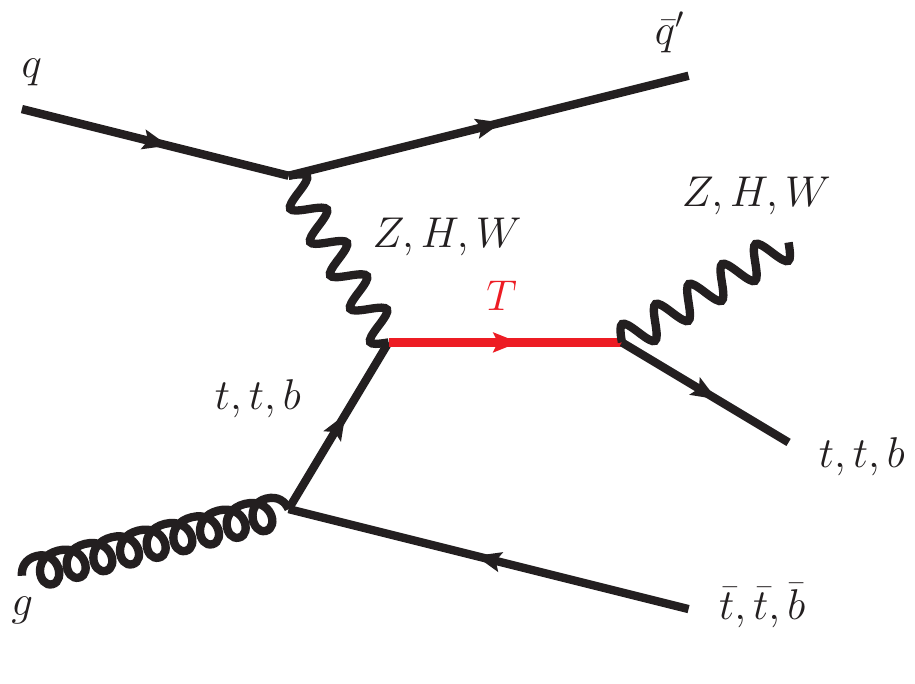}}
\end{center}
\caption{Representative diagrams of (a) pair production and (b) single production of a vector-like $T$ quark.}
\label{fig:vlq_proc}
\end{figure}

\begin{table*}[tp!]
\begin{center}
\caption{Production and decay channels explored by the different VLQ searches discussed in this section.}
\begin{tabular}{l c c c}
\hline
\hline
Search & Production mode & Decay channel & Section \\\hline
Hadronic $T$ search~\cite{EXOT-2019-07} & Single & $T\to Ht$ & \ref{sec:vlqHad} \\
Hadronic $B$ search~\cite{EXOT-2019-04} & Single & $B\to Hb$ & \ref{sec:vlqHad} \\
Multilepton (single)~\cite{EXOT-2020-01} & Single & $T\to Zt$ & \ref{sec:vlqLep} \\
Multilepton (pair)~\cite{EXOT-2018-58} & Pair & $TT\to Zt Vt$, $BB\to Zb Vb$, $V=W,Z,H$& \ref{sec:vlqLep} \\
High \met~\cite{EXOT-2019-08} & Pair & $T\to Vt$ or $B\to Vb$, $V=W,Z,H$ & \ref{sec:vlqMet} \\
Lepton and jets~\cite{EXOT-2018-52} & Single & $T\to Ht$, $T\to Zt$ & \ref{sec:vlqJets} \\
\hline
\hline
\end{tabular}
\label{tab:vlqStates}
\end{center}
\end{table*}

\subsubsection{Hadronic searches}
\label{sec:vlqHad}

Single production of $T$ and $B$ quarks is explored in fully hadronic final states, and the two analyses share similar characteristics and challenges. The $T$ search~\cite{EXOT-2019-07} focuses on the decay channel $T\rightarrow H t$, where the Higgs boson decays into a pair of \bquarks. The $B$ search~\cite{EXOT-2019-04} focuses on the decay channel $B\rightarrow H b$ and the same Higgs decay is considered. In both cases, the VLQ invariant mass can be reconstructed explicitly by selecting the Higgs boson and the corresponding heavy quark, and it is the search variable used in both analyses. The analyses also share the use of large-$R$ jets to reconstruct and identify the Higgs boson, implementing substructure techniques based on the mass of the jet, the presence and kinematics of \btagged variable-$R$ track-jets inside the large-$R$ jet and the $\tau_{21}$ variable~\cite{Thaler:2010tr}, which is a relative measure of whether the jet has a two-body or one-body structure.

The $B$ analysis uses a four-region variation of the method based on the presence of a forward jet in the event (typical of single VLQ production) and the $b$-tagging state of the \bjet chosen to reconstruct the $B$ candidate. Correlations are considered by studying events outside the Higgs mass window for the large-$R$ jets chosen as Higgs candidates. Finally, a reweighting method is implemented event-by-event between the control regions used to estimate the background; this allows further correction for kinematic effects produced by correlations between the variables considered in the construction of the ABCD regions.

The $T$ analysis uses an extension of the ABCD method that brings the number of regions to 81, building a $9\times 9$ selection matrix. The different regions correspond to the different tagging states of the two leading large-$R$ jets in the event, i.e.\ whether these jets are Higgs candidates or not, whether they are top-tagged or not, and whether they contain \btagged jets or not. The signal region corresponds to events with one top-tagged jet, one Higgs candidate jet, and at least a total of three \btagged jets inside those two jets. The additional regions are used in the background estimation itself, in the normalization of the subdominant $\ttbar$ background, which is estimated from MC samples in the SR, and to evaluate the correlations between the different selection criteria and take them into account in the final background estimation.

The statistical analysis in the two searches is based on the reconstructed invariant mass of the VLQ candidate and finds good agreement between the data and SM prediction in the signal region of each search.

\subsubsection{Multilepton searches}
\label{sec:vlqLep}

Events with opposite-sign same-flavour (OSSF) leptons are investigated in searches for pair-produced $T$ or $B$ quarks~\cite{EXOT-2018-58} and singly produced $T$ quarks~\cite{EXOT-2020-01}. In all cases, the relevant decay channel involves a $Z$ boson decaying into a pair of leptons, $T\rightarrow Zt$ or $B\rightarrow Zb$. Additional leptons can appear in the final state through the leptonic decays of top quarks. Both analyses optimize two search channels independently, one with exactly two OSSF leptons (electrons or muons) and the other with two OSSF leptons and at least one additional lepton. The OSSF lepton pair is used to reconstruct the $Z$ boson candidate in both channels of both analyses, and it is required to have a mass close to the mass of the $Z$ boson.

The single-production search defines regions by using the number of forward jets, the number of $b$-tagged jets, the number of top-tagged jets,\footnote{These are jets reclustered with a variable $R$ parameter value~\cite{Krohn:2009zg} to reconstruct more-boosted or less-boosted top candidates.} and the number of leptons defining the channel. In the 2-lepton channel, events are assigned to the signal region if they have at least one forward jet, one \btagged jet, and one top-tagged jet. Other combinations define control regions to help constrain the major background processes, $Z$+jets, $VV$, and \ttbar production, which are modelled using MC samples. A kinematic reweighting is used to further constrain the $Z$+jets background. In the 3-lepton channel, the signal region is defined by selecting events with at least one \btagged jet and at least one forward jet. Additional cuts are applied to the azimuthal angle $\Delta\phi$ between the reconstructed $Z$ boson and the non-OSSF lepton or between the $Z$ boson and the leading \btagged jet in the event. Control regions are defined by inverting or removing some of those requirements and selections to help constrain the $VV$ and $\ttbar+X$ processes (including $\ttbar+V$, $\ttbar\ttbar$, and $\ttbar WW$ processes), which produce the dominant backgrounds for this channel. The statistical analysis is performed on the transverse momentum $\pt(\ell\ell)$ of the OSSF leptons, simultaneously fitting the five CRs and two SRs, and good agreement between the background prediction and data is seen in Figure~\ref{fig:multilep_single}.

\begin{figure}[tb!]
\begin{center}
\subfloat[]{\includegraphics[width=0.4\textwidth]{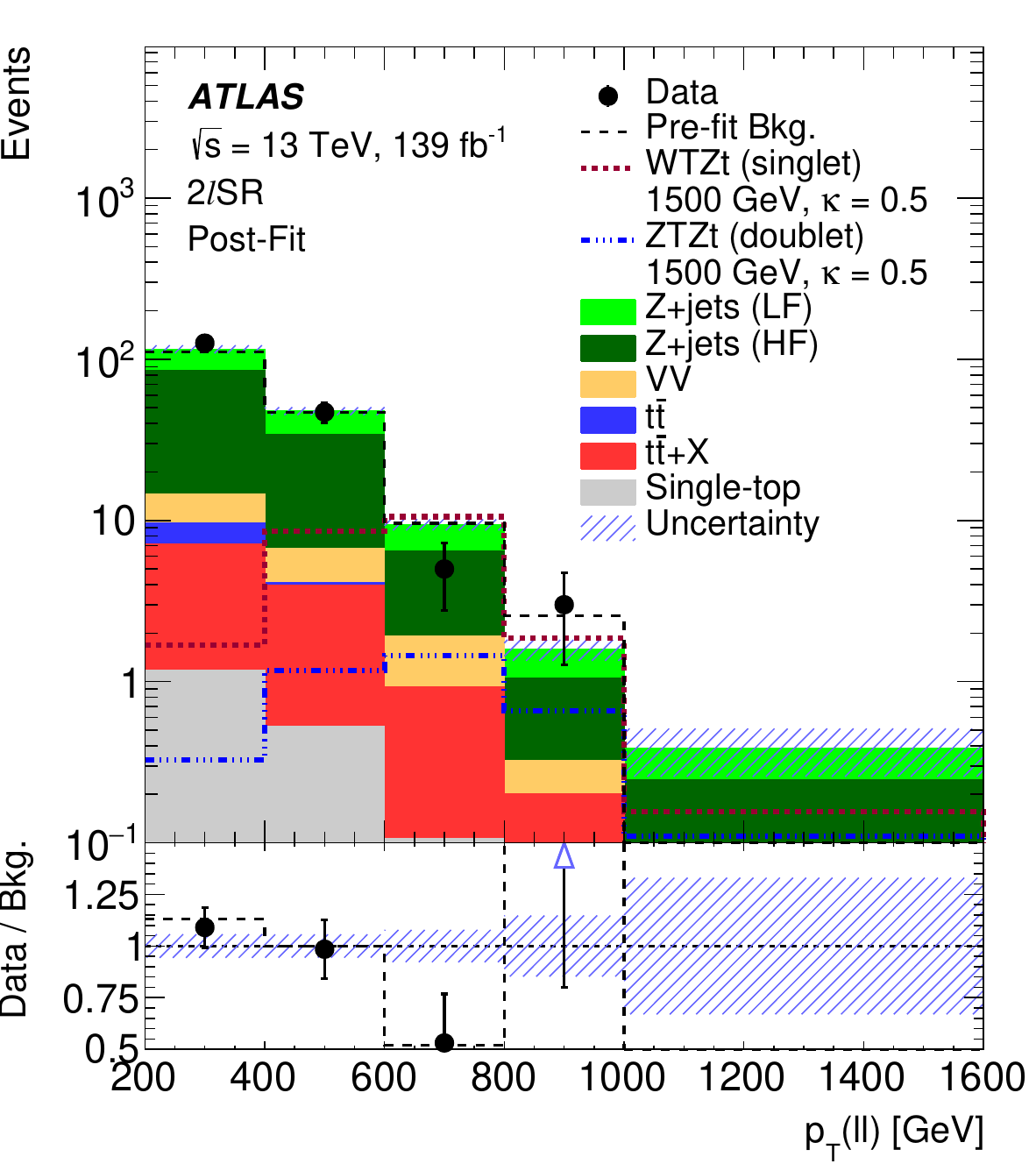}}
\qquad
\subfloat[]{\includegraphics[width=0.4\textwidth]{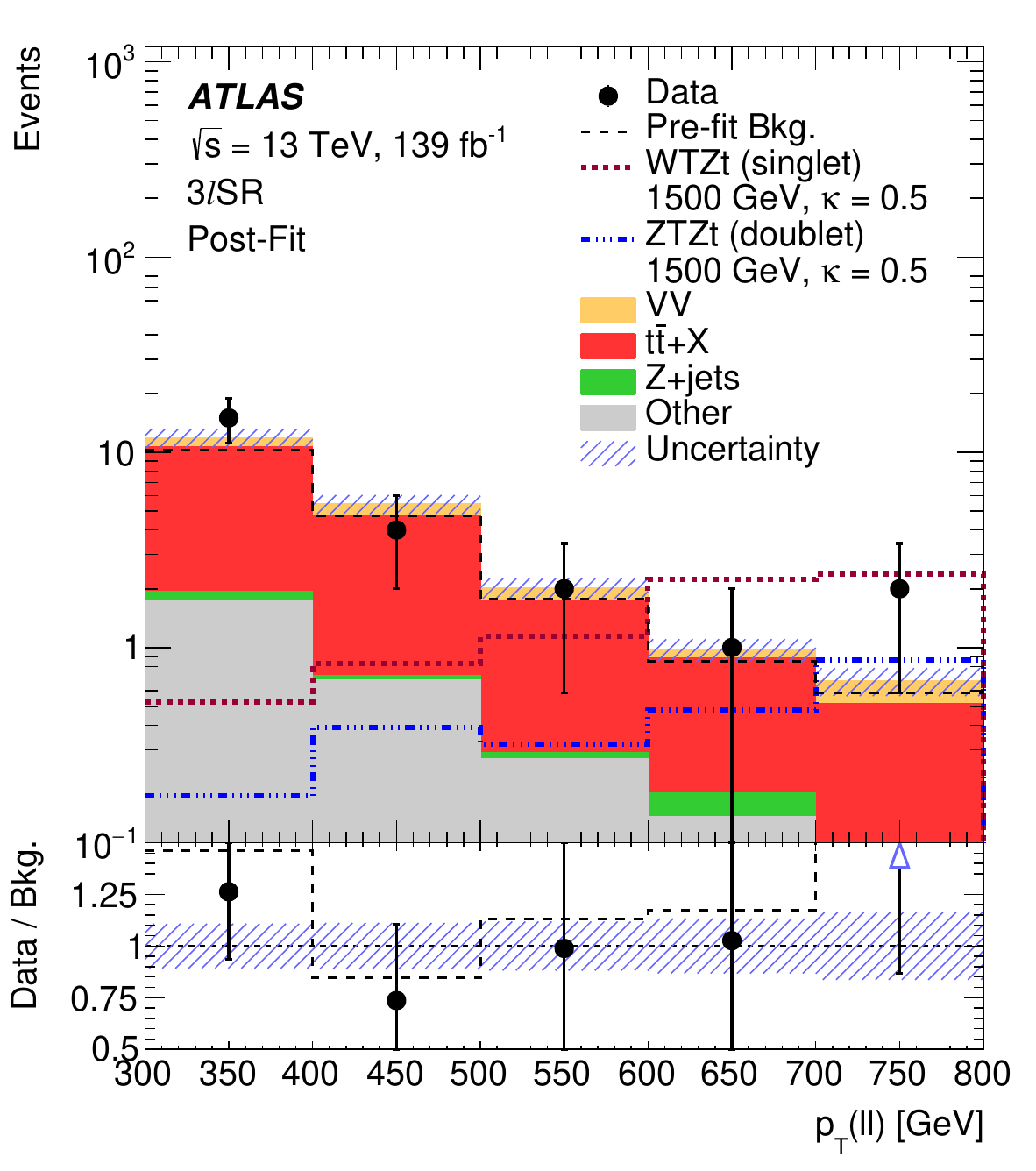}}
\end{center}
\caption{Distributions of $\pT(\ell \ell)$ in the signal region of the (a) 2-lepton channel and (b) 3-lepton channel of the VLQ search for singly produced $T$ quarks~\cite{EXOT-2020-01}. The distributions are shown after a background-only fit, with examples of expected signal contributions overlaid.}
\label{fig:multilep_single}
\end{figure}

The pair-production analysis has many more possible final states than the single-production analysis. One of the VLQs is always expected to decay into a leptonically decaying $Z$ boson and a heavy quark to provide the OSSF leptons. However, the other VLQ can decay into various bosons: $W$, $H$, or $Z$. To deal with this type of final state, a multiclass boosted-object tagger (MCBOT) is employed to classify reclustered (RC) jets as $V$-tagged, $H$-tagged, or top-tagged. This selection aims to take advantage of the large number of jets and consider the possibility of an invisibly decaying $Z$ boson. Events are also required to have at least one $b$-tagged jet. For the two channels, SRs are defined using the number of \btagged jets and slightly different definitions of the \HT variable, which are based on all reconstructed objects relevant to that particular channel. Once the SRs are defined, the MCBOT is used to further classify them according to the tagging status of the RC jet in the event. In total, 19 SRs and 3 CRs are defined and fitted simultaneously in the statistical analysis, with different variables closely related to the mass or transverse mass of the leptonic VLQ being used in each of them for the final fit. All background processes, dominated by $Z$+jets in the 2-lepton channel and $VV$ and $\ttbar+X$ in the 3-lepton channel, are modelled using MC samples. The background yields in all regions after the background-only fit are shown in Figure~\ref{fig:multilep_pair}, and good agreement with data is seen in all signal regions.

\begin{figure}[tb!]
\begin{center}
\includegraphics[width=0.49\textwidth]{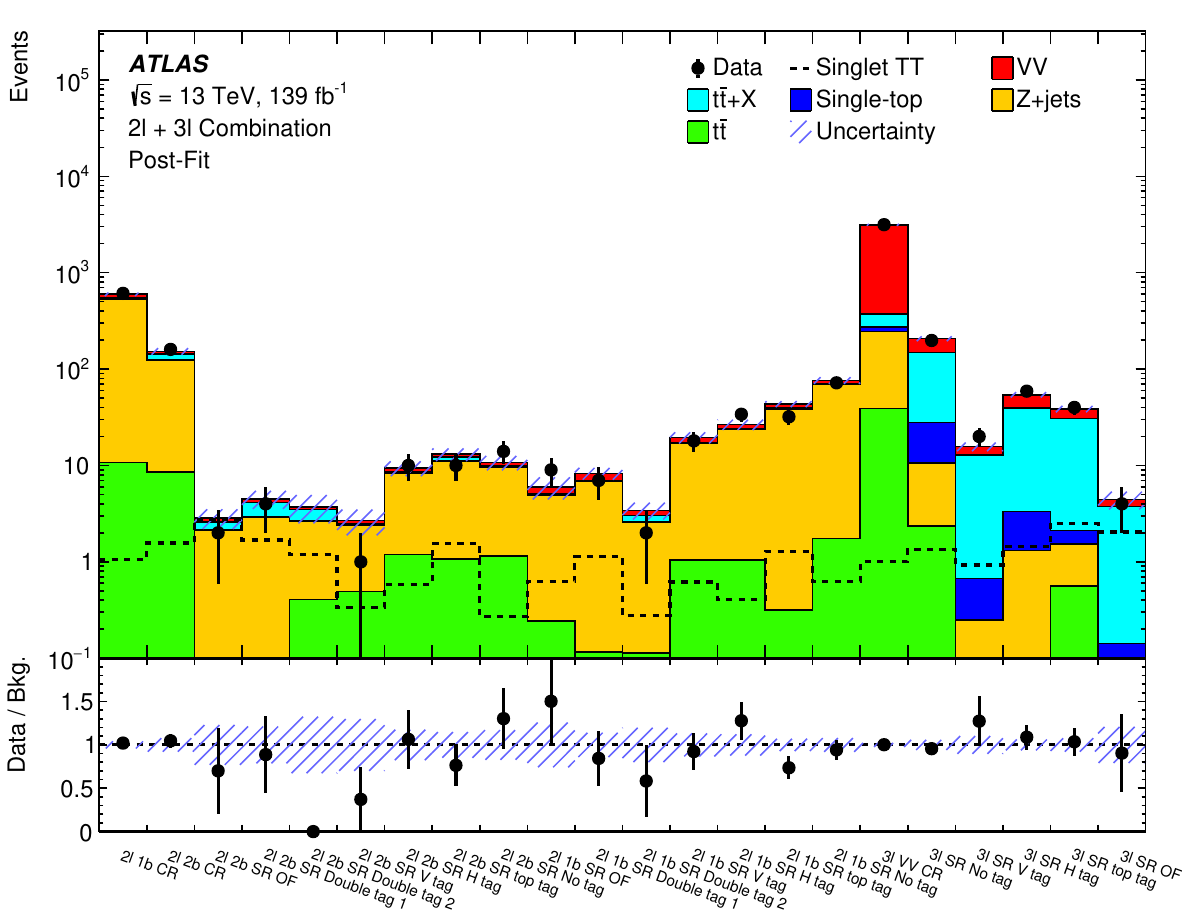}
\end{center}
\caption{Summary of the data and background yields in all analysis regions from the 2-lepton and 3-lepton channels of the search for pair-produced VLQs~\cite{EXOT-2018-58} after a background-only fit. The expected yields for a benchmark signal are also shown overlaid.}
\label{fig:multilep_pair}
\end{figure}

\subsubsection{Searches in events with large missing transverse momentum}
\label{sec:vlqMet}

Pair production of VLQs is also investigated in events with a large amount of \met~\cite{EXOT-2019-08}. This type of search is particularly sensitive to VLQ decays with a $Z$ or $W$ boson in the final state, which can provide a significant amount of \met through the presence of neutrinos and are relevant to the search for pair-produced $B$, $T$, and $X$. Events are required to have $\met > 250$~\GeV, exactly one lepton (electron or muon) and at least four jets, including at least one \btagged jet. Additional requirements on the azimuthal distance between jets and \met and on the $m_\mathrm{T}(\ell,\met)$ variable are also imposed. The latter helps to reject $W$+jets events, which form one of the major backgrounds in this analysis.  Another important background, \ttbar, is reduced by using the $am_\mathrm{T2}$ variable. A training region is defined by selecting events with large $am_\mathrm{T2}$ and $m_\mathrm{T}(\ell,\met)$ values and one large-$R$ jet. Three additional regions obtained by inverting some of those requirements are used to assist in modelling the $W$+jets, single-top, and \ttbar backgrounds. The third region is used to obtain a kinematic event-by-event reweighting of the \ttbar background, applied to correct the MC modelling of this process in the signal region. The first two are used directly in the final fit as control regions. A neural network based on 13 kinematic variables from multiple objects in the event is trained and used to classify events in the training region. High scores define the signal region, while low scores are used as a final control region. The statistical analysis is performed by simultaneously fitting the signal region and three control regions, with all backgrounds modelled using MC samples. The post-fit distributions in the signal region and one of the control regions after a background-only fit are shown in Figure~\ref{fig:vlq_met} and exemplify the excellent agreement between the data and the expected background.

\begin{figure}[tb!]
\begin{center}
\subfloat[]{\includegraphics[width=0.4\textwidth]{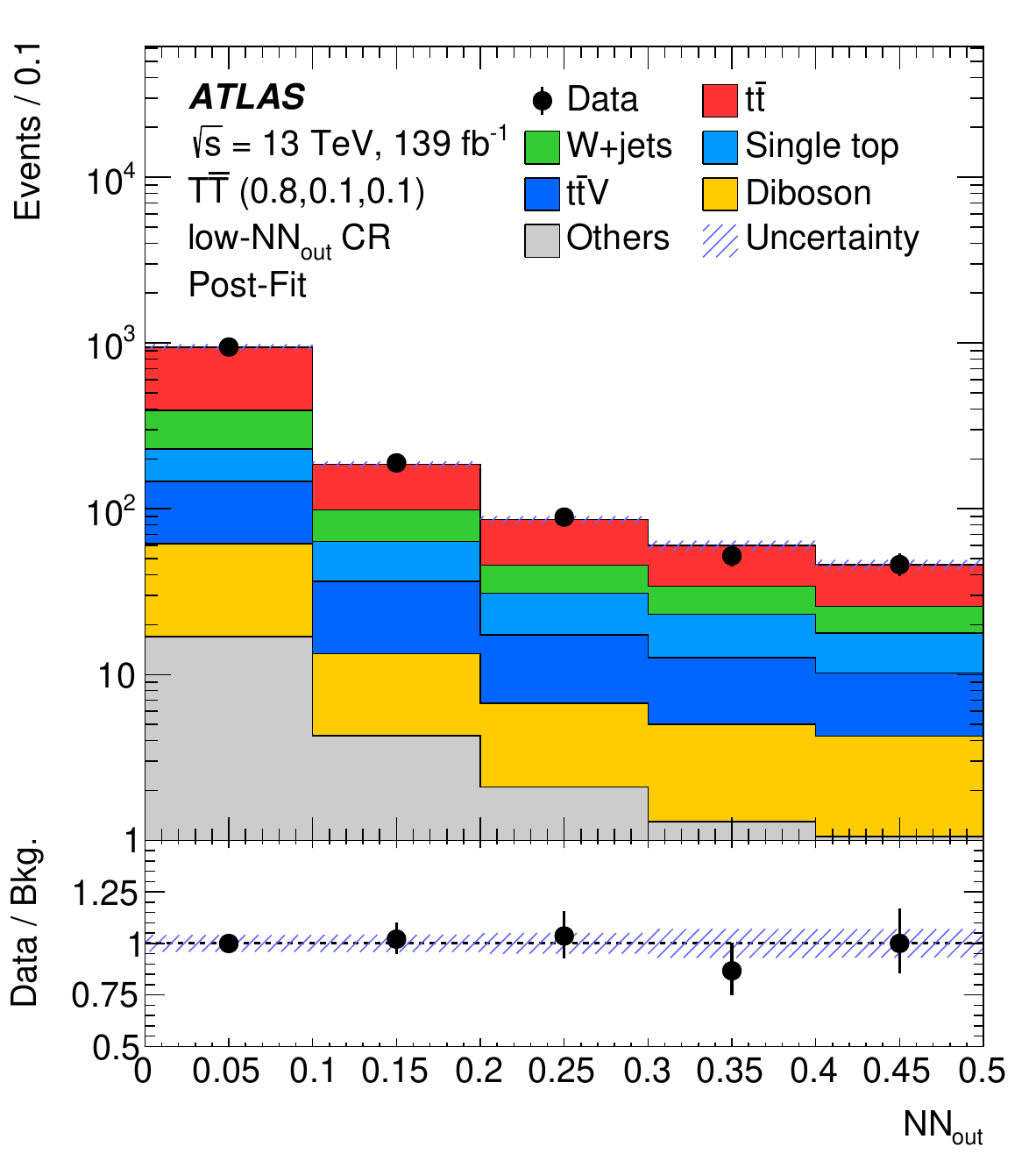}}
\qquad
\subfloat[]{\includegraphics[width=0.4\textwidth]{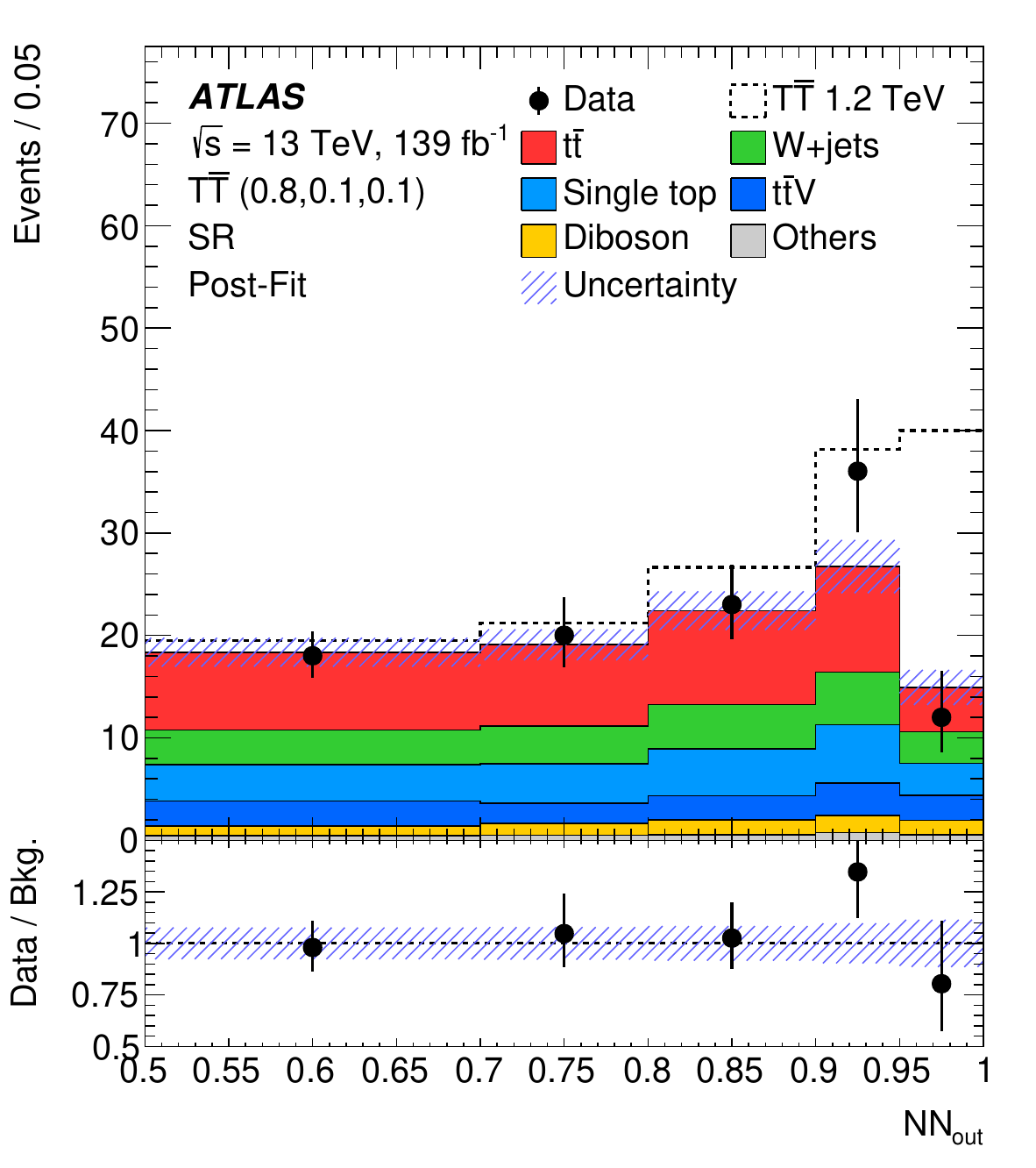}}
\end{center}
\caption{Distributions of the neural network score for (a) one of the control regions and (b) the signal region used in the search for VLQs in events with large \met~\cite{EXOT-2019-08}. Distributions are shown after a background-only fit, with a benchmark signal overlaid.}
\label{fig:vlq_met}
\end{figure}

\subsubsection{Searches in events with one lepton and multiple jets}
\label{sec:vlqJets}

Before hadronization, many of the final states that appear in single $T$ production contain a large number of quarks, top quarks, and $b$-quarks, and a heavy boson ($H$ or $Z$). When the heavy boson decays hadronically (into a pair of \btagged jets or other jets) and one of the top quarks decays leptonically, a signature of a single lepton and a large number of jets (from three to six) becomes particularly powerful in selecting and identifying this type of process. To cover the largest possible number of different subprocesses involved in single $T$ production, a large number of dedicated regions with different jet and hadronically decaying boson multiplicities are investigated~\cite{EXOT-2018-52}. The background from multijet production is suppressed by placing requirements on \met and the transverse mass of the lepton and \met system. Events must also have at least three jets, with one of them \btagged.  RC jets are used to identify hadronically decaying boosted top quarks, Higgs bosons, and $V$ ($W$ or $Z$) bosons. The mass of each RC jet and its number of subjets are used to tag it as one of the possible hadronic resonances. Top-quark candidates decaying semileptonically are identified by combining the lepton, the \met, and one of the \btagged jets in the event and imposing kinematic constraints on the combination. A total of 22 signal regions are defined for different multiplicities of jets, \btagged jets, leptonically decaying top quarks, top-tagged RC jets, Higgs-tagged RC jets, and $V$-tagged RC jets. All of the SRs are also required to have one forward jet. Additional regions without a forward jet are used to help with background modelling in the final fit. All background processes are modelled using MC samples. The multijet background prediction is improved by normalizing it to data in a multijet-enriched region, while a dedicated kinematic reweighting is used to improve the \ttbar and $W$+jets modelling, which is known to underestimate the data at high jet multiplicities and/or high \pt. The statistical analysis in the 22 SRs and 2 CRs uses $m_\mathrm{eff}$, which
shows strong signal discrimination power. Good agreement between the SM prediction and the data in all the fit regions is seen in Figure~\ref{fig:vlq_multijet}.

\begin{figure}[tb]
\begin{center}
\includegraphics[width=0.49\textwidth]{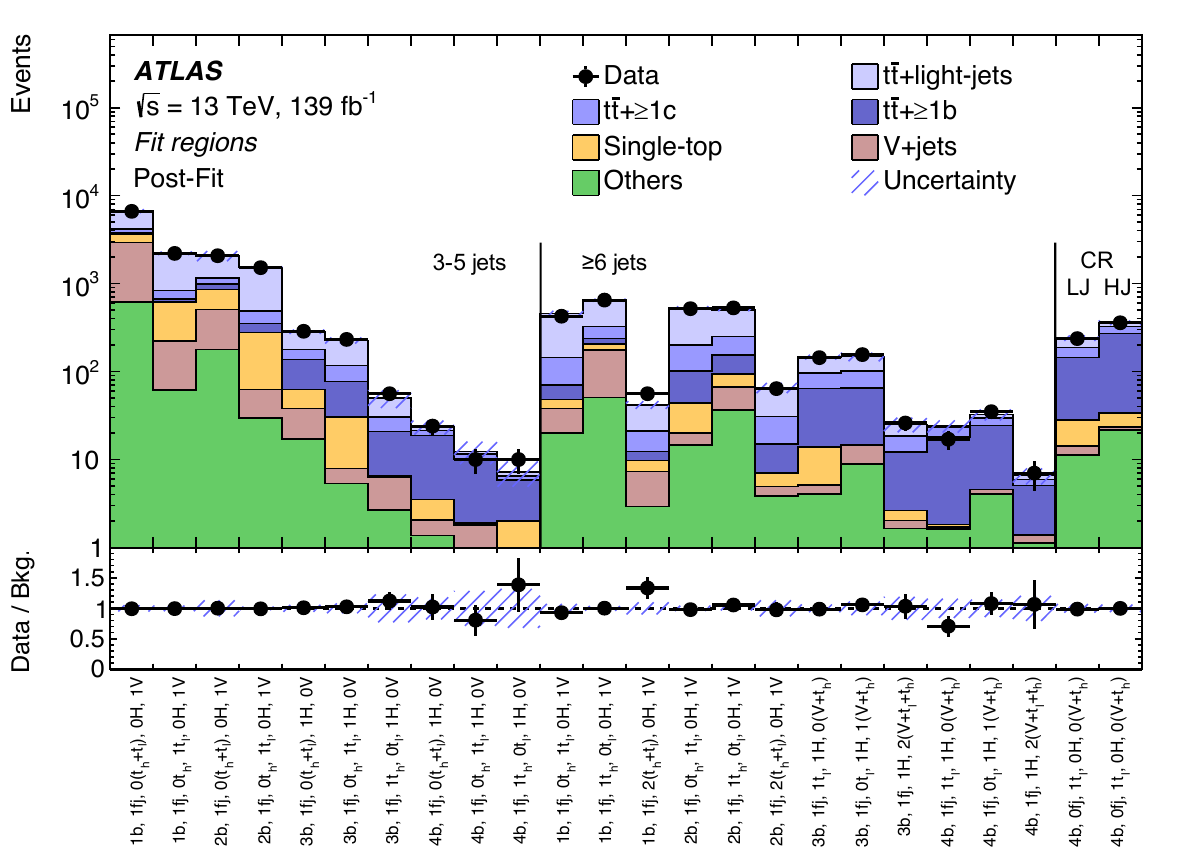}
\end{center}
\caption{Summary of the data and background yields in all analysis regions used in the search for singly produced $T$ VLQs in final states with one lepton and multiple jets~\cite{EXOT-2018-52} after a background-only fit.}
\label{fig:vlq_multijet}
\end{figure}

\subsubsection{Limits on VLQ pair production}

The two searches looking for pair production of VLQs have complementary sensitivity. Mass limits are set in a two-dimensional plane of the possible branching ratios (BR) of $T$ and $B$ decays and are shown in Figure~\ref{fig:vlqPairLimits}. The limits corresponding to the singlet and different doublet hypotheses are shown in each plot. For a $T$ quark, the multilepton search is generally more sensitive, but the \met-based search has better sensitivity in the top left corner of the BR plane. They have very similar sensitivities for the singlet and doublet models, excluding masses below about 1.25~\TeV and 1.41~\TeV, respectively. For a $B$ quark, the two searches behave very differently and are more sensitive in opposite regions of the BR plane. The \met-based search has a higher mass limit in the singlet scenario, excluding masses up to 1.33~\TeV. The doublet scenarios in the two searches correspond to different VLQ combinations and thus cannot be compared directly. Masses below ${\sim}1$~\TeV are excluded for any BR combination of pair-produced $B$ and $T$ quarks.

\begin{figure}[tb]
\begin{center}
\subfloat[]{\includegraphics[width=0.5\textwidth]{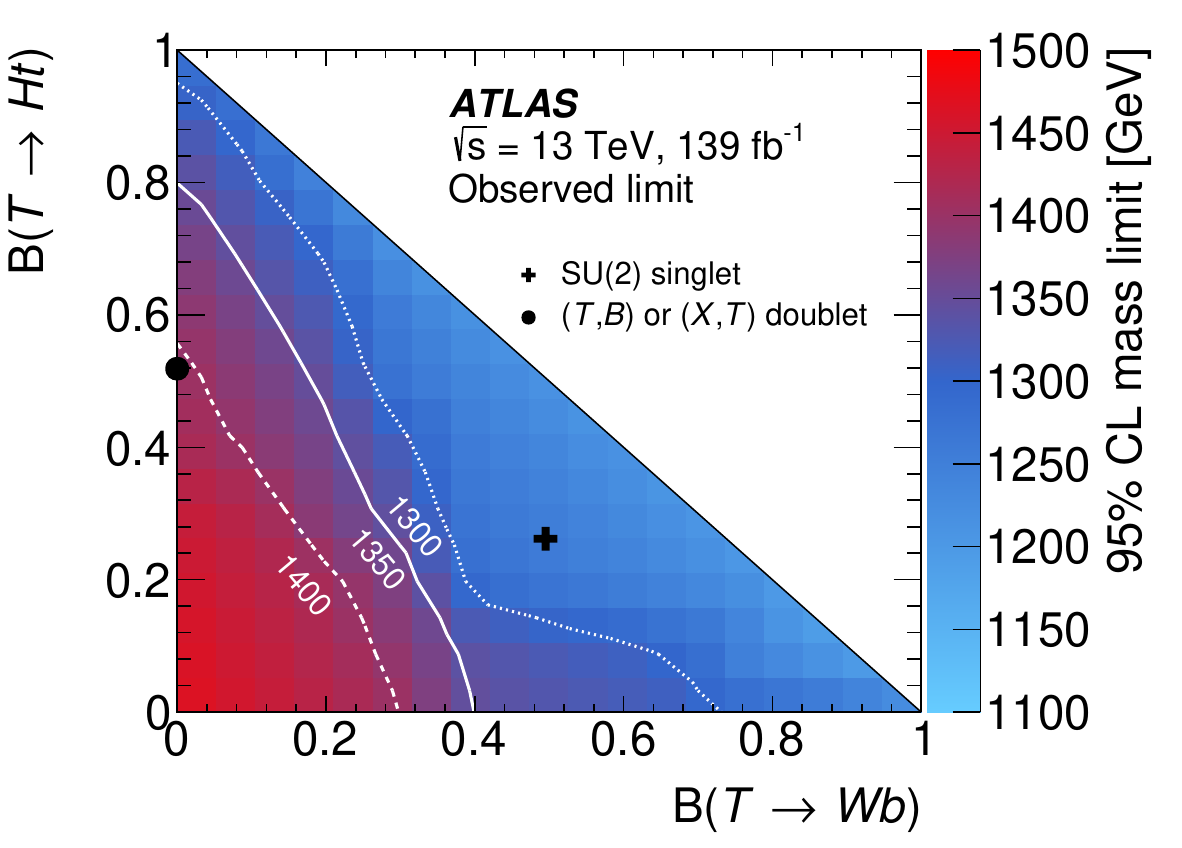}}
\subfloat[]{\includegraphics[width=0.5\textwidth]{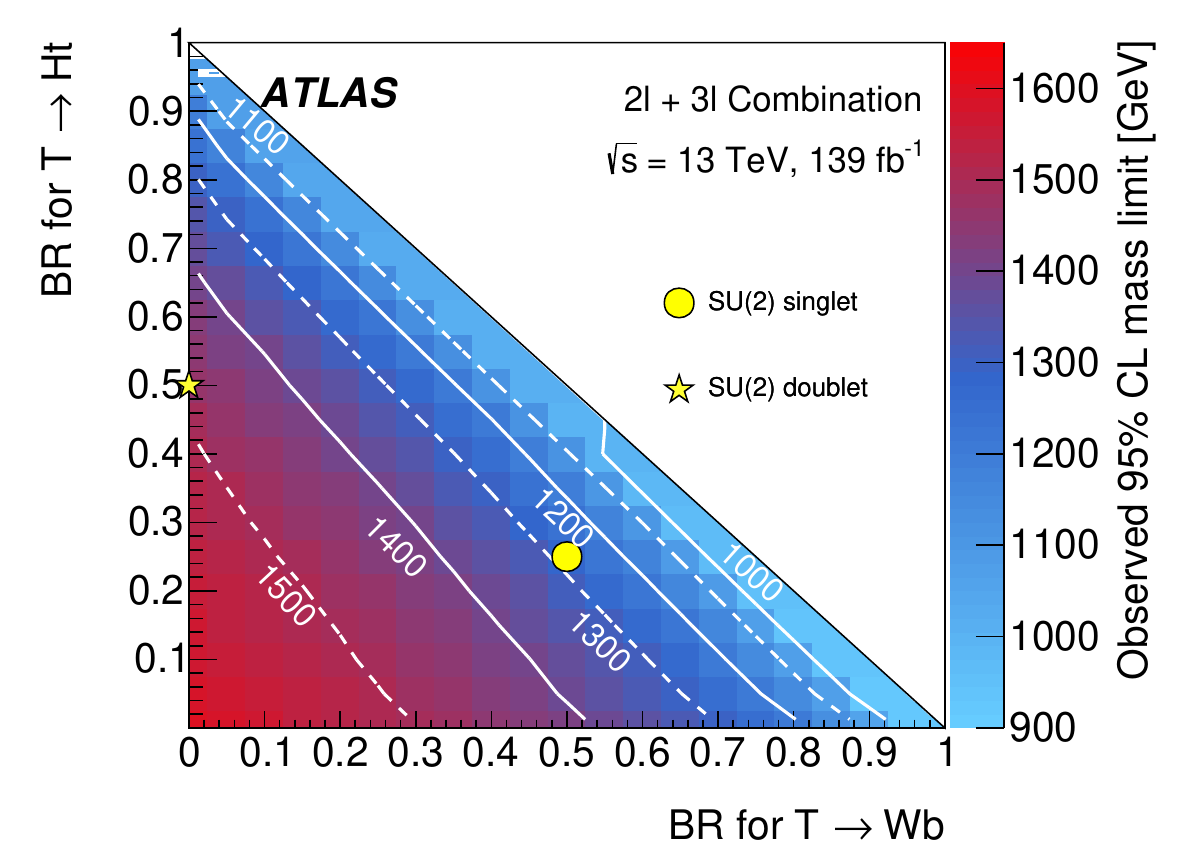}}\\
\subfloat[]{\includegraphics[width=0.5\textwidth]{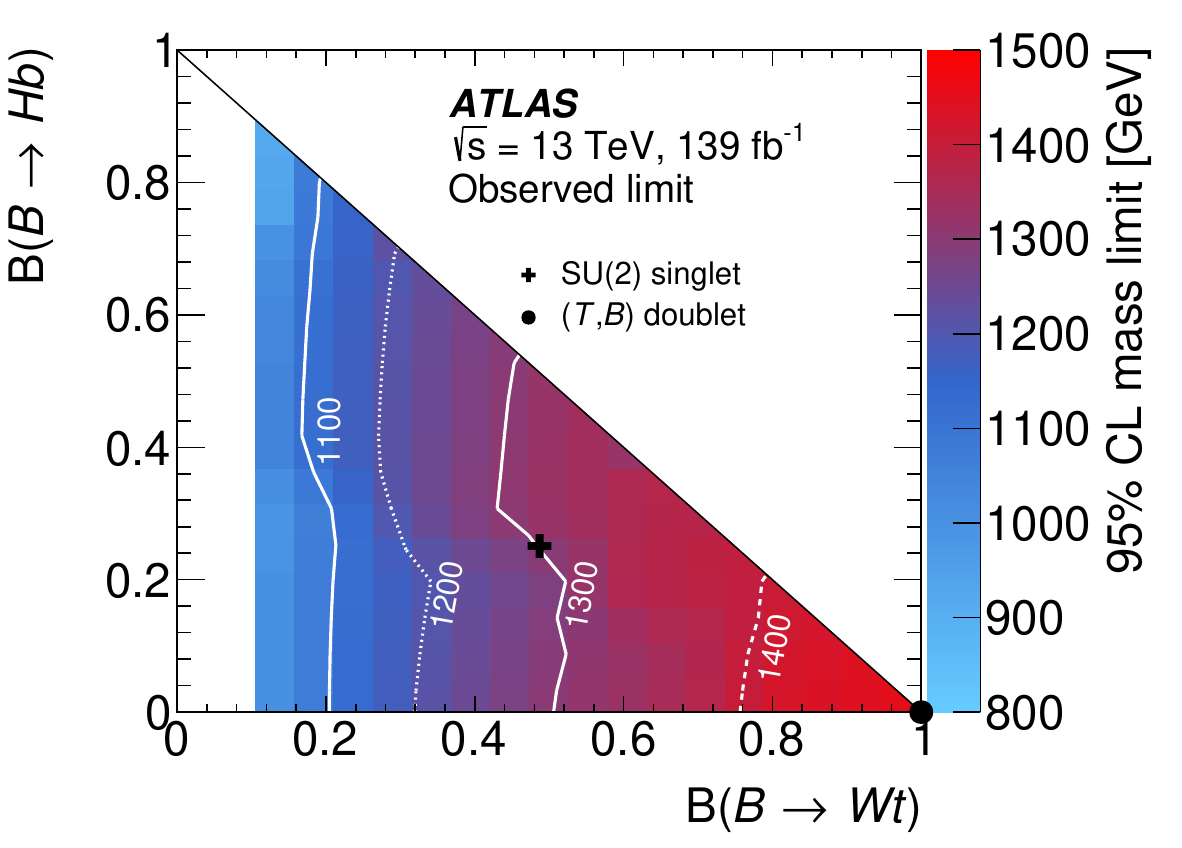}}
\subfloat[]{\includegraphics[width=0.5\textwidth]{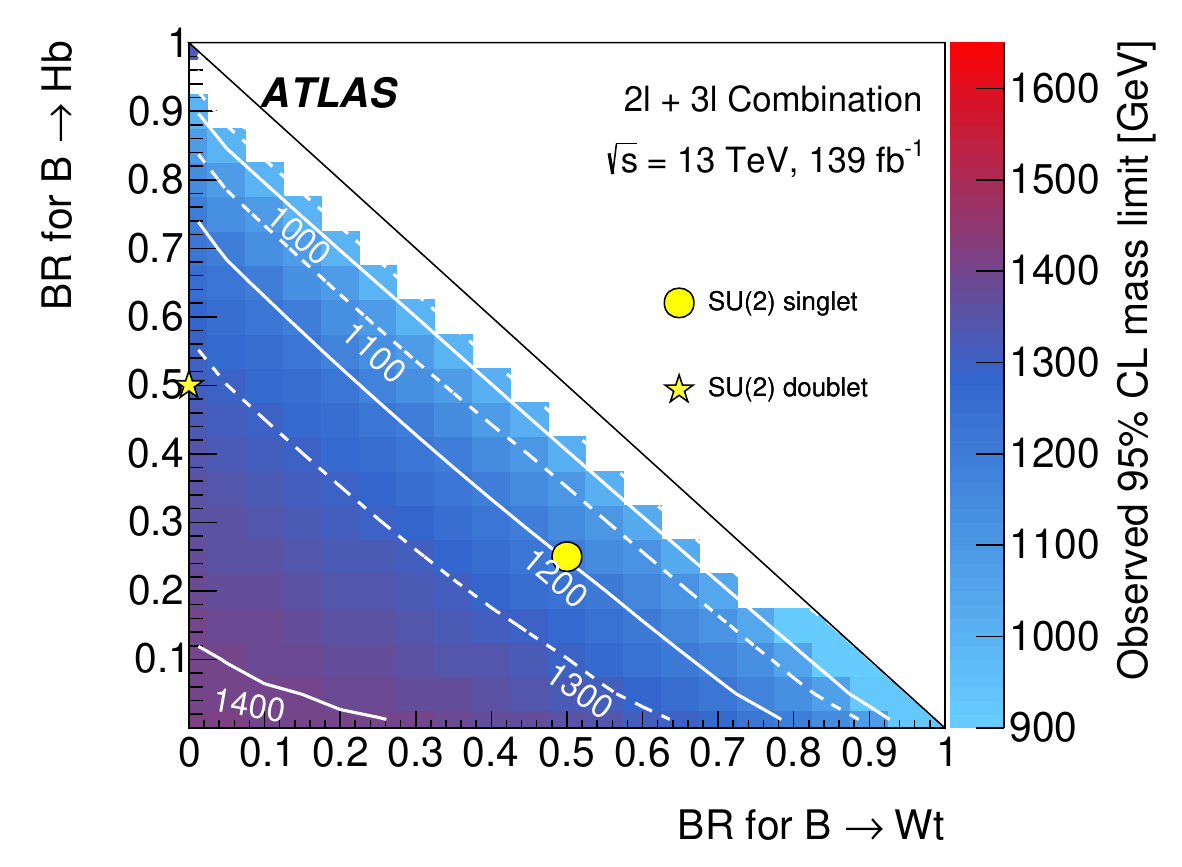}}
\end{center}
\caption{Observed lower limits on the $T$ and $B$ masses in the BR plane for pair production of (a,b) $TT$ and (c,d) $BB$. Results from (a,c) the search for VLQs in events with large \met~\cite{EXOT-2019-08} and (b,d) the search for pair-produced VLQs in multilepton final states~\cite{EXOT-2018-58} are included. The doublet hypothesis considered in (b,d) is $(T, B)$.}
\label{fig:vlqPairLimits}
\end{figure}

In addition to these two analyses using the full \RunTwo dataset, seven VLQ searches for pair-produced VLQs in a partial \RunTwo dataset (recorded between 2015 and 2016) were interpreted together in a single statistical combination~\cite{EXOT-2017-17}, which improved on the individual results.

\subsubsection{Limits on single VLQ production}

Limits on single $T$ production are set for various coupling values as a function of the mass of the VLQ. Limits on the coupling as a function of the $T$ mass are shown in Figure~\ref{fig:vlqSingleSinglet} and Figure~\ref{fig:vlqSingleDoublet} for the singlet and $(T, B)$ or $(B, Y)$ doublet hypotheses, respectively. Of the three searches, the multilepton analysis has the strongest sensitivity for the full range of masses considered for any multiplet hypothesis. In the semileptonic analysis, the limit for high masses reaches lower coupling values because of a slight deficit of data observed in some signal regions. Only the hadronic $B$ decay search in ATLAS considers single $B$ production, and the corresponding limit plots are also shown in Figure~\ref{fig:vlqSingleSinglet} and Figure~\ref{fig:vlqSingleDoublet}. The hadronic $T$ decay analysis did not consider the doublet hypothesis and, therefore, is not included in Figure~\ref{fig:vlqSingleDoublet}.

\begin{figure}[tb]
\begin{center}
\subfloat[]{{\includegraphics[width=0.49\textwidth]{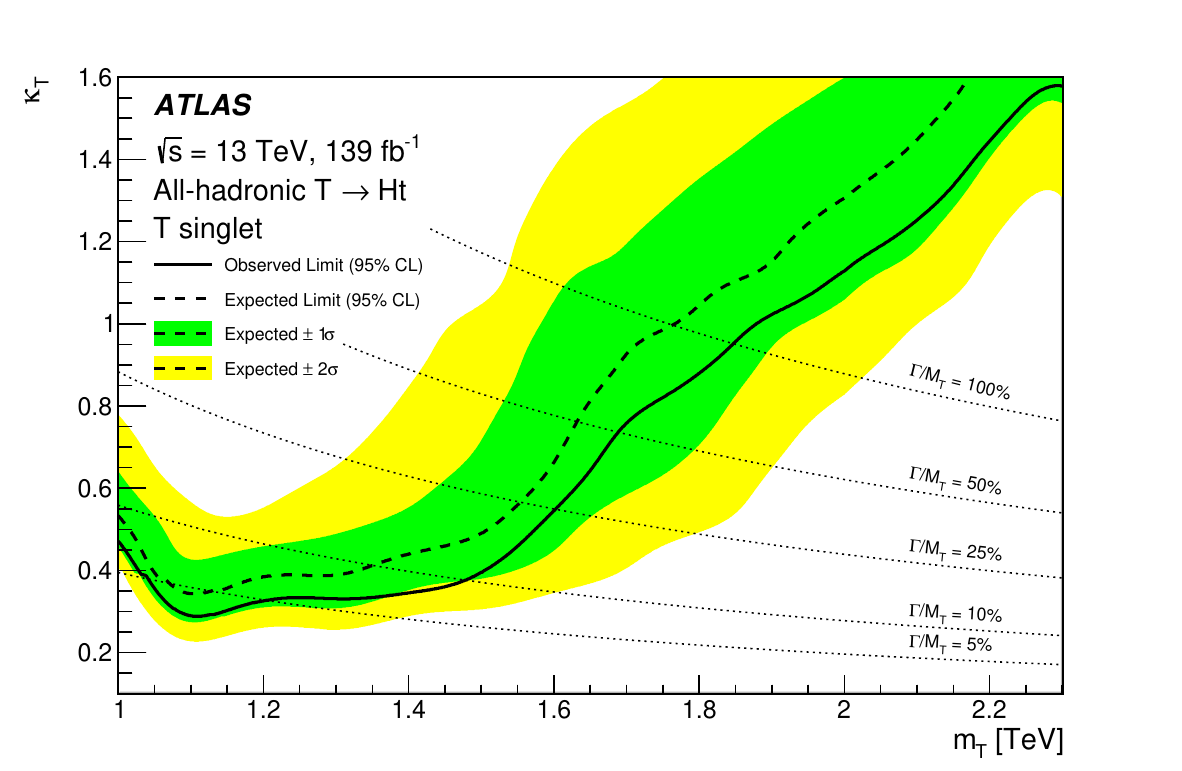}}}
\qquad
\subfloat[]{\includegraphics[width=0.44\textwidth]{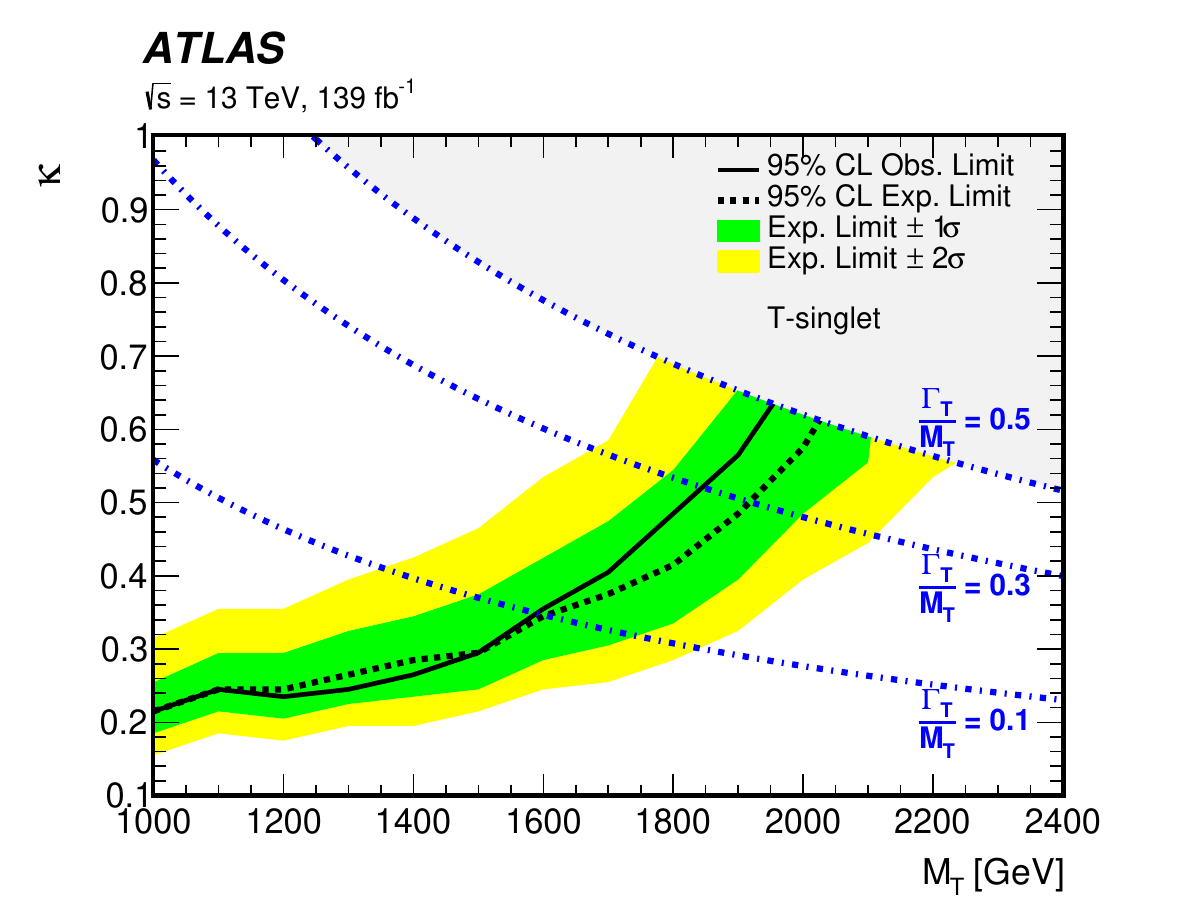}}\\
\qquad
\subfloat[]{\includegraphics[width=0.42\textwidth]{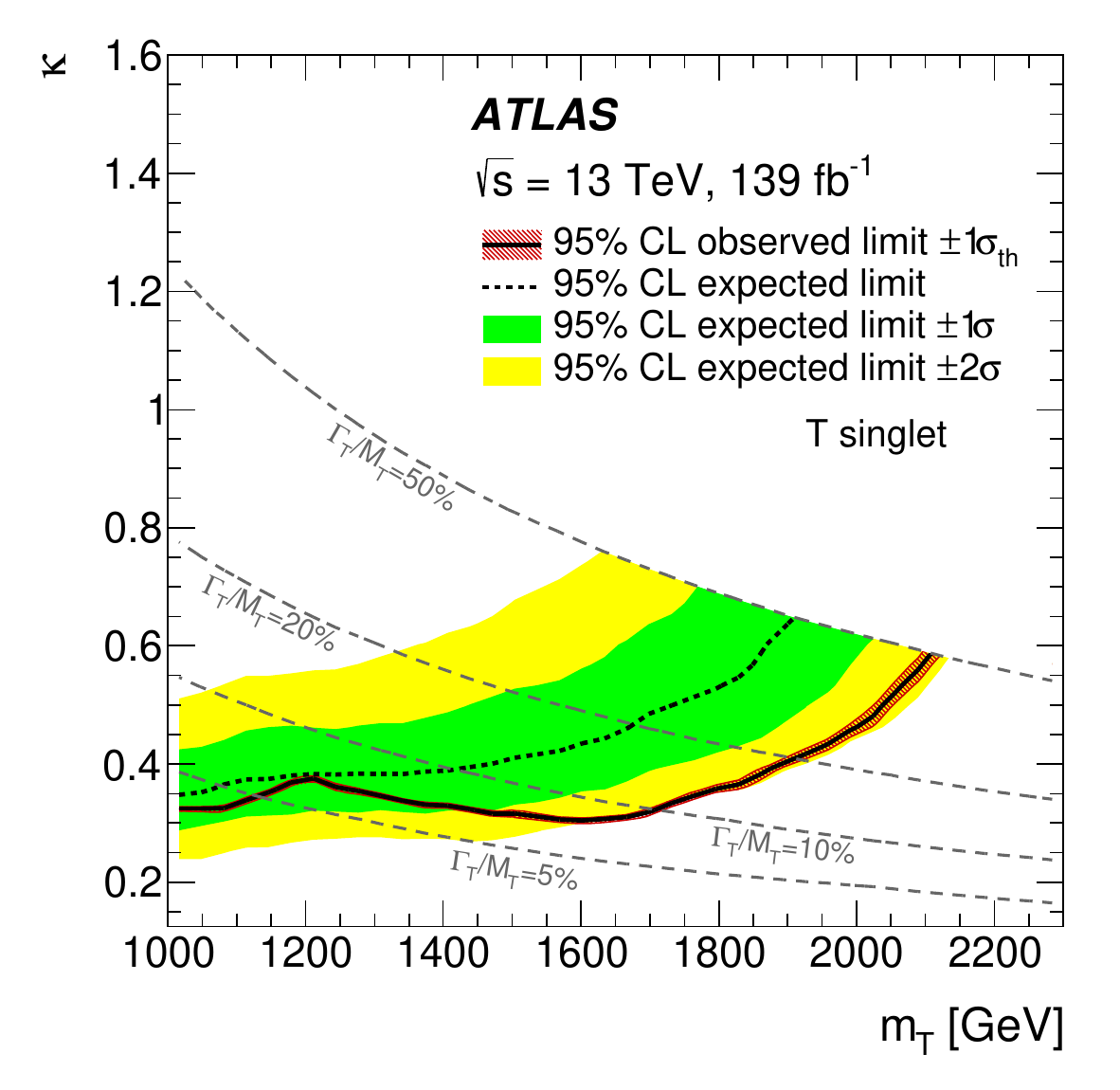}}
\qquad
\subfloat[]{\raisebox{0.05\height}{\includegraphics[width=0.47\textwidth]{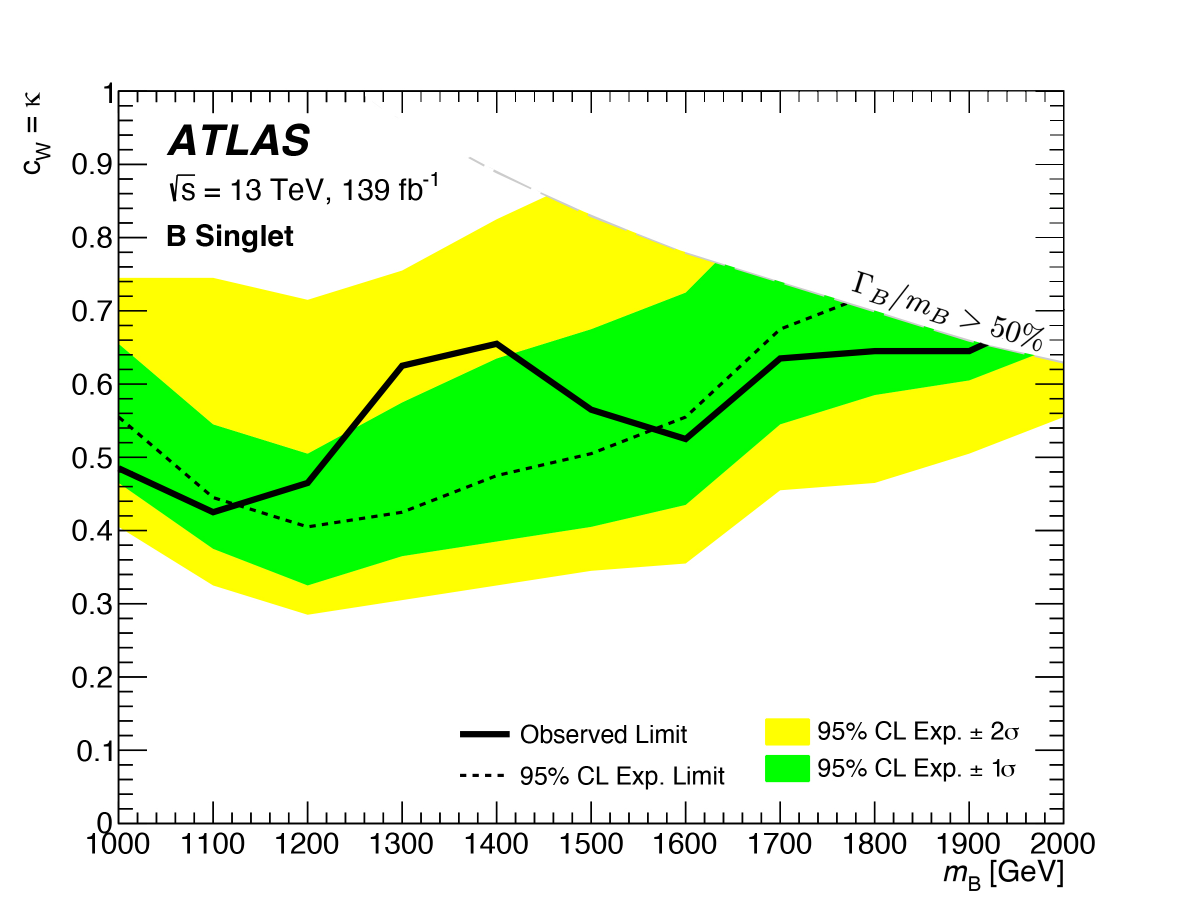}}}
\end{center}
\caption{Observed and expected upper limits on (a,b,c) single $T$ and (d) single $B$ production as a function of the mass of the VLQ for a singlet hypothesis. Results from searches for (a) singly produced $T$ in hadronic final states~\cite{EXOT-2019-07}, (b) singly produced $T$ in multilepton final states~\cite{EXOT-2020-01}, (c) singly produced $T$ in final states with one lepton and multiple jets~\cite{EXOT-2018-52} and (d) singly produced $B$ in hadronic final states~\cite{EXOT-2019-04}, are included. The dashed curves represent contours of constant relative VLQ width.}
\label{fig:vlqSingleSinglet}
\end{figure}

\begin{figure}[tb]
\begin{center}
\subfloat[]{\raisebox{0.0\height}{\includegraphics[width=0.53\textwidth]{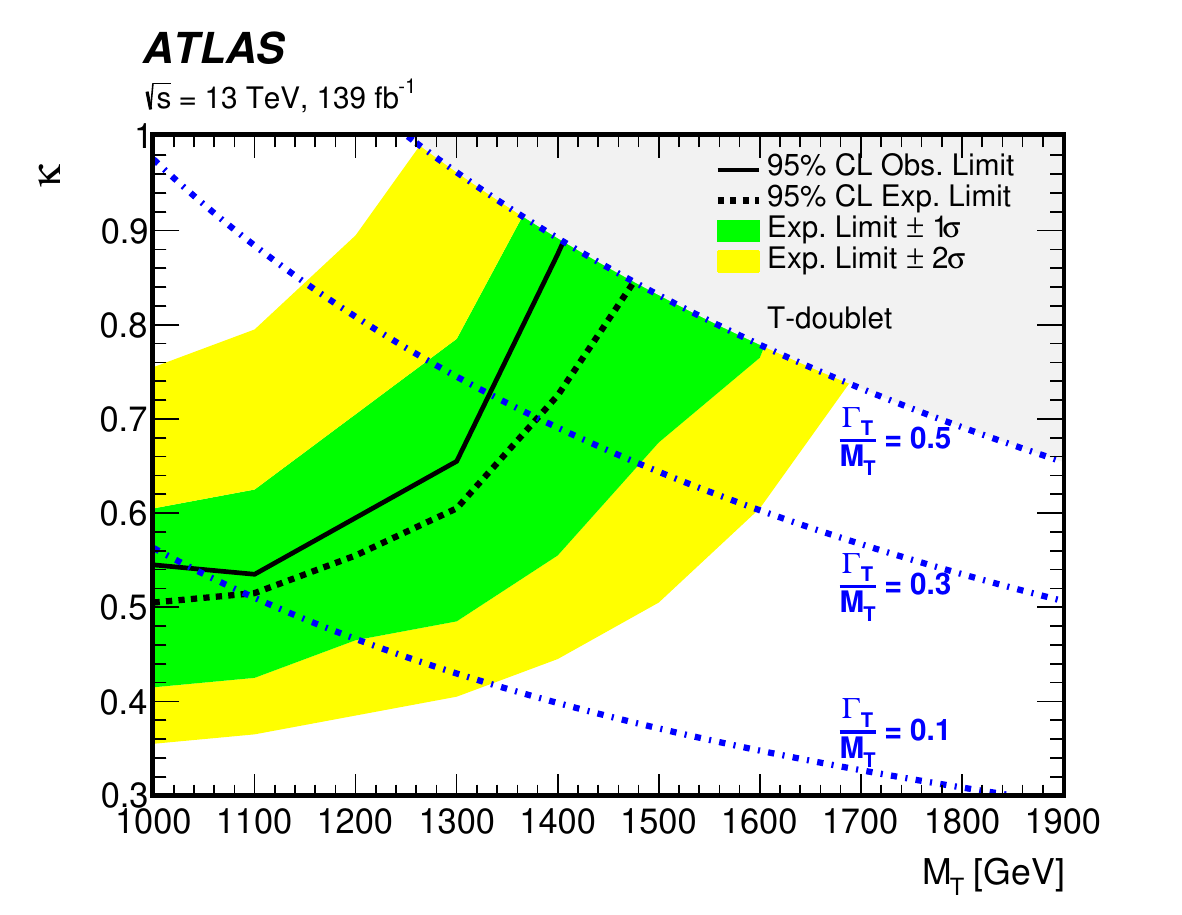}}}
\subfloat[]{\includegraphics[width=0.4\textwidth]{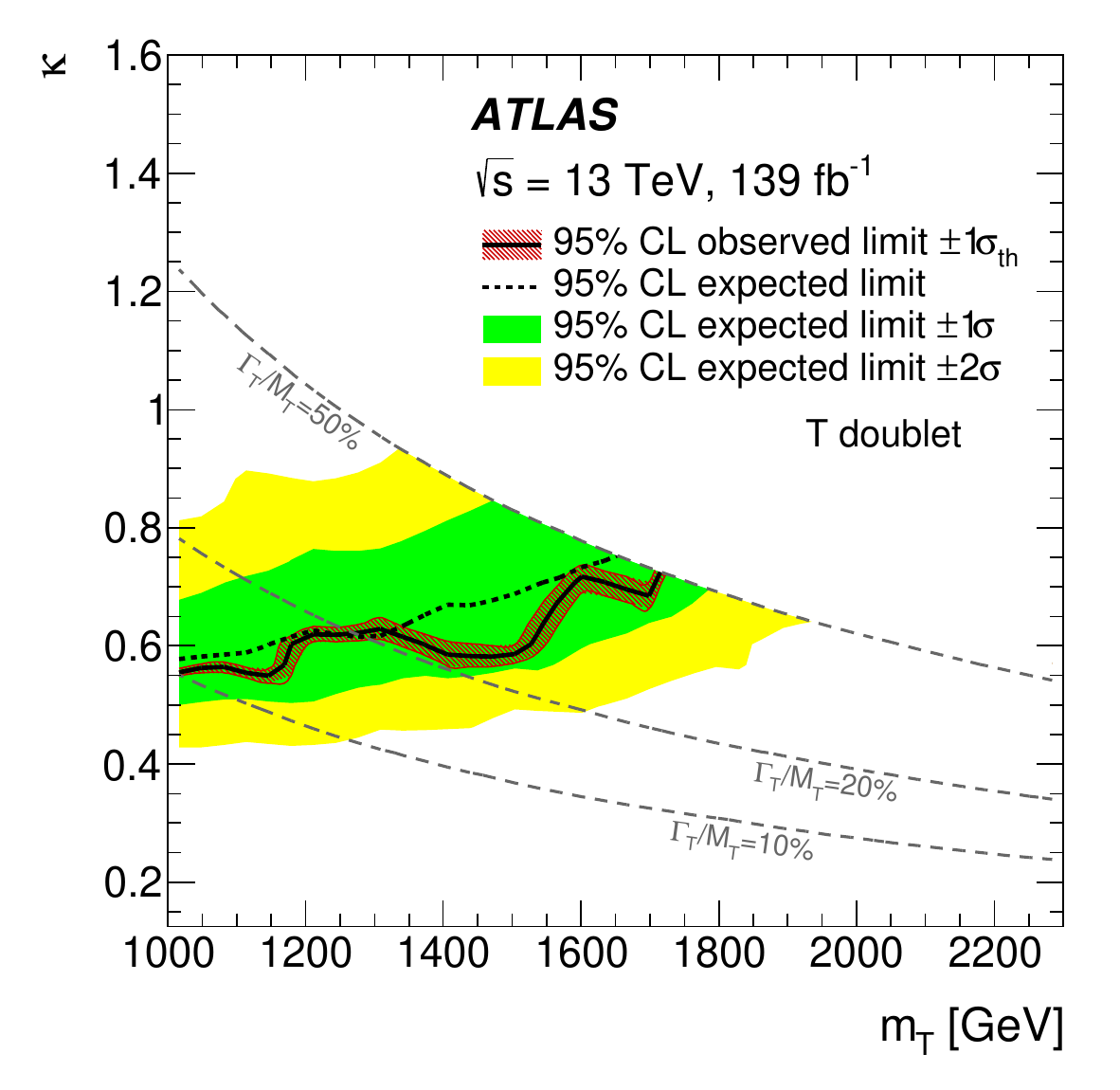}}\\
\subfloat[]{\includegraphics[width=0.5\textwidth]{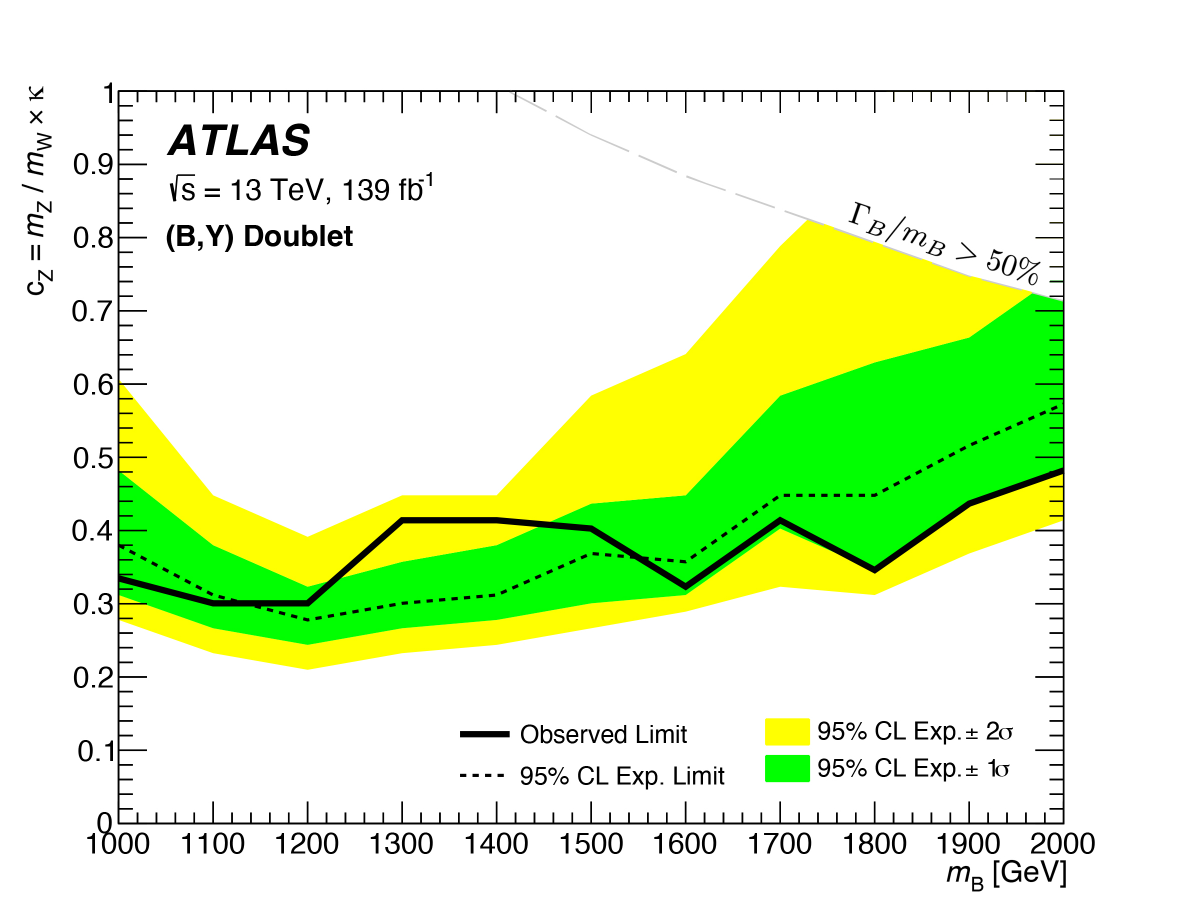}}
\end{center}
\caption{Observed and expected upper limits on (a,b) single $T$ and (c) single $B$ production as a function of the mass of the VLQ for a doublet hypothesis. Results from searches for (a) singly produced $T$ in multilepton final states~\cite{EXOT-2020-01}, (b) singly produced $T$ in final states with one lepton and multiple jets~\cite{EXOT-2018-52} and (c) singly produced $B$ in hadronic final states~\cite{EXOT-2020-01} are included. The dashed curves represent contours of constant relative VLQ width. The doublet hypothesis considered in (a,b) is $(T, B)$.}
\label{fig:vlqSingleDoublet}
\end{figure}
\FloatBarrier


%
\section{Leptoquarks}
\label{sec:lq}

One striking feature of the SM is the similar structure of the lepton and quark sectors, which, a priori, could have been very different. One possibility that could lie behind the similarity is the presence of a BSM symmetry connecting the two sectors. With this idea in mind, many theories introduce new particles, called leptoquarks (LQs)~\cite{Pati:1974yy,Dimopoulos:1979es}, with both baryon and lepton number. Theories dealing with unification~\cite{Georgi:1974sy}, supersymmetric models with R-parity violation~\cite{Barbier:2004ez}, and models with composite fermions~\cite{Gripaios:2014tna} are some areas that typically consider adding such particles. Having baryon and lepton numbers makes them special, as it confers on them the ability to convert quarks into leptons and vice versa. LQs have fractional charge and are triplets of the strong interaction, with other SM quantum numbers, such as their weak isospin representation or their spin, varying between theories.

Because of their unique properties, they can mediate processes that violate lepton-flavour universality and were proposed as a possible explanation for anomalies observed in $B$-meson decays~\cite{BaBar:2013mob, Belle:2010tvu, LHCb:2023zxo}. A scalar LQ can also explain the discrepancy between the measured and predicted values of the muon's  anomalous magnetic moment $(g-2)$~\cite{Aoyama:2020ynm,Muong-2:2006rrc}.

Because they interact strongly, LQs can be pair-produced at the LHC with large cross sections, which explains why this has been the most studied production mode. This production mode is independent of the coupling of the LQ to leptons and quarks but is suppressed for high LQ masses. Single LQ production, in which the LQ coupling to leptons and quarks appears explicitly, becomes more copious than LQ pair production at high LQ mass and coupling values and has also been investigated. Non-resonant LQ production is also possible and has been considered in some searches. Illustrative diagrams of the different modes of LQ production are shown in Figure~\ref{fig:lqProd} for a specific choice of LQ coupling and decay chain.

\begin{figure}[tb]
\begin{center}
\subfloat[]{\includegraphics[width=0.25\textwidth]{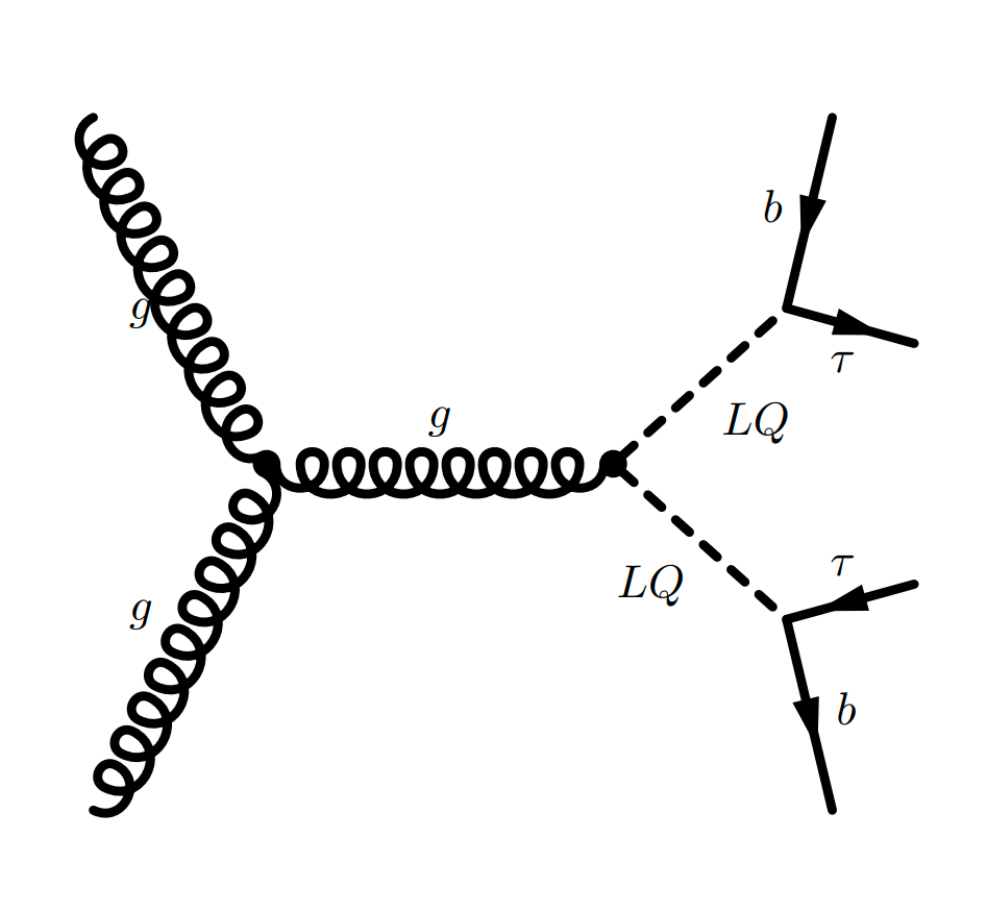}}
\subfloat[]{\includegraphics[width=0.25\textwidth]{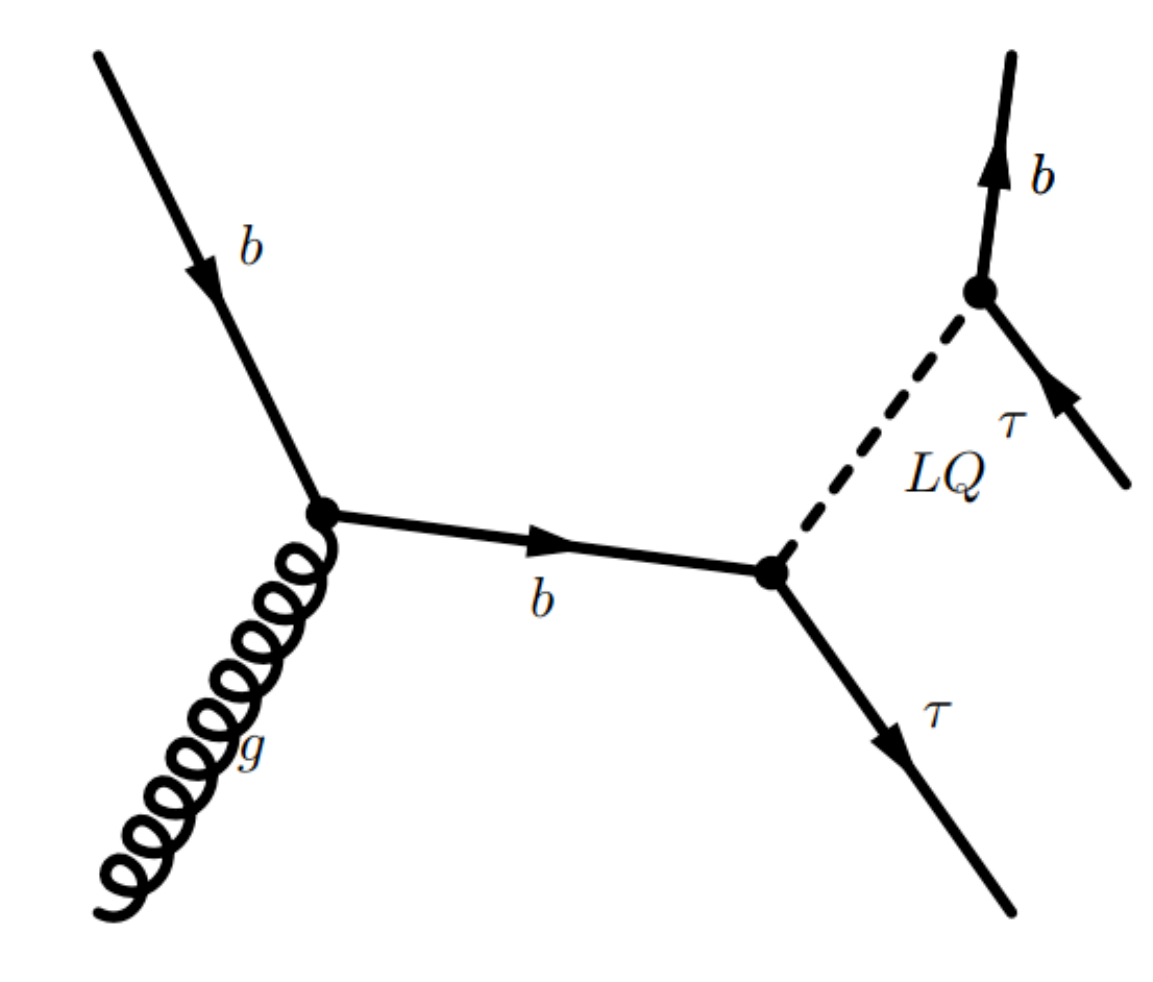}}
\subfloat[]{\includegraphics[width=0.25\textwidth]{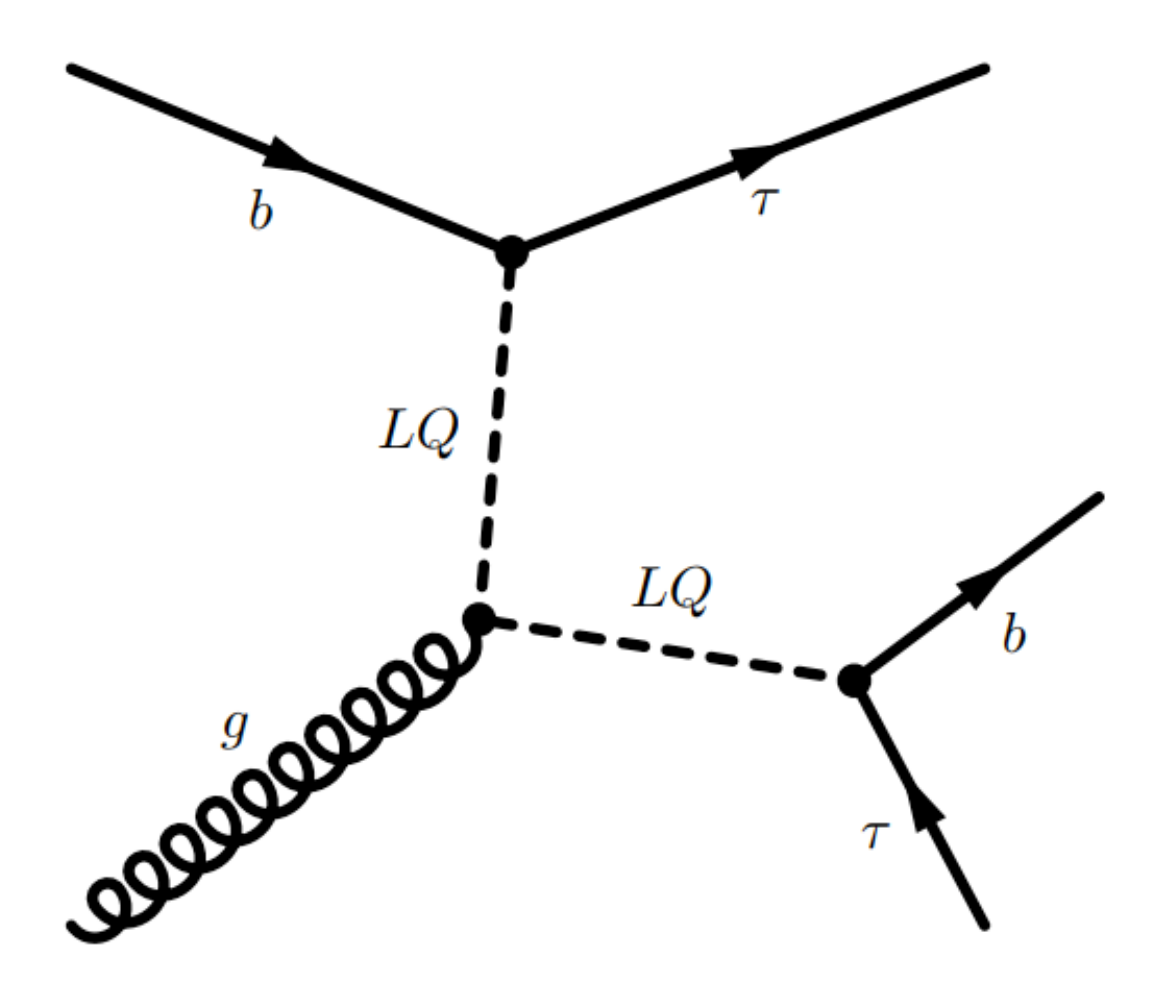}}
\subfloat[]{\includegraphics[width=0.25\textwidth]{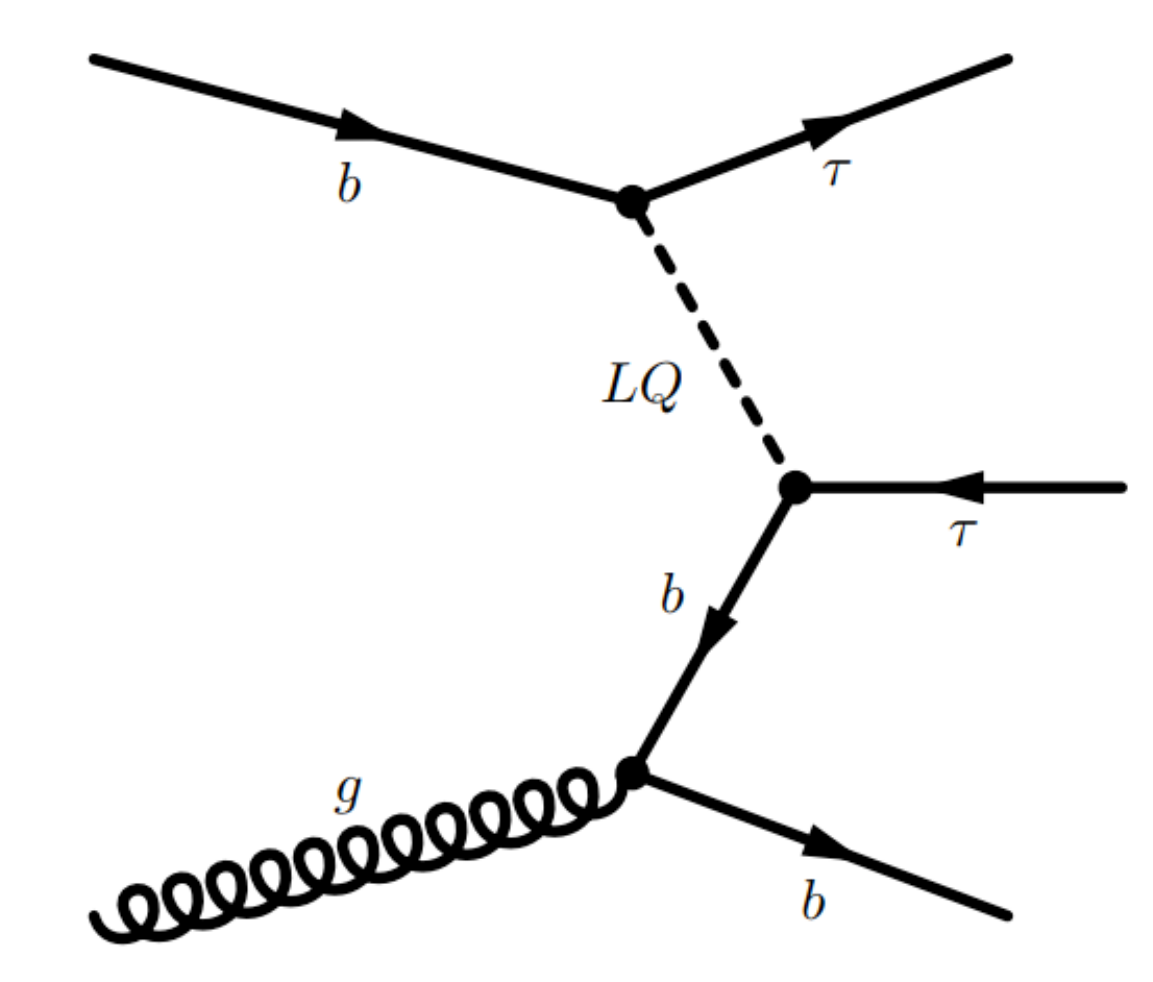}}
\end{center}
\caption{Illustrative Feynman diagrams of (a) LQ pair production, (b,c) single LQ production and (d) non-resonant LQ production.}
\label{fig:lqProd}
\end{figure}

Initially, searches for LQs in ATLAS focused on models in which they would only couple with quarks and leptons of the same generation, with special emphasis on the third-generation LQ that would couple to bottom quarks, top quarks, $\tau$-leptons, and $\nu_\tau$-neutrinos. However, the possibility of off-diagonal couplings, and therefore cross-generational quark--lepton conversions, was investigated in an attempt to explain the $B$-meson and muon $g-2$ results~\cite{Greljo:2021xmg,Cornella:2019hct}.

This section starts by discussing the third-generation searches performed by ATLAS in final states with top quarks~\cite{EXOT-2019-15,SUSY-2018-12} and bottom quarks~\cite{EXOT-2022-39,EXOT-2021-15,SUSY-2018-34,SUSY-2019-18}, and finishes by discussing the cross-generational searches with light leptons~\cite{EXOT-2020-08,EXOT-2019-12,EXOT-2019-19,EXOT-2019-13}  or charm quarks~\cite{EXOT-2020-18} in the final state. Results are discussed in the context of scalar LQs in this report, but interpretations of vector-like LQs are also included in some of the analyses referenced. The following subsections distinguish between an up-type LQ ($LQ^{u}$, with a charge of $2/3$) and a down-type LQ ( $LQ^{d}$, with a charge of $-1/3$).

\subsection{Searches for LQs with exclusive third-generation couplings}

This search type focuses on LQ models in which the LQ couples exclusively or primarily to the third generation of SM fermions ($LQ_{3}^{u,d}$ for up-type and down-type LQs respectively). They deal with final states with multiple top quarks, bottom quarks, and $\tau$-leptons.

\subsubsection{Down-type LQs with exclusive third-generation couplings}

For scenarios with down-type LQs decaying exclusively into third-generation fermions, two decays are possible: $t \tau$ and $b \nu_{\tau}$. Both have been investigated in ATLAS in three distinct analyses.

The $t\tau t\tau$~\cite{EXOT-2019-15} analysis defines seven orthogonal channels to search for the pair production of LQs, selecting events according to their light-lepton and jet multiplicities, and always requiring the presence of a hadronically decaying $\tau$-lepton (\tauhad). Additional kinematic variables such as the transverse momentum of the leptons or jets, the missing transverse momentum, and invariant mass combinations among the different objects are used in every channel to separate the SRs from CRs used to help in the modelling of the various SM backgrounds. Additional control regions are built from selections without a reconstructed \tauhad. Irreducible backgrounds from SM processes are modelled using MC samples, with kinematic reweighting applied to $\ttbar$ processes, one of the major background components. Contributions from fake \tauhad and non-prompt light leptons are estimated using simulations, with normalization corrections obtained in dedicated regions. In total, 15 CRs and 7 SRs are fitted simultaneously in the statistical analysis, with $m_\text{eff}$ being used as the discriminating variable in the SRs. Figure~\ref{fig:tttautau} shows good overall agreement between data and MC simulation.

\begin{figure}[tb]
\begin{center}
\subfloat[]{\raisebox{0.1\height}{\includegraphics[width=0.53\textwidth]{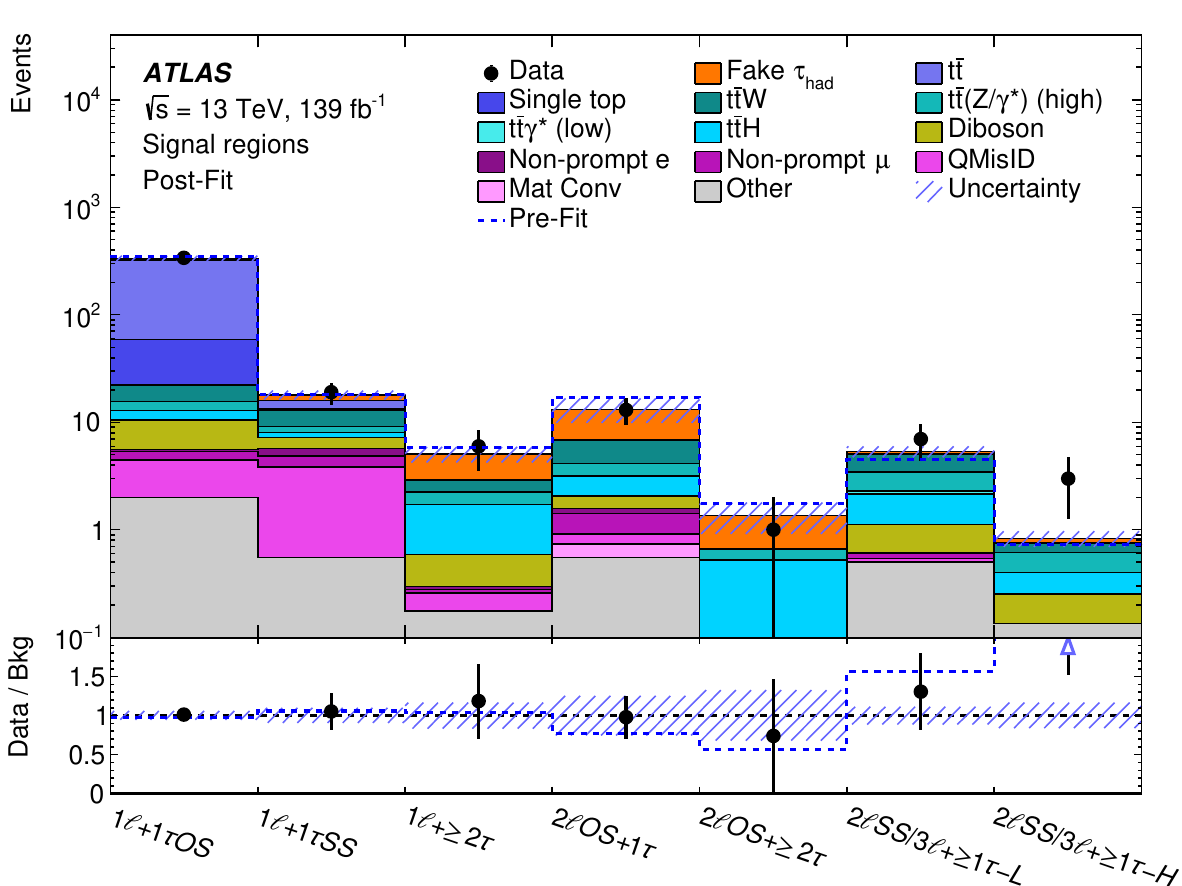}}}
\qquad
\subfloat[]{\includegraphics[width=0.42\textwidth]{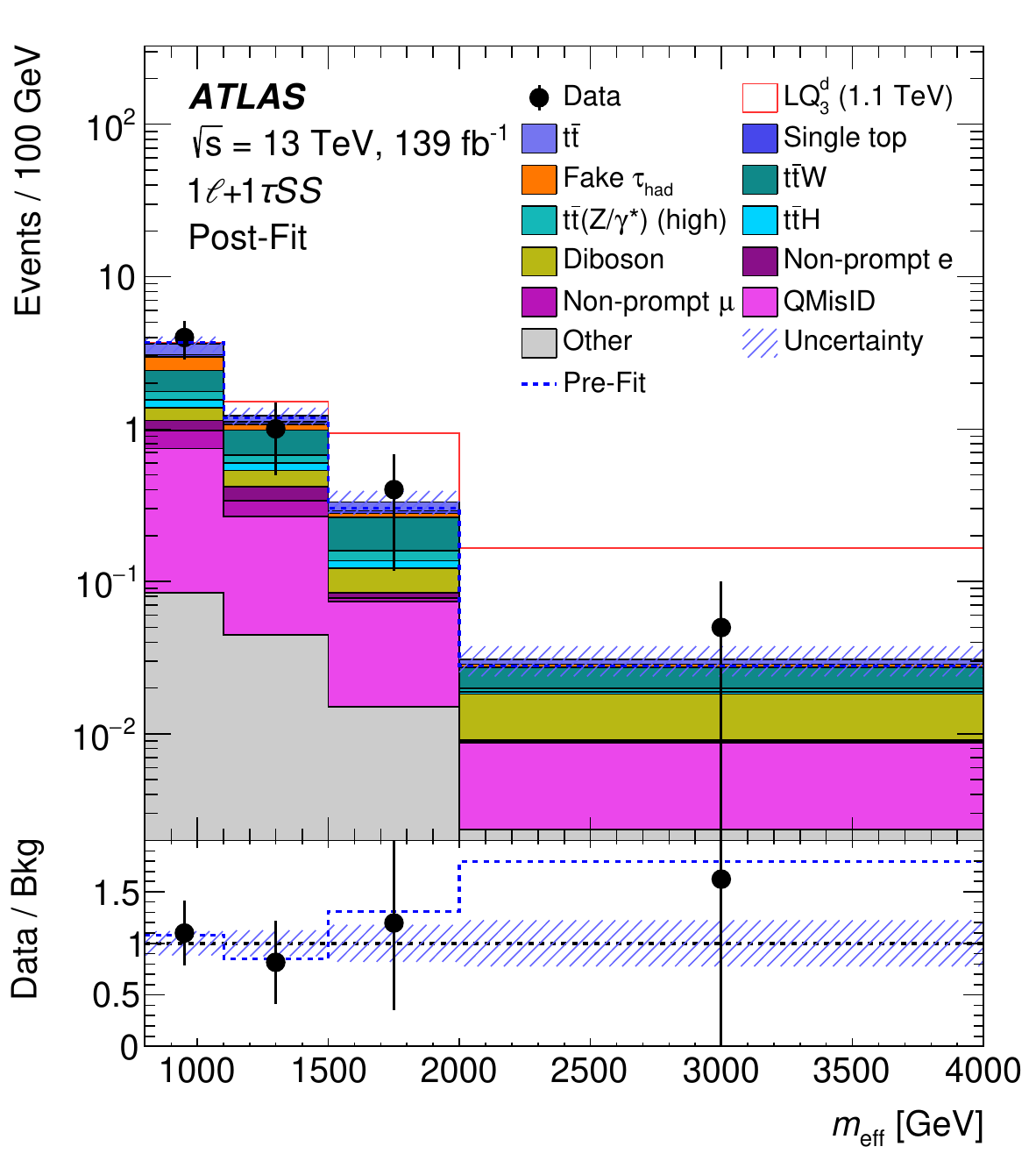}}
\end{center}
\caption{(a) Event yields in all signal regions and (b) the distribution of $m_\text{eff}$ in one of the signal regions used in the search for $LQLQ\rightarrow tt\tau\tau$~\cite{EXOT-2019-15}. The distributions are shown after a background-only fit.}
\label{fig:tttautau}
\end{figure}

Two searches have been conducted in ATLAS to explore pair production of LQs decaying in the $b\nu_{\tau}$ channel. They both select events with a large amount of \met but focus on different final states. One of them~\cite{SUSY-2018-34} focuses on a symmetrical final state in which both LQs decay in the same channel. The final state has two \btagged jets in addition to significant \met. The other~\cite{SUSY-2019-18} considers the asymmetrical case in which one LQ decays into $b\nu_{\tau}$ while the other decays into $t\tau$. Compared to the symmetrical search, this search has a final state with one $\tau$-lepton and additional jets.

In the symmetrical case, events with large \met ($> 250$~\GeV), no light leptons, and between two and four jets are selected to form a signal region (SRA). The leading and subleading jets in the event must be \btagged, and no additional \btagged jets are allowed. Additional selection criteria are imposed on $m_{bb}$, $m_\mathrm{eff}$ and $m_\mathrm{CT}$. A second signal region (SRB) has a relaxed requirement on the $b$-tagging configuration, in which the leading jet is required not to be \btagged. Instead of using the set of kinematic criteria used in SRA, a BDT is employed to define the signal region by using many kinematic variables constructed from the jets in the event and the \ptmiss. Two control regions, one associated with each SR, are also defined by selecting events with two opposite-sign same-flavour (OSSF) light leptons to help model the main SM background, $Z$+jets. Additional signal regions are defined with different jet and \btagged jet criteria, but because they target other BSM particles, they are not discussed here. The statistical analysis relies on a simultaneous fit of the two SRs and process templates provided by MC samples for signal and SM backgrounds. The fit uses the event yield in the CR, the $m_\mathrm{CT}$ distribution in SRA, and the BDT score in SRB, revealing good agreement between data and background as shown in Figure~\ref{fig:susy_down}(a).

In the asymmetrical case, a single SR with large \met ($> 280$~\GeV), at least two \btagged jets, no light leptons, and exactly one \tauhad is used to look for pair production of LQs. Additional requirements are imposed on the \pt of the \tauhad, on the scalar sum of the transverse momenta, $S_\mathrm{T}$, computed from the \tauhad and the two leading jets, and on $m_\mathrm{T}(X,\met)$ where $X$ is either one of the two leading jets or the $\tau$-lepton. Two CRs which help in modelling some of the important SM backgrounds, \ttbar and single-top production, are defined by inverting or relaxing some of the additional requirements. Two more CRs are defined for events with two \tauhad.  A corresponding di-$\tau$ SR region is defined to target other BSM particles but is not discussed further. The statistical analysis uses the event yield in the CR and the \pt distribution of the $\tau$-lepton in the SR. All signal and background templates are obtained from simulated events. A background-only fit, shown in Figure~\ref{fig:susy_down}(b), finds good agreement between data and the expected background.

\begin{figure}[tb]
\begin{center}
\subfloat[]{\includegraphics[width=0.45\textwidth]{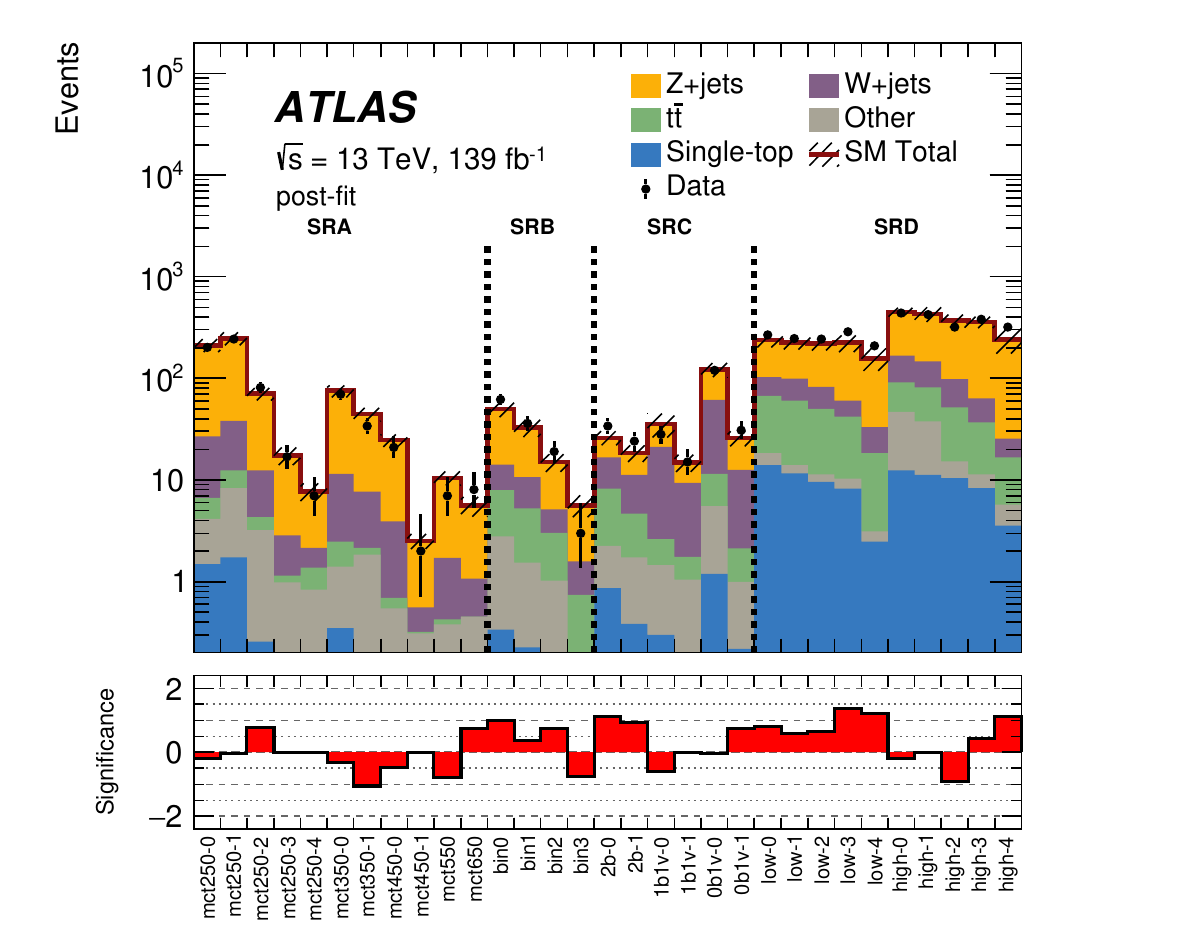}}
\qquad
\subfloat[]{\raisebox{0.1\height}{\includegraphics[width=0.5\textwidth]{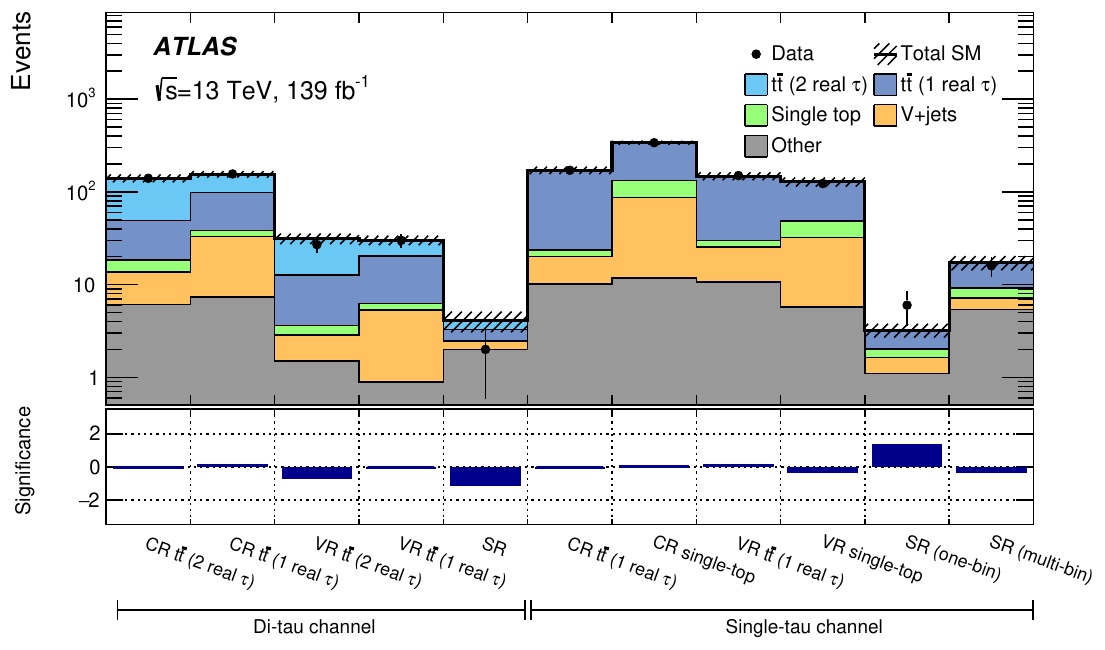}}}
\end{center}
\caption{Event yields in (a) all signal regions used in the search for $LQLQ\rightarrow b b \nu \nu$~\cite{SUSY-2018-34} and (b) all regions used in the search for $LQLQ \to tb\tau\nu$ ~\cite{SUSY-2019-18} after a background-only fit. Only (a) SRA and SRB and (b) regions in the single-$\tau$ channel are used while searching for LQ production.}
\label{fig:susy_down}
\end{figure}

\subsubsection{Up-type LQs with exclusive third-generation couplings}

Two decays, into $b \tau$ and $t \nu$, are allowed for an up-type LQ with exclusive third-generation couplings. Three ATLAS analyses tackle this configuration explicitly, one focusing on pair production of LQs decaying symmetrically in the $b \tau$ channel~\cite{EXOT-2021-15}, another focusing on single production of a LQ decaying the same way~\cite{EXOT-2022-39}, and a third  focusing on pair production of LQs decaying symmetrically into $t \nu$~\cite{SUSY-2018-12}. The asymmetrical search~\cite{SUSY-2019-18} described in the previous section can also be used to search for up-type LQs.

The $b \tau b \tau$ analysis selects events with either one \tauhad or two oppositely charged \tauhad, building two distinct channels. In the single-$\tau$ channel, events must have a light lepton (electron or muon) with charge opposite to that of the \tauhad. Both channels require the presence of at least two jets and one or two \btagged jets. An additional requirement is imposed on $m_{\tau\tau}$, reconstructed using the MMC technique, to help reduce the $Z$+jets background contribution. Finally, requirements on $S_\mathrm{T}$ (the scalar sum of the transverse momenta of the objects in the event) and \met are imposed to reduce the contamination from multijet backgrounds. Signal regions are built using a parameterized neural network (PNN), parameterized as a function of the  hypothesized LQ's mass. The PNN uses either six or seven variables, depending on the channel, selected from amongst many kinematic, multiplicity, and angular variables. The dominant $Z$+jets and \ttbar backgrounds are estimated using MC samples with normalization corrections derived in a \ttbar-rich CR. In the SR, fake-$\tau$ contributions from \ttbar events are estimated from simulated \ttbar events after applying data-driven corrections obtained in a fake-rich CR. A small but non-negligible fake contribution from multijet events is estimated using the fake-factor method. The PNN score is used in the statistical analysis, and the two SRs are fitted simultaneously. Reasonable agreement between data and the SM prediction is obtained after a background-only fit, with a slight deficit of data events observed in the di-$\tau$ channel. The result of the background fit in the di-$\tau$ region is shown in Figure~\ref{fig:third_up_post}.

The $t t \nu \nu$  analysis looks for a pair of LQs in final states with a large amount of \met ($ > 250$~\GeV) and a pair of hadronically decaying top quarks. Besides this \met requirement, SRs are built using events with exactly zero leptons (electrons, muons, or $\tau$-leptons) and at least four jets, two of them \btagged. The leading RC jet is also required to have a mass higher than 120~\GeV, indicating the presence of a hadronic top-quark decay. Two regions (SRA and SRB) are built according to the value of $m_{\mathrm{T},\chi^2}$~\cite{Lester:1999tx} which makes them sensitive to different mass hypotheses. Additional cuts on variables related to the \btagged and RC jets, and their relative positions and kinematic properties, are employed in both regions to increase the sensitivity. Finally, each region is divided into three categories according to the mass of the subleading RC jet. Five CRs built with selections that contain light leptons are used to help in the modelling of the different backgrounds, dominated by $Z\rightarrow \nu \nu$ production. Additional signal and control regions are defined to look for other BSM particles but are not discussed here. The statistical analysis simultaneously fits the event yields in the six SRs and five CRs. A background-only fit shows good agreement between data and the SM prediction, with a slight excess of data in one of the signal regions, as seen in Figure~\ref{fig:third_up_post}.

\begin{figure}[tb]
\begin{center}
\subfloat[]{\includegraphics[width=0.53\textwidth]{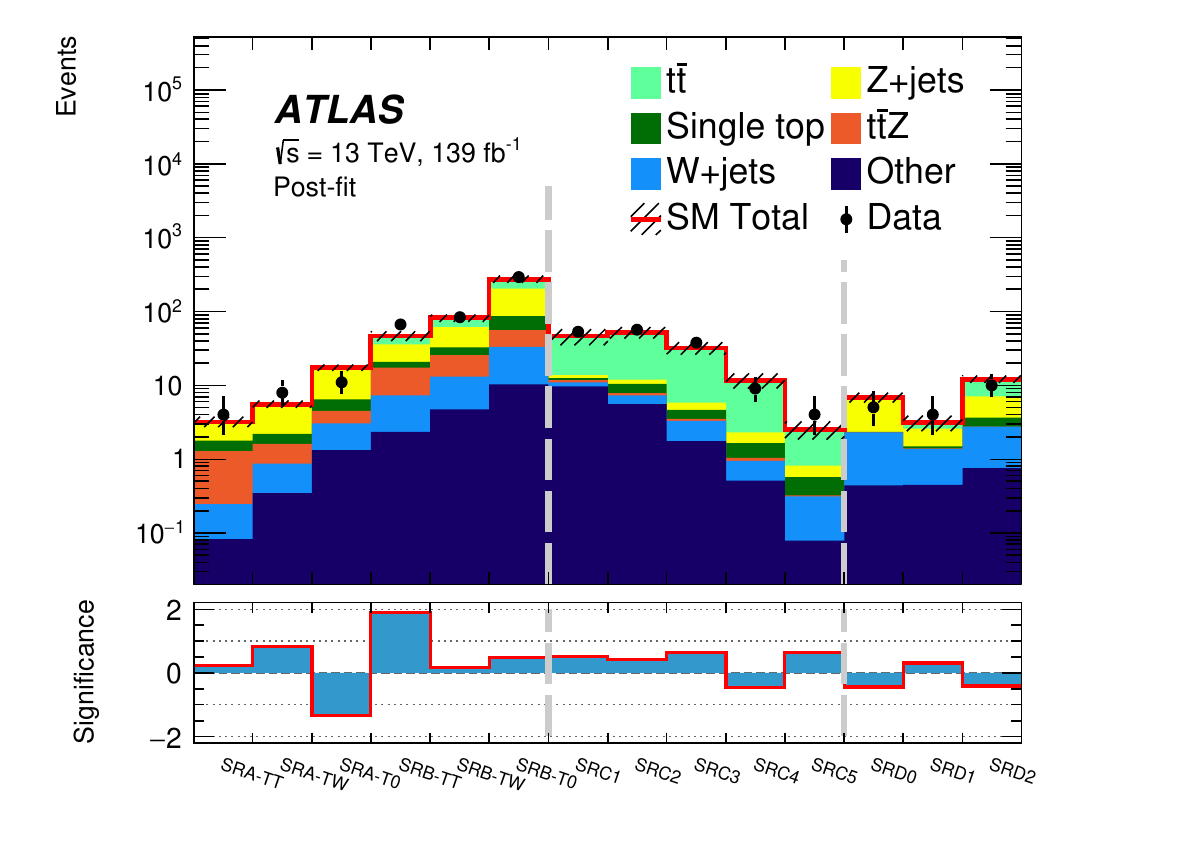}}
\qquad
\subfloat[]{\includegraphics[width=0.42\textwidth]{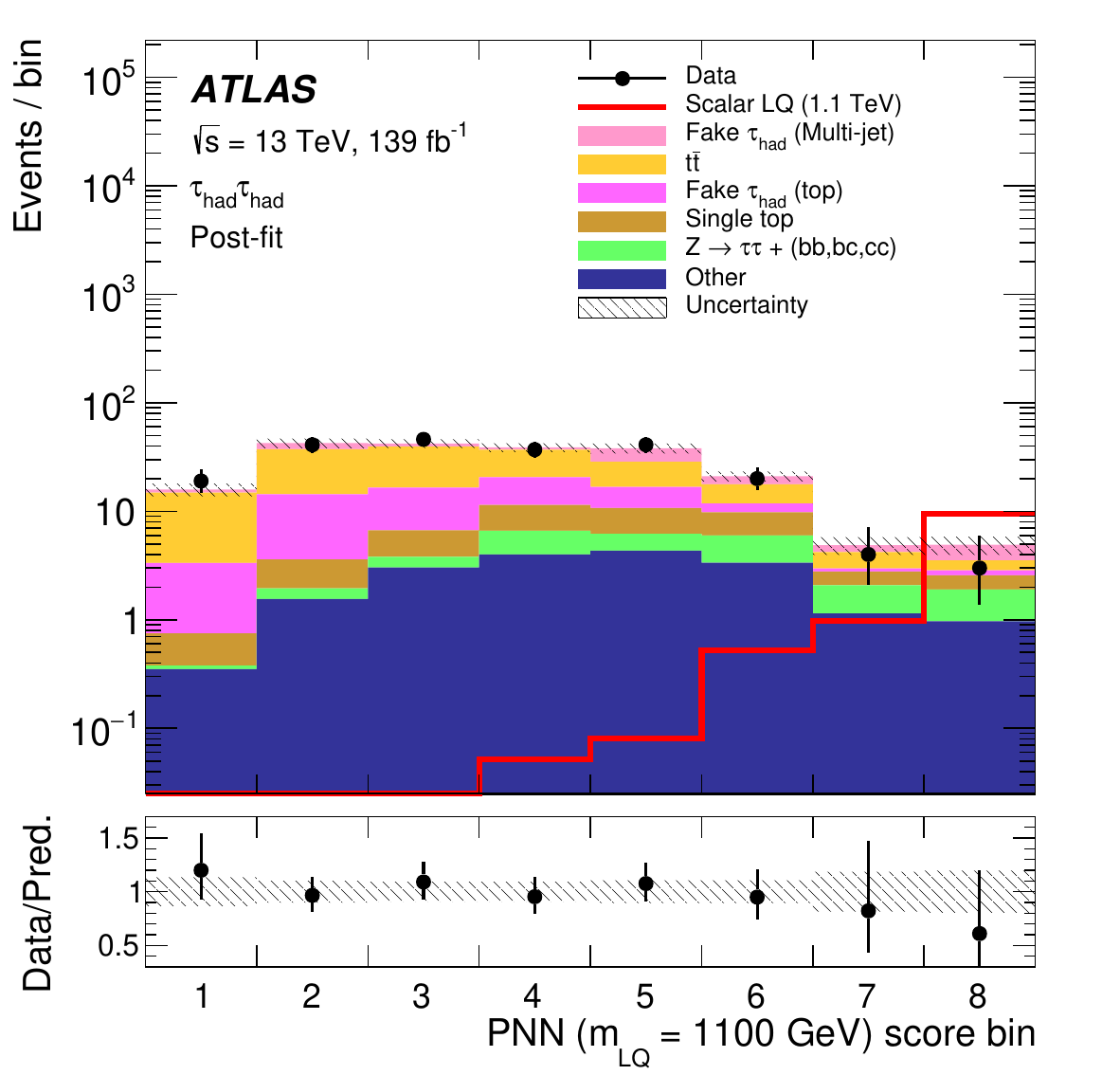}}
\end{center}
\caption{(a) Event yields in all signal regions used in the search for $LQLQ\to tt\nu\nu$~\cite{SUSY-2018-12} and (b) the PNN distribution in the di-$\tau$ signal region used in the search for $LQLQ\to bb\tau\tau$~\cite{EXOT-2021-15} after a background-only fit. (a) Only SRA and SRB are used to search for LQ production.}
\label{fig:third_up_post}
\end{figure}

The single-production analysis proceeds similarly to the pair-production one. It separates events into two channels, one with two \tauhad of opposite charge and the other with one \tauhad and one isolated light lepton (electron or muon) of opposite charge. Both channels define a signal region by requiring at least one \btagged jet and imposing additional criteria based on the invariant mass of the visible decay products of the two $\tau$-leptons and the angular separation between the light lepton and the \met. These criteria aim to reduce the $Z\rightarrow\tau \tau$ and \ttbar backgrounds. Each signal region is divided into two \pt ranges, $\pt < 200$~\GeV and $\pt > 200$~\GeV, making four signal regions. In the single-$\tau$ channel, the \ttbar and single-top backgrounds are modelled using MC samples with a data-driven correction derived in a dedicated CR. In the SR, fake-$\tau$ contributions from top-quark processes are estimated similarly to the pair-production analysis, using a region rich in fakes. A final contribution from multijet events faking $\tau$-leptons is estimated using a data-driven fake factor. In the di-$\tau$ channel, defining a top-quark-process-rich CR close to the SR, as is done in the single-$\tau$ channel, is very complex, and the modelling is corrected using a kinematic reweighting based on $S_\mathrm{T}$, which is computed from the two \tauhad and the leading-\pT $b$-jet. Background from $Z\rightarrow\tau\tau$ is estimated from MC simulations, with a scale factor obtained from a CR with only one \tauhad. Background from multijet events is estimated using a data-driven fake-factor method. The statistical analysis employs the $S_\mathrm{T}$ distributions in the four signal regions. Reasonable agreement is found between data and the SM background, with a minor excess of data observed in the high-\pt signal region for high $S_\mathrm{T}$.

\subsubsection{Limits on LQs with exclusive third-generation couplings}

In the absence of significant excesses in data, limits are obtained for the pair production of scalar leptoquarks as a function of the mass of the LQ for different values of the BR to $b\tau$ or $t\tau$ for down-type or up-type leptoquarks respectively. Two-dimensional limits on the production of up-type and down-type scalar leptoquarks as a function of the mass of the LQ and its BR are shown in Figure~\ref{fig:lq_3rd_limits}. In addition to the limits obtained in the individual searches, the result obtained after a statistical combination~\cite{EXOT-2020-27} of those searches is also shown in all cases. Two of the searches described in this section~\cite{EXOT-2021-15,SUSY-2019-18} also include interpretations regarding the pair production of vector-type LQs. Lower limits on the mass are between 200 and 400~\GeV higher than the corresponding mass limits for scalar LQs in the same BR scenario. Searches with visible decays in the final state are more sensitive to higher values of the BR, while searches with neutrinos in the final state become important for lower values.

The analysis focusing on singly produced LQs~\cite{EXOT-2022-39} sets limits on different models of vector-like LQs. Despite its focus on single production, it considers the three possible production modes in its interpretation: non-resonant, single, and pair production. In the Yang--Mills scenario, the observed lower limits on the LQ mass are between 1.58~\TeV and 2.05~\TeV, depending on the coupling. The lower limits in the minimal-coupling scenario are between 1.35~\TeV and 1.99~\TeV. Finally, the lower limits for a scalar LQ scenario with charge $(4/3)e$ are between 1.28~\TeV and 1.53~\TeV. Figure~\ref{fig:lq_singly} summarizes those limits.

\begin{figure}[tb]
\begin{center}
\subfloat[]{\includegraphics[width=0.45\textwidth]{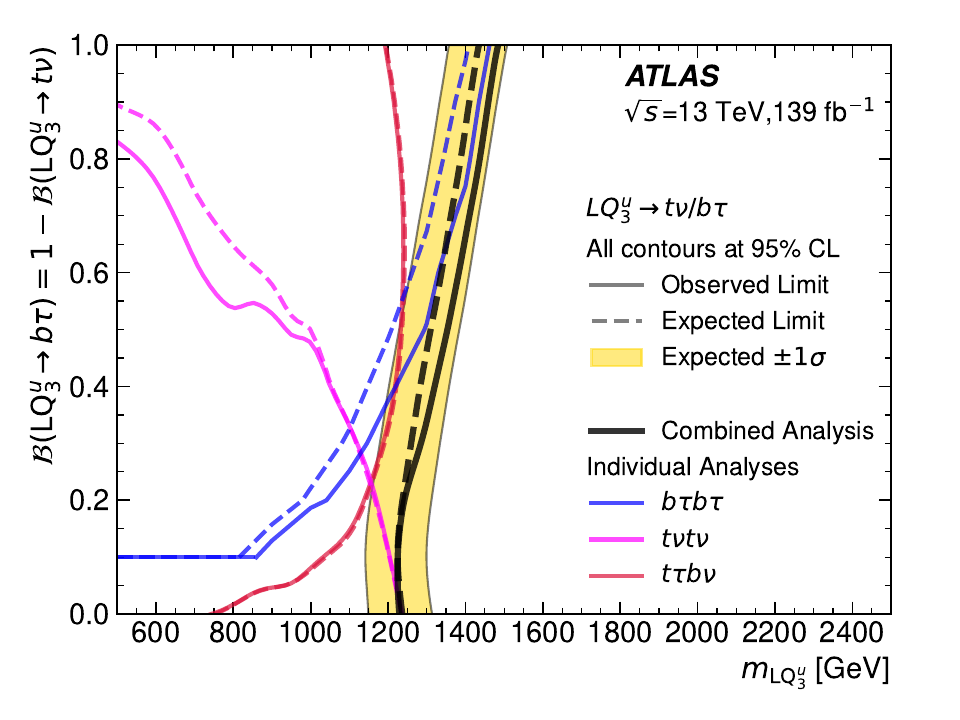}}
\qquad
\subfloat[]{\includegraphics[width=0.45\textwidth]{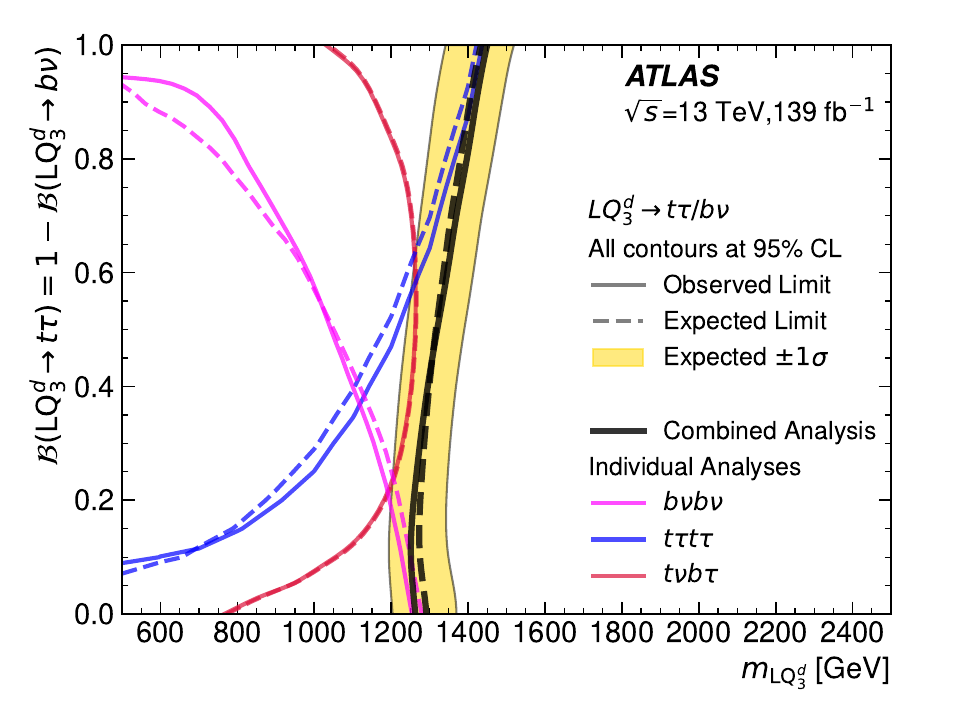}}
\end{center}
\caption{Two-dimensional limits (a) on the production of up-type scalar leptoquarks with exclusive third-generation couplings as a function of the mass of the LQ and the BR to $b \tau$ from the search for (blue) $LQLQ\to b b \tau \tau$~\cite{EXOT-2021-15}, (red) $LQLQ\to b t \tau\nu$~\cite{SUSY-2019-18}, (pink) $LQ\to t\nu t\nu$~\cite{SUSY-2018-12} and (black) their statistical combination~\cite{EXOT-2020-27} and (b) on the production of down-type scalar leptoquarks with exclusive third-generation couplings as a function of the mass of the LQ and the BR to $t \tau$ from the search for (blue) $LQLQ\to t \tau t \tau $~\cite{EXOT-2019-15}, (red) $LQLQ\to t \tau b \nu$~\cite{SUSY-2019-18}, (pink) $LQLQ\to b \nu b \nu$~\cite{SUSY-2018-34} and (black) their statistical combination.}
\label{fig:lq_3rd_limits}
\end{figure}

\begin{figure}[tb]
\begin{center}
\subfloat[]{\includegraphics[width=0.45\textwidth]{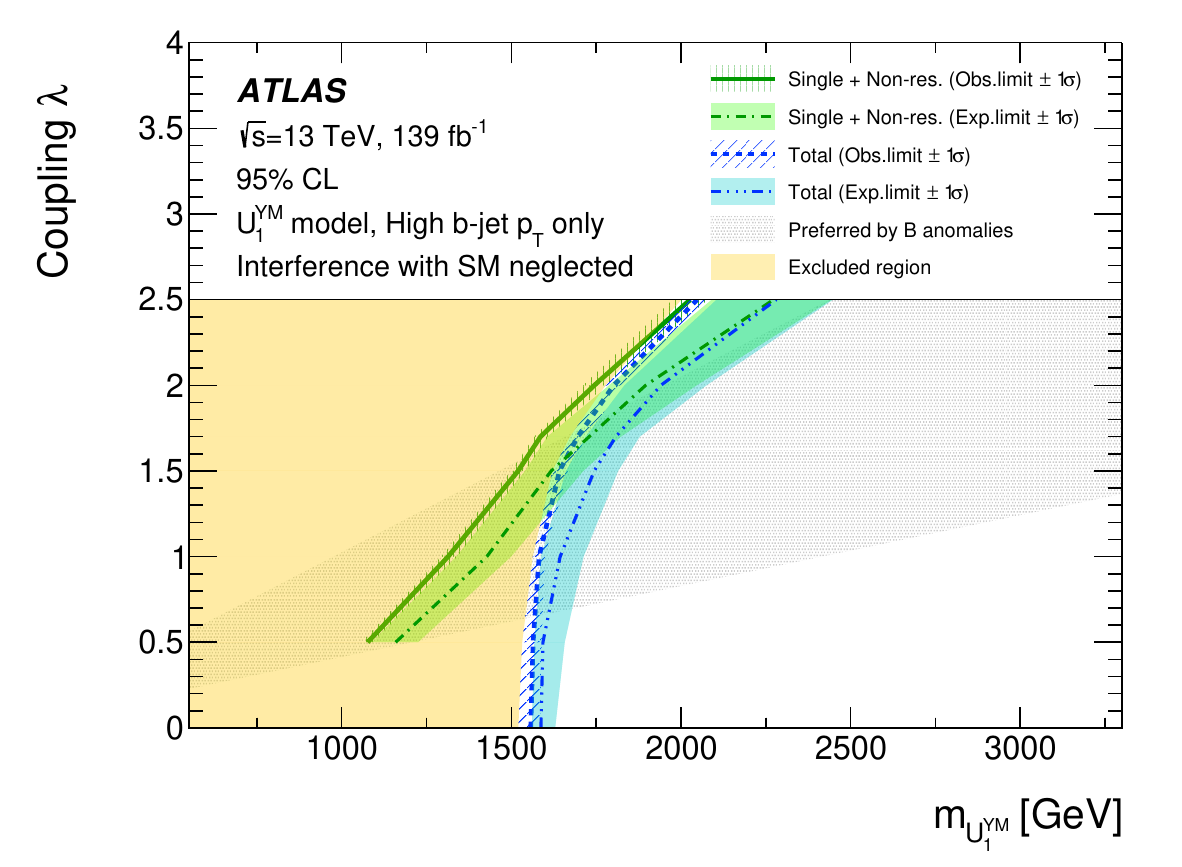}}
\qquad
\subfloat[]{\includegraphics[width=0.45\textwidth]{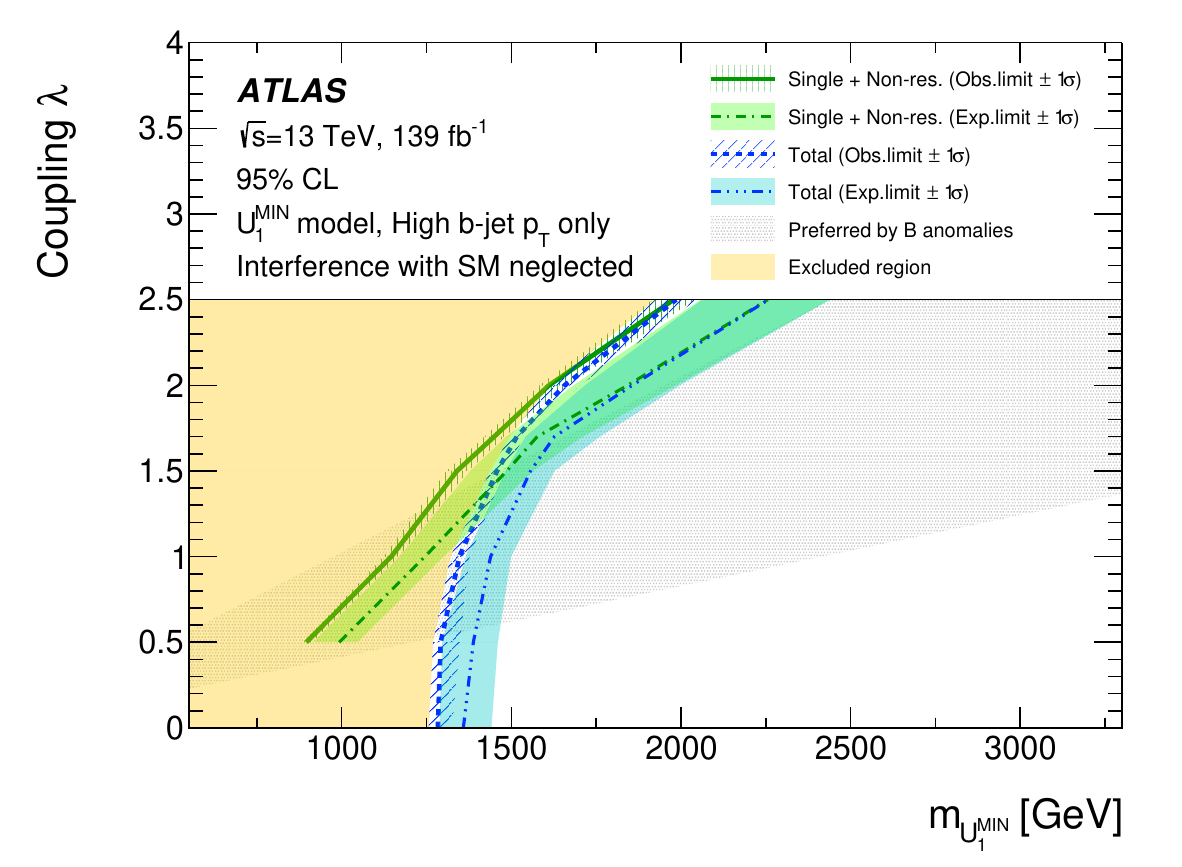}}\\
\subfloat[]{\includegraphics[width=0.45\textwidth]{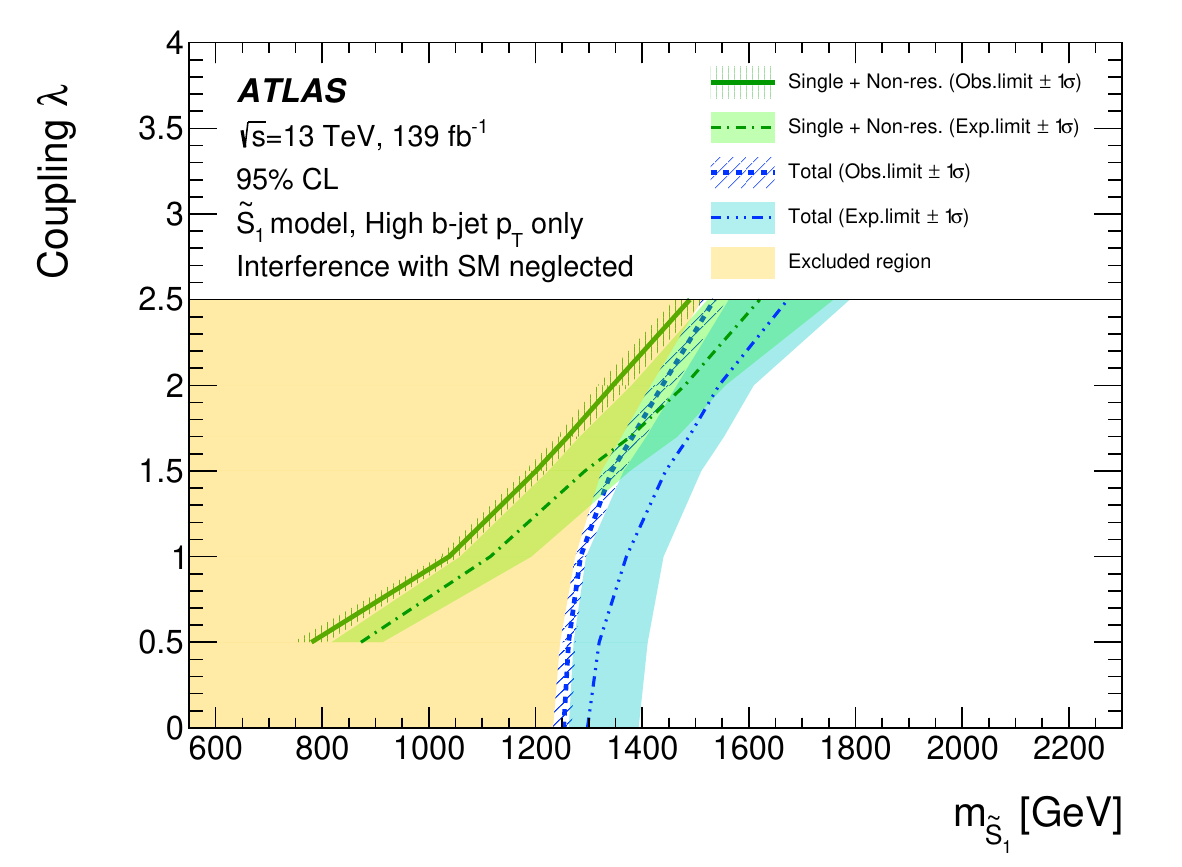}}
\end{center}
\caption{Two-dimensional limits for single plus non-resonant production of (a) vector-like LQs in the Yang--Mills scenario, (b) vector-like LQs in the minimal-coupling scenario and (c) scalar LQs. Although the search~\cite{EXOT-2022-39} focuses on singly produced LQs, its interpretation includes pair production in the \enquote{Total} limits.}
\label{fig:lq_singly}
\end{figure}

\subsection{Searches for LQs with cross-generational couplings}

These searches explore the possibility of an LQ mediating an interaction that connects quarks and leptons from different generations ($LQ_\mathrm{mix}^{u,d}$ for up-type and down-type LQ, respectively). Most deal with final states with multiple light leptons (electrons or muons) and multiple bottom or top quarks. The other cross-generational possibility (light quarks and $\tau$-leptons) is also investigated in one of the searches described. These cross-generational couplings are particularly relevant in models attempting to explain $B$-meson and muon $g-2$ results. Electron and muon selections are considered independently, duplicating the SRs and CRs for use in separate statistical analyses.

\subsubsection{Up-type LQs with cross-generational couplings}
\label{sec:upLQcross}

Two searches for pair-produced up-type LQs with cross-generational couplings have been performed in ATLAS. One is aimed at final states with exactly two light leptons~\cite{EXOT-2019-13} and the other is aimed at final states with one light lepton~\cite{EXOT-2019-12}.

For the two-lepton analysis, events with exactly two OSSF light leptons (electrons or muons) and at least two jets are selected. Additional requirements on the invariant mass and \pt of the dilepton system suppress background from Drell--Yan and $Z$ boson production. LQ candidates are identified from the different possible pairs of lepton+jet by choosing the two pairs closest in $m_{\ell j}$. The maximum and minimum $m_{\ell j}$ in the event is used to construct the $m^\mathrm{asym}$ variable, which helps to reduce the SM background further. Events with low $m^\mathrm{asym}$ form a single signal region, while the rest constitute a sideband region (SB) that is used as a CR in the statistical analysis. A second CR is defined to help in the modelling of the \ttbar background by selecting events with two opposite-sign opposite-flavour leptons. Two independent statistical analyses are defined, one targeting $LQ\rightarrow b\ell$ decays and the other targeting $LQ\rightarrow c\ell$ decays. For the former decay, the SR and SB are split into three subregions according to the number of \btagged jets in the event (zero, one, or at least two). For the latter decay, events in the SR and SB are separated into those with zero tags, at least one $c$-tag, and at least one $b$-tag. Events with a \btagged jet and a $c$-tagged jet are placed in the $c$-tagged category. Both statistical analyses use the distribution of the average lepton+jet invariant mass of the two LQ candidates in the event, and all SM and signal processes are modelled using MC samples. The data and MC simulation agree well, as shown in Figure~\ref{fig:lq_mix_up}(a) for one of the electron selection's SRs.

In the one-lepton analysis, events with exactly one lepton (electron or muon), a significant amount of \met, and at least four jets are selected. This very general selection can capture events from a variety of up- and down-type LQ decay modes ($LQ\to t\nu$, $LQ\to b\ell$, $LQ\to t\ell$ and $LQ\to b\nu$). Events with at least one \btagged jet, large $m_\mathrm{T}(\ell,\met)$, and high $am_\mathrm{T2}$ are assigned to a training region. In this region, a neural network (NN) is trained to separate signal and background events, using 15 input variables related to the objects in the event, including the lepton flavour. An SR is defined using the events with a high NN score, while the lower score region is kept as a CR. Two additional CRs are defined to help in modelling the single-top and $W$+jets backgrounds with selections based on the presence of two \btagged jets and lower $m_\mathrm{T}(\ell,\met)$ respectively. A kinematic reweighting of the MC prediction of the \ttbar background is implemented before the statistical analysis, with corrections obtained in a \ttbar-rich region defined using $am_\mathrm{T2}$. The statistical analysis uses the distribution of the NN score in the SR and the overall event yield in each CR, with all signal and background processes modelled using MC samples. The data and SM prediction agree well after a background-only fit, as shown in Figure~\ref{fig:lq_mix_up}(b) for the electron selection's SR.

\begin{figure}[tb]
\begin{center}
\subfloat[]{\includegraphics[width=0.47\textwidth]{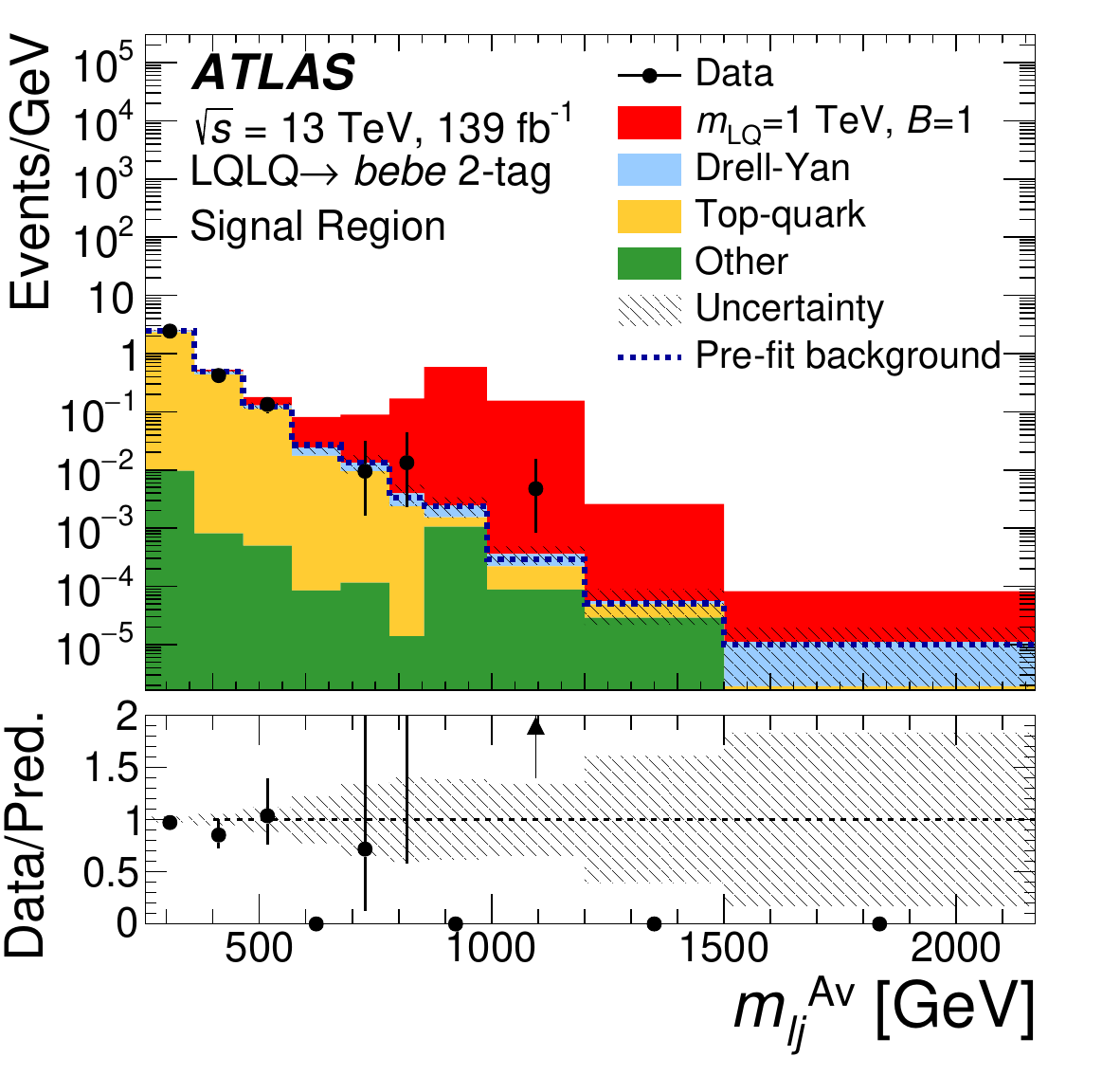}}
\qquad
\subfloat[]{\includegraphics[width=0.43\textwidth]{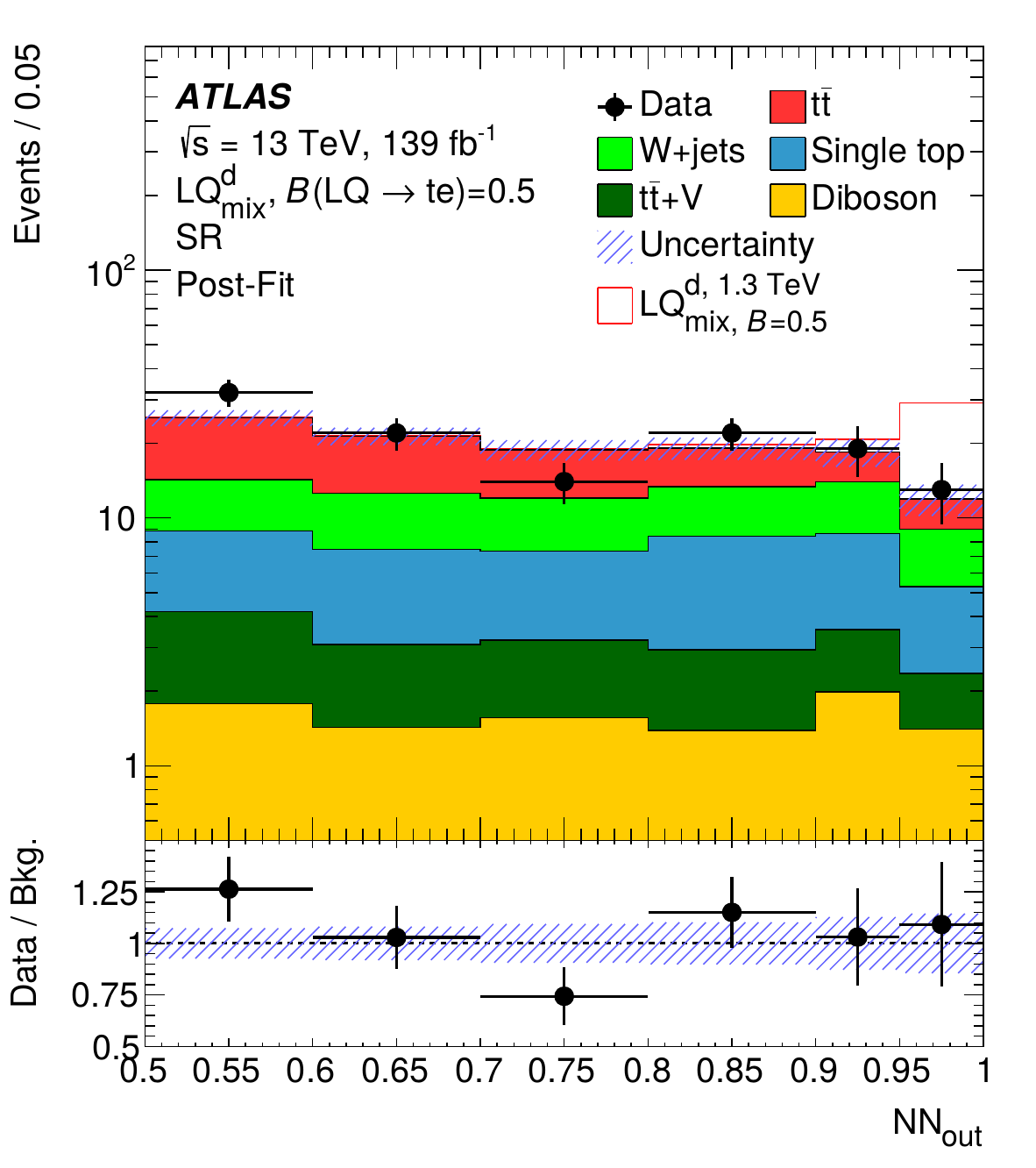}}
\end{center}
\caption{(a) Average invariant mass of the lepton+jet system in one of the SRs from the search for up-type LQs in final states with two light leptons~\cite{EXOT-2019-13} and (b) NN score in the SR from the search for LQs in final states with one light lepton~\cite{EXOT-2019-12} after a background-only fit to data.}
\label{fig:lq_mix_up}
\end{figure}

\subsubsection{Down-type LQs with cross-generational couplings}

Three ATLAS searches consider the scenario of a down-type LQ with cross-generation couplings and are defined according to the number of light leptons present in the final state. The one-lepton search~\cite{EXOT-2019-12} is described in Section~\ref{sec:upLQcross}, as the same search is also used for up-type scenarios. A separate two-lepton search is, however, introduced~\cite{EXOT-2019-19} to take into account the presence of top quarks, and a recent multilepton analysis~\cite{EXOT-2020-08} investigates events with two, three or four light leptons in the final state arising from the pair production of LQs and the $LQ\rightarrow t\ell$ decay channel.

The dilepton analysis builds a single SR with events containing exactly two OSSF light leptons, at least two large-$R$ jets, and large \mll, which is used to suppress the background from $Z$ boson production. A CR defined to help in the modelling of $Z$+jets production is built with events having lower \mll values. A second control region with OS different-flavour pairs of leptons is built to help model the \ttbar background. A BDT is used in the SR to further separate signal and background, with its input including variables related to the substructure and kinematics of the large-$R$ jets. The BDT is parameterized using the masses of the different signal hypotheses, which are included in the training. The BDT score in the SR and the overall number of events in each CR are used in the statistical analysis, with all signal and background templates modelled using MC samples. A background-only fit to data, shown in Figure~\ref{fig:lq_mix_down}(a) for all fitted regions, reveals good agreement between data and MC simulation.

The multilepton analysis is divided into three distinct channels, defined by the lepton multiplicity in each event: two same-sign light leptons (2$\ell$SS), three light leptons ($3\ell$), or at least four leptons ($4\ell$). Only the $3\ell$ and $4\ell$ channels contain signal regions, while the $2\ell$ channel is used entirely to define CRs to help in modelling the \ttbar and single-top backgrounds. Additional CRs are defined in the $3\ell$ channel to help model other SM backgrounds, such as $\ttbar+Z$ or diboson production. In the $3\ell$ channel, events are classified as being in the signal region if they have at least two jets, at least one \btagged jet, and a pair of opposite-sign same-flavour leptons that have a mass far from the $Z$ boson mass window, which helps to suppress SM background containing a $Z$ boson. Events are also required to have a large ($>200$~\GeV) invariant mass $m_{\ell\ell}^{\text{min}}$, which corresponds to the smallest invariant mass among of all two-lepton combinations and has the ability to separate background and signal processes very efficiently. A procedure to identify electrons with incorrect charge assignment that originate from internal photon conversions or photon conversions in matter is implemented through a BDT discriminant. Electrons in the signal region are required to pass a conversion veto, while events with electrons not passing the veto are assigned to one of two additional CRs. The SR is split in two, according to the flavour of the OSSF pair of leptons. In the $4\ell$ channel, events are classified following criteria very similar to those in the $3\ell$ channel, except for the conversion veto, which is not applied. The SR is split in two, according to the lepton flavour with higher multiplicity. For events with two electrons and two muons, the leading lepton determines the SR. Two independent statistical analyses are defined, one with only the electron SRs and one with only the muon SRs. All considered backgrounds and signals, except that from electrons with misassigned charge, are estimated using MC samples. The predicted diboson, non-prompt-lepton, and \ttbar backgrounds from simulation are improved using data-driven corrections obtained from dedicated CRs. The estimated background from electrons with misassigned charge is fully data-driven. The statistical analyses use the $m_\mathrm{eff}$ distribution in all SRs, in the $3\ell$ CR, and in one of the CRs of the $2\ell$ channel. In the rest of the CRs, the overall number of events is used. The data and background prediction agree reasonably well in all SRs, with small data excesses in the electron SRs. This is shown in Figure~\ref{fig:lq_mix_down}(b) for one of those regions.

\begin{figure}[tb]
\begin{center}
\subfloat[]{\raisebox{0.1\height}{\includegraphics[width=0.52\textwidth]{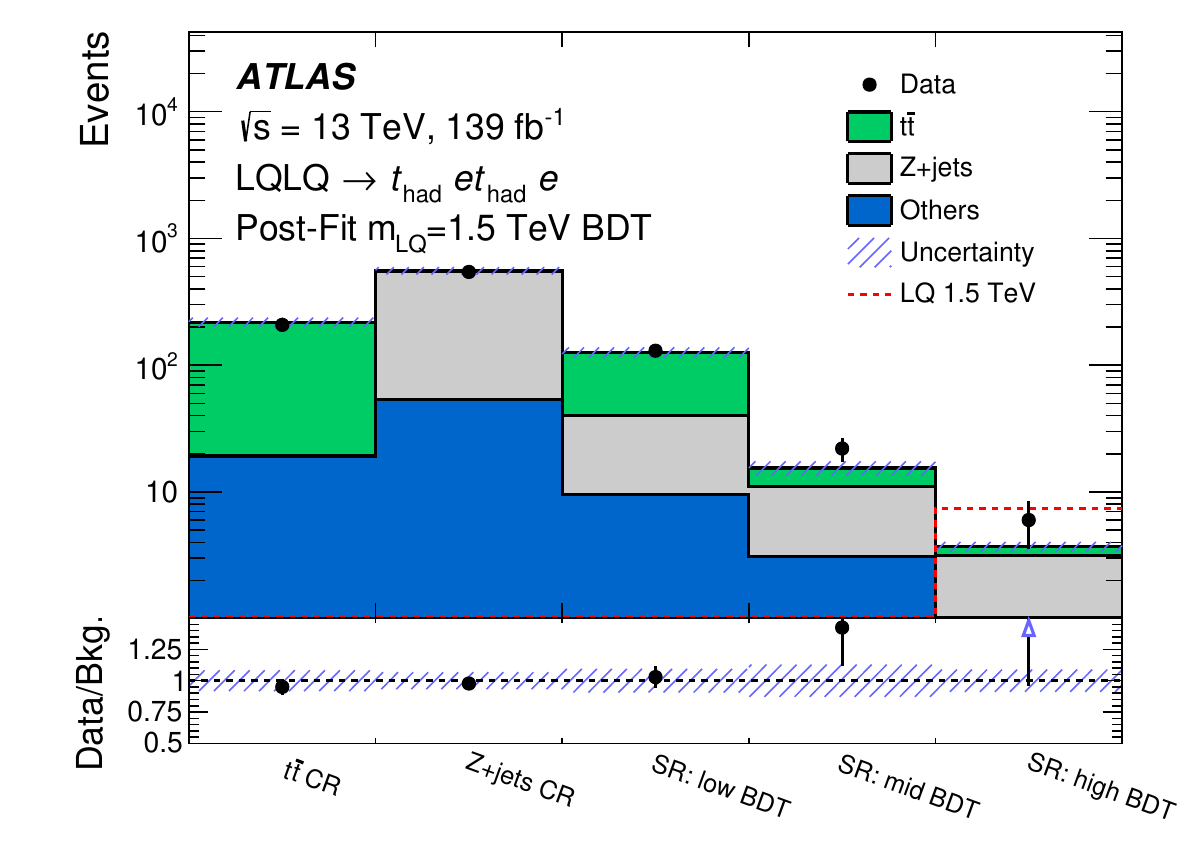}}}
\qquad
\subfloat[]{\includegraphics[width=0.4\textwidth]{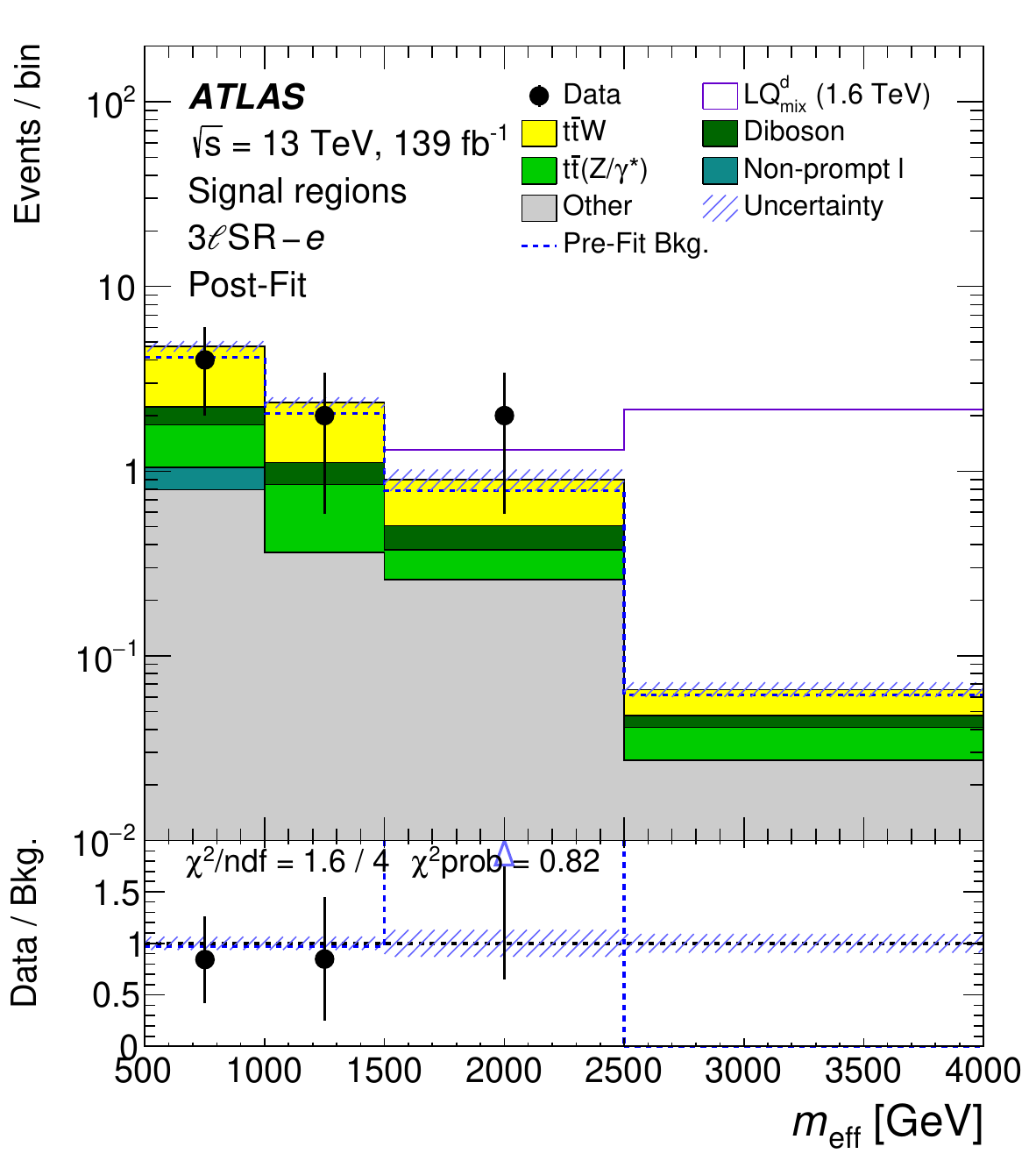}}
\end{center}
\caption{(a) Distribution of the event yields in all fitted regions of the electron selection used in the search for down-type LQs in final states with two light leptons~\cite{EXOT-2019-19} and (b) the effective mass in one of the SRs used in the search for down-type LQs in multilepton final states~\cite{EXOT-2020-08} after a background-only fit to data.}
\label{fig:lq_mix_down}
\end{figure}

The only ATLAS analysis considering cross-generational couplings without light leptons focuses on LQs decaying into a $\tau$-lepton and a $c$-quark~\cite{EXOT-2020-18}. Events are selected for a SR if they have two reconstructed hadronically decaying $\tau$-leptons and at least two reconstructed jets. Events with any light leptons are vetoed. The two \tauhad are required to be well separated and have opposite charges. To suppress the $Z\rightarrow \tau \tau$ background, $m_{\tau\tau}^{\text{col}}$ is required to be larger than 110~\GeV. The visible momentum fraction for the two $\tau$ candidates, i.e.\ the ratio of the visible \pt to the estimated total \pt of their decay products, is used to reduce the background from fake-$\tau$ sources. Two CRs are defined and used in the statistical analysis: a CR defined for lower values of $m_{\tau\tau}^{\text{col}}$ to help model the $Z\rightarrow \tau \tau$ background and another one using a single-$\tau$ selection in order to improve the modelling of background from single-top and \ttbar production. The statistical analysis uses the $S_\mathrm{T}$ distributions in the SR and the two CRs with all signal and background processes modelled using MC samples, except for the fake background due to jets misidentified as \tauhad, which is estimated using a fake-factor method. The data is in good agreement with the SM prediction in the SR after a background-only fit.

\subsubsection{Limits on LQs with cross-generational couplings}

In the absence of any significant excess in data, limits are obtained for the pair production of scalar leptoquarks with cross-generational couplings as a function of the mass of the LQ for different values of the branching ratio to $b\ell$ or $t\ell$ for down-type or up-type leptoquarks respectively. Two-dimensional limits on the production of up-type and down-type scalar leptoquarks as a function of the mass of the LQ and its BR are shown in Figure~\ref{fig:lq_mix_limits}. In addition to the limits obtained in the individual searches, the result obtained after a statistical combination~\cite{EXOT-2020-27} of those searches is also shown in all cases. Results are shown separately for muon and electron final states. Searches with fully visible final states reach higher in mass but lose sensitivity rapidly as the BR approaches zero. Searches that consider semi-invisible final states are less sensitive but obtain more consistent limits as a function of the BR.

The $\tau c$ analysis sets limits on the pair-production of LQs that decay exclusively into a $\tau$-lepton and $c$-quark as a function of the LQ mass. LQs with masses below 1.3~\TeV are excluded for the model considered.

\begin{figure}[tb]
\begin{center}
\subfloat[]{\includegraphics[width=0.45\textwidth]{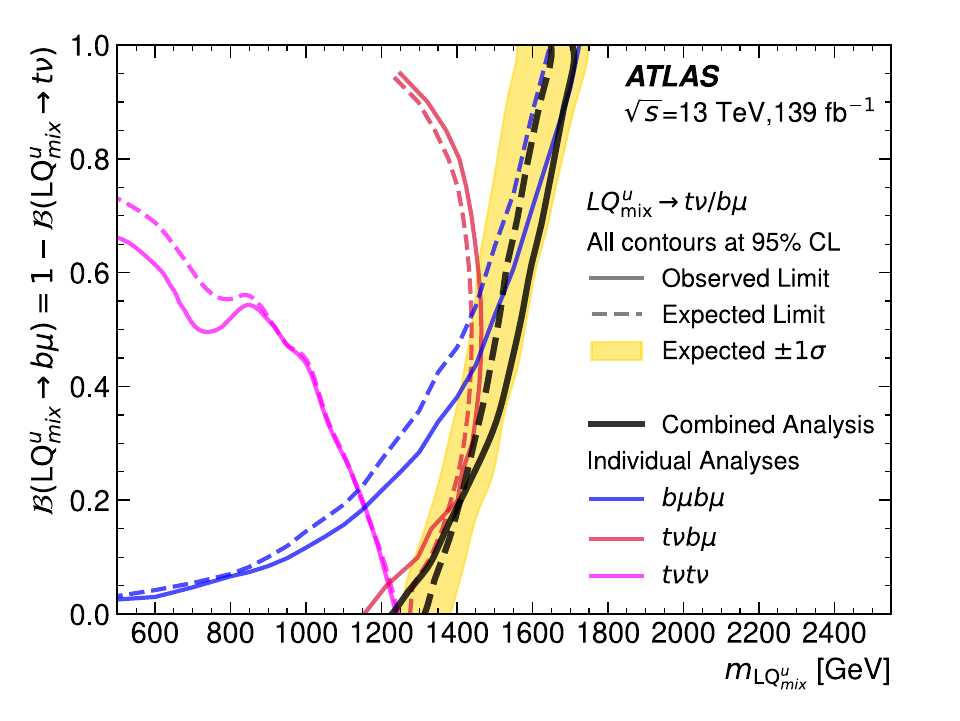}}
\qquad
\subfloat[]{\includegraphics[width=0.45\textwidth]{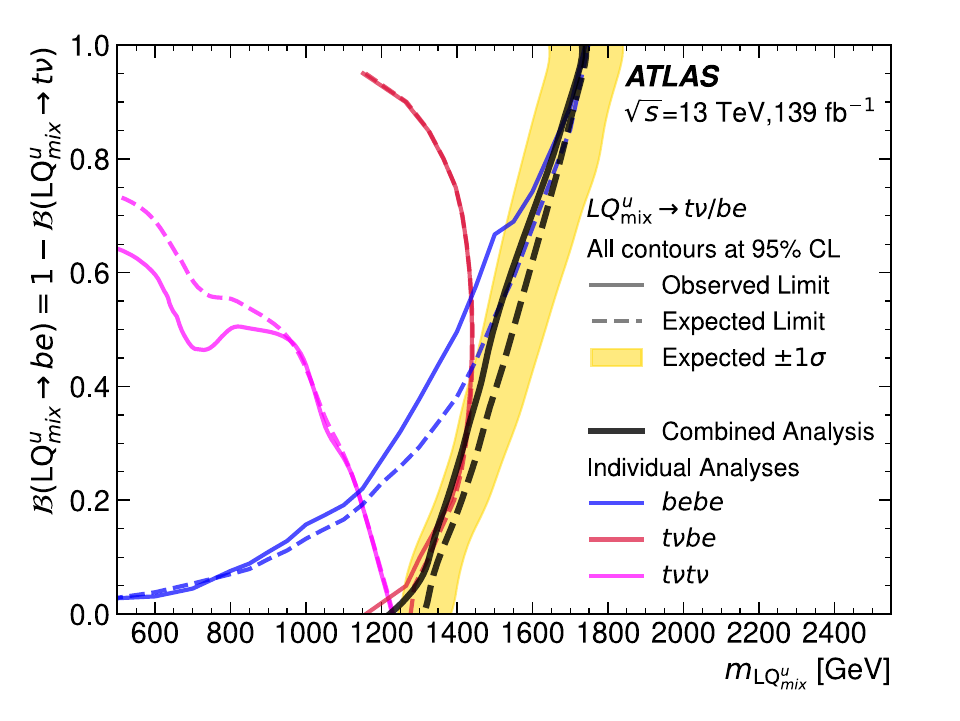}}\\
\subfloat[]{\includegraphics[width=0.45\textwidth]{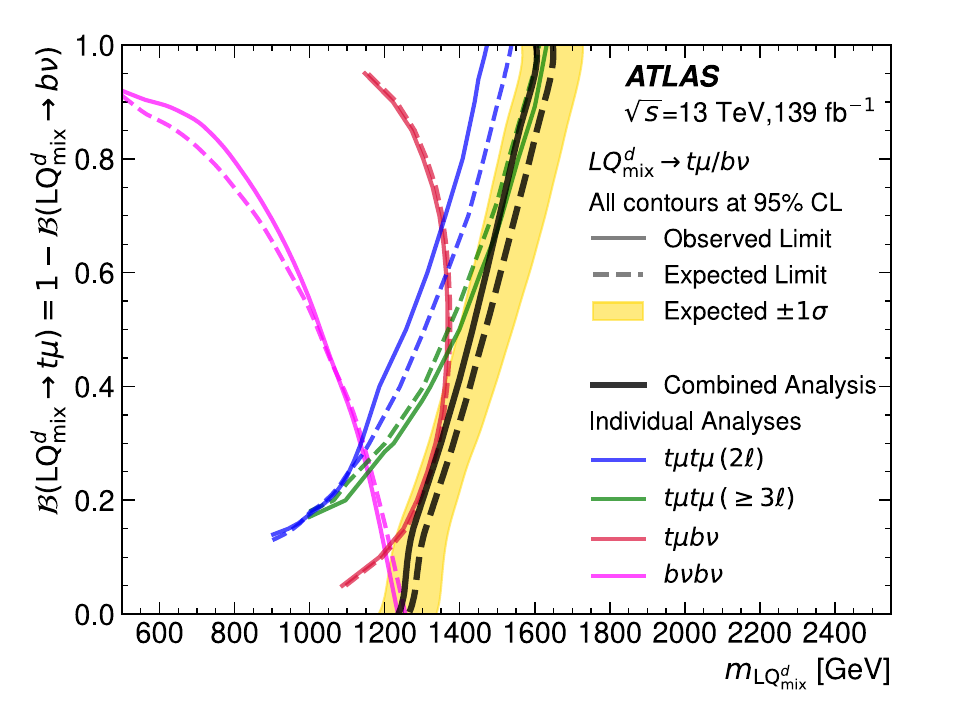}}
\qquad
\subfloat[]{\includegraphics[width=0.45\textwidth]{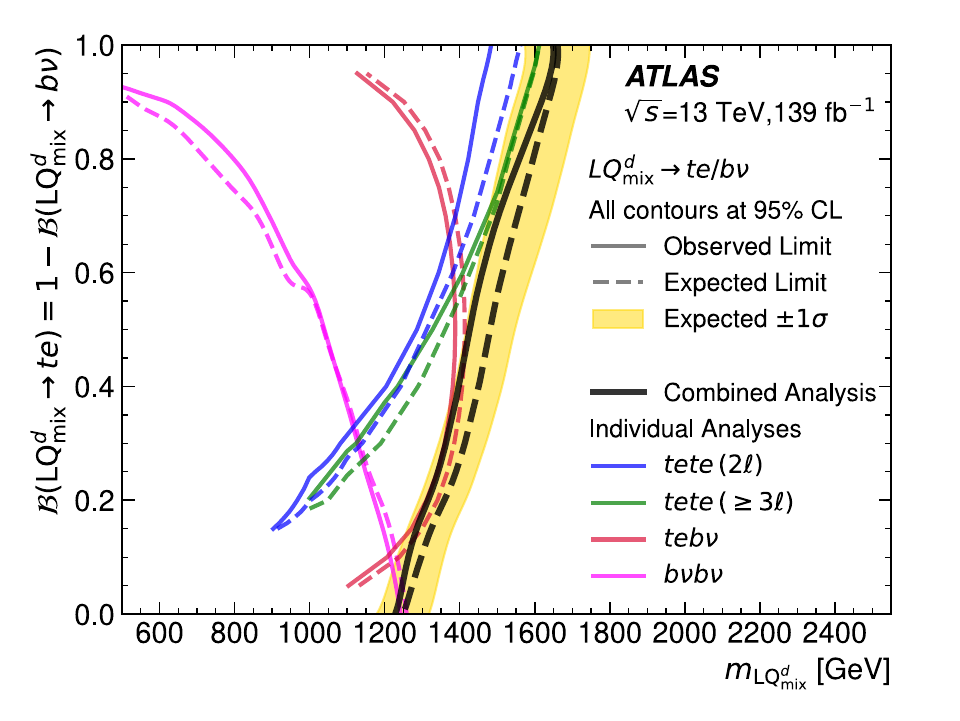}}
\end{center}
\caption{Two-dimensional limits (a,b) on the production of up-type scalar leptoquarks with cross-generational couplings as a function of the mass of the LQ and the BR to $b \ell$ from the search of up-type LQs in final states with (blue) two leptons~\cite{EXOT-2019-13}, (red) only one lepton~\cite{SUSY-2019-12}, (pink) no leptons~\cite{SUSY-2018-34} and (black) their statistical combination~\cite{EXOT-2020-27}, and (c,d) on the production of down-type scalar leptoquarks with cross-generational couplings as a function of the mass of the LQ and the BR to $t\ell$ from the search of down-type leptoquarks in final states with (blue) two leptons~\cite{EXOT-2019-19}, (green) multiple leptons~\cite{ATLAS-CONF-2022-052}, (red) only one lepton~\cite{SUSY-2019-12}, (pink) no leptons~\cite{SUSY-2018-34} and (black) their statistical combination~\cite{EXOT-2020-27}. Electron and muon final states are considered separately.}
\label{fig:lq_mix_limits}
\end{figure}


%
\section{Additional vector bosons}
\label{sec:gauge}

Multiple BSM theories predict the existence of new vector bosons, which are naturally predicted in theories with an extended gauge sector. A typical historical benchmark is the Sequential Standard Model (SSM)~\cite{Altarelli:1989ff}, in which the new \Zprime boson has the same fermion couplings as its SM counterpart: it can hence be searched for as a resonance in the invariant mass distribution of fermion pairs in a variety of leptonic and hadronic final states (the same can also apply to a new SSM \Wprime boson). A $\Zprime_{\psi}$ vector boson can also be predicted by an E$_6$-motivated Grand Unification model~\cite{PhysRevD.34.1530}. Other models predict \Zprime bosons which decay into top quarks with a large branching ratio, such as the colour-singlet boson predicted in the topcolour-assisted-technicolour model~\cite{Hill:1993hs,Hill:1994hp}, making the search for a resonance in the $m_{tt}$ invariant mass spectrum particularly interesting. The \Zprime and \Wprime can also be the nearly mass-degenerate heavy vector triplet (HVT) members of a new SU(2) gauge group~\cite{Pappadopulo:2014qza}. This scenario includes a wider range of phenomena because the new bosons can decay into diboson final states. The SSM also appears as a particular benchmark within the HVT coupling space.

These resonances can be searched for in two-body final states or in production modes with other associated jets or leptons, or in the case of a new vector colouron~\cite{Chivukula:1996yr} (a massive version of the gluon, predicted in gauge extensions of QCD), in a pair of dijet resonances. If the new vector boson's mass is higher than the $pp$ collision centre-of-mass energy, its effects can still be seen and then interpreted in terms of contact interactions, which are the focus of other dedicated searches. All of these searches are reviewed in this section, and the limits set on the \Zprime and \Wprime bosons are summarized in Table~\ref{tab:Vprime}. Some other, specific models of new vector bosons are covered elsewhere in this report: the left--right symmetric model were already discussed in Section~\ref{sec:leptons}, while lepton-flavour-violating \Zprime decays will be discussed in Section~\ref{sec:lfv}, and a simplified model of dark-matter production through a \Zprime, in Section~\ref{sec:dm}.

\subsection{Resonant searches}
\label{sec:gauge_resonant}

\subsubsection{Two-body final states with quarks or leptons}
\label{sec:resqqll}

The dijet resonance analysis discussed in Section~\ref{sec:qstar} can also be used to search for either a SSM \Zprime or \Wprime. A two-$b$-tags channel is also defined for the \Zprime search, which differs from the previously described $b^*$ search only by requiring one extra $b$-tagged jet. The best constraints on these SSM gauge bosons, which assume the same fermion couplings as their SM counterparts, however come from the search in the leptonic final states described below, as shown in Table~\ref{tab:Vprime}.

In the $\Zprime\to\ell\ell$ search~\cite{EXOT-2018-08}, where $\ell=\{e,\mu\}$, the $m_{\ell\ell}$ spectrum above 225~\GeV is probed after selecting events with at least two same-flavour (SF) leptons.\footnote{If more than two leptons of the same-flavour are found, the highest-\pT ones are used. If both an $ee$ pair and a $\mu\mu$ pair are found, the $ee$ pair is kept because it has better mass resolution and signal efficiency.} The muon channel also requires the two muons to be of opposite sign, a requirement which is not made in the electron channel because of the larger charge misidentification rate at high \pT and the fact that a wrong charge assignment in the electron channel would not impact the calorimetric measurement of the electron's momentum. The lower bound on $m_{\ell\ell}$ is imposed to avoid the $Z$ boson peak region, because the background estimate is (as in the dijet resonance search) obtained by fitting a functional form to the data in the SR, in this case $f_{\ell\ell}(m_{\ell\ell})=f_{\mathrm{BW},Z}(m_{\ell\ell})(1-x^c)^b x^{\Sigma_{i=0}^{3} p_i\log(x)^i}$, where $x=m_{\ell\ell}/\sqrt{s}$, $b$ and $p_i$ are free parameters with independent values for the $ee$ and $\mu\mu$ channels, $c$ is fixed to 1 (for $ee$) or 1/3 (for $\mu\mu$), and $f_{\mathrm{BW},Z}(m_{\ell\ell})$ is a Breit--Wigner function with the mass and width of the $Z$ boson. Uncertainties in the MC background modelling are propagated into the estimation of the spurious signal, which is taken as a systematic uncertainty of the background estimation. The distribution of $m_{\ell\ell}$ in the electron channel data is compared with the background fit and some example signals in Figure~\ref{fig:vprimelep}(a). No significant deviation of the data from the background fit is observed. The same also applies to the muon channel, which is, however, slightly less sensitive due to poorer signal resolution.

\begin{figure}[tb]
\begin{center}
\subfloat[]{\includegraphics[width=0.54\textwidth]{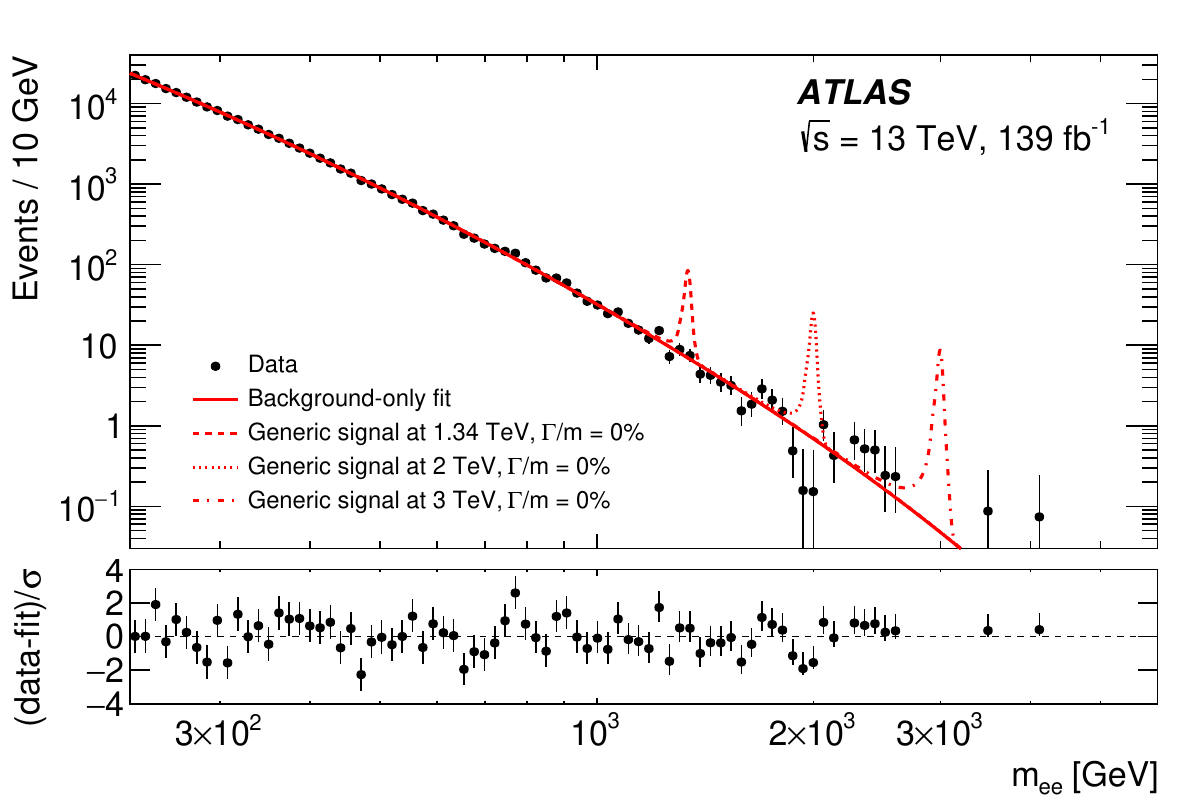}}
\qquad
\subfloat[]{\includegraphics[width=0.35\textwidth]{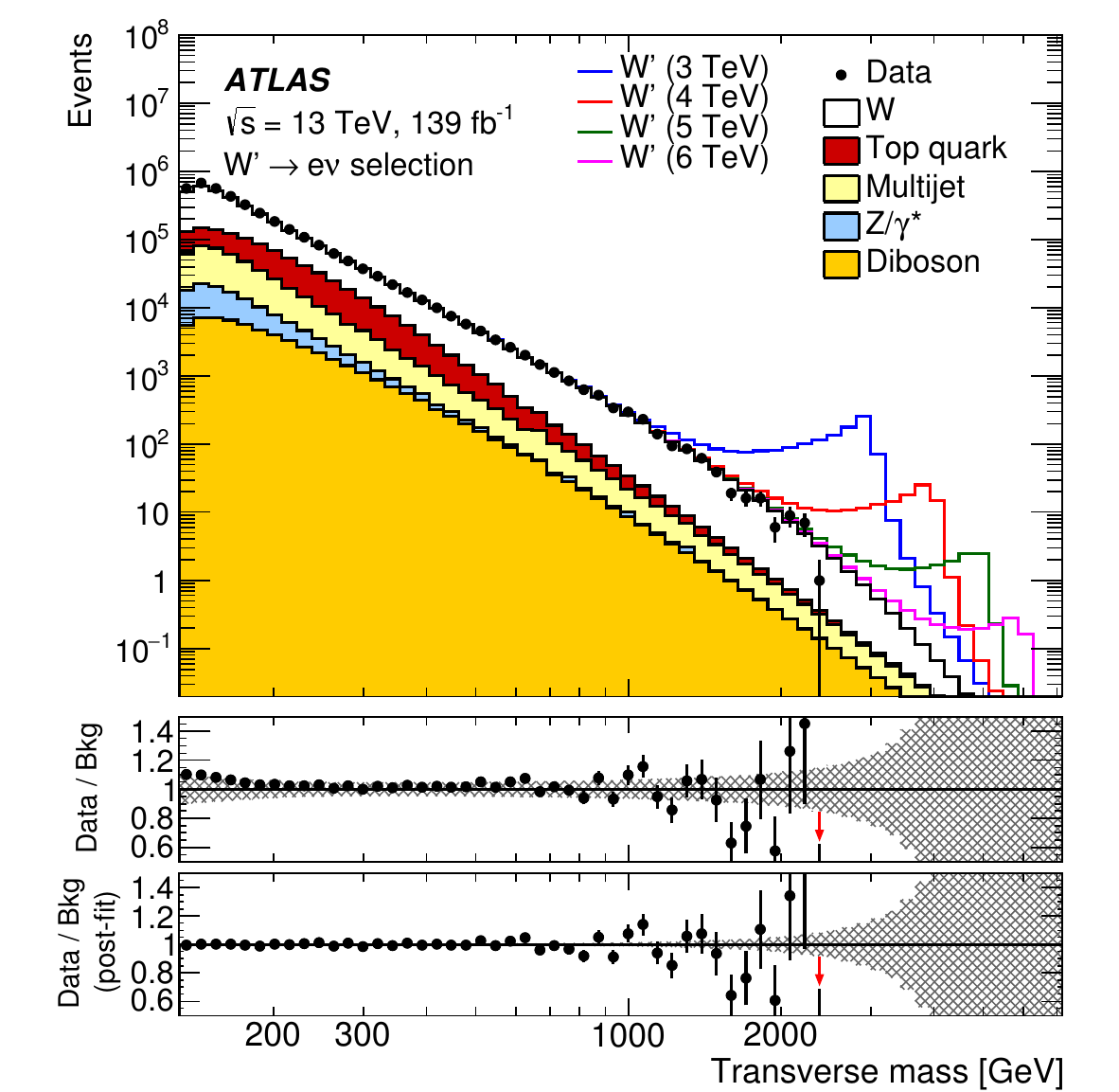}}
\end{center}
\caption{(a) Dielectron invariant mass spectrum used in the search for a \Zprime resonance~\cite{EXOT-2018-08} compared to the fitted background template and some generic signal examples, and (b) transverse mass of the electron--\met system in the \Wprime search~\cite{EXOT-2018-30} compared to the background expectation and example \Wprime signals.}
\label{fig:vprimelep}
\end{figure}

In the $\Wprime\to\ell\nu$ search~\cite{EXOT-2018-30}, the presence of exactly one electron or muon is required in the SR, along with a large value of \met and of $\mT(\ell,\met)$, which is used as the final discriminating variable. The background is dominated by the Drell--Yan (DY) production of $W$ bosons, but also contains some contributions from \ttbar, single-top, diboson, $Z/\gamma^*$ and multijet processes. The backgrounds involving true leptons are evaluated using MC simulations; in the \ttbar and diboson cases, to compensate for the limited number of events at high \mT, the smoothly falling \mT distribution from the MC simulation is fitted using functional forms and extrapolated to high values. The contribution from fake leptons from multijet processes, more important in the electron channel, is evaluated with a matrix method.
No significant deviation of the data from the fitted background's \mT spectrum is observed, neither in the electron channel (shown in Figure~\ref{fig:vprimelep}(b)) nor in the muon channel, whose sensitivity is, as in the \Zprime search, slightly worse than in the electron channel due to poorer resolution.

Analyses with $\tau$-leptons in the final state can be important in the search for new \Wprime or \Zprime bosons and are particularly relevant in scenarios with enhanced third-generation couplings. A search for $\Wprime\to\tau\nu$ using the full Run 2 dataset was very recently made public~\cite{EXOT-2018-37}. The data agree with the expected background’s $\mT(\tauhadvis,\met)$ distribution within uncertainties, which are dominated by the number of data events for masses above 2~\TeV. The search for a $\Zprime\to\tau\tau$ has only been performed with a partial Run~2 dataset~\cite{HIGG-2016-12} recorded between 2015 and 2016.

Another analysis performs a generic search for simple two-body resonances and is based on anomaly detection techniques~\cite{EXOT-2022-07}. This approach considers resonances decaying into a pair of jets (which could be \btagged) or a combination of one jet and one lepton (or photon). A simple preselection of events with at least one jet and at least one lepton is performed, and an unsupervised machine-learning method based on an autoencoder is used to find events with kinematic properties that are likely to be different from the bulk of preselected data events. A total of nine independent statistical analyses using different two-body invariant mass combinations are used to look for excesses above the SM background, which in each case is obtained using a parametric fit. No significant excess is found in any of the combinations, with the largest deviation being seen at a mass of 4.9~\TeV when using the $m_{j\mu}$ statistical analysis. The sensitivity of this search is poorer than in dedicated approaches but it remains interesting due to its very general, model-independent approach, which can guide more sensitive, dedicated searches.

\subsubsection{Diboson final states}
\label{sec:dibosonres}

Six searches in ATLAS consider final states with two bosons, and all of them can be interpreted in the context of additional vector gauge bosons, either \Zprime or \Wprime. They can be classified into two subclasses with common characteristics: final states without ($VV$) or with ($VH$) Higgs bosons, where $V$ is a $W$ or $Z$ boson. In addition to vector gauge bosons, several analyses in ATLAS look for scalar resonances in diboson final states such as $H\gamma$, $Z\gamma$ or $ZZ$; these analyses are included in a different report~\cite{HDBS-2023-15}.

The two $VH$ searches consider a final state with a Higgs boson decaying into two \bquarks: a fully hadronic final state~\cite{HDBS-2018-11} and a leptons+jets final state~\cite{HDBS-2020-19}.

The fully hadronic analysis considers events with no light leptons and two large-$R$ jets, well separated in rapidity. The leading (highest-\pt) jet is considered to be the Higgs candidate, while the subleading jet is considered to be the $V$ boson candidate. Two independent channels are built, aimed at final states with either a $W$ or a $Z$ boson. The jets are identified (tagged) as coming from a Higgs, $W$ or $Z$ boson according to their mass, the number of tracks, and the $D_2$ variable~\cite{Larkoski:2015kga}. The latter is defined as a ratio of two-point to three-point energy correlation functions based on the jet's constituents and is expected to peak at values below one for heavy hadronic resonances. In each channel, events with one tagged candidate $V$~boson and one tagged candidate $H$~boson are categorized into SRs. Two independent categories are built in each channel according to the number (one or two) of $b$-tagged jets associated with the large-$R$ Higgs candidate jet. The invariant mass of the two large-$R$ jets is used in the statistical analyses. In both categories and in both channels, the background in the signal regions is obtained using a data-driven method. The initial template is obtained from a third category with no $b$-tagged jet associated with the Higgs candidate. The final template in the 1-tag and 2-tag categories is obtained using a combination of a multidimensional kinematic reweighting, based on a boosted decision tree (BDT), and normalization corrections. The BDT is trained on data from a dedicated CR with relaxed identification criteria for the boson candidates. The same region is used to obtain the normalization correction between the 0-tag category and the 1-tag or 2-tag categories. Finally, the estimated background is smoothed using a functional-form fit. Two independent statistical analyses are done, one for each $V$ boson final state associated with the equivalent $V'$ hypothesis. Agreement in both channels and both categories is good after a background-only fit.

The leptons+jets analysis uses two complementary selections aimed at different \pt regimes: a resolved selection with at least two jets, one of them \btagged, and a merged selection with a large-$R$ jet with at least one associated \btagged variable-$R$ track-jet. Both selections are divided according to the number of \btagged jets or associated \btagged variable-$R$ track-jets, respectively, into 1-tag and 2-tag categories. The pair of \btagged jets or the \btagged jet and the leading non-\btagged jet (for the 2-tag and 1-tag categories, respectively) are chosen in the resolved selection as the Higgs candidate and are required to have an invariant mass close to the Higgs boson mass. In the merged selection, the Higgs candidate is chosen as the leading large-$R$ jet that has least one associated \btagged jet and a mass close to that of the Higgs boson. If an event passes both the resolved and merged selection, it is classified as resolved. Events from the two selections are then classified into six categories by looking at the number of light leptons in the event: zero, one, or two. The 0-lepton channel requires a large amount of \met and imposes additional kinematic selections based on the relationships between the \met and the jets in the event. The 1-lepton channel requires events to have one light lepton and a significant amount of \met and imposes additional selections sensitive to the presence of a $W$~boson in the event, e.g.\ on the $m_\text{T}(\ell,\met)$ variable. Finally, the 2-lepton channel requires an event to have two same-flavour leptons, and additional selections, designed to select events with a $Z$~boson in the final state, are imposed on the dilepton system. For each combination of the number of leptons and number of \btagged jets, and separately for the merged and resolved categories, a signal region is built by imposing a very tight selection around the Higgs boson mass. In contrast, events with slightly lower or higher \mH values are kept as control regions in the 0-lepton and 1-lepton channels. In the 2-lepton resolved category, a CR is defined using events with two different flavour leptons, while the 2-lepton merged category has no corresponding CR. Two independent statistical analyses are built to search for events consistent with either a $Z'$ or a $W'$ hypothesis, with the former using the 0-lepton and 2-lepton channels and the latter using the 0-lepton and 1-lepton channels. The invariant mass of the $VH$ candidate is used in the 0- and 2-lepton channels, while the transverse mass of the $VH$ candidate is used in the 1-lepton channel. All templates for background and signal are obtained from MC samples, with the small multijet background in some regions obtained using a data-driven template method~\cite{HIGG-2018-04}. Agreement between data and the SM background expectation after a background-only fit is found to be good in all regions.

In the absence of significant excesses, cross-section limits are set as a function of the mass of the new gauge boson (\Wprime or \Zprime). Such limits from an HVT signal hypothesis are shown in Figure~\ref{fig:vprimeVH} for the quark--quark annihilation (qqA) production mode. The two searches have similar sensitivity for the $W'$ case, but the leptons+jets search dominates across the whole mass range for a $Z'$ hypothesis.

\begin{figure}[tb]
\begin{center}
\subfloat[]{\includegraphics[width=0.45\textwidth]{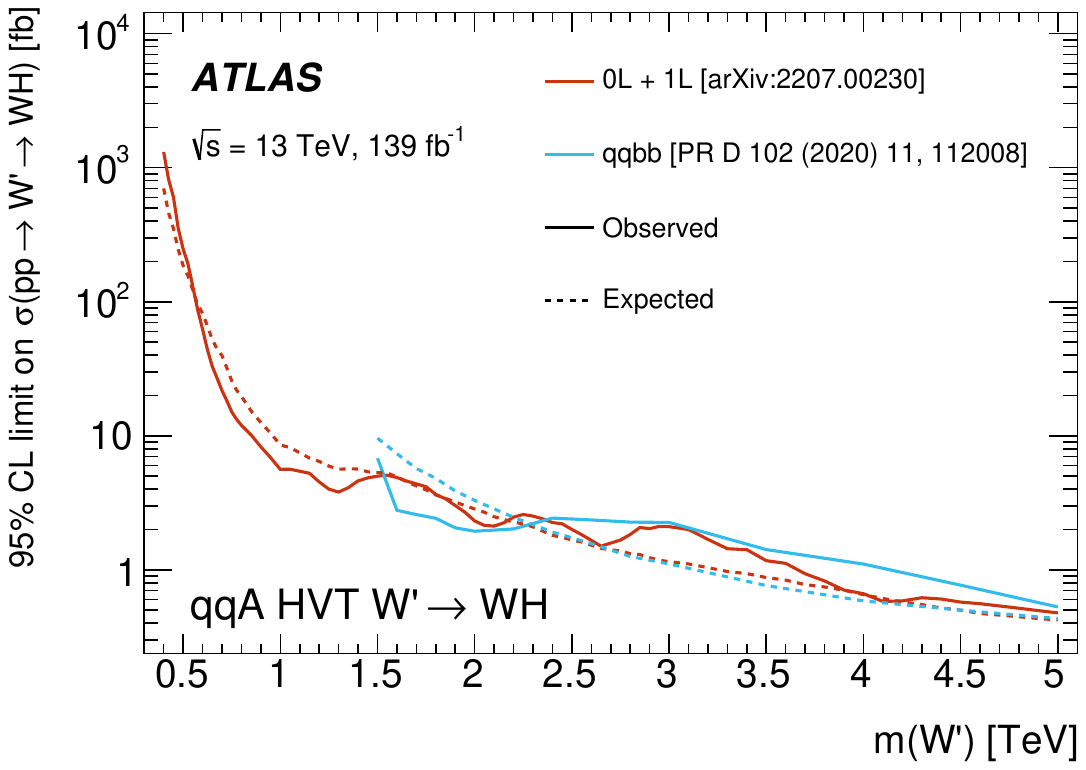}}
\subfloat[]{\includegraphics[width=0.45\textwidth]{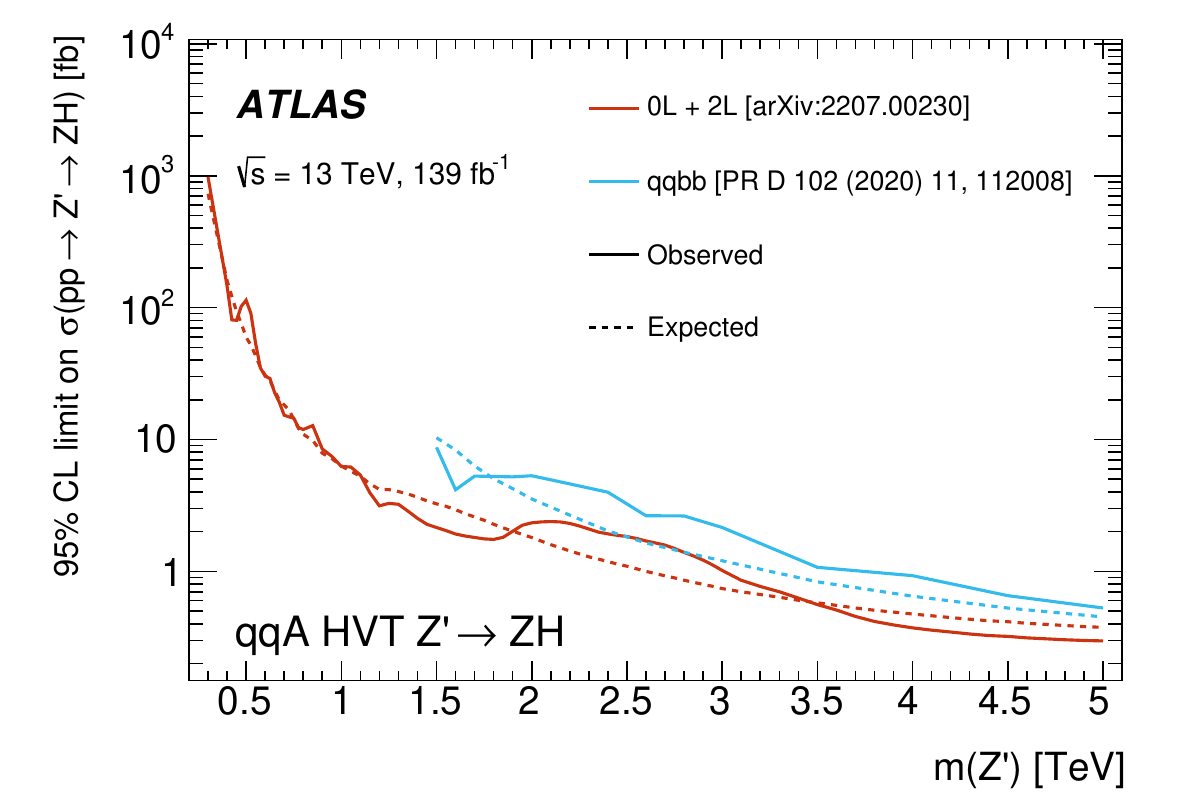}}
\end{center}
\caption{Upper limits on the production cross section of new gauge bosons, (a) \Wprime or (b) \Zprime decaying into a $VH$ final state from the (red) leptons+jets~\cite{HDBS-2020-19} and (blue) fully hadronic~\cite{HDBS-2018-11} analyses. The HVT model is used to define the signal hypothesis.}
\label{fig:vprimeVH}
\end{figure}

The four $VV$ searches cover the large variety of $WW$ and $WZ$ final states, since each boson can decay either hadronically or leptonically.

The semileptonic $VV$ search~\cite{HDBS-2018-10} considers three separate channels, named according to the number of leptons: zero, one, or two. The 0-lepton channel targets the $ZV$ final state and requires events to have a large amount of \met and no leptons in addition to at least one large-$R$ jet or two small-$R$ jets. The 1-lepton channel, aimed at the $WV$ final state, has the same jet requirements but selects events with a significant amount of \met and exactly one lepton. Finally, the 2-lepton channel, also dedicated to the $ZV$ final state, requires a pair of same-flavour leptons with an \mll value close to the $Z$ boson mass. Each channel considers the two $V$ hypotheses separately. The hadronic boson candidate is identified using the leading large-$R$ jet, with a mass window close to the $V$ boson mass considered. If no large-$R$ jet candidate is found, an attempt is made to reconstruct a hadronic candidate using pairs of small-$R$ jets. The dijet system formed from the two leading jets is used as a candidate if its \mjj is consistent with that of the $V$ boson considered. The selected events are classified into a large number of signal regions using three criteria. The first criterion uses the $D_2$ variable to separate events according to the reconstruction quality of the hadronic $V$ candidate. The second criterion aims to separate vector-boson fusion (VBF) events from events produced by other processes and uses a recurrent neural network to separate events into these two categories. The final criterion is based on the number of variable-$R$ track-jets associated with the hadronic $V$ candidate. A total of 40 SRs are defined across the three channels. CRs rich in \ttbar, $W$+jets, and $Z$+jets background processes are defined in the 0-lepton,1-lepton, and 2-lepton channels, respectively. The statistical analysis utilizes either the diboson invariant mass or the transverse mass of the $VV$ final state (for the 0-lepton channel only) in all SRs and CRs, with most background and signal templates obtained using MC samples. The multijet background, which is non-negligible in the 1-lepton channel, is estimated using a template method. The data are found to be in good agreement with the background expectation. This search has better sensitivity than any other ATLAS analysis when considering new gauge bosons decaying into $VV$ final states. The other three analyses~\cite{HDBS-2018-31,HDBS-2018-19,ATLAS-CONF-2022-066}, considering fully leptonic or fully hadronic final states, are included in the summary plots below but not described further.

With no significant excesses, cross-section limits are set as a function of the mass of the new gauge boson ($W'$ or $Z'$). Limits for an HVT signal hypothesis are shown in Figure~\ref{fig:vprimeVV} for the two possible final states, $WZ$ and $WW$, and separately for the two possible production modes, VBF and qqA. The semileptonic search described in this section dominates the sensitivity for most of the mass range considered.

\begin{figure}[tb]
\begin{center}
\subfloat[]{\includegraphics[width=0.45\textwidth]{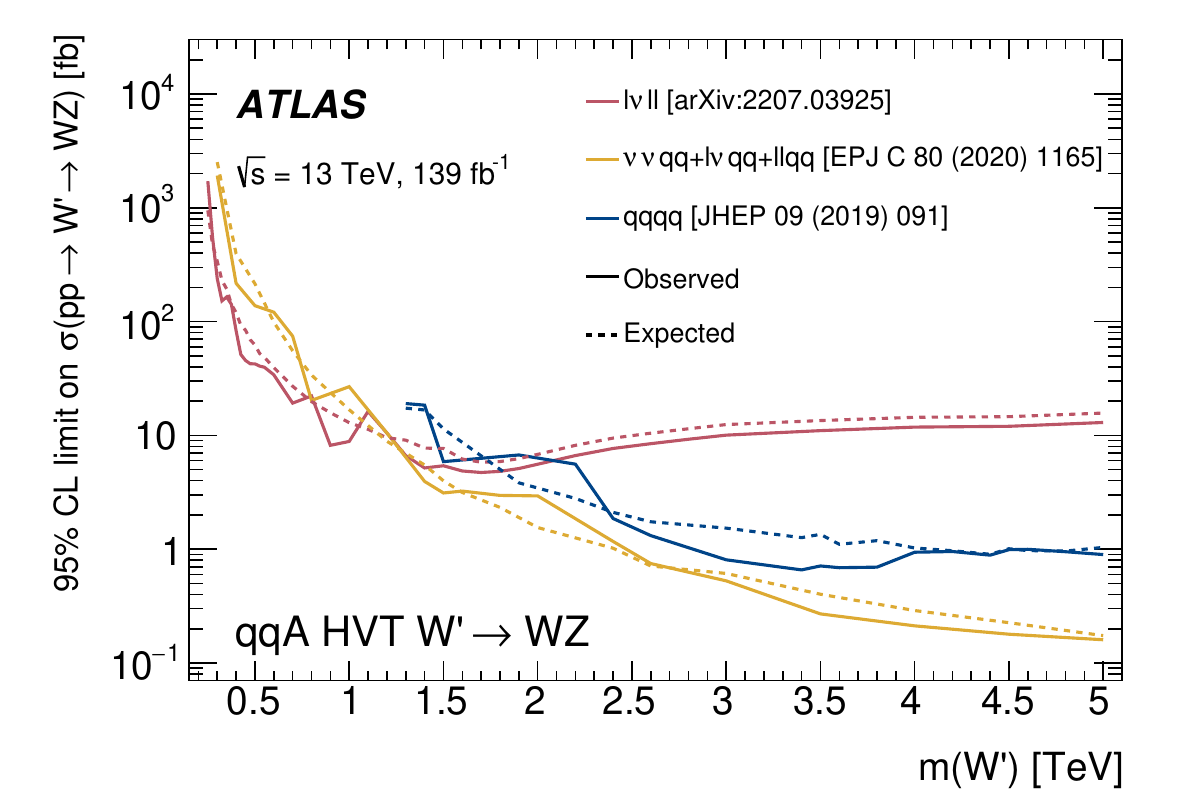}}
\subfloat[]{\includegraphics[width=0.45\textwidth]{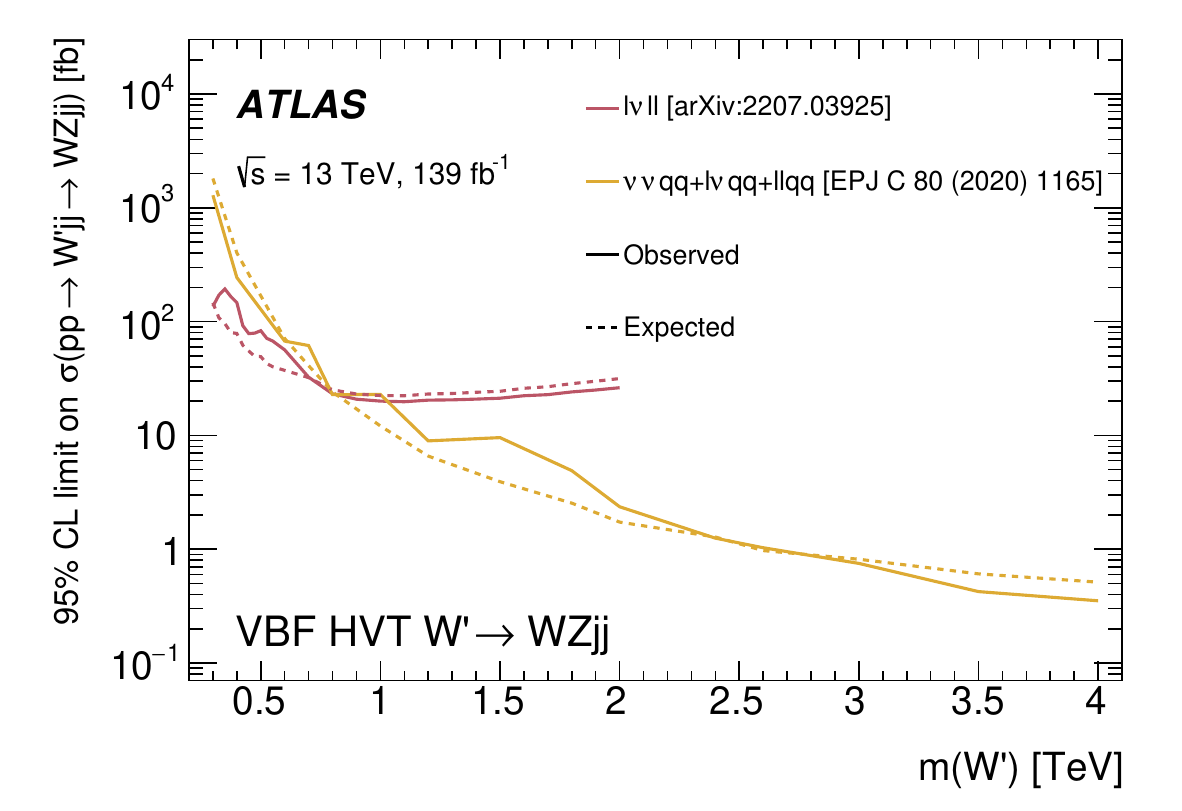}}\\
\subfloat[]{\includegraphics[width=0.45\textwidth]{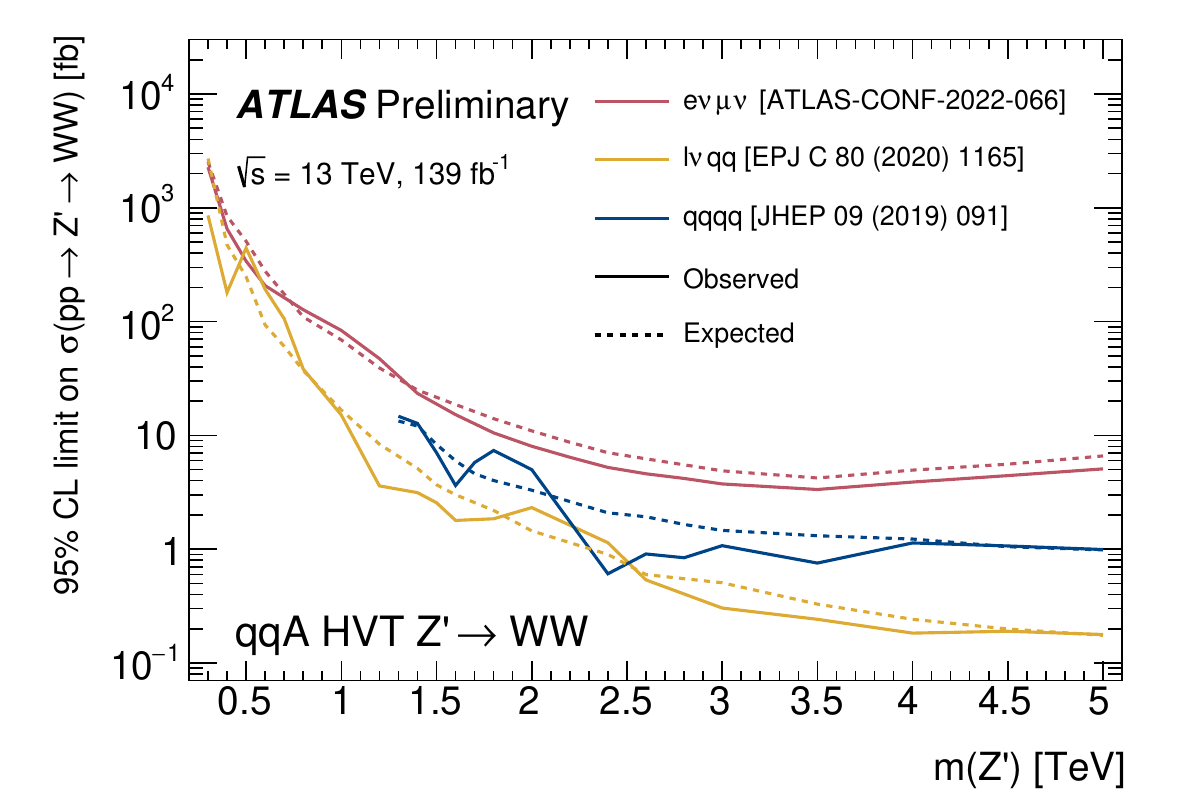}}
\subfloat[]{\includegraphics[width=0.45\textwidth]{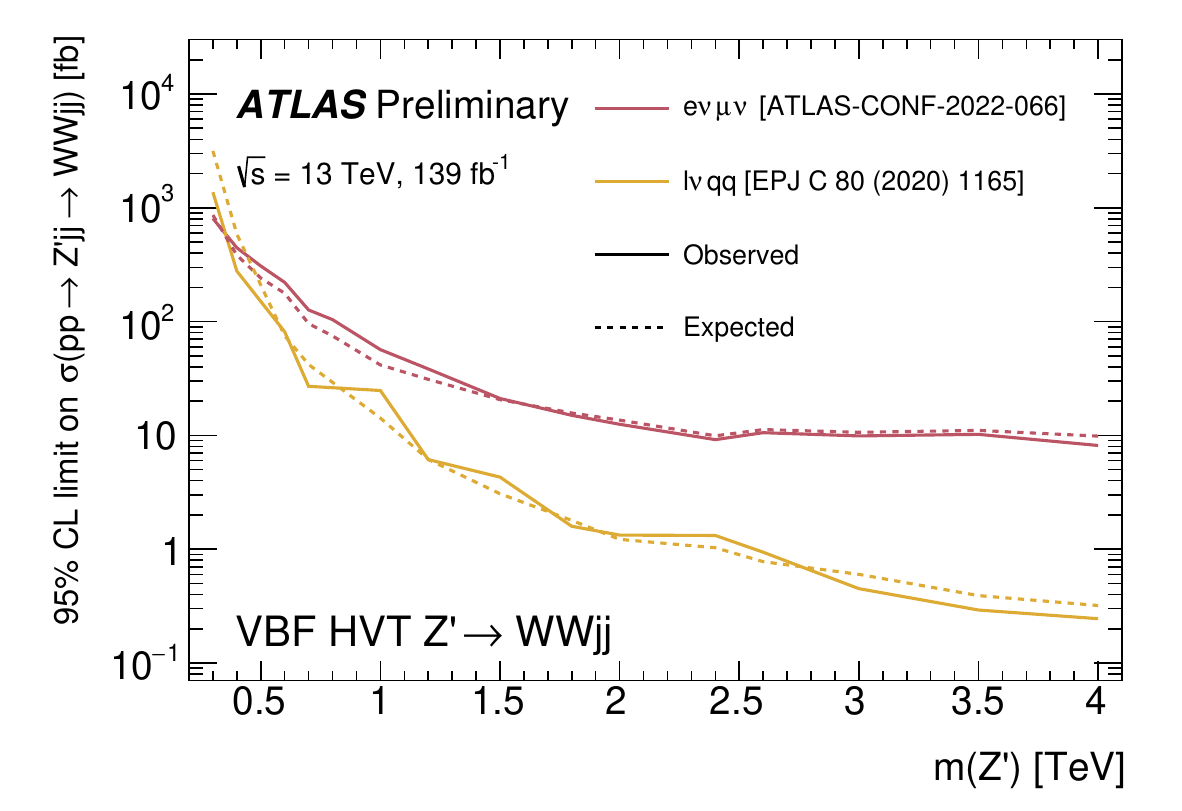}}
\end{center}
\caption{Upper limits on the production cross section of new gauge bosons, (a,b) \Wprime or (c,d) \Zprime decaying into a (yellow) $VV$ semileptonic final state~\cite{HDBS-2018-10}, (red) leptonic final state~\cite{HDBS-2018-19,ATLAS-CONF-2022-066}, or (blue) hadronic final state \cite{HDBS-2018-31}. The HVT model is used to define the signal hypothesis. The qqA and VBF production modes are considered separately.}
\label{fig:vprimeVV}
\end{figure}

\subsubsection{Final states with heavy quarks}
\label{sec:ttortbres}

Searches with top quarks in the final state are particularly relevant for models with preferred couplings to the third generation. A hadronic $\Zprime\to\ttbar$~\cite{EXOT-2018-48} search is employed to set limits on a leptophobic \Zprime from a topcolour-assisted-technicolour model, which falls in that category. This search selects events with two top-tagged large-$R$ jets, which are required to be back-to-back in the ($\eta, \phi$) plane. A selection based on the rapidity separation between the two jets is imposed in order to reject events with large rapidity separation, which are dominated by $t$-channel events from multijet production and represent a significant background. The selected events are separated into two signal regions, according to the number of \btagged variable-$R$ track-jets (one or two) associated with the selected large-$R$ jets. The statistical analysis employs the invariant mass of the two leading large-$R$ jets as the discriminating variable. The background distribution is obtained in~situ in each signal region using a parametric function fitted to data. A background fit shows good agreement between data and MC simulation in both regions, as can be seen in Figure~\ref{fig:ttbar}.

\begin{figure}[tb]
\begin{center}
\subfloat[]{\includegraphics[width=0.45\textwidth]{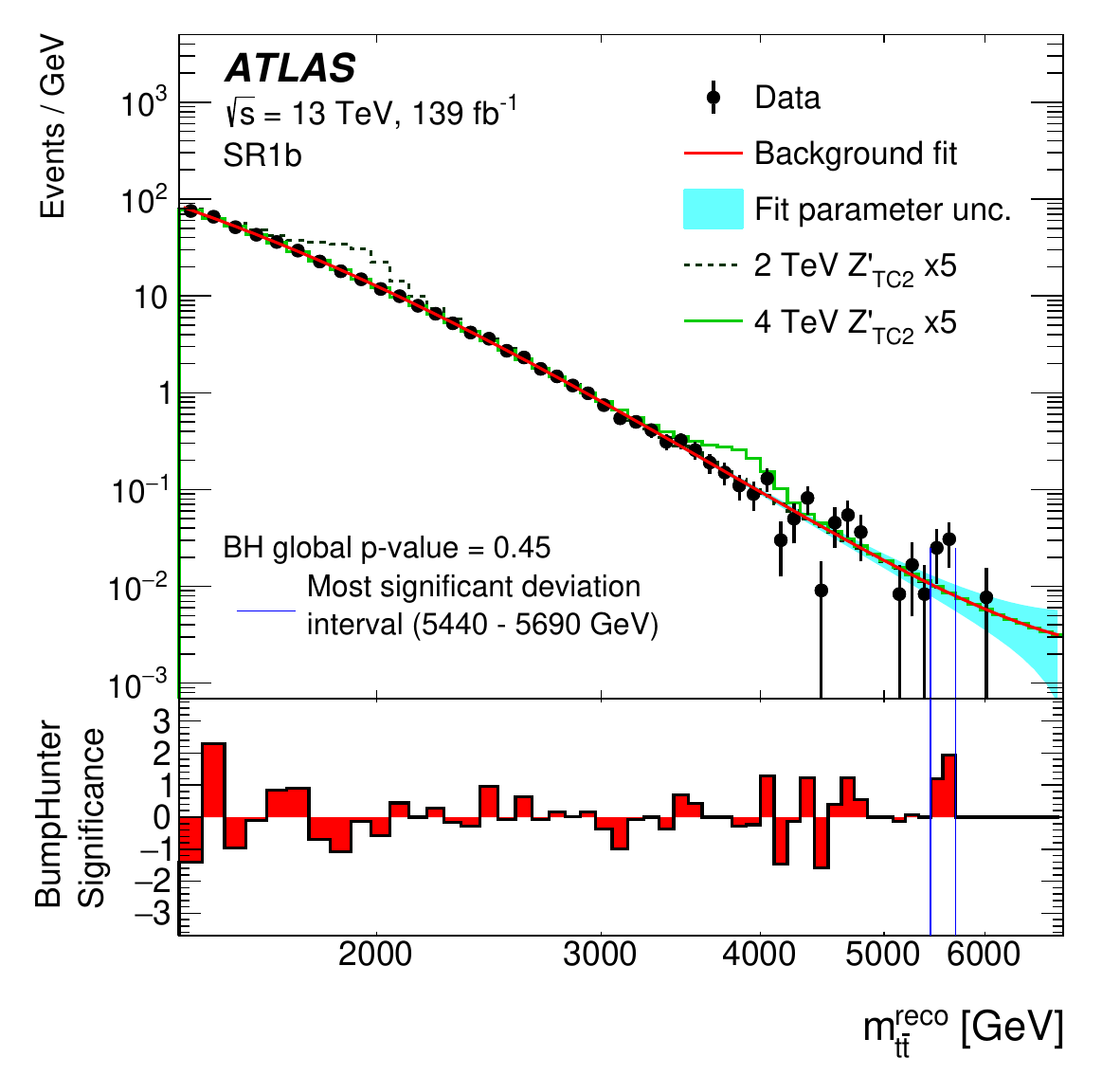}}
\subfloat[]{\includegraphics[width=0.45\textwidth]{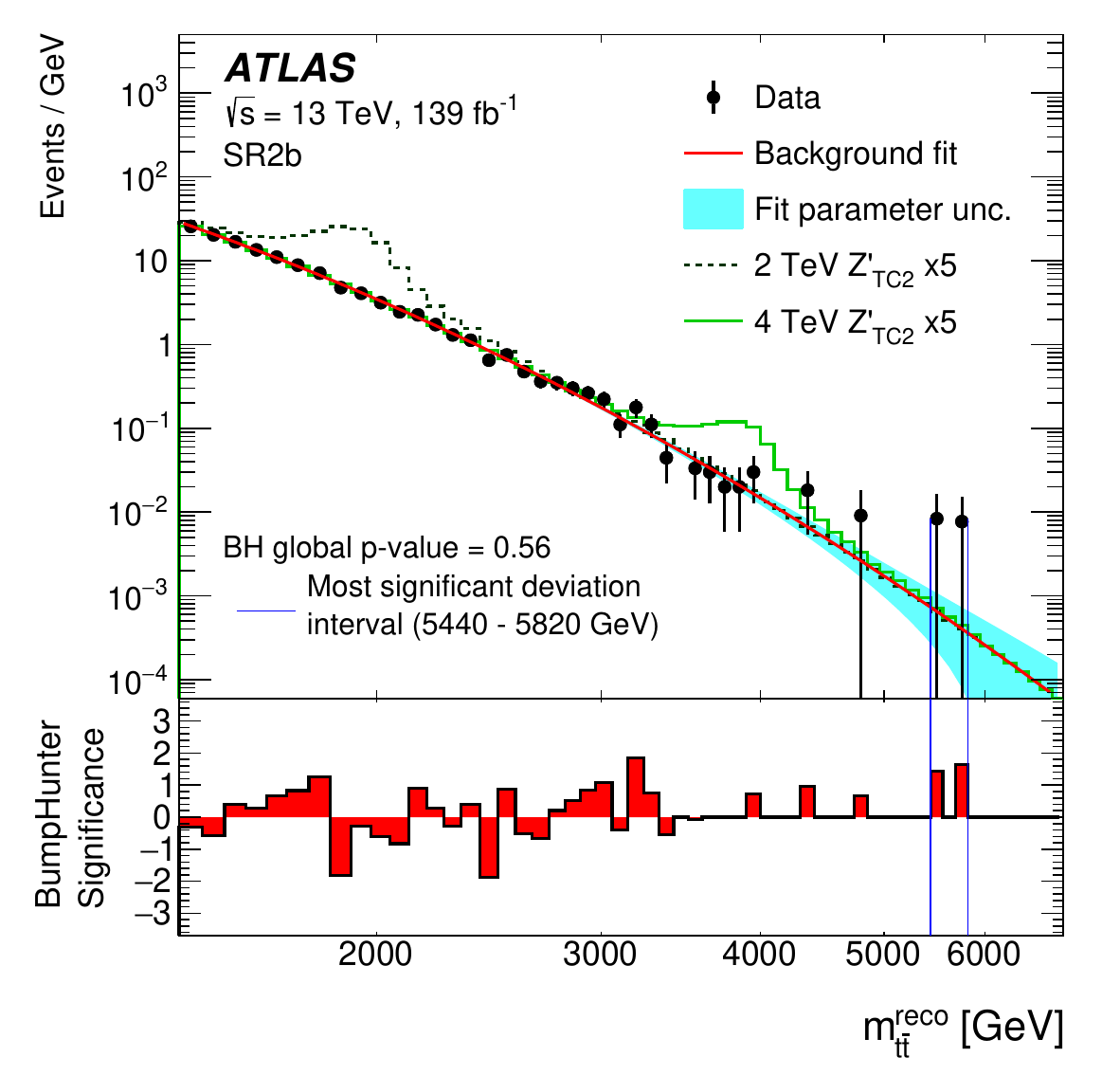}}
\end{center}
\caption{Distributions, in the two SRs of the $\Zprime\to\ttbar$ analysis~\cite{EXOT-2018-48}, of the invariant mass of the two leading large-$R$ jets. The fit of the parametric function to the data is shown in both regions, together with the distributions of two benchmark signal hypotheses with different masses.}
\label{fig:ttbar}
\end{figure}

The same argument can be used to justify a \Wprime search in the $\Wprime\to tb$ final state~\cite{EXOT-2021-36}, which considers both the hadronic (0-lepton) and lepton+jets (1-lepton) channels. The 0-lepton channel selects events with exactly one top-tagged large-$R$ jet back-to-back with a \btagged jet. Such a pair is used to reconstruct the \Wprime boson. Events are classified into two categories according to the presence of additional \btagged jets inside the top-tagged large-$R$ jet. A total of three SRs are defined, with 12 additional regions (6 per category), built by relaxing or inverting the $b$-tagging or top-tagging criteria utilized to estimate the main background, from QCD multijet production. This background is estimated using an ABCD method.
The $\Wprime\to tb$ search uses an eight-region variant, with two SRs per category. The 1-lepton channel selects events with exactly one lepton (electron or muon), a significant amount of \met, two or three additional jets, and exactly one or two \btagged jets. Events are classified according to the numbers of jets and \btagged jets. Additional kinematic selections are imposed to reduce the major backgrounds (\ttbar and $W$+jets) and define four SRs, one per category. Additional cuts based on the same variables are used to build CRs and VRs to control or validate those major backgrounds. The \Wprime candidate is reconstructed using the lepton, the \met, and a combination of jets that satisfy constraints based on the top quark and $W$ boson masses. The final statistical analysis combines the two channels, with seven SRs and two CRs. Using the invariant mass of the \Wprime candidate as the discriminating variable, a background fit shows good agreement between the data and the SM background prediction. Limits are obtained for a leptophobic \Wprime model in which the coupling of the \Wprime to quarks is a free parameter. Both the left-handed and right-handed \Wprime chiralities are considered. Two-dimensional limit plots in the coupling vs mass plane are shown in Figure~\ref{fig:wprime}.

\begin{figure}[tb]
\begin{center}
\subfloat[]{\includegraphics[width=0.45\textwidth]{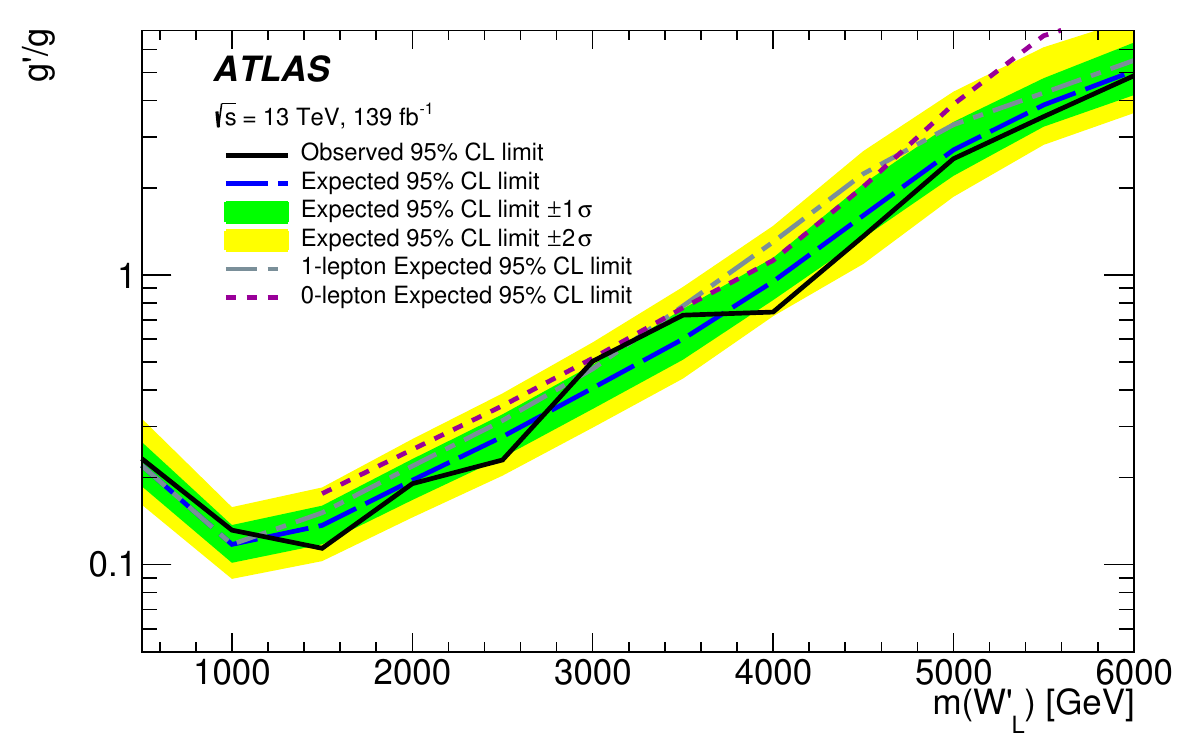}}
\subfloat[]{\includegraphics[width=0.45\textwidth]{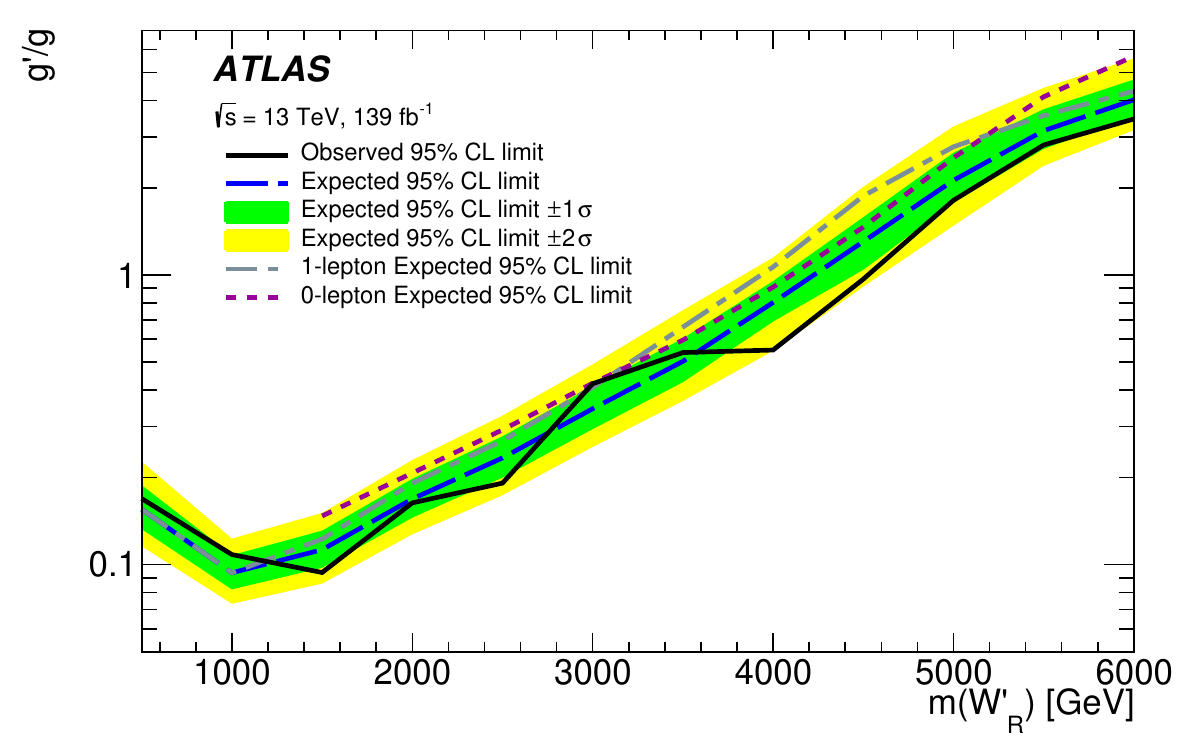}}
\end{center}
\caption{Two-dimensional limits in the mass vs coupling plane for the production of a \Wprime decaying into $tb$~\cite{EXOT-2021-36}, for a (a) left-handed and (b) right-handed hypothesis.}
\label{fig:wprime}
\end{figure}

A second $\Zprime\to\ttbar$ search is done to address a more extreme case in which the \Zprime couples exclusively to top quarks~\cite{EXOT-2022-14}. In this context, the \Zprime is produced primarily through associated production with two additional top quarks, thus composing a $\ttbar\Zprime\to\ttbar\ttbar$ final state. Depending on the exact parameters of the model, final states of the type $tj\Zprime$ and $tW\Zprime$ can also be relevant and are considered in the analysis. This search selects events with exactly one lepton (electron or muon), at least two large-$R$ jets, with the two leading ones used to reconstruct the \Zprime candidate, and at least two additional small-$R$ jets not overlapping the selected large-$R$ jets. Events must also have at least two \btagged jets, which can overlap the large-$R$ jets or be counted as additional jets. Six regions with at least three \btagged jets and classified according to the numbers of jets and \btagged jets are used as SRs. The background in each SR is estimated using a data-driven method. The shape of the initial background template is obtained using a parametric function fit in a region with two \btagged jets and two small-$R$ jets, and extrapolated into the SR using factors obtained from MC simulations. The invariant mass of the reconstructed \Zprime candidate is used in the statistical analysis, and a background-only fit shows good agreement with the data in each of the SRs, as can be seen in Figure~\ref{fig:ttbar2} for the two most sensitive SRs.

\begin{figure}[tb]
\begin{center}
\subfloat[]{\includegraphics[width=0.45\textwidth]{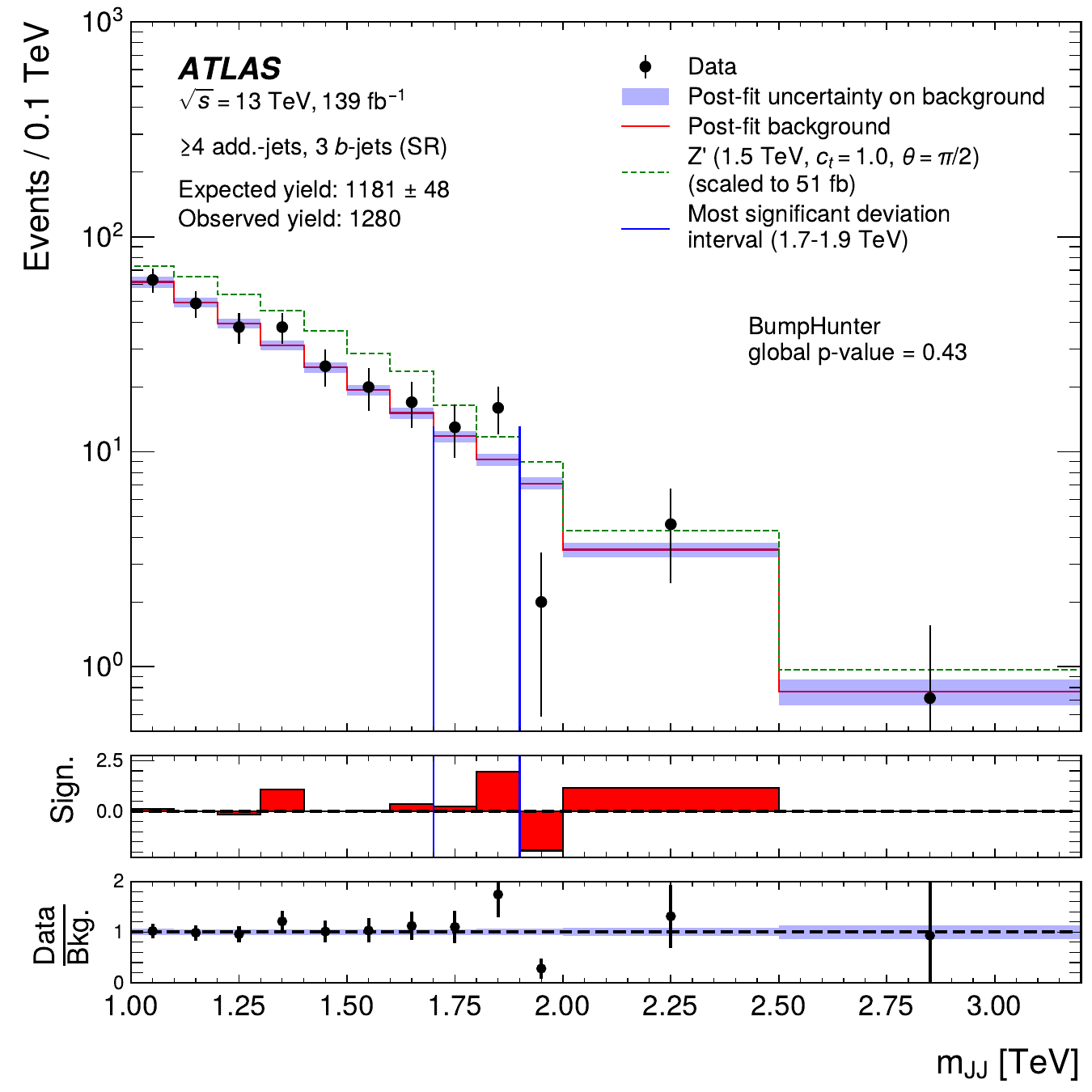}}
\subfloat[]{\includegraphics[width=0.45\textwidth]{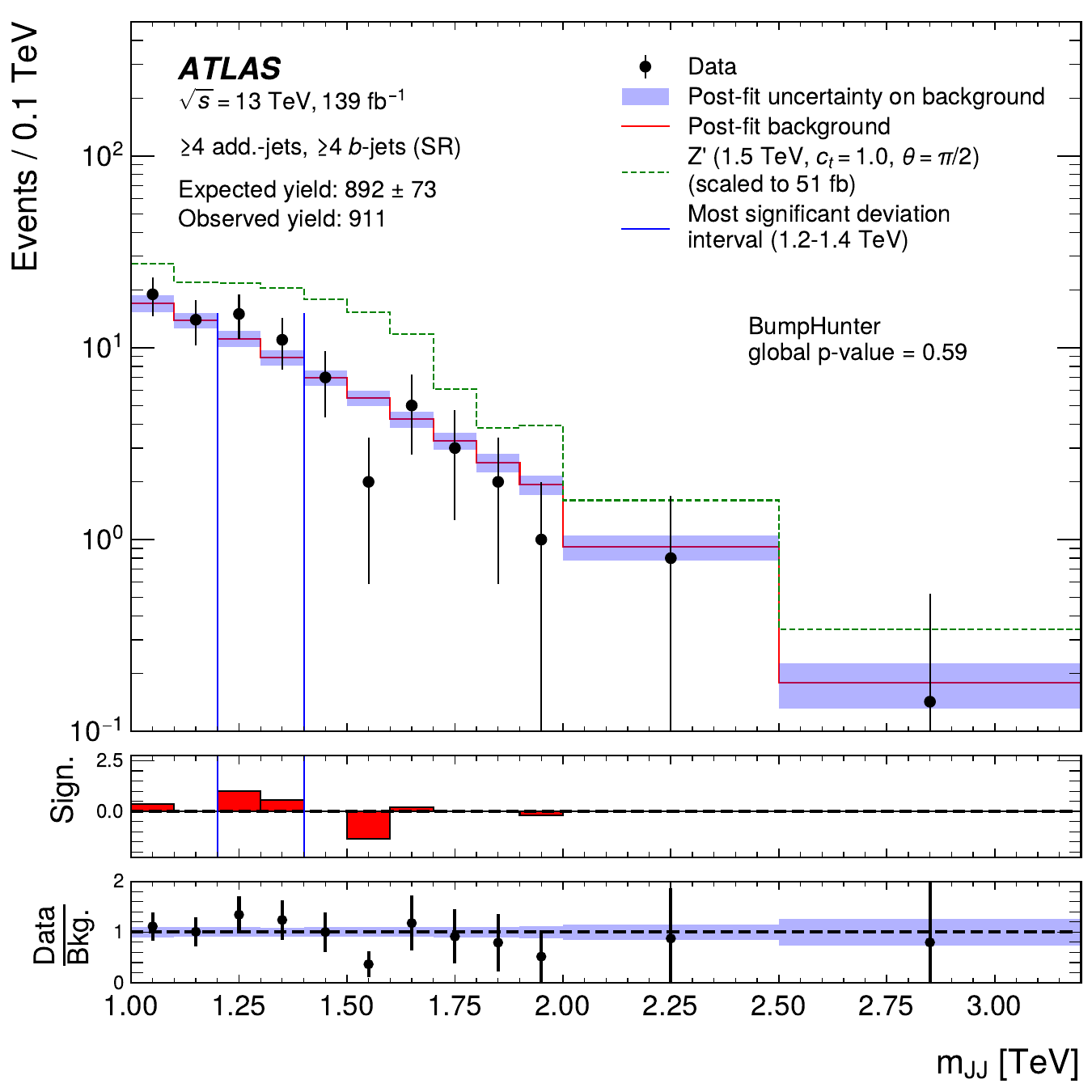}}
\end{center}
\caption{Distributions, in the two most sensitive SRs of the $\ttbar\Zprime\to\ttbar\ttbar$ analysis~\cite{EXOT-2022-14}, of the invariant mass of the two leading large-$R$ jets. They are shown after a background-only fit to data. The distribution for one benchmark signal hypothesis is also shown.}
\label{fig:ttbar2}
\end{figure}

A third and final \Zprime analysis considers an associated production scenario similar to the one in the previous search, in this case involving \bquarks~\cite{EXOT-2018-09}. The analysis selects events with at least three \btagged jets, with asymmetric \pt cuts, corresponding to the $bb\Zprime\to b\bar{b}b\bar{b}$ topology. Additional kinematic cuts on the rapidity separation between the two leading jets help to differentiate between signal events and the main background, coming from QCD multijet production. The background estimate is obtained in~situ in the SR using the functional decomposition method~\cite{Edgar:2018irz} and shows good agreement between the data and the SM background estimate. The search sets limits using a model that includes lepton-universality violation (LUV) and third-generation exclusive couplings in the quark sector.

\subsubsection{Many-body final states}

Although simple final states are the target of many searches, more complicated configurations are useful in looking for \Zprime or \Wprime bosons in specific models where simpler searches do not have enough sensitivity. The $\mu\mu\Zprime\to4\mu$ analysis~\cite{HDBS-2018-57} targets one such scenario, in which a \Zprime couples exclusively to second- and third-generation leptons. Because of this, they cannot be produced in the usual way and appear in final states with multiple leptons. An example of such a production mode and corresponding final state can be seen in Figure~\ref{fig:zprimemuons}(a). The search selects events with four muons with kinematics compatible with producing a pair of muons with a large invariant mass ($Z_1$) and an extra pair of opposite-sign muons ($Z_2$), which follow the expected signature. Both pairs of muons are used to look for heavy resonances with relatively high-mass \Zprime signals appearing as a peak in the invariant mass distribution of $Z_1$ and low-mass ones appearing as a peak in the invariant mass distribution of $Z_2$. The background is estimated using a  fake-factor method.
A parameterized deep neural network (pDNN)~\cite{Baldi:2016fzo} is used to further separate signal and background and to define the final signal regions, with the final cut depending on the signal hypothesis. Agreement between the data and the SM prediction is excellent for most of the mass range, and is shown in Figures~\ref{fig:zprimemuons}(b) and~\ref{fig:zprimemuons}(c) for two different mass hypotheses.

\begin{figure}[tb]
\begin{center}
\subfloat[]{\includegraphics[width=0.3\textwidth]{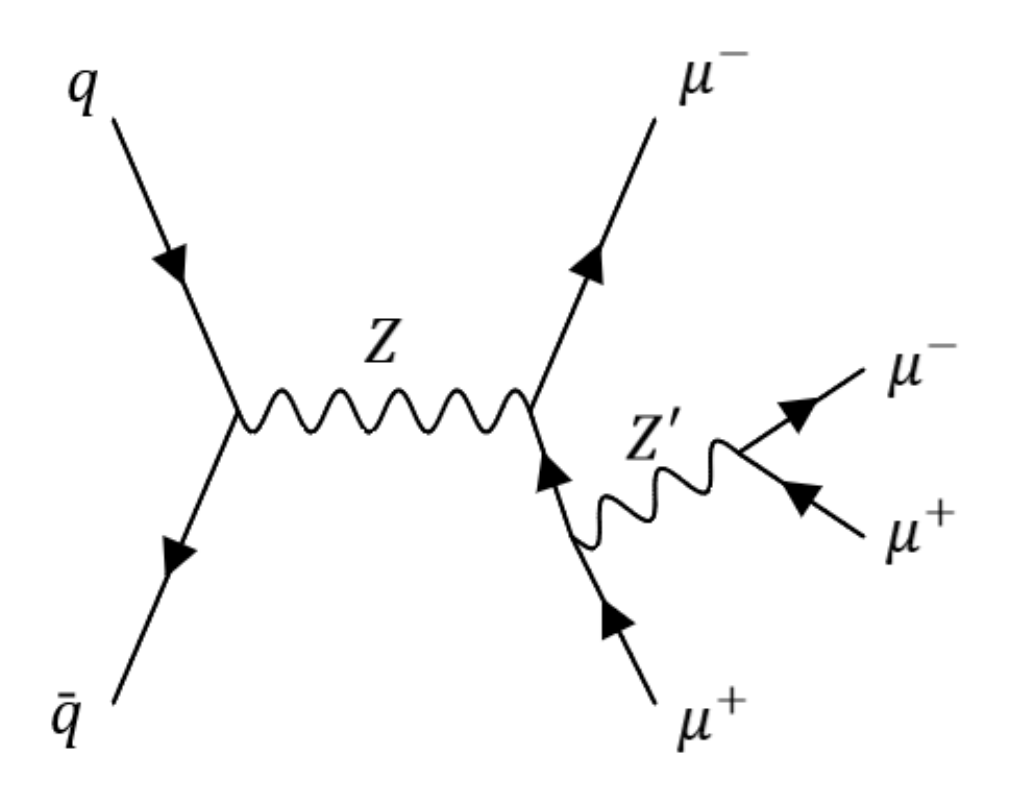}}
\qquad
\subfloat[]{\includegraphics[width=0.45\textwidth]{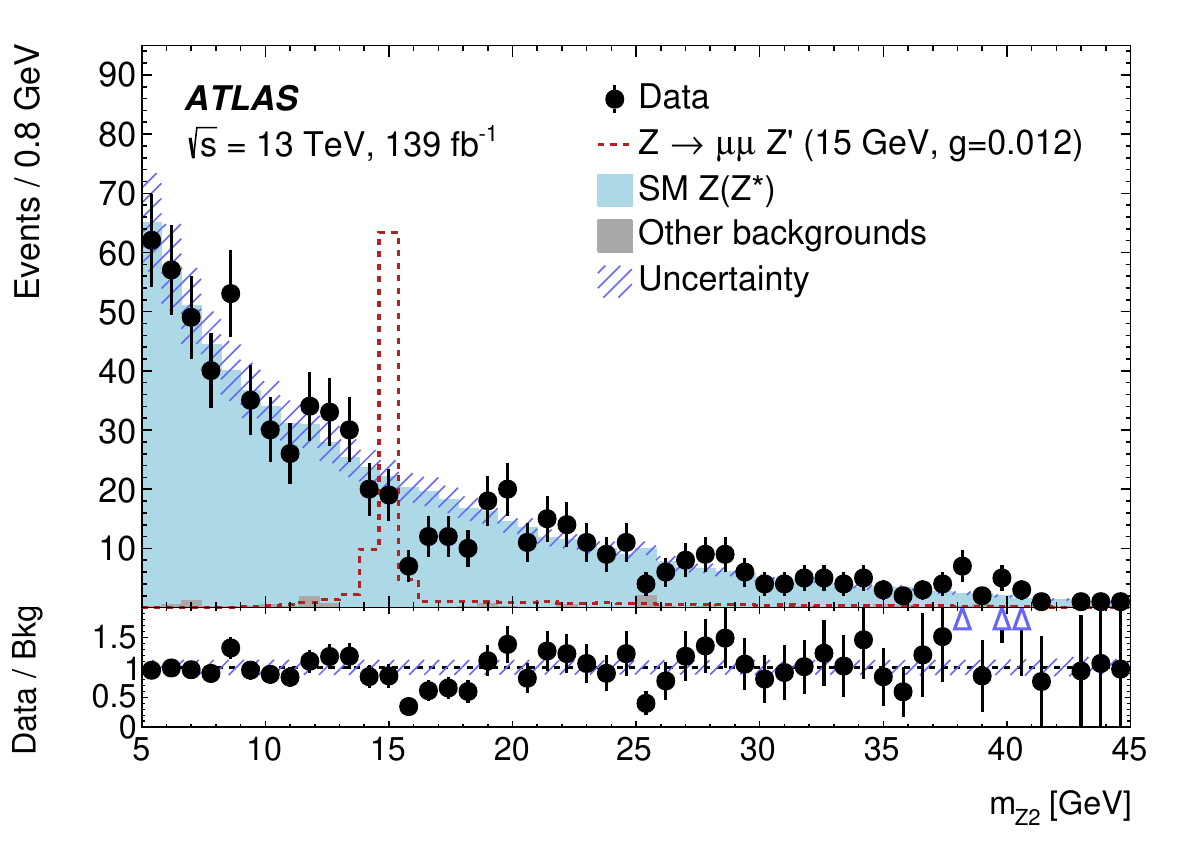}}
\qquad
\subfloat[]{\includegraphics[width=0.45\textwidth]{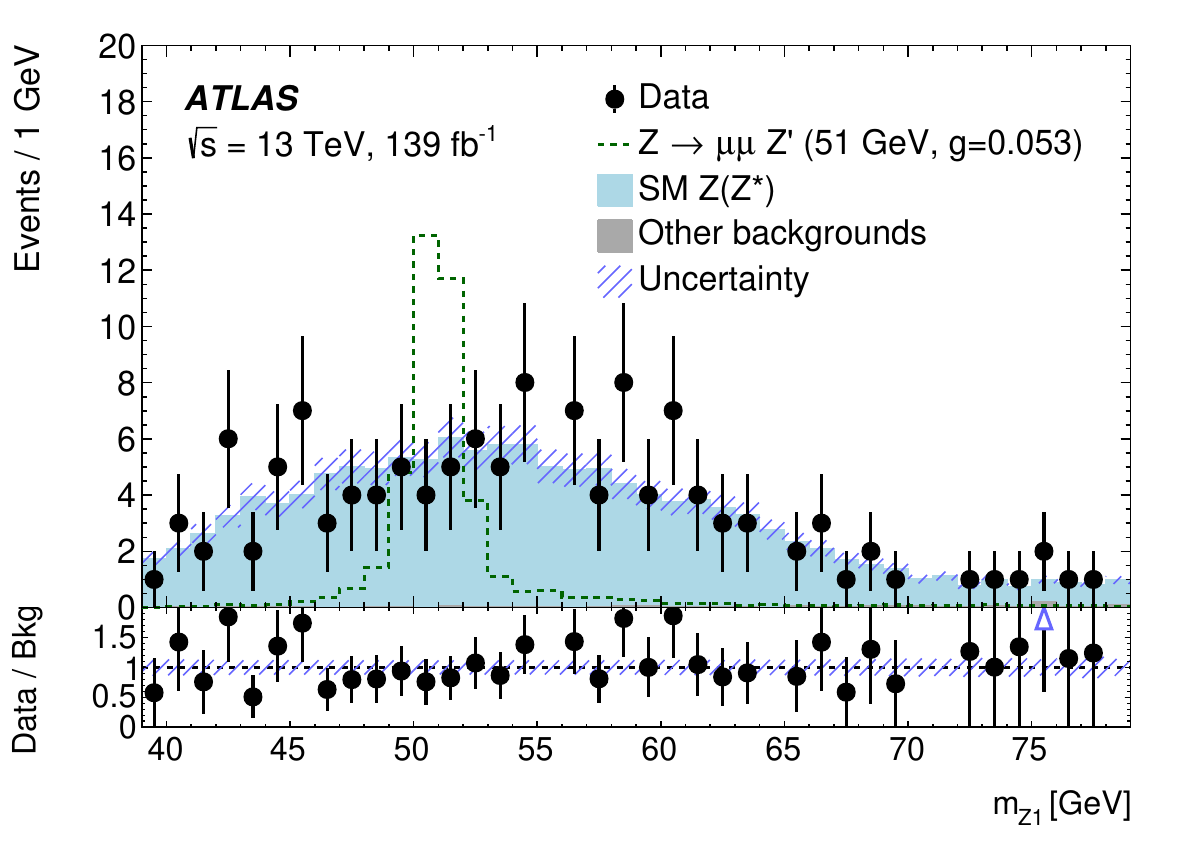}}
\end{center}
\caption{(a) Feynman diagram of \Zprime production through radiation in a Drell--Yan process, and the (b) $m_{Z_{2}}$ and (c) $m_{Z_{1}}$ spectra from the $\mu\mu\Zprime\to4\mu$ analysis~\cite{HDBS-2018-57} with signal examples at 15~\GeV and 51~\GeV, respectively.}
\label{fig:zprimemuons}
\end{figure}

Another scenario in which simple final states may prove insufficient arises when dealing with complex decay chains that follow the production of a new \Zprime or \Wprime boson. More than two objects coming from the heavy resonance must be considered in this context. The analysis described in Ref.~\cite{EXOT-2020-15} constructs a single SR by selecting events with at least one isolated lepton (electron or muon) and a pair of jets. The background is then estimated in~situ by using a parametric function to fit the data. Four fully independent statistical analyses are defined for four different invariant masses: $m_{jj\ell}$, $m_{jj\ell\ell}$, $m_{jb\ell}$, and $m_{bb\ell}$.
The largest deviation of the data from the fitted background is found in the $m_{jj\ell}$ analysis at a mass aound 1.3~\TeV and has a local significance of 3.5$\sigma$ (1.5$\sigma$ global), but otherwise the data agrees well with the predicted SM background. Figure~\ref{fig:multibody}(a) shows one of the relevant distributions.  A similar analysis tackles the simpler scenario of a dijet resonance, arising from a \Wprime or \Zprime, produced in association with a light charged lepton~\cite{EXOT-2018-32}. It follows a similar strategy and looks for excesses in the \mjj distribution of such events. In this case, no significant disagreement with the SM prediction is found.

A fourth analysis of multi-body final states, which considers fully hadronic scenarios with two pairs of jets arising from a single heavy resonance, was also completed recently~\cite{EXOT-2022-18}. This analysis reconstructs events with two pairs of jets that are required to have similar invariant masses, corresponding to a scenario of the type $Y\rightarrow XX \rightarrow jjjj$, and uses a single SR. Additional kinematic requirements are imposed to reduce the multijet background. Two different distributions are used for the statistical analysis: the average dijet invariant mass and the invariant mass of the tetra-jet system. A total of six independent statistical analyses are defined for those two variables and three ranges of the $\alpha$ parameter, defined as the ratio of those masses. The background is obtained by fitting a parametric function directly to the data in the SR in those six scenarios, and good agreement is found. One of the distributions is shown in Figure~\ref{fig:multibody}(b).

\begin{figure}[tb]
\begin{center}
\subfloat[]{\includegraphics[width=0.45\textwidth]{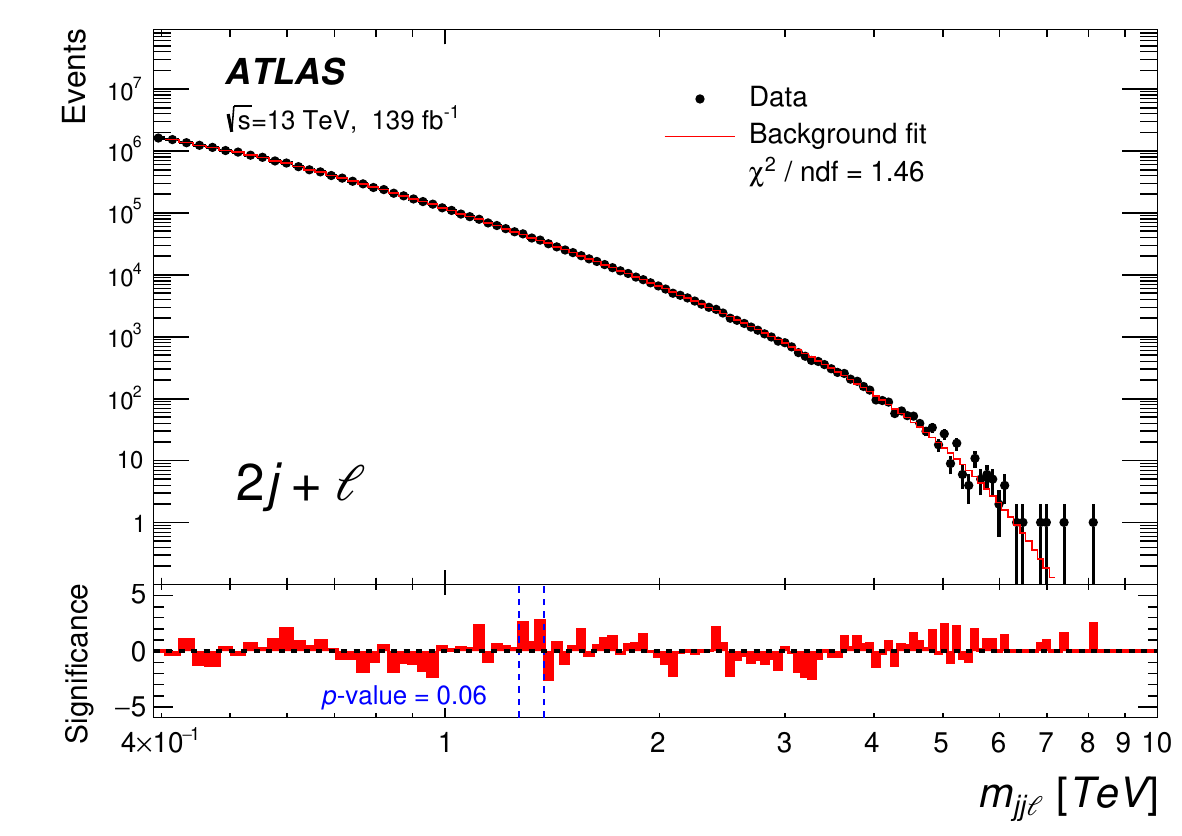}}
\subfloat[]{\includegraphics[width=0.45\textwidth]{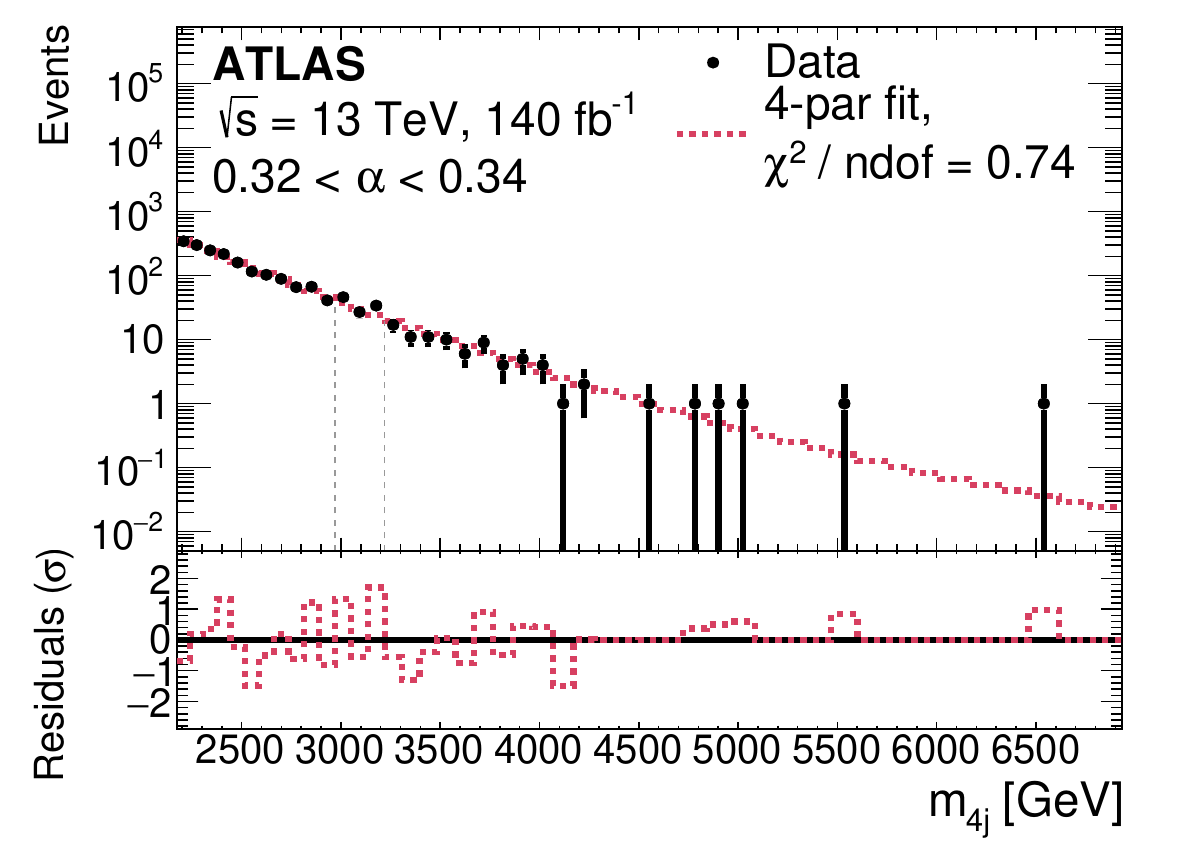}}
\end{center}
\caption{Distributions of two of the relevant invariant masses in the (a) multi-body search~\cite{EXOT-2020-15} and (b) the $Y\rightarrow XX \rightarrow jjjj$ analysis~\cite{EXOT-2022-18}. The background estimate obtained using a parametric fit is shown in both cases.}
\label{fig:multibody}
\end{figure}

\subsubsection{Summary of lower limits on the masses of \Zprime and \Wprime resonances}

For most of the analyses described in this section, limits are set on the production of \Zprime or \Wprime bosons. Table~\ref{tab:Vprime} summarizes the lower limits placed on the mass in the various models.

\renewcommand{\arraystretch}{1.2}
\begin{table*}[htp!]
\begin{center}
\caption{95\% CL lower mass limits obtained in various analyses for different models predicting new heavy vector bosons $\Wprime$ and \Zprime, whose decay modes and branching ratios are determined by the model or the model parameters  (see the text for more details).  }
\begin{tabular}{l c c c}
\hline
\hline
$V'$ & Analysis final state & Observed lower limit on $m_{V'}$ [$\text{T\electronvolt}$] & Section \\\hline
$\Zprime_\mathrm{SSM}$			& $bb$			& 2.7 ~\cite{EXOT-2019-03} & \ref{sec:resqqll} \\
& $ee+\mu\mu$	& 5.1 ~\cite{EXOT-2018-08} & \ref{sec:resqqll} \\\hline
$\Wprime_\mathrm{SSM}$		& $qq$			& 4.0 ~\cite{EXOT-2019-03} & \ref{sec:resqqll} \\
& $e\nu+\mu\nu$	& 6.0 ~\cite{EXOT-2018-30} & \ref{sec:resqqll} \\
& $\tau\nu$		& 5.0 ~\cite{EXOT-2018-37} & \ref{sec:resqqll} \\\hline
$\Zprime_{\psi}$				& $ee+\mu\mu$	& 4.5 ~\cite{EXOT-2018-08} & \ref{sec:resqqll} \\\hline
$\Zprime_\mathrm{TC2}$			& $\ttbar$			& 3.9 ~\cite{EXOT-2018-48} & \ref{sec:ttortbres} \\\hline
$\Zprime_\mathrm{LUV}$ & $b\bar{b}b\bar{b}$ & \phantom{0}1.45 ~\cite{EXOT-2018-09} & \ref{sec:ttortbres}  \\\hline
$\Wprime_{R} (g'/g=1.0)$ & $tb$ & 4.6  ~\cite{EXOT-2021-36} &  \ref{sec:ttortbres} \\\hline
$W'_\mathrm{HVT}$ (model A)	& $WZ\to XXqq$		& 3.9 ~\cite{HDBS-2018-10} & \ref{sec:dibosonres} \\\hline
$W'_\mathrm{HVT}$ (model B)	& $WZ\to XXqq$		& 4.3 ~\cite{HDBS-2018-10} & \ref{sec:dibosonres} \\\hline
$W'_\mathrm{HVT}$ (model C)	& $WZ\to \ell\nu \ell\ell$   & 0.34 ~\cite{HDBS-2018-19} & \ref{sec:dibosonres} \\\hline
$Z'_\mathrm{HVT}$ (model A)	& $WW\to \ell\nu qq$   & 3.5 ~\cite{HDBS-2018-10} & \ref{sec:dibosonres} \\\hline
$Z'_\mathrm{HVT}$ (model B)	& $WW\to \ell\nu qq$   & 3.9 ~\cite{HDBS-2018-10} & \ref{sec:dibosonres} \\\hline
\hline
\end{tabular}
\label{tab:Vprime}
\end{center}
\end{table*}
\renewcommand{\arraystretch}{1.0}

\subsection{Contact interactions}
\label{sec:gauge_contact}

Searches complementary to those for resonances, and looking for broad excesses in the tails of invariant mass distributions, are performed as a test for a four-fermion contact interaction (CI) approximating new physics at a scale $\Lambda$ with coupling $g^*$. The value of $\Lambda$ can be much higher than the $pp$ collision centre-of-mass energy, extending the sensitivity of the LHC to mass scales well beyond its direct reach.

In the dilepton final state~\cite{EXOT-2019-16}, the same selection criteria as the ones reported for the resonant search (see Section~\ref{sec:resqqll}) are used, and the background is fitted with the same functional form, but only considering the data in a low \mll region which forms the CR. The fitted function is then used to extrapolate to the expected number of background events in a single inclusive bin at high \mll which forms the SR. The boundaries of the CR and the SR are optimized for sensitivity for each of the four scenarios considered: a CI with constructive or destructive interference with the SM, in the $ee$ and $\mu\mu$ channels. An example is shown in Figure~\ref{fig:CI}(a) for the $ee$ constructive interference channel, where a slight but non-significant excess is seen. Lower limits are set on $\Lambda$ for various assumptions about the chirality of the quarks and leptons in the $qq\ell\ell$ interaction. Combining the two lepton channels, they are as high as 35.8 (28.8)~\TeV in the constructive (destructive) interference scenario.

A more specific $bs\ell\ell$ CI, inspired by lepton-flavour anomalies~\cite{Afik:2018nlr}, is the target of another search~\cite{EXOT-2018-16}, with independent $ee$ and $\mu\mu$ spectra drawn for the cases in which there is no $b$-tagged jet or exactly one $b$-tagged jet in the event, the latter being shown in Figure~\ref{fig:CI}(b) for the dimuon case. The background in this case is estimated using CRs, built at lower \mll values for the $Z$+jets background and by selecting two \btagged jets for the top-related backgrounds. As in the leptonic \Wprime search discussed in Section~\ref{sec:resqqll}, some fit-based extrapolations are needed to populate the higher-mass bins of the background samples.  The fake leptons are estimated using a matrix method. Values of $\Lambda/g^*$ smaller than 2.0 (2.4)~\TeV are excluded in the $ee$ ($\mu\mu$) channel.

\begin{figure}[tb]
\begin{center}
\subfloat[]{\includegraphics[width=0.45\textwidth]{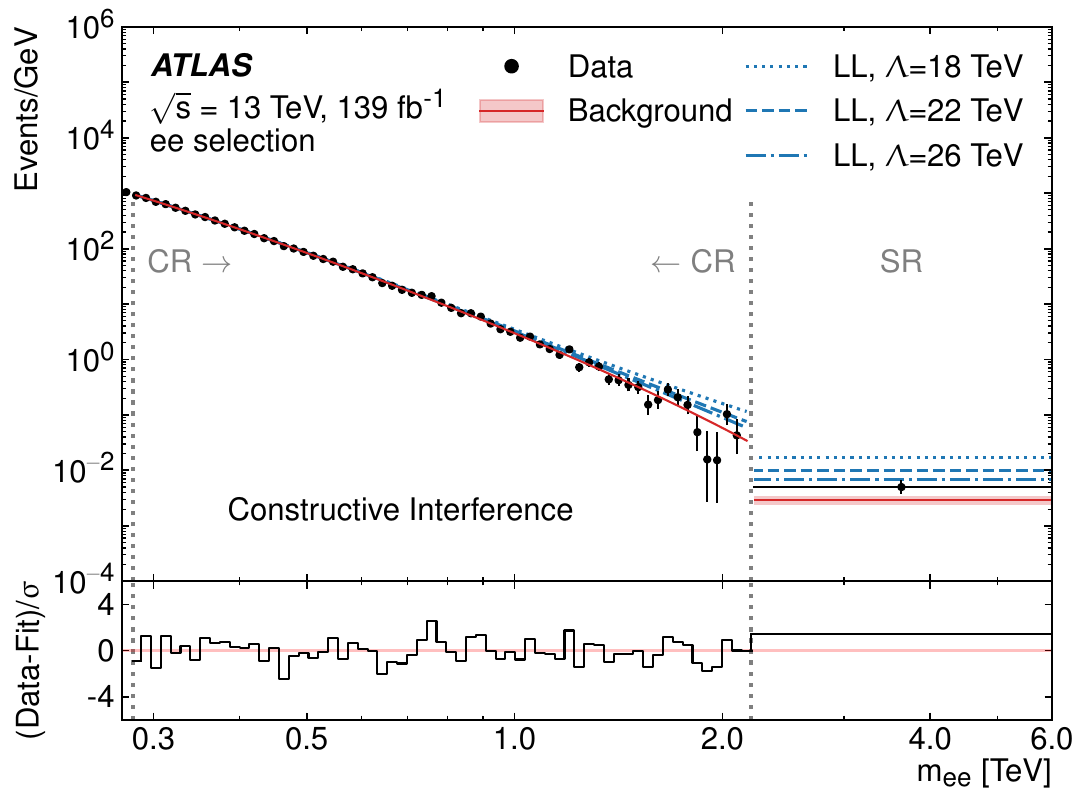}}
\subfloat[]{\includegraphics[width=0.55\textwidth]{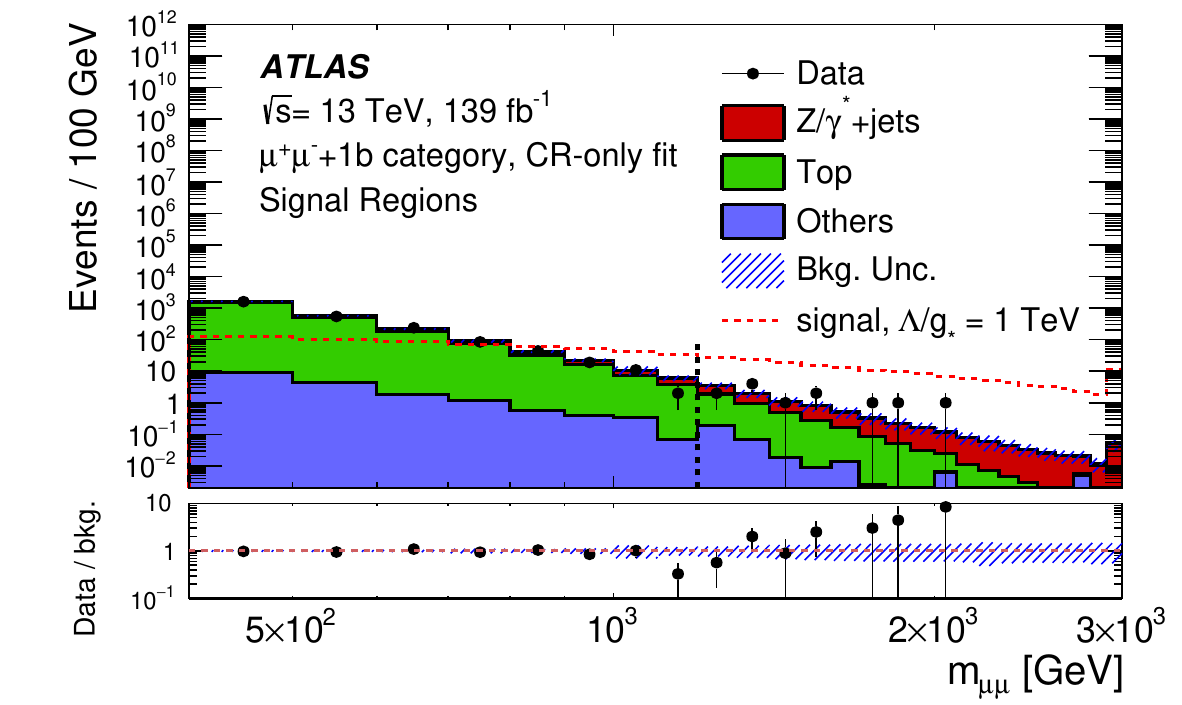}}
\end{center}
\caption{Distribution of \mll in the CR (where the data is fitted) and the SR (where the background estimate is an extrapolation from the fit) of the $qq\ell\ell$ CI search~\cite{EXOT-2019-16}, shown for (a) the $ee$ channel with constructive interference, and (b) the $bs\ell\ell$ search~\cite{EXOT-2018-16} in the $\mu\mu$ channel with one \btagged jet, where the region to the right of the dashed vertical line needs some fit-based extrapolation for the top background. }
\label{fig:CI}
\end{figure}

A similar approach can be followed to construct a non-resonant hadronic search~\cite{EXOT-2016-21}, which was completed using a partial \RunTwo dataset recorded between 2015 and 2016.



%
\section{Charged-lepton flavour violation}
\label{sec:lfv}

As discussed in Section~\ref{sec:leptons}, neutrinos oscillate, indicating that lepton flavour violation (LFV) does occur in nature. However, no LFV has ever been seen for processes involving charged leptons, even though no fundamental principles forbid it. Charged-lepton flavour transitions mediated by neutrino oscillations should occur, but their predicted rate is negligibly small in the SM (e.g.\ ${\cal B}(Z\to e\mu)<4\times10^{-60}$).

\subsection{Lepton flavour violation in $Z$ boson decay}
\label{sec:lfvz}
In some BSM theories, LFV rates can be significantly enhanced by interactions involving new particles such as heavy neutrinos~\cite{Illana:2000ic}. It is therefore interesting, given the abundance of $Z$ bosons produced at the LHC, to look for LFV decays of the $Z$ boson as a probe for new physics, as was done in the $Z\to e\mu$ channel~\cite{EXOT-2018-35} and the $Z\to e\tau$ and $Z\to\mu\tau$ channels, with the $\tau$-lepton decaying leptonically~\cite{EXOT-2020-28} or hadronically~\cite{EXOT-2018-36}.
The limits obtained in these channels are summarized in Table~\ref{tab:lfv}.

In the $e\mu$ final state~\cite{EXOT-2018-35}, the invariant mass of the $Z$ boson candidate can be reconstructed fully from the visible leptons: the search can thus look for a narrow signal peak in the invariant mass distribution of the oppositely charged leptons, restricted to the range $70<m_{e\mu}<110$~\GeV. The main background comes from leptonic decays of $\tau$-leptons in the $Z\to\tau\tau$ process, misidentification of a muon as an electron in $Z\to\mu\mu$ events, or dileptonic decays in \ttbar or diboson production. While the \ttbar background is reduced by rejecting events with a $b$-tagged jet, jets with $\pT>60$~\GeV, or $\met>50$~\GeV, the main background reduction strategy is to use a BDT to select events in the signal region. This BDT is based on the \pT of the leading jet (when there is one), the $\met$, and the transverse momentum of the electron--muon system, $\pT^{e\mu}$. Control regions with same-flavour or same-sign lepton pairs are also used to help constrain some of the background contributions. The resulting invariant mass distribution in the SR is shown in Figure~\ref{fig:lfv}(a), where the data is seen to agree well with the background predictions.
A binned fit of this distribution gives an $e\mu$ branching ratio value that is consistent with zero within uncertainties, the systematic uncertainties being dominated by the statistical uncertainty of the simulated $Z\to\tau\tau$ and $Z\to\mu\mu$ processes.

\begin{figure}[b]
\begin{center}
\subfloat[]{\includegraphics[width=0.36\textwidth]{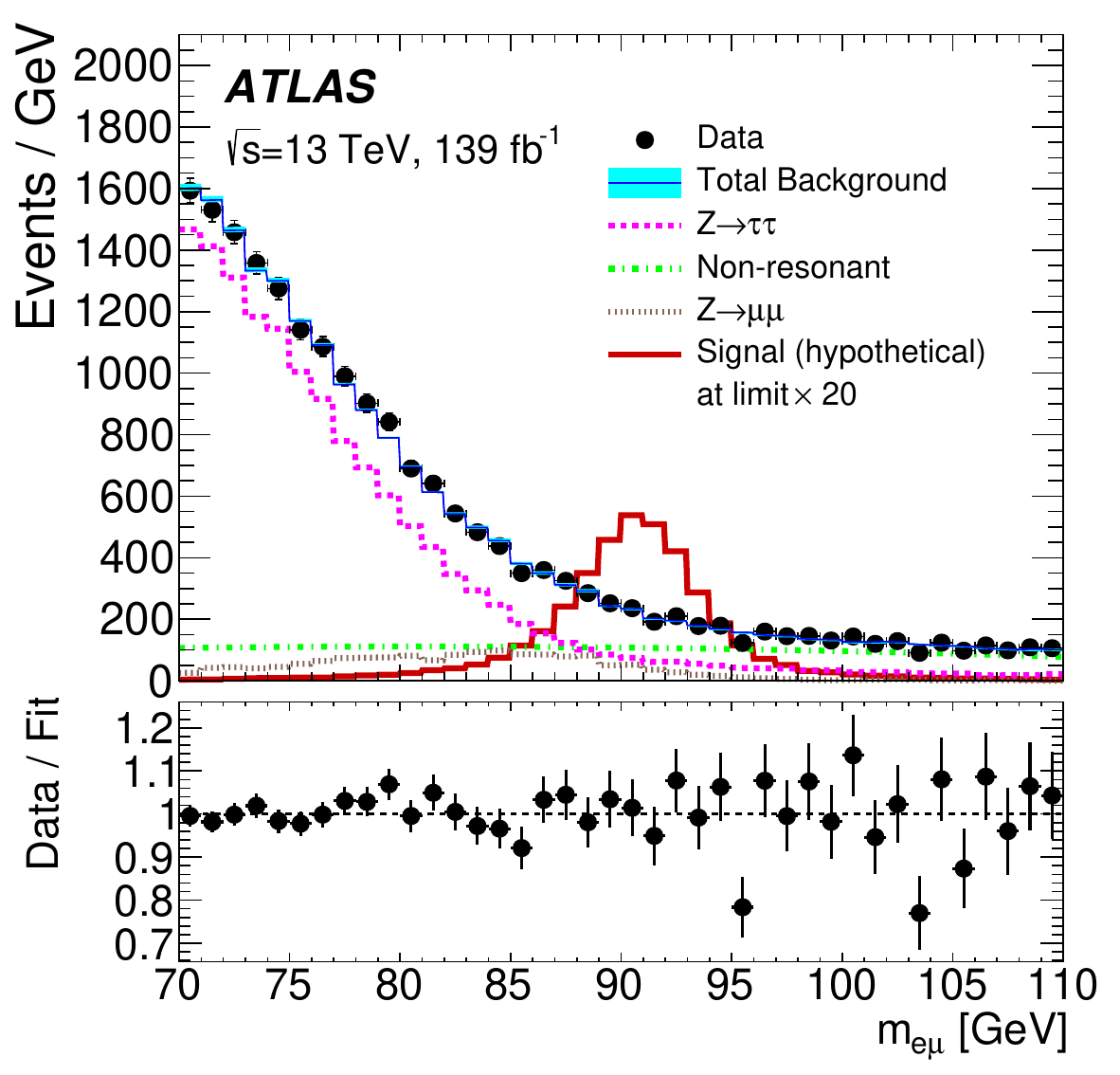}}
\subfloat[]{\includegraphics[width=0.3\textwidth]{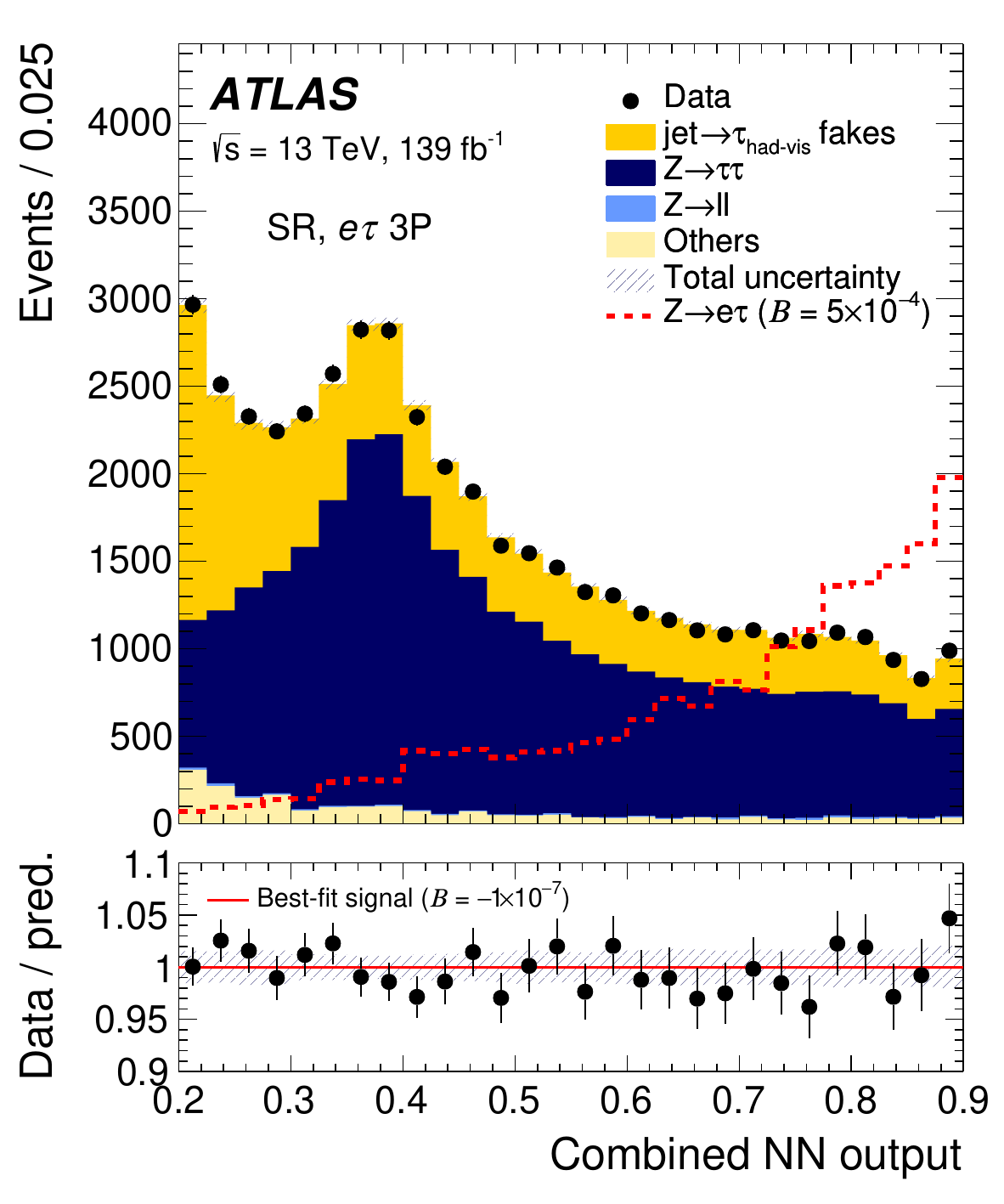}}
\subfloat[]{\includegraphics[width=0.3\textwidth]{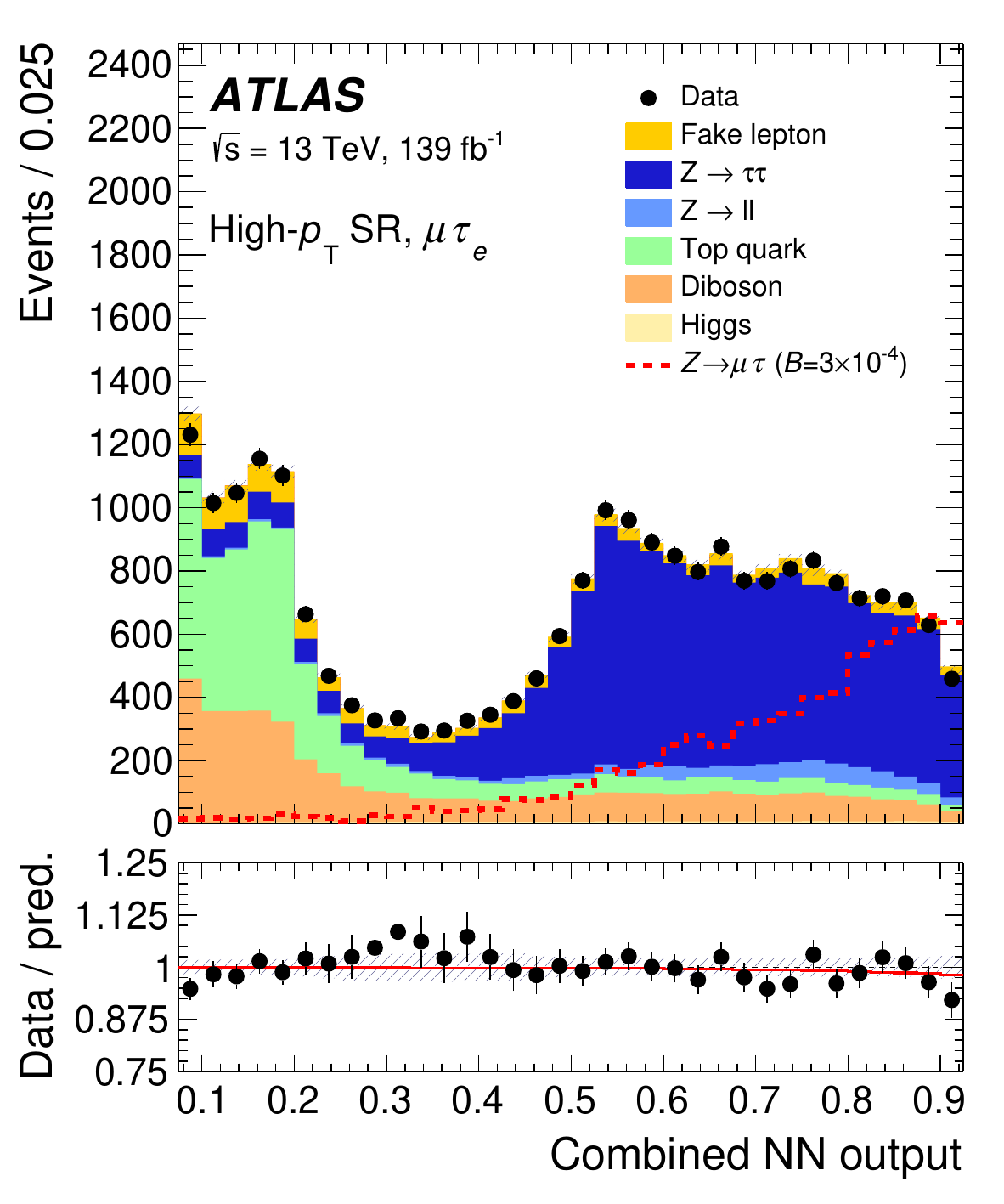}}
\end{center}
\caption{Observed distribution of the final discriminating variables compared to the expected background and example LFV $Z$ signals: (a) the $e\mu$ invariant mass in the $Z\to e\mu$ search~\cite{EXOT-2018-35}, and the combined NN score in the (b) $e\tau$(3P) SR of the $Z\to\ell\tauhad$ search~\cite{EXOT-2018-36} and in the (c) $\mu\tau_e$ SR of the $Z\to\ell\tau_{\ell'}$ search~\cite{EXOT-2020-28}.}
\label{fig:lfv}
\end{figure}

In the $\ell\tauhad$ channel~\cite{EXOT-2018-36}, where $\ell=e,\mu$ and the $\tauhad$ can be either one-pronged (1P) or three-pronged (3P), the $\ell$ and $\tauhad$ are required to have opposite electric charges.
The value of $\mT(\tauhadvis,\met)$ is required to be less than 35~\GeV to remove $Z\to\tau\tau$ and $W$+jets backgrounds, $m_{\ell\tauhadvis}$ must be above 60~\GeV to be compatible with a $Z$ boson decay, and no $b$-tagged jets must be found in the events to reject top-quark-related backgrounds. After these selections, three NN classifiers for each of the four channels ($(\ell=e,\mu)\times(\tauhad=\text{1P,\,3P})$) are used to discriminate further against $W$+jets, $Z\to\tau\tau$ and $Z\to\ell\ell$ background processes, respectively. They use low-level inputs such as $\ell$ and \tauhadvis kinematic variables and the \met, but also higher-level inputs which are properties of the $\ell{-}\tauhadvis{-}\met$ system, such as $m_{\ell\tau}^\mathrm{col}$. To estimate the remaining $Z$+jets background in the SR, two CRs are used: one for the $Z\to\tau\tau$ background (built by reversing the \mT and $m_{\ell\tauhadvis}$ requirements and requiring $m_{\ell\tau}^\mathrm{col}$ to be compatible with the $Z$ boson mass), and one for the $Z\to\ell\ell$ background (built by reversing the corresponding NN output requirement). A fake-factor method is used to estimate the remaining background in which a $\tauhad$ is falsely identified. The distribution of the final discriminating variable in the $e\tau$(3P) SR, the combined NN score, is shown in Figure~\ref{fig:lfv}(b) along with an example expected signal. Since the data does not deviate significantly from the expected background in the SR, limits are set on the LFV $Z$ boson branching ratios, combining them with the \RunOne results~\cite{HIGG-2015-09}.

These results can be improved significantly by also examining the $\ell\tau_{\ell'}$ channel~\cite{EXOT-2020-28}, in which the leptonic $\tau_{\ell'}\to\ell'+2\nu$ decay is considered instead. In this search, the two light leptons $\ell$ and $\ell'$ are required to have different flavours and opposite electric charges and no $\tauhad$ candidate must be found, in order to remain orthogonal to the $\ell\tauhad$ channel. As in the hadronic channel, a $b$-jet veto is imposed, and similar variables, based on the leading and subleading leptons ($\ell_0$ and $\ell_1$), are used to reject backgrounds: $\mT(\ell_1,\met)<35$~\GeV and $m_{\ell_0\ell_1}>40$~\GeV. Furthermore, $|\Delta\phi(\ell_0,\met)|>1$ is required and, in the $\mu\tau_e$ channel, the electron \pT reconstructed from its track must be compatible with the transverse energy reconstructed from its cluster, i.e.\ $\pT^\mathrm{track}/\ET^\mathrm{cluster}<1.1$, to reject muons faking electrons. The $m_{\ell\tau}^\mathrm{col}$ variable is also used here, built with the same collinear assumptions for the $2\nu$ system as made for the single $\nu$ in the hadronic channel. The main background comes from dileptonic $Z\to\tau\tau\to\ell\ell'+4\nu$, with some contribution from $\ttbar$ and diboson production. Three NN classifiers are built to discriminate against these, and their combined score\footnote{The combined score is computed as $1-\sqrt{(1/3)\Sigma_{i=1}^{3}(1-\mathrm{NN}_{i})^2}$, where NN$_i$ are the individual scores.}
is used to select the events in the SR. A $Z\to\tau\tau$ CR is built by reversing its corresponding NN score, and a fake-factor method is used to estimate misidentified-lepton background. The combined NN score used as the final discriminant in the $\mu\tau_e$ channel is shown in Figure~\ref{fig:lfv}(c), where it is compared with the predicted background and an LFV signal. Since no excess in data is seen in this search, limits are set on the $Z$ LFV branching ratios, combining them with the hadronic channel limits.

\begin{table*}[tb]
\begin{center}
\caption{95\% CL upper limits on the LFV branching ratio of the $Z$ boson, in the various $\ell\ell'$ channels.}
\begin{tabular}{l c}
\hline
\hline
Channel & Upper limit on  ${\cal B}(Z\to \ell\ell')$  \\\hline
$e\mu$~\cite{EXOT-2018-35} & $2.62\times10^{-7}$  \\\hline
$e\tau$ (\tauhad~\cite{EXOT-2018-36} and $\tau_{\ell'}$~\cite{EXOT-2020-28} channels combined) & $5.0\phantom{0}\times10^{-6}$ \\\hline
$\mu\tau$ (\tauhad~\cite{EXOT-2018-36} and $\tau_{\ell'}$~\cite{EXOT-2020-28} channels combined) & $6.5\phantom{0}\times10^{-6}$ \\
\hline
\hline
\end{tabular}
\label{tab:lfv}
\end{center}
\end{table*}

\subsection{Lepton flavour violation in the decay of a heavy \Zprime}
A new gauge boson \Zprime, as introduced in Section~\ref{sec:gauge}, could also have LFV decays. The three final states $e\mu$, $e\tauhad$ and $\mu\tauhad$ were therefore probed at high mass in a search for such a signal~\cite{EXOT-2019-20}, by requiring that the invariant mass of different-flavour, opposite-sign leptons be above 600~\GeV and that these leptons be back-to-back in the transverse plane. Since the \tauhad in this search are more boosted than the ones in the $Z$ boson LFV search discussed in the previous section, the collinear approximation used to assign a four-momentum to the neutrino from the $\tau$ decay is even more justified here and is used to reconstruct the invariant mass, significantly improving its resolution.

The irreducible background in each channel is dominated by \ttbar and $WW$ production. In the SR, the \ttbar background is reduced by rejecting events with $b$-tagged jets, a criterion which is reversed to build a CR for this background. A $WW$ CR is built by reversing the $\Delta\phi(\ell,\ell')$ selection in the $e\mu$ channel. In the channels involving $\tauhad$, such $WW$ CRs would suffer from too large a contamination from fake leptons and are not used. Instead, the correction factor $k$ that needs to be applied to the $WW$ simulation in a given $\tauhad$ channel is extrapolated from the one obtained in the $e\mu$ channel, by multiplying it by the ratio of the correction factors obtained in the \ttbar CRs: $k^{e(\mu)\tauhad}_{WW}=k^{e\mu}_{WW} \times k^{e(\mu)\tauhad}_{\ttbar}/k^{e\mu}_{\ttbar}$, this extrapolation being applicable due to lepton-flavour universality. The reducible fake-lepton backgrounds are estimated using either a matrix method in the $e\mu$ channel or dedicated CRs enriched in $W$+jets events in the $e\tauhad$ and $\mu\tauhad$ channels.

In each of the three channels, the total uncertainty is dominated by the statistical precision, and a binned profile-likelihood fits is performed on  $m_{\ell\ell'}$. Since the data agree with the expected background, limits are set on the \Zprime mass, assuming an SSM benchmark augmented with one non-zero LFV $\ell\ell'$ coupling at a time, which is taken to be the same as the corresponding $\ell\ell$ coupling. The observed (expected) lower limits placed on $m_{\Zprime}$ are 5.0 (4.8), 4.0 (4.3) and 3.9 (4.2)~\TeV for decay into $e\mu$, $e\tau$, and $\mu\tau$, respectively.



%
\section{Hidden sectors leading to long-lived neutral particles}
\label{sec:hidden}

In models with a hidden (or \emph{dark}) sector, new particles which are neutral under the SM gauge groups are introduced; these particles can potentially interact between themselves via new interactions in which the SM particles are neutral. The SM and dark particles thus exist in parallel and can only communicate through a mediator (or \emph{portal}): this mediator can, for example, be the SM Higgs boson or a new scalar, pseudoscalar, vector or axial-vector particle carrying a double SM--dark charge (i.e.\ a \emph{bi-fundamental} mediator) or mixing with a SM particle with small probability. These models can often offer candidates to explain the nature of dark matter, which is the focus of Section~\ref{sec:dm}. The present section instead focuses on hidden sectors for which the very small coupling to the SM through the portal of interest leads to neutral long-lived particles (LLPs) which decay at some point in the detector. This  leads to very distinctive signatures which are sometimes more akin to detector noise or beam-induced background than to SM processes from the $pp$ collisions. Decays of the Higgs boson into such LLPs are predicted in multiple BSM theories, such as some models with neutral naturaleness~\cite{Burdman:2006tz,Curtin:2015fna}. Here, neutral LLPs produced through a Higgs or Higgs-like portal are the subject of four different searches: for dark-photon jets~\cite{EXOT-2019-05, EXOT-2022-15}, for a long-lived pseudoscalar via displaced decay vertices in the inner detector~\cite{EXOT-2018-57}, and for a long-lived scalar via displaced jets in the calorimeter~\cite{EXOT-2019-23} or in the muon spectrometer~\cite{EXOT-2019-24}.

\subsection{Search for dark-photon jets}
\label{sec:dpj}
The long-lived particle can be a light dark photon ($\gamma_\text{d}$) which would have a small but non-zero value of the parameter $\epsilon$ for mixing with the SM photon. It would therefore decay into leptons and light quarks with branching ratios and a lifetime determined by its mass and $\epsilon$. They would be pair-produced through a Higgs boson decay, either directly as in the Hidden Abelian Higgs Model (HAHM)~\cite{Curtin:2014cca}, or through a more complex chain involving other dark-sector particles as in the Falkowski--Ruderman--Volansky--Zupan (FRVZ) model~\cite{Falkowski:2010cm,Falkowski:2010gv} (see Figure~\ref{fig:dpj}(a)). Due to their small mass, dark photons would be highly boosted: their decays would be seen as jet-like structures composed of a collimated group of fermions, called dark-photon jets. Three Higgs boson production modes are considered and combined: gluon--gluon fusion (ggF) and $WH$~\cite{EXOT-2019-05}, and VBF~\cite{EXOT-2022-15}.

If the displaced $\gamma_\text{d}$ decays into a pair of muons, it can be seen as a muonic dark-photon jet ($\mu$DPJ), i.e.\ collimated stand-alone muon spectrometer (MS) tracks which are not matched to a prompt muon and are not close to a jet.  Cosmic-ray muons in time coincidence with a $pp$ collision can be an important source of background for $\mu$DPJs: a DNN is trained to discriminate between simulated signal and a cosmic-ray-enriched dataset collected during empty bunch crossings, using as inputs the MS track position, direction and timing. A displaced $\gamma_\text{d}$ decaying into electrons or quarks can be seen, for an appropriate range of lifetimes, as a calorimeter DPJ (caloDPJ), i.e.\ a jet with an unusually low EMF, the fraction of its total energy that is deposited in the EM calorimeter. To select caloDPJs efficiently, the cleaning selections which are usually applied to jets are relaxed, e.g.\ by removing the usual JVT or EMF criteria. Other selections are applied, such as a timing requirement to remove cosmic-ray muons or beam-induced background (BIB). Multijet events can also mimic caloDPJs: a \enquote{QCD} NN is therefore trained on MC signal and background events, using as inputs the shape of the calorimeter energy deposits associated with the jet. These inputs are also used to train another NN against a BIB-enriched dataset, collected with a dedicated trigger. The reconstruction of $\mu$DPJs and caloDPJs is illustrated in Figure~\ref{fig:dpj}(b).

Events without leptons are then selected for the ggF~\cite{EXOT-2019-05} and VBF~\cite{EXOT-2022-15} categories, while events with one lepton are used for the $WH$ category~\cite{EXOT-2019-05}. In the ggF and VBF SRs, dedicated $\mu$DPJ triggers are used in combination with \met (VBF) or  low-EMF (ggF) triggers, while lepton triggers are used for $WH$. In the ggF SRs, two DPJs are required. Besides the use of the NNs mentioned above, selections are made on the azimuthal separation $\Delta\phi_\mathrm{DPJs}$ of the DPJs and their timing, and they are required to be ID-isolated: the scalar sum of the \pT  ($\Sigma\pT$) of all inner-detector (ID) tracks found close to the DPJ direction must be small. In the $WH$ SRs, the events are required to have at least a minimum amount of \met and \mT,
and no $b$-tagged jet. Events are further separated into those having only one caloDPJ (in which case the \mT requirement is increased
to reduce the $W$+jets background), two or more caloDPJs, or a mix of caloDPJ and $\mu$DPJ. Besides the use of the QCD NN, timing, low-JVT and low-width selections on the caloDPJ, an upper limit is placed on the minimum $\Delta\phi(\mathrm{DPJ},\met)$ value.
In the VBF SRs, the special production topology is selected by requiring a pair of highly energetic jets separated by a significant gap in $\eta$, resulting in a large value of \mjj, besides requiring some \met and vetoing on the presence of leptons or \btagged jets. This reduces the SM background relative to the other channels, allowing selected events to have a minimum of only one DPJ, which must be ID-isolated and central, and either be neutral (for a $\mu$DPJ) or have a good NN score (for a caloDPJ).

For each SR, an ABCD method is used to estimate the background. Possible signal leakage from the SR into the other three regions is accounted for by a simultaneous fit of the signal and background in all regions. In the ggF SRs, the planes are formed by the larger $\Sigma\pT$ value of the two DPJs, and either $\Delta\phi_\mathrm{DPJs}$ or the QCD NN score, the latter also being used for the caloDPJ VBF search. In the $WH$ SRs, they are based on $\min{\Delta\phi(\mathrm{DPJ},\met)}$ and the minimum QCD NN score. For the VBF $\mu$DPJ search, the ABCD plane uses the charge of the $\mu$DPJ versus its ID isolation.

In all SRs, the data are compatible with the expected background within statistically dominated uncertainties. Limits on the FRVZ model obtained by combining the ggF, $WH$ and VBF categories are shown in Figure~\ref{fig:dpj}(c) in the $\epsilon$ versus $m_{\gamma_\text{d}}$ plane for various $H\to2\gamma_\text{d}$ branching ratios: for $m_{\gamma_\text{d}}<2m_\mu$, the sensitivity drops because $\mu$DPJs do not contribute anymore, while above this range the structures seen in the displayed limits depend on the $\gamma_\text{d}$ branching ratios, as decays into QCD resonances significantly alter the sensitivity. The figure also shows the complementarity of that search to a previous \RunOne ATLAS search for prompt dark photons~\cite{EXOT-2014-09}, which was able to exclude larger couplings, and also to the jet+\met analysis that is discussed in Section~\ref{sec:monoj}, which is able to rule out lower couplings at intermediate masses for a large branching ratio, and to non-ATLAS searches~\cite{Ilten:2018crw} which exclude lower masses/couplings when some assumptions are made about the interactions.

\begin{figure}[tb]
\begin{center}
\subfloat[]{{\includegraphics[width=0.4\textwidth]{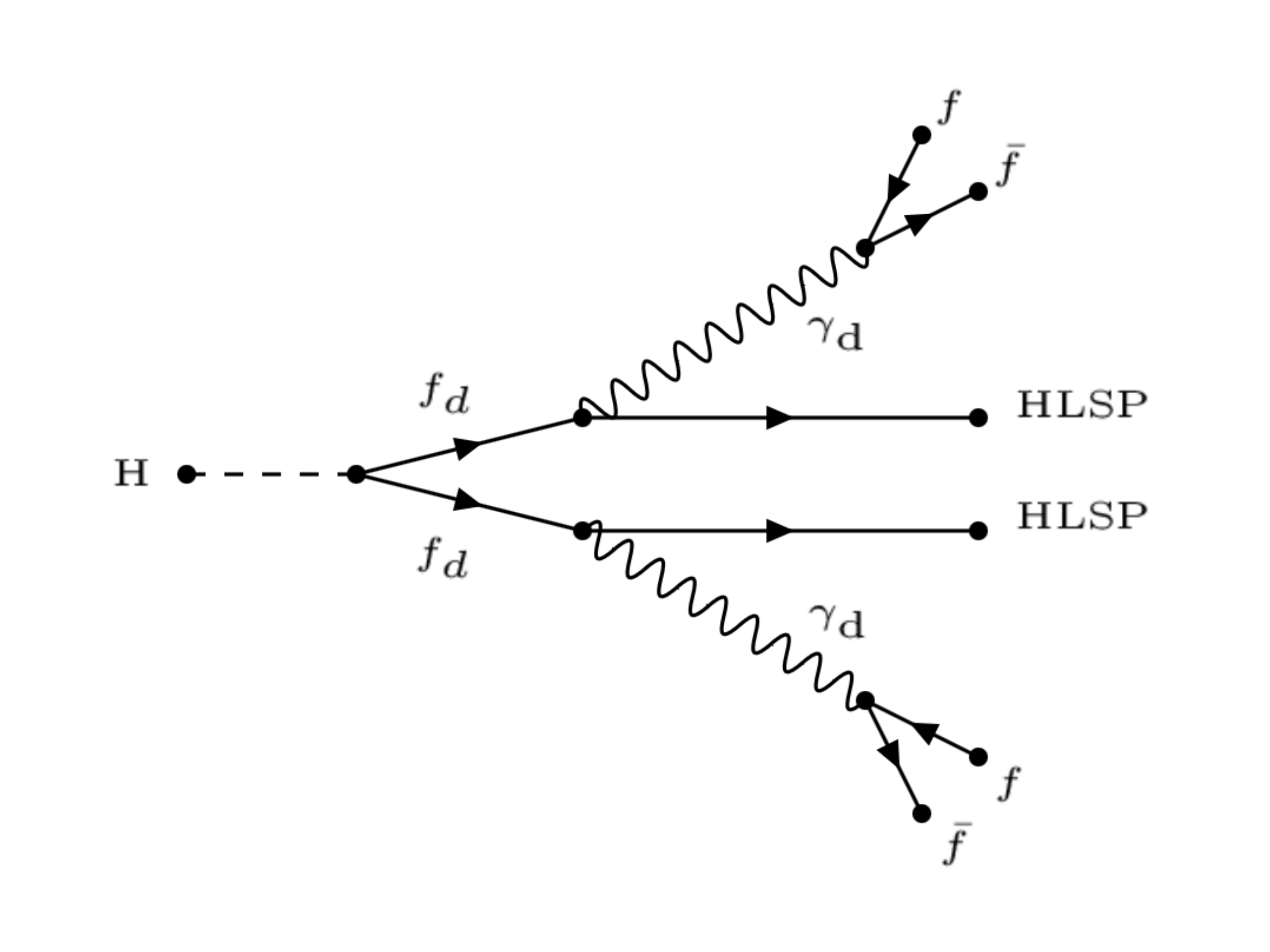}}}
\qquad
\subfloat[]{\includegraphics[width=0.5\textwidth]{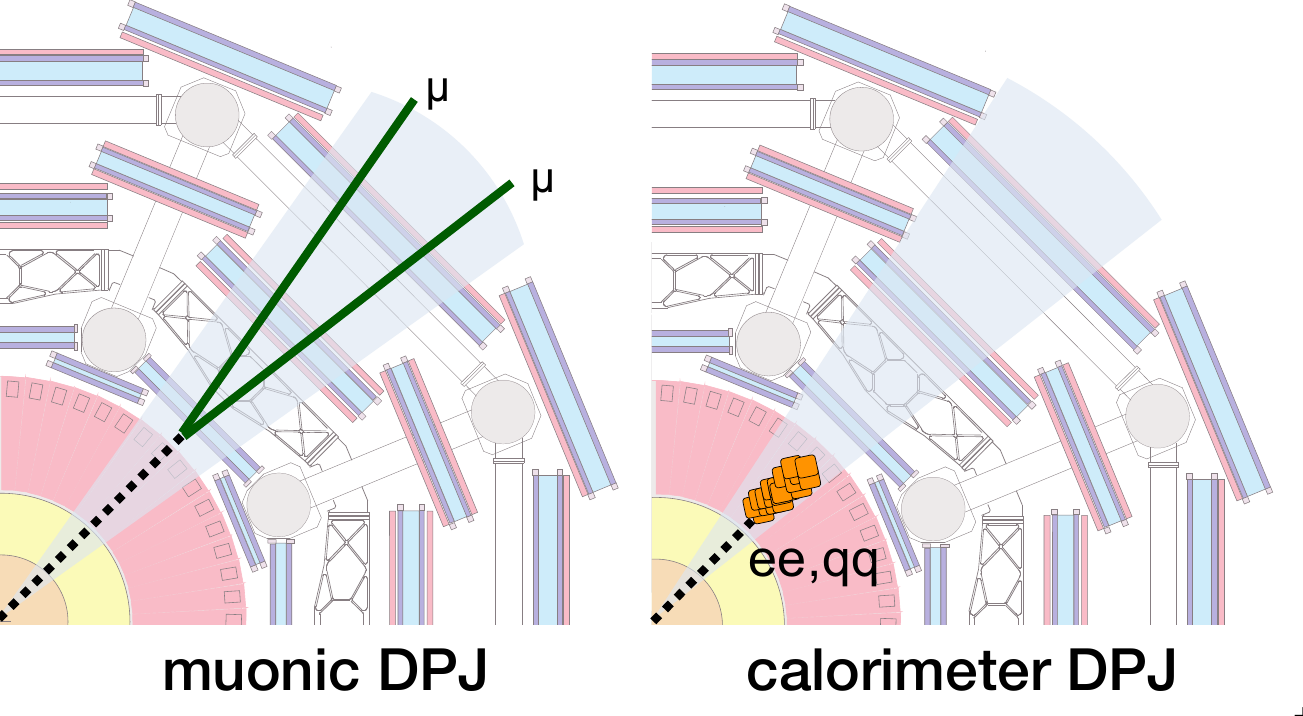}}
\qquad
\subfloat[]{\includegraphics[width=0.5\textwidth]{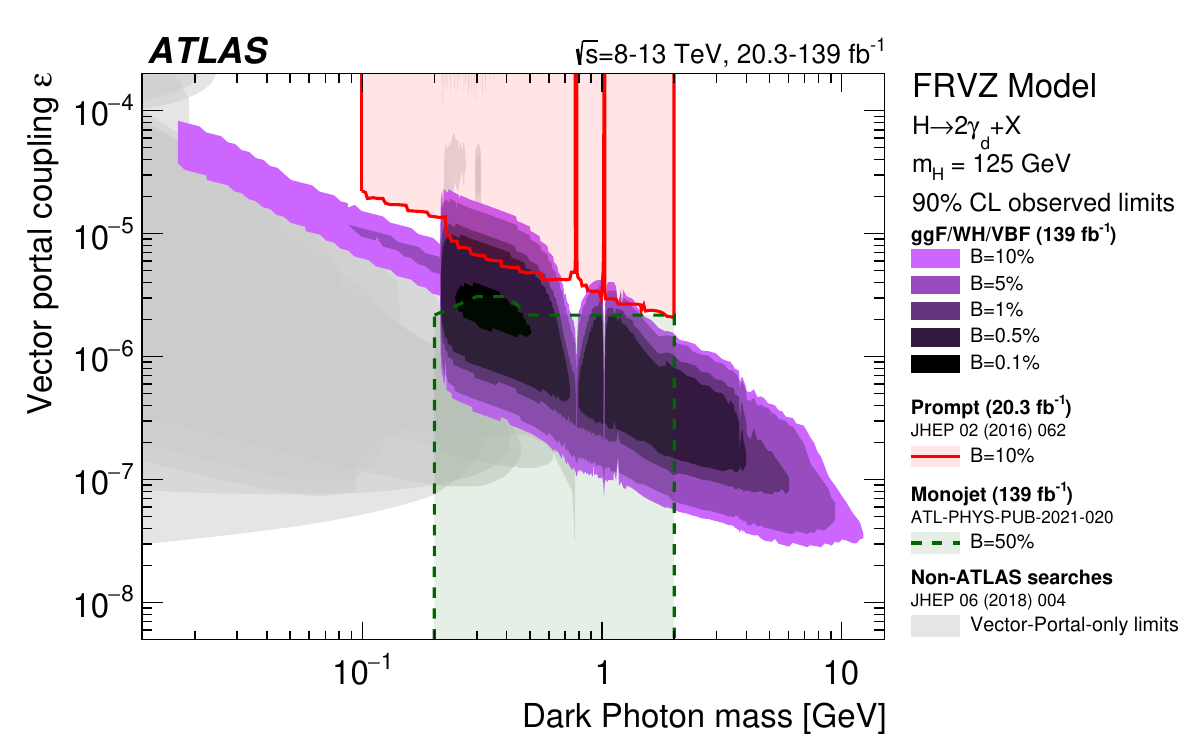}}
\end{center}
\caption{(a) Example of dark-photon production and decay in
the FRVZ model, where $f_\text{d}$ are dark-sector fermions and HLSP is the hidden lightest stable particle. (b) Illustrations of $\mu$DPJ (left) and caloDPJ (right) reconstruction. (c) Exclusion at 90\% CL in the plane of the kinetic mixing parameter $\epsilon$ versus the $\gamma_\text{d}$ mass for different values of the $H\to2\gamma_\text{d}$ branching ratio obtained in the search for dark-photon jets~\cite{EXOT-2019-05,EXOT-2022-15} in the context of the FRVZ model, compared to other exclusions (see the text).}
\label{fig:dpj}
\end{figure}

\subsection{Searches for long-lived scalars or pseudoscalars in the decay of a Higgs-like boson}

\subsubsection{Looking for decays in the inner detector}

This search~\cite{EXOT-2018-57} focuses on a simplified model in which two new long-lived pseudoscalars $a$ are produced in the decay of the Higgs boson and then decay via the $a\to b\bar{b}$ mode\footnote{This decay should be dominant when kinematically accessible if $a$ mixes with the Higgs boson and inherits its Yukawa couplings.} within the inner-detector volume, leading to DVs. In this model, the proper lifetime and mass of $a$ are free parameters. The $Z(\to\ell\ell)H$ production channel is investigated because the two leptons offer a highly efficient trigger and selection strategy. Searches for prompt $a$ decays were also pursued, using a partial \RunTwo dataset, and are reviewed in another report.%

Standard and LRT tracks (see Section~\ref{sec:typeI}) are used to reconstruct the DVs. Tracks which are too loosely associated with the DVs are removed and some quality requirements are applied to the DVs: they must be within the inner tracker but not within known detector material, have a good $\chi^2/n_\mathrm{dof}$ vertex fit, have at least three tracks including at least one with transverse impact parameter $|d_0|>3$~mm, and be within $\Delta R = 0.6$ of one of the four leading jets. Vertices due to random track crossings are suppressed by requiring $m/\Delta R_\mathrm{max}>3$~\GeV, where $m$ is the vertex mass and $\Delta R_\mathrm{max}$ is the maximal angular separation found between any given track and the combined momentum of the remaining vertex constituents when removing this track.

Events are selected by requiring two OS same-flavour leptons with an invariant mass compatible with the $Z$ boson, at least two jets, and at least two DVs matched to different jets. Furthermore, one of these two jets must have low track activity: the ratio of the \pT of all its geometrically-matched prompt tracks to its total \pT must be low, and it must be mostly geometrically matched to tracks which have a low probability of being compatible
with any $pp$ collision PV candidate $i$, based on a measurement of $\pT^{\mathrm{jet~trk} \in i} / \pT^{\mathrm{all~jet~trk}}$.
A CR is defined by reversing the DV multiplicity requirement; it is used to compute a probability for a jet to be matched to a DV, as a function of the jet kinematics and properties, and this probability is used to estimate the number of background events in the SR, which is predicted to be $1.30\pm0.08\,\text{(stat.)}\pm0.27\,\text{(syst.)}$. Zero events are observed in the SR, and limits on the branching ratio of $H\to a a \to b\bar{b}b\bar{b}$ are set as a function of the proper lifetime of $a$ for various $a$ mass scenarios, as shown in Figure~\ref{fig:Haabbbb}(a).

\subsubsection{Looking for decays in the calorimeters}

Searches are also conducted for a hidden sector which is connected to the SM via a heavy scalar boson $\Phi$ which decays into two long-lived scalar particles $s$. While $\Phi$ can be the Higgs boson, the search also considers $\Phi$ masses ranging from 60~\GeV to 1~\TeV and $s$ masses from 5 to 475~\GeV, with $s$ decaying into SM fermions. In this model the couplings of $s$ to SM fermions are determined by the Higgs boson's Yukawa couplings through mixing, so $s$ decays preferentially into the heaviest SM fermion pair which is kinematically accessible, thus usually favouring hadronic decays. The signal models considered assume gluon--gluon fusion (ggF) production of $\Phi$. An illustration of a $\Phi\to ss$ production with the long-lived $s$ decaying either in the muon spectrometer or the hadronic calorimeter is shown in Figure~\ref{fig:Haabbbb}(b) in the transverse plane of the detector.

A search for $\Phi\to ss$ with $s$ decaying mainly in the calorimeters is performed~\cite{EXOT-2019-23}, each $s$ being reconstructed as a jet which is narrow, trackless and has a low EMF value, a signature which is similar to that of the calorimetric dark-photon jets (caloDPJs) previously discussed in Section~\ref{sec:dpj}. Two SRs labelled low-\ET or high-\ET are designed, targeting a value of $m_\Phi$ below or above 200~\GeV, respectively.

The events are selected using a low-EMF jet trigger and are required to have at least two jets after applying, like in the dark photon $(\gamma_\text{d}$) search, a modified cleaning algorithm where, in this case, the jet EMF requirement is removed. Furthermore, the jets must be trackless, ensured by requiring $\Sigma \Delta R_\mathrm{min}\mathrm{(jet,tracks)}>0.5$, where the  sum runs overs selected jets which have $\pT>50$ GeV and $\Delta R_\mathrm{min}\mathrm{(jet,tracks)}$ is the angular distance between the jet and the closest PV-associated track with $\pT>2$~\GeV.

These jets are then tagged using two complex NNs (low-\ET or high-\ET) which are trained using three samples: a sample of MC signal events (low-\ET or high-\ET), a sample of SM multijet events taken from a dataset of events passing a jet trigger but failing the low-EMF trigger, and a BIB sample, collected with a dedicated trigger. The inputs to these per-jet NNs are low-level jet information concerning the jet-associated tracks, calorimeter-energy topological clusters, muon-spectrometer track segments, and general kinematics; each NN outputs three scores to classify these jets as either signal-like, multijet-like or BIB-like.

At event level, a BDT is then trained to further discriminate between BIB and signal events, separately for the low-\ET and high-\ET cases, using as input variables not only the per-jet NN scores but also event-level kinematics such as the angular distance between the signal jet candidates, or $\HT^\mathrm{miss}/\HT$ where $\HT^\mathrm{miss}$ is the magnitude of the vectorial sum of the jet \pT while \HT is the scalar sum of the jet \pT. A selection on the BDT output is then made along with other selections, including one on the jet timing, to further suppress the BIB and ensure that the only significant remaining background consists of multijet events. Once this is done, the low-\ET and high-\ET SRs are defined by final selections on $\HT^\mathrm{miss}/\HT$, the EMF of the signal jets, their \pT, and the product of the two highest per-jet NN signal scores.

An ABCD method is used to estimate the background, using $\Delta R_\mathrm{min}\mathrm{(jet,tracks)}$ and the low-\ET or high-\ET BDT event score as the uncorrelated variables. After the fit to the four regions, the 22 (23) observed data events in the low-\ET (high-\ET) SR agree well with the $18.8\pm3.5$ $(20.6\pm4.0)$ estimated background events. Limits are then placed on the cross section times branching ratio of the mediator's decay $\Phi\to ss$ as a function of the $s$ lifetime for various masses of $\Phi$ and $s$, an example of which is shown in Figure~\ref{fig:Haabbbb}(c) for the case in which $\Phi$ has a mass of 60~\GeV.

\subsubsection{Looking for decays in the muon spectrometer}
A search was also performed to look for $s$ decaying in the muon spectrometer~\cite{EXOT-2019-24}. These decays would be seen as DVs in the MS.  A dedicated algorithm is used to reconstruct the MS DVs~\cite{PERF-2013-01}. In the MS, MDT chambers consist of two sets (called multilayers) of three of four layers of drift tubes. Hits in the multilayers can form track segments, and segments coming from the two multilayers can form tracklets. Displaced vertices are formed from clusters of three tracklets in the barrel, or four in the endcaps.

After passing a dedicated MS-based trigger, events are selected by requiring at least one MS DV which must be geometrically matched to the MS cluster found by the trigger. If two trigger MS clusters are found, two MS DVs must be matched. Requirements are also made on the position in $\eta$ and the transverse decay radius $L_{xy}$ of the DV in order to reduce the background, which is dominated by \emph{punch-through} jets, i.e.\ jets which are not fully contained in the calorimeter volume and create tracks in the MS.  Since signal events are expected to have many more hits than a reconstructed MS DV in a background event, the number of hits found in the MDT and the RPC or TGC in a cone
around the DV must be high.
To reduce the background, the DVs
are also required to be isolated from any significant activity in the ID,
and from any large-EMF jets.
In the SR, at least two isolated DVs must be found, well separated in $\Delta R$.

The remaining background is estimated in a data-driven way. This background is dominated by events in which there are two isolated DVs, but only one MS cluster found by the trigger. These events are estimated by counting the number of events in a CR in which there is only one isolated DV and it is matched to the only MS cluster, and weighting them by the probability to find another, unassociated DV which is unmatched to any MS cluster. This probability is measured in a dataset selected with a zero-bias trigger, by dividing the total number of unmatched DVs by the total number of events. Background events in which there are two DVs matched to two MS clusters are also estimated in a data-driven way, but give a much smaller contribution.  In the SR, the expected background is $0.32\pm0.05$ events and zero events are observed. Limits are set in the same planes as for the calorimeter-based search, as shown in Figure~\ref{fig:Haabbbb}(c). The MS search excludes smaller branching ratios than the calorimeter-based search, but the calorimeter-based search probes lower lifetimes, as expected.

\begin{figure}[tb]
\begin{center}
\subfloat[]{\includegraphics[width=0.43\textwidth]{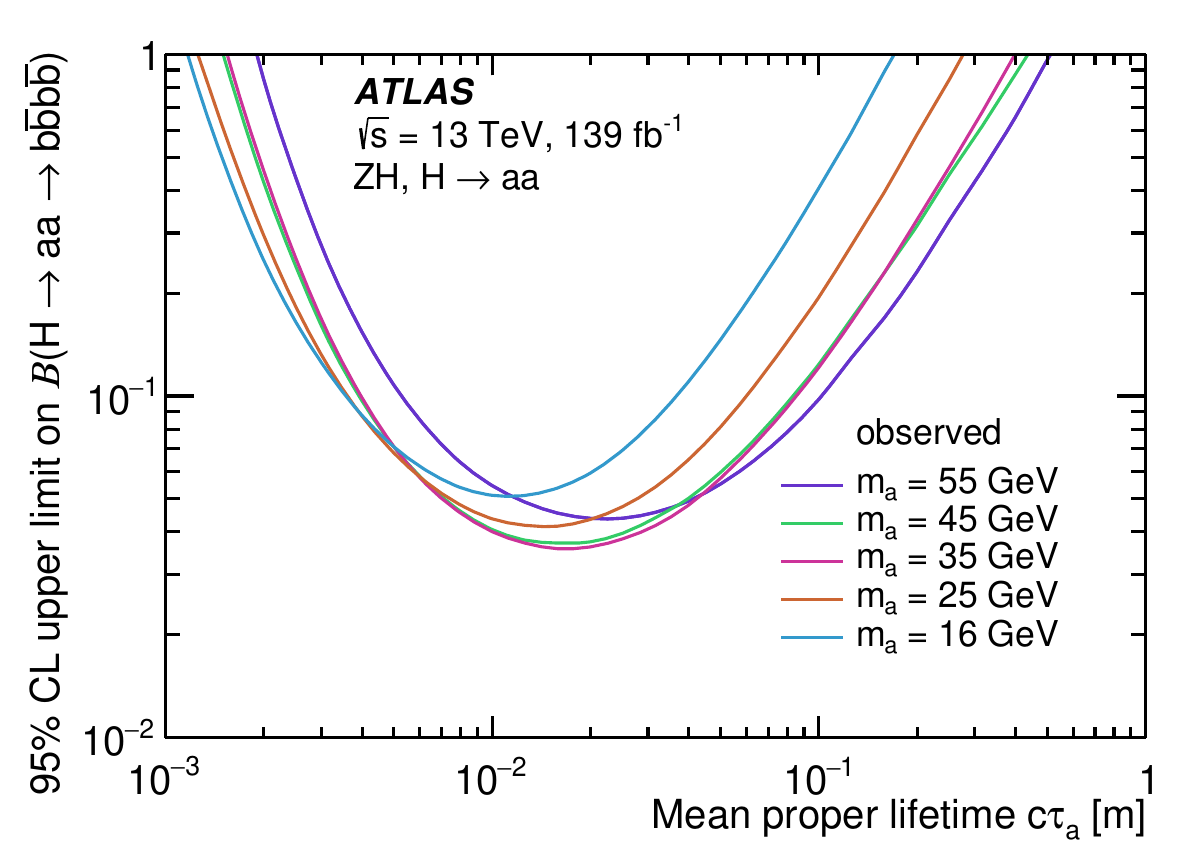}}\\
\qquad
\subfloat[]{\includegraphics[width=0.35\textwidth]{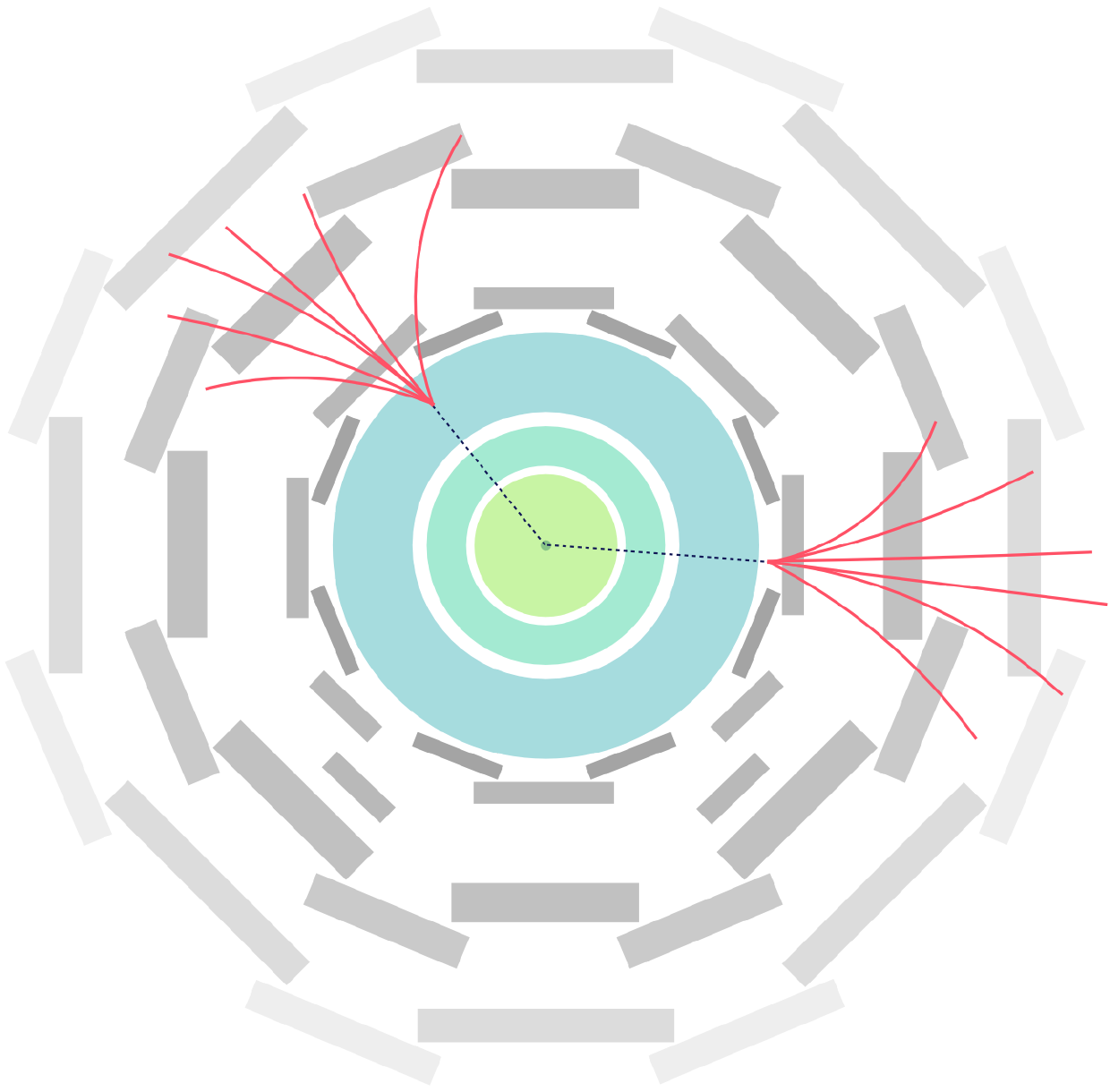}}
\qquad
\subfloat[]{\raisebox{0.2\height}{\includegraphics[width=0.55\textwidth]{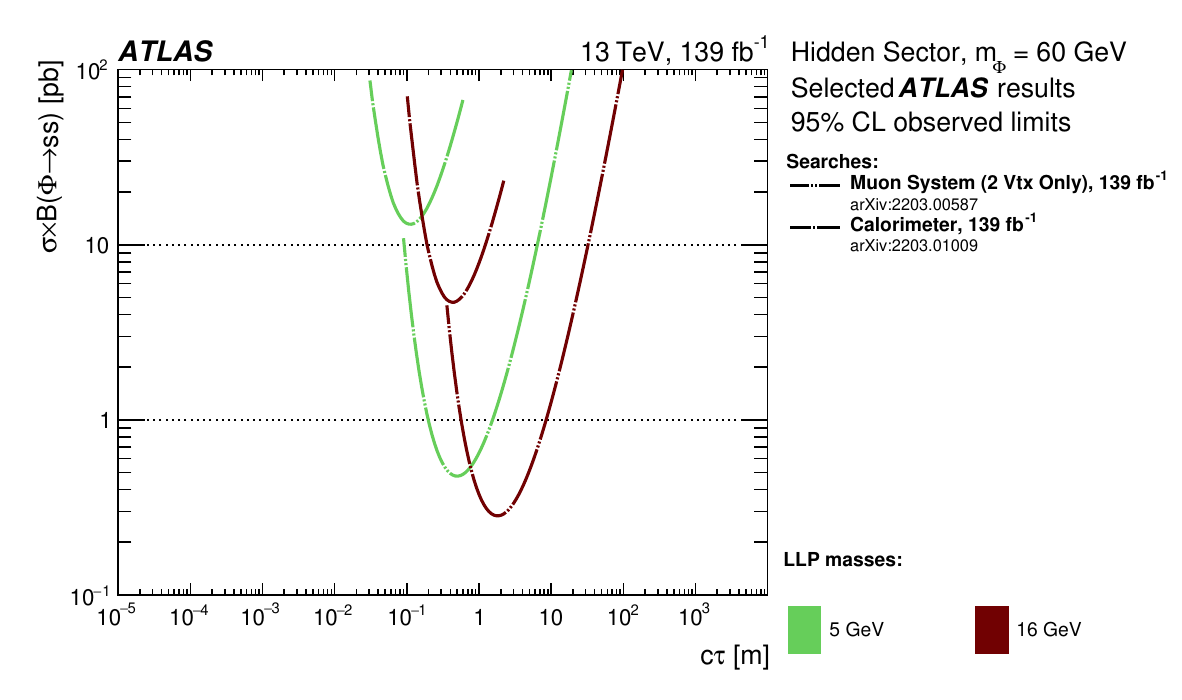}}}
\end{center}
\caption{(a) Upper limit on the branching ratio of $H\to a a \to b\bar{b}b\bar{b}$ obtained as a function of the proper lifetime of $a$ for various mass hypotheses in the displaced vertex analysis~\cite{EXOT-2018-57}. (b) Example of an LLP pair production followed by their decay, with one LLP decaying in the MS and the other one at the end of the hadronic calorimeter, with no additional detector activity. (c) Observed upper limit on the cross section times branching ratio of the mediator's decay $\Phi\to ss$ as a function of the $s$ lifetime for a $\Phi$ mass of 60~\GeV and two $s$ masses as obtained by looking for decays in the calorimeters~\cite{EXOT-2019-23} and the muon spectrometer~\cite{EXOT-2019-24}. For clarity, parts of the exclusion curves outside the most sensitive region are omitted. }
\label{fig:Haabbbb}
\end{figure}

\FloatBarrier



%
\section{Dark-matter candidates}
\label{sec:dm}

The nature of dark matter (DM), whose existence is supported by a variety of astrophysical and cosmological measurements~\cite{Corbelli:1999af,Rubin:1980,Begemann:1991,Hinshaw:2012aka,Akrami:2018vks,Trimble1987,Bertone2005,Feng2010}, remains one of the biggest puzzles in modern physics.
Should DM be a particle interacting weakly with SM particles, it could be possible to detect it in various ways: directly, via elastic scattering of the local DM by nuclei or electrons in a low-background detector~\cite{Aalbers:2022dzr,LZ:2022lsv,Akerib:2016vxi,PICO:2019vsc,PandaX-II:2021nsg,Cui:2017nnn,XENON:2022ltv,XENON:2023sxq,PhysRevLett.122.141301,PhysRevLett.123.241803,DarkSide-50:2022qzh,Angloher:2015ewa,Lai:2023qub,Amaudruz:2017ekt,Agnese:2017njq,Agnese:2017jvy}; indirectly, by the detection of their annihilation or decay products in the universe~\cite{Acharyya:2023ptu,HESS:2022ygk,MAGIC:2021mog,CTA:2020qlo,HAWC:2017udy,Aartsen:2018mxl,IceCube:2021xzo,Fermi-LAT:2016uux,TheFermi-LAT:2017vmf}; or by producing them at colliders, which is the way explored here.

Weakly interacting massive particles (WIMPs), often denoted by $\chi$, are a class of DM candidates of particular interest for the ATLAS experiment. With a DM particle mass close to the electroweak scale and an interaction strength with SM particles of the order of the weak interaction's strength, they could be readily pair-produced at the LHC. They are also cosmologically interesting as indicated by the so-called WIMP miracle~\cite{Steigman:1984ac}:  with these mass and coupling scales, the right relic density can easily be achieved via a freeze-out mechanism in the early universe.

Once pair-produced in proton--proton collisions the WIMPs, since they interact only weakly and are stable, would only be detectable through the presence of missing transverse momentum: thus, to be detected, some visible particles must also be produced in the interaction in order to measure this \pt imbalance. These visible particles can come from a chain of decays ending in the WIMP particles, such as in R-parity-conserving supersymmetric scenarios~\cite{Martin:1997ns,Farrar:1978xj,Goldberg:1983nd,Ellis:1983ew}, or from initial-state radiation (ISR) in a simplified model of DM pair production, to name but two examples.

This section focuses on searches guided by simplified models of DM, produced through a vector, axial-vector, pseudoscalar or Higgs portal, largely following the work of the DM Forum/LHC DM Working Group~\cite{Abercrombie:2015wmb,Boveia:2016mrp,Albert:2017onk,LHCDarkMatterWorkingGroup:2018ufk}. While in Run 1 most DM searches were based on an effective field theory approach in which the DM candidates are produced in pairs through a contact interaction, this approach is not valid when the typical momentum transfer in the collisions reaches the scale of the interaction; the simplified models considered in Run 2 alleviate this validity issue. The constraints placed on these models by some of the DM searches introduced here are also compared with those from direct-detection experiments. A more exotic model, in which DM is a composite stable particle of a strongly interacting hidden sector is also discussed. A stable WIMP which can be a DM candidate often features in supersymmetric models that conserve R-parity; a thorough review of ATLAS searches for supersymmetry can be found elsewhere~\cite{SUSY-2023-10}.

\subsection{Vector or axial-vector portal}
\label{sec:dmV}
In order to search for DM at the LHC, some interactions between DM particles and SM particles must be assumed. Here, the focus is on a simplified model in which the DM candidate is a Dirac fermion and the DM--SM interactions are due to a new U(1) symmetry under which they are charged. Five free parameters are considered: the mass of the new vector ($\Zprime_\mathrm{V}$) or axial-vector ($\Zprime_\mathrm{A}$) mediator, $m_\Zprime$, the mass of the DM candidate, $m_\chi$, and the coupling of the \Zprime to the quarks, $g_q$, to the DM candidate, $g_\chi$, and to the charged and neutral leptons $g_\ell$, where the couplings $g_q$ and $g_\ell$ are assumed to be universal in flavour. The width of the mediator is taken to be the minimum width allowed given the couplings and masses.

The searches looking for such V/A-mediated DM pair production then rely on the presence of an ISR gluon~\cite{EXOT-2018-06}, photon~\cite{EXOT-2018-63}, or $Z$ boson~\cite{HIGG-2018-26}, as shown in Figure~\ref{fig:jetmet}(a). While the first two searches are presented in this section, the $Z$ boson one is instead discussed in Section~\ref{sec:Zmet}, as its impact in the Higgs-portal model is more important than in this ISR-based model. Since $g_q$ must be non-zero for the LHC to produce DM, and because $g_\ell$ could also be non-zero, the resonant searches (see Figure~\ref{fig:jetmet}(b)) presented in Section~\ref{sec:gauge} also put interesting constraints on these models, as shown in Section~\ref{sec:complDM}.

\subsubsection{The jet+\met analysis}
\label{sec:monoj}
Since gluon emission dominates ISR production, the jet+\met search~\cite{EXOT-2018-06} has a very broad range of applicability which also probes other DM models such as the one in the invisible-Higgs search (see Section~\ref{sec:hinv}), or exotic models related to gravity (see Section~\ref{sec:gravity}), amongst others. The analysis, based on a \met trigger, requires the presence of at least one high-\pT small-$R$ jet along with a large value of \met.
Although this analysis is often dubbed \enquote{mono-jet}, this is not a faithful description as the selection does allow up to three additional jets
to allow for
extra radiation, and to reduce the associated modelling systematic uncertainties. The selected jets are required be well separated from the \met direction to reduce the background due to mismeasured multijet events, with a larger separation required in events with the least \met, which are more problematic.
Finally, events are vetoed if they contain identified electrons, photons, muons or $\tau$-leptons. To increase the sensitivity, the SR is binned in \met.

The main SM background comes from \Znunujets, along with a significant contribution of \Wlnujets events in which the lepton is not identified.
Much smaller contributions come from \Zlljets, top-related and diboson processes; the multijet and non-collision backgrounds are determined in a fully data-driven way, but are found to contribute at a level of at most 1.2\% in the lowest \met bin.
Five CRs are built
by reversing the lepton veto, requiring exactly one electron or muon with an \mT value compatible with a $W$ boson,
for the two \Wlnujets CRs and the combined \ttbar CR, or exactly two electrons or muons with an \mll value compatible with the $Z$ boson
for the \Zlljets CRs. By either vetoing or requiring events with a \btagged jet, the \Wlnujets CRs are separated from the \ttbar CR. Further selections based on the \met and jets are applied in the $1e$ CRs to suppress the multijet background.
Finally, the \met distribution of these backgrounds in the SR, arising from neutrinos or unidentified charged leptons, is mimicked by computing an \met proxy ($\pT^{\mathrm{recoil}}$), effectively treating the charged leptons as invisible.

Instead of using the $Z$+jets and $W$+jets CRs separately to constrain their respective backgrounds, the uncertainties are reduced by using the four $W$ and $Z$ CRs  simultaneously to constrain the main \Znunujets background. This is made possible by a careful study of the correlations between the QCD corrections to the $Z$+jets and $W$+jets processes, with dedicated high-order QCD and electroweak parton-level predictions~\cite{Lindert2017} provided separately for $W$+jets, \Zlljets and \Znunujets as a function of the vector boson's \pT. These are used to reweight the \SHERPA[2.21] $Z$+jets and $W$+jets MC event samples, leading to a good description of their respective and relative $\pT^{\mathrm{recoil}}$ distributions. Because of this, the simultaneous binned-likelihood fit of the five CRs needs only three background normalization factors to adjust the MC-based expectations to the data across all bins: a common one for the $W$+jets and $Z$+jets backgrounds, one for the \ttbar background and one for the single-top background. This advanced background estimation method leads to much reduced systematic uncertainties, at the level of 1.5\%--4.2\%, depending on the bin. The dominant remaining uncertainties, besides those in the $V$+jets predictions, are those related to the electron, muon and jet identification and reconstruction efficiencies. The resulting $\pT^{\mathrm{recoil}}$ distribution in the SR after a simultaneous fit to all regions is shown in Figure~\ref{fig:jetmet}(c). Good overall agreement between the data and the background prediction is obtained, so limits can be set, as is shown in Section~\ref{sec:complDM}.

\begin{figure}[tb]
\begin{center}
\subfloat[]{\raisebox{0.35\height}{\includegraphics[width=0.19\textwidth]{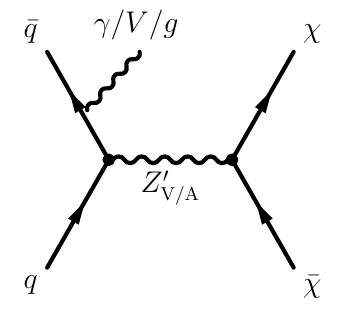}}}
\qquad
\subfloat[]{\raisebox{0.35\height}{\includegraphics[width=0.19\textwidth]{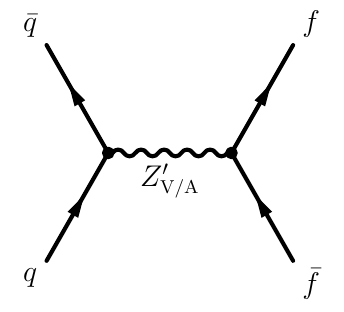}}}
\qquad
\subfloat[]{\includegraphics[width=0.52\textwidth]{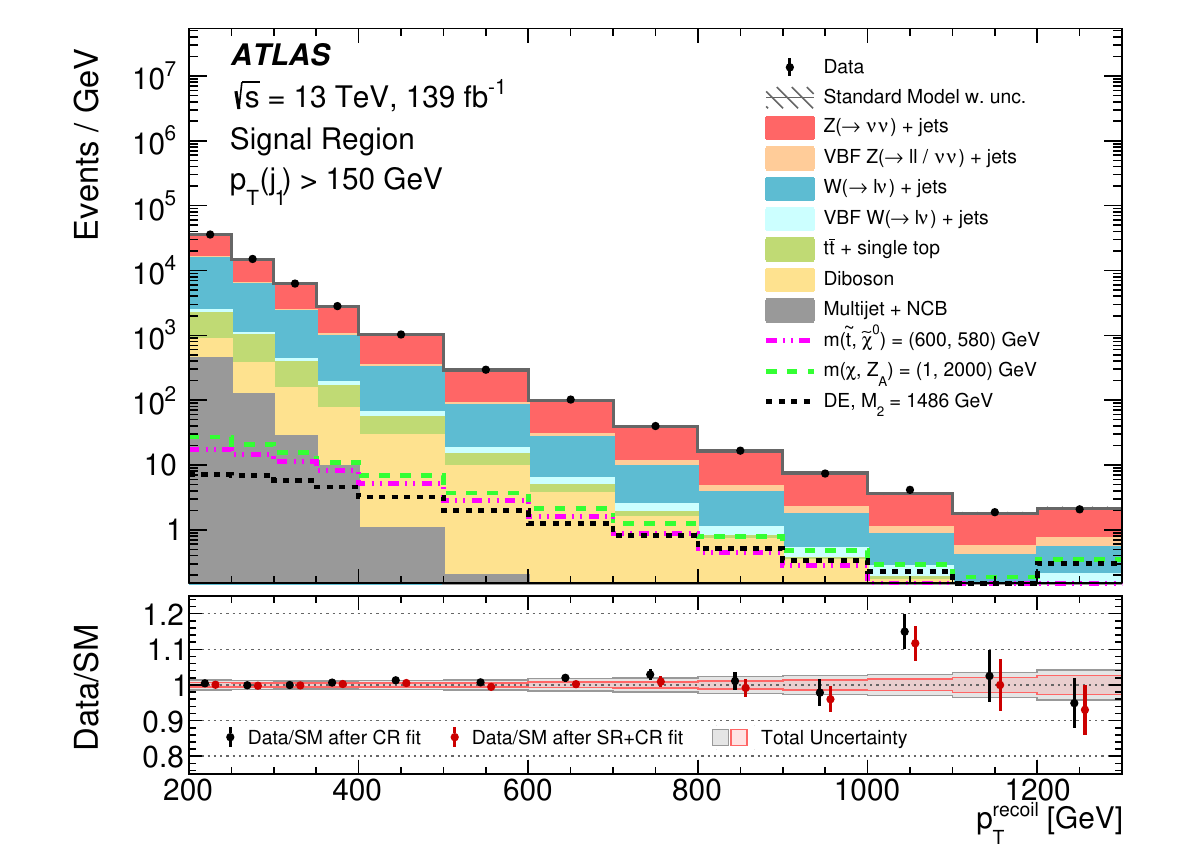}}
\end{center}
\caption{Production, through a V/A portal in the $s$-channel, of (a) a DM pair along with ISR or (b) a fermion pair, and (c) measured distribution of $\pT^{\mathrm{recoil}}$ in the SR of the jet+\met search~\cite{EXOT-2018-06} compared to the expected background and some exotic-model signals (including an A-portal DM model with $(g_q,g_\chi,g_\ell)=(0.25,1,0)$).}
\label{fig:jetmet}
\end{figure}

\subsubsection{The $\gamma{+}\met$ analysis}
\label{sec:monogam}

This analysis~\cite{EXOT-2018-63} looks for a final state in which there is a well-identified high-\pT photon, on which the trigger is based, and a large and significant amount of \met in a direction well separated from the photon (and an eventual extra jet). The photon must loosely point back to the PV to suppress the non-collision background, and a veto on leptons ($e,\mu,\tau$) is imposed. The SR is binned in \met to improve the sensitivity. The main backgrounds stem from a $Z(\to\nu\nu)$ boson, or a $W(\to\ell\nu)$ boson where the lepton is missed, produced in association with a photon. Smaller contributions are also expected from $\gamma$+jet events, especially in the lower \met bin, and from electrons or jets falsely identified as photons. While the fake-photon backgrounds are estimated in a fully data-driven way, $V\gamma$ CRs are constructed by inverting the lepton veto, and a $\gamma$+jet CR is defined mainly by reversing the \met requirement. Good agreement between data and the expected background is seen in the SR, within uncertainties dominated by the statistical precision.

\subsubsection{Complementarity of ATLAS searches}
\label{sec:complDM}

The results of the $X{+}\met$ and resonant searches are complementary for the V/A mediators~\cite{ATL-PHYS-PUB-2022-036}, as can be seen in Figure~\ref{fig:sumDMV}, which shows the observed limits obtained for the individual analyses in the $m_\chi$ versus $m_\Zprime$ plane for four benchmarks as recommended in Ref.~\cite{Albert:2017onk}. The first two benchmarks are for an A mediator which is either leptophobic, $(g_q,g_\chi,g_\ell)=(0.25,1,0)$, or has a small lepton coupling, $(g_q,g_\chi,g_\ell)=(0.1,1,0.1)$, while the two last benchmarks are for a V mediator which is leptophobic, $(g_q,g_\chi,g_\ell)=(0.25,1,0)$, or has a much smaller coupling to leptons, $(g_q,g_\chi,g_\ell)=(0.1,1,0.01)$. The lepton-coupling scenarios differ between the A and the V models, because in the case of a pure vector, the lepton coupling can be much smaller than the quark coupling, if for example the mediator couples to quarks and DM at tree level but couples to leptons at loop-level through mixing with SM gauge bosons.

For sufficiently large couplings of the \Zprime to quarks or to leptons, the resonant searches are able to exclude the parameter space up to mediator masses of around 3.6~\TeV, as the branching ratio of the \Zprime to visible states dominates the sensitivity. When these couplings are lowered while maintaining a relatively large DM coupling, the region excluded by constraints from resonant searches starts to shrink, mainly covering the parameter space for which the mediator cannot decay on-shell into DM, thus making the visible final state much more likely. For the $\Zprime\to qq$ scenarios, the lowest mediator masses are excluded by previous dijet resonant searches performed on a partial \RunTwo dataset; they avoided the jet trigger threshold of the high-mass dijet resonance search by either doing the analysis at trigger level~\cite{EXOT-2016-20} or requiring the presence of an additional ISR object in order to satisfy the trigger~\cite{EXOT-2018-05,EXOT-2017-01}. Conversely, the $X{+}\met$ searches cover the region where the on-shell decay into DM is possible, and in that case are also able to cover very low mediator masses, which is difficult for the resonant searches because of triggering limitations. However, the $X{+}\met$ searches also depend on the value of $g_q$, as can be seen by comparing the different coupling scenarios, because the production cross section of the \Zprime depends on that coupling.

\begin{figure}[tb]
\begin{center}
\subfloat[]{\includegraphics[width=0.45\textwidth]{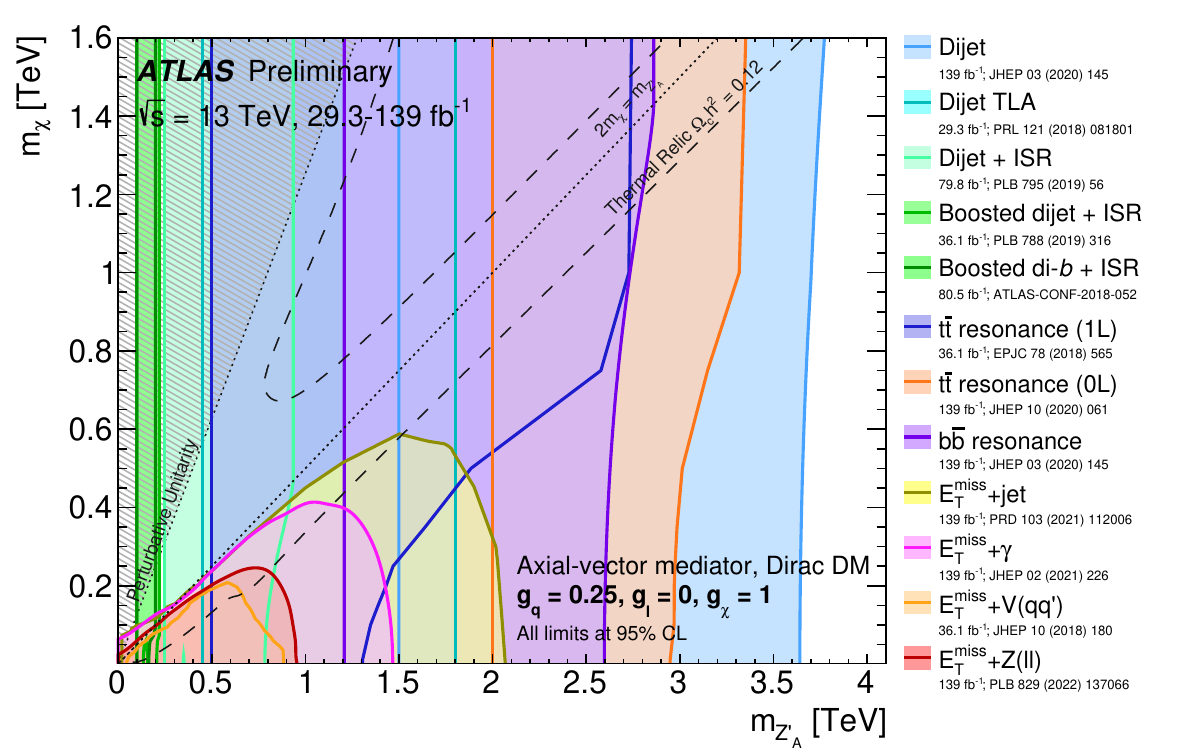}}
\qquad
\subfloat[]{\includegraphics[width=0.45\textwidth]{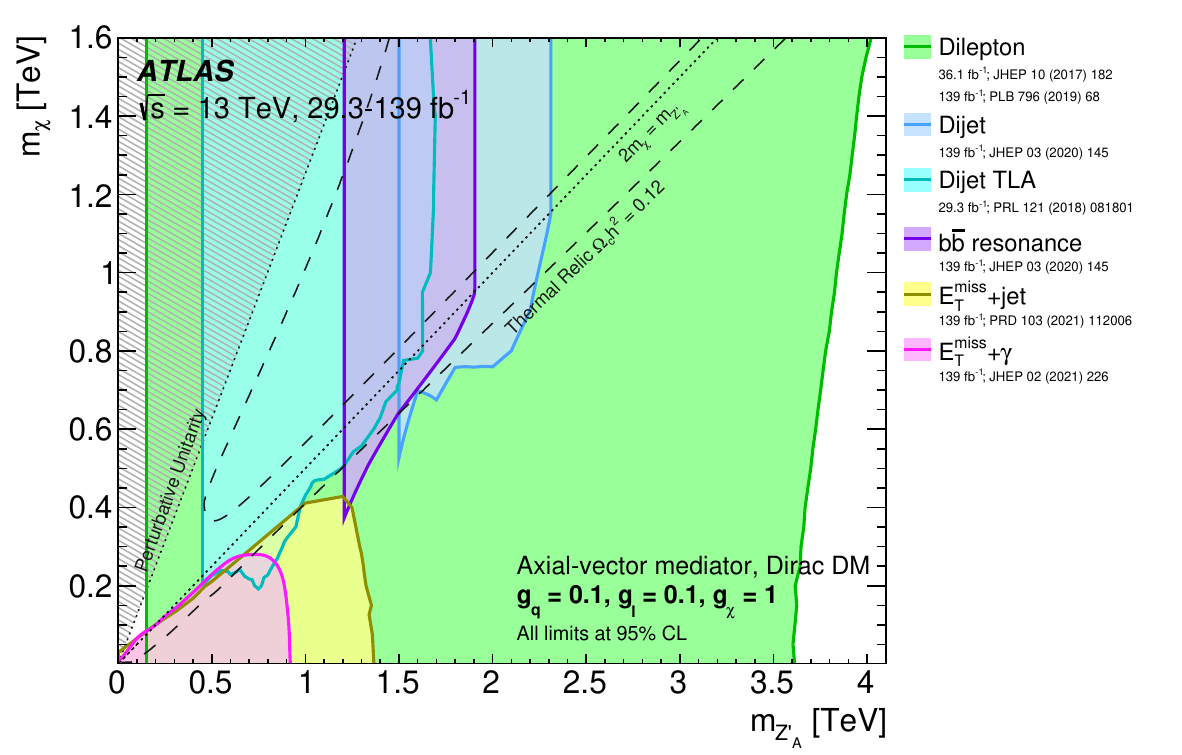}}
\qquad
\subfloat[]{\includegraphics[width=0.45\textwidth]{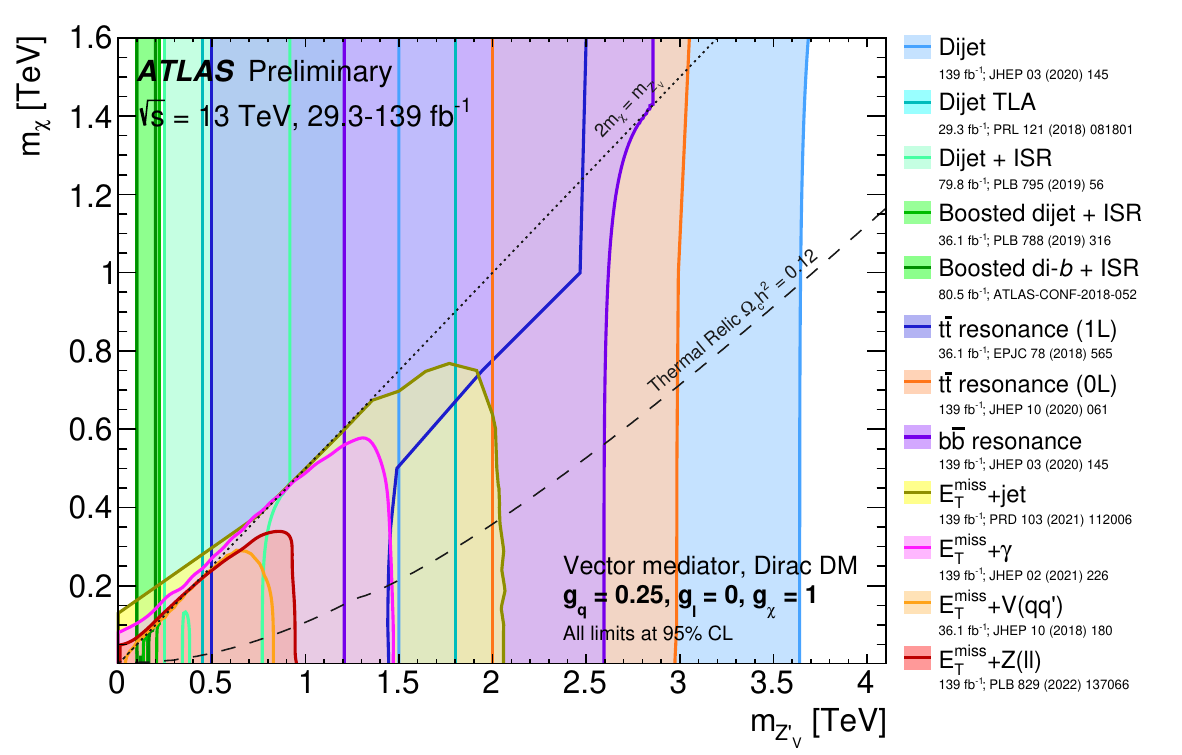}}
\qquad
\subfloat[]{\includegraphics[width=0.45\textwidth]{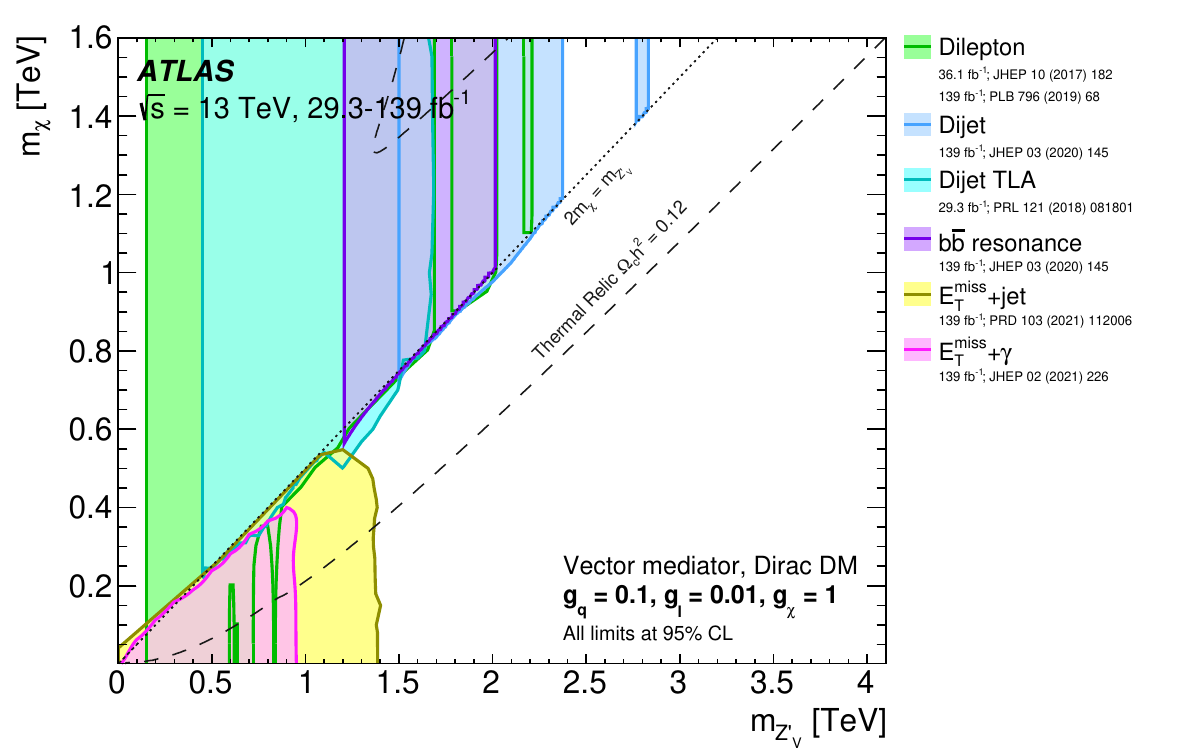}}
\end{center}
\caption{Regions in the $m_\chi$ versus $m_\Zprime$ plane that are excluded at 95\% CL by resonant and $X{+}\met$ searches, for the four coupling scenarios described in the text. Dashed curves labelled \enquote{thermal relic} correspond to parameters consistent with a DM density of $\Omega h^2=0.12$ as computed in MadDM~\cite{Albert:2017onk,Backovic:2015tpt} when assuming a standard thermal history. (c) For the parameter space above the line, or (a,b,d) in between these lines, $\Omega h^2<0.12$. The dotted line indicates the kinematic threshold where the mediator can decay on-shell into DM. }
\label{fig:sumDMV}
\end{figure}

\subsection{Vector portal with a dark Higgs boson}
\label{sec:dmVdarkH}
The simplified model described in the previous section can be extended with a dark Higgs boson $s$ to generate the DM mass through Yukawa interactions~\cite{Duerr:2017uap}. If $m_s<m_\chi$, a new annihilation channel into SM particles can open, relaxing the relic density constraints shown in the previous section. In this two-mediator model~\cite{Duerr:2016tmh}, a Majorana DM candidate is considered and there are two additional dark-Higgs free parameters: $m_s$ and the mixing angle with the SM Higgs boson, $\theta$. This model can lead to an $s{+}\met$ final state with $s\rightarrow WW/ZZ$ at high enough $m_{s}$, as shown in Figure~\ref{fig:dmVdarkH}(a). The signal models considered
assume $m_{\chi}= 200$~\GeV (to forbid the $s\rightarrow\chi\chi$ decay for the $m_s$ considered), along with $\sin{\theta}=0.01$, $g_\chi=1.0$ and $g_q=0.25$.  While the $g_q$ and $g_\chi$ values used here are excluded by the resonant \Zprime searches in the simplified model shown in the last section, this does not lessen the interest of looking for the unique $s{+}\met$ signature of this more complete model, as the couplings can be varied to relax the constraints.

Two final states are explored: a fully hadronic $WW/ZZ$ final state~\cite{EXOT-2018-40}, not described further here, and a more sensitive semileptonic $WW$ final state~\cite{EXOT-2020-04}. In both analyses, boosted dark-Higgs decays can lead to multi-prong large-$R$ jets. A track-assisted reclustering (TAR) is used~\cite{ATL-PHYS-PUB-2018-012}, in which the resolution of the reclustered jet substructure variables is improved by using information from ID tracks which are matched to the small-$R$ jet constituents.

In the semileptonic channel, events must have exactly one lepton with a large value of $\mT(\ell,\met)$,
a large and significant amount of \met
and no \btagged jets, to suppress the \ttbar backgound. A $W$ boson candidate must then be found with a mass compatible with that boson, and which must not be too far away from the lepton. In the merged SR, this candidate is given by a two-pronged TAR jet, and in the resolved SR, by a high-\pT system of two small-$R$ jets.
The discriminating variable in both SRs is the minimum possible reconstructed mass of the dark-Higgs boson candidate, assuming that the charged lepton is massless. Dedicated CRs are used to estimate the dominant $W$+jets and subdominant \ttbar backgrounds, built by reversing the lepton--$W$ angular separation requirement or the $b$-tagging veto.

As no significant excess above the expected background is found, limits are set on the model parameters in the two analyses, as shown in Figure~\ref{fig:dmVdarkH}(b); the semileptonic channel sets the most stringent limits.

\begin{figure}[tb]
\begin{center}
\subfloat[]{\raisebox{0.3\height}{\includegraphics[width=0.24\textwidth]{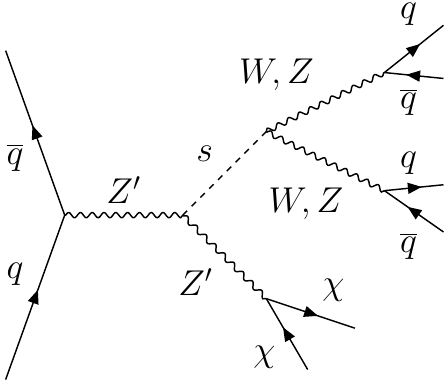}}}
\qquad
\subfloat[]{\includegraphics[width=0.49\textwidth]{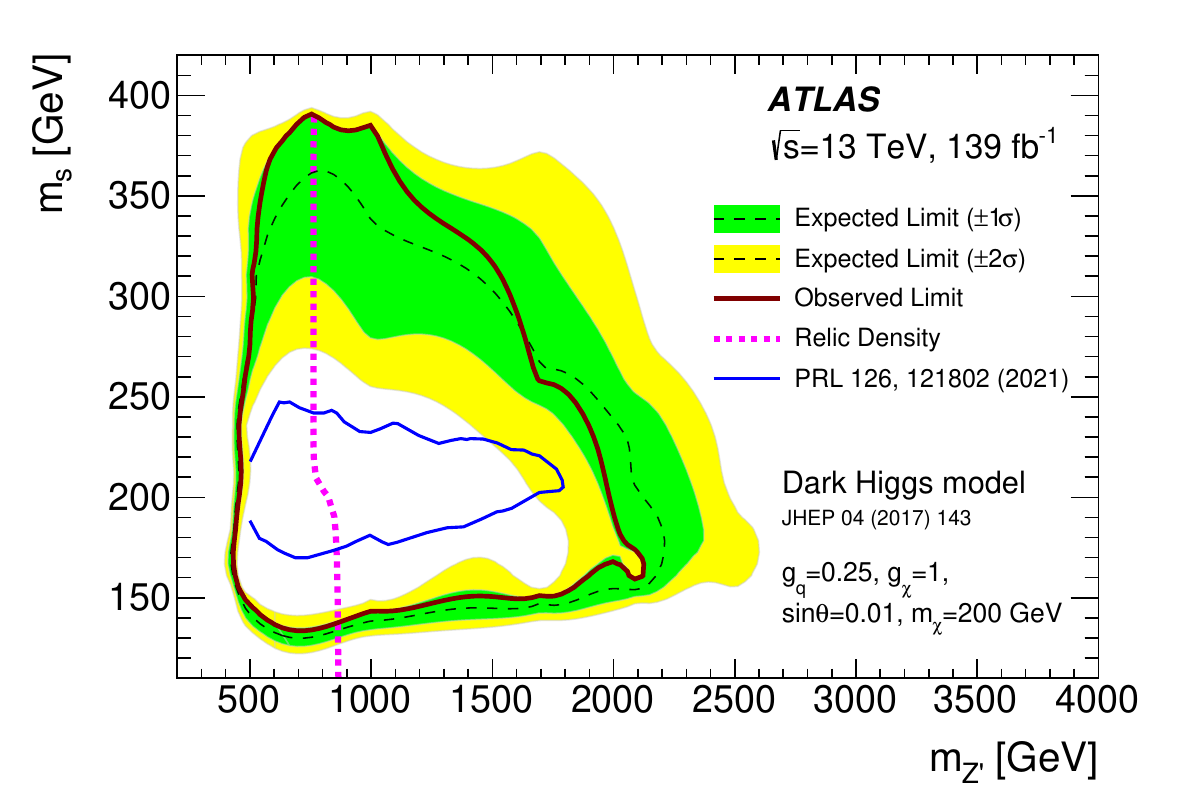}}
\end{center}
\caption{(a) Production of a $WW/ZZ+\met$ final state in the two-mediator model comprising a \Zprime and a dark Higgs $s$, and (b) corresponding exclusion contours set by (red line) the semileptonic~\cite{EXOT-2020-04} and (blue line) fully hadronic~\cite{EXOT-2018-40} channels. In (b), the $m_\Zprime$ values above the dashed relic density line correspond to a DM overabundance.}
\label{fig:dmVdarkH}
\end{figure}

\subsection{Higgs portal}
\label{sec:hinv}

Another interesting avenue to explore is whether the Higgs boson, the last piece of the SM discovered by the ATLAS and CMS Collaborations in 2012~\cite{HIGG-2012-27,CMS-HIG-12-028}, can act as a portal between DM and the SM via either Yukawa-type couplings or other mechanisms~\cite{Antoniadis:2004se,ArkaniHamed:1998vp,Datta:2004jg,Kanemura:2010sh,Djouadi:2011aa,Djouadi:2012zc,Shrock:1982kd,Choudhury:1993hv,Eboli:2000ze,Davoudiasl:2004aj,Godbole:2003it,Ghosh:2012ep,Belanger:2013kya,Curtin:2013fra}. An exciting signature of this type of interaction would be the decay of the Higgs boson into a pair of DM particles, if kinematically allowed, leading to an \emph{invisible} Higgs decay. As long as there is a visible final-state object, this invisible Higgs decay can be searched for in various Higgs boson production modes: VBF, associated production with a weak boson ($VH$) or a \ttbar pair ($\ttbar H$), or even ggF if one relies on ISR, as shown in Figure~\ref{fig:Hinv}. This section covers these various searches, in increasing order of expected sensitivity, and their combination.

All of the presented results assume that the Higgs boson production cross section is the one predicted by the SM~\cite{deFlorian:2016spz, Lidnert:2018kmw, Djouadi:2018kms, Bonetti:2018ukf, Dulat:2018rbf, Harlander:2018yio, Cacciari:2018dksz} and that the Higgs boson mass is 125~\GeV.
The invisible Higgs decay is mimicked by using the SM $H\rightarrow Z^{*}Z\rightarrow 4\nu$ process (which has a 0.1\% branching ratio in the SM).

\begin{figure}[tb]
\begin{center}
\subfloat[]{\includegraphics[height=0.13\textwidth]{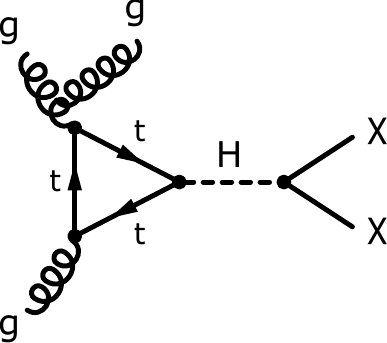}}
\qquad
\subfloat[]{\includegraphics[height=0.13\textwidth]{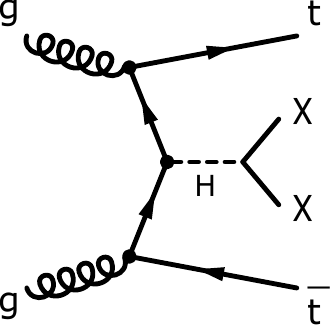}}
\qquad
\subfloat[]{\includegraphics[height=0.12\textwidth]{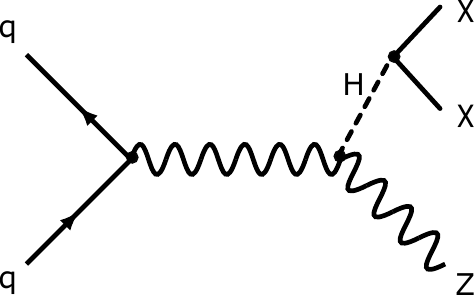}}
\qquad
\subfloat[]{\includegraphics[height=0.13\textwidth]{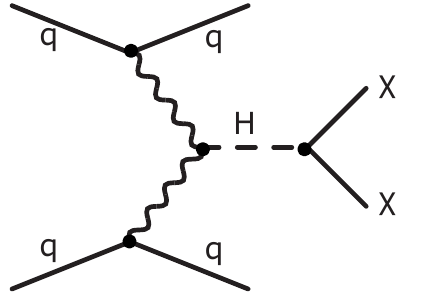}}
\qquad
\subfloat[]{\includegraphics[height=0.13\textwidth]{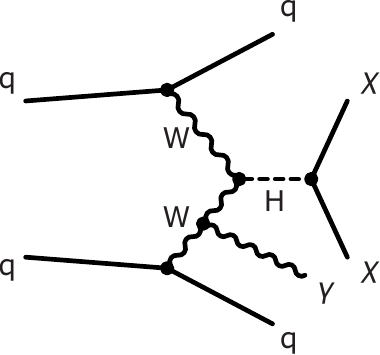}}
\end{center}
\caption{The various diagrams leading to final states which are probed in the search for a DM Higgs portal: (a) ggF with an ISR gluon, (b) \ttbar associated production, (c) $Z$-associated production, (d) VBF, and (e) VBF with a photon.}
\label{fig:Hinv}
\end{figure}

\subsubsection{Gluon--gluon fusion}
\label{sec:jetmet}
As ggF is the main Higgs boson production channel at the LHC, it would seem natural to start an invisible-Higgs search with this channel. However, as in the V/A simplified model, this search must rely on an ISR jet against which the Higgs boson decay products recoil as shown in Figure~\ref{fig:Hinv}(a), leading to the jet+\MET final state~\cite{EXOT-2018-06} already discussed in Section~\ref{sec:monoj}.  This analysis places a 95\% CL observed (expected) upper limit of $0.34$ ($0.39_{-0.11}^{+0.16}$) on the branching ratio for an invisibly decaying Higgs boson, where the sensitivity is mainly driven by the lower \met region.

\subsubsection{Production in association with a \ttbar pair}
\label{sec:ttmet}

In order to fully exploit the data, even the $\ttbar H$ production mode can be used; in spite of having the smallest production cross section, it competes favourably with ggF production when combining~\cite{SUSY-2019-12} the various final states from the two top decays, which can contain zero~\cite{SUSY-2018-12} , one~\cite{SUSY-2018-07} or two~\cite{SUSY-2018-08} leptons. It can also be used in searches motivated by other DM models, as shown in Section~\ref{sec:2hdmA}.

The 0-lepton search is described in Section~\ref{sec:lq} as it is also used as a leptoquark search. However, the DM search combination~\cite{SUSY-2019-12} extends it with three additional SRs which are able to improve the acceptance for events with lower \met or lower-momentum objects by relying on a combination of \met and \btagged jet triggers. This improves the expected sensitivity to invisible Higgs decay by 12\%. The combined observed (expected) upper limit on the Higgs-to-invisible branching ratio is 0.95 ($0.52^{+0.23}_{-0.16}$).

In the 1-lepton channel~\cite{SUSY-2018-07}, events must have one electron or muon, at least four small-$R$ jets, of which two must be \btagged, sizeable and significant \met,
and large $\mT(\ell,\met)$.
An algorithm based on a variable jet radius is used to catch both the highly boosted and less boosted hadronically decaying top-quark candidates.
In order to remove background coming from dileptonic $\ttbar$ decays in which one lepton is not reconstructed, a variable called \emph{topness}~\cite{Graesser:2012qy} is used, based on top quark and $W$ boson mass constraints and considering an invisible lepton candidate.
The SR is split into four bins in $\Delta\phi(\ptmiss,\ell)$ and the
dominant dileptonic $\ttbar$ and $\ttbar+Z$ backgrounds are estimated through dedicated CRs, built by either reversing the topness requirement, or requiring more leptons.
This search is expected to be less sensitive than the 0-lepton channel, but its observed limit is found to be slightly more stringent: at 95\% CL, the observed (expected) upper limit on the branching ratio is 0.74 ($0.80^{+0.40}_{-0.26}$).

The most sensitive $\ttbar H$ channel is the 2-lepton one~\cite{SUSY-2018-08}, which requires two opposite-sign leptons (electrons or muons) %
along with at least one \btagged jet and a significant amount of \met. %
Furthermore, the azimuthal angle $\Delta\phi_\mathrm{boost}(\ptmiss,\mathbf{p}_\mathrm{T}^\mathrm{boost})$ must be smaller than 1.5, where $\mathbf{p}_\mathrm{T}^\mathrm{boost}$ is defined as the vectorial sum of the \ptmiss and the \pt of the leptons. The events are then separated depending on whether they contain same-flavour leptons (incompatible with the $Z$ boson mass), or different-flavour leptons. The discriminating variable, in which the SR is binned, is
$m_\mathrm{T2}$.
The main backgrounds after all requirements, \ttbar and $\ttbar+Z$ events, are estimated using dedicated CRs, based on $e\mu$ events at lower values of  $m_\mathrm{T2}$ or on the presence of three leptons, respectively. The data agree well with the background expectations, as shown in Figure~\ref{fig:ttHZMET}(a): the observed (expected) upper limit on the Higgs-to-invisible branching ratio is 0.36 ($0.40^{+0.18}_{-0.12}$) at 95\% CL.

Combining the three $\ttbar H$ channels increases the sensitivity further: the combined expected limit is $0.30^{+0.13}_{-0.09}$, with an overall uncertainty dominated by the statistical precision of the data.
The observed value of 0.38 agrees well with the expected limit.

\subsubsection{Production in association with a $Z$ boson}
\label{sec:Zmet}

The $Z(\to\ell\ell)H$ channel  puts even more stringent constraints on the Higgs-to-invisible branching ratio~\cite{HIGG-2018-26}: the leptonic trigger allows the analysis to probe \met values which are not as high as in the jet+\met analysis, thus increasing the sensitivity of the search to this particular signal, and the larger production cross section and relatively clean final state provide better sensitivity than the $\ttbar H$ channel. This analysis is  also sensitive to other DM signals (see Sections~\ref{sec:dmV} and \ref{sec:2hdmA}).

Events are selected by requiring exactly two opposite-sign electrons or muons with an invariant mass compatible with a $Z$ boson.
Since the leptons should recoil against an invisible Higgs boson, their angular separation should not be too large,
and the events should have a large and significant \met.
In order to increase the sensitivity further, the SR uses a BDT which is based on eight kinematic variables.
After this selection, the dominant background comes from the $ZZ$ process, followed by $WZ$, $Z$+jets and smaller non-resonant backgrounds ($WW$, \ttbar, single-top and $Z\rightarrow\tau\tau$). Three CRs are used: an $e\mu$ CR to constrain the non-resonant backgrounds, a 4$\ell$ CR to estimate the $ZZ$ contribution, and a 3$\ell$ CR to estimate the $WZ$ background.
The $Z$+jets contribution is taken from MC simulation, but verified in a validation region reversing the requirement on $S_{\met}$. The resulting BDT output distribution is shown in Figure~\ref{fig:ttHZMET}(b).

The best-fit Higgs-to-invisible branching ratio is found to be $(0.3\pm9.0)\%$ where the uncertainty is dominated by the $ZZ$ modelling uncertainties and the jets/\met-related experimental uncertainties. The observed 95\% CL upper limit of 0.19 set on the branching ratio coincides with the expected limit.

\subsubsection{Vector-boson fusion}

The most sensitive channel in the search for a DM Higgs portal is VBF production~\cite{EXOT-2020-11}, leading to a VBF jets+\met final state. While this final state differs from the ggF jet+\met one by the VBF characteristics of the jets, many of the analysis techniques for the two analyses are similar.
After requiring a large \met value and imposing a veto on leptons and photons, the analysis exploits the VBF signature. In this topology, the two leading jets are usually in opposite hemispheres of the detector, $\eta^\mathrm{j1}\cdot\eta^\mathrm{j2}<0$, and are more forward, leading to a large separation in pseudorapidity
and a large invariant mass.
Unlike the jets in a multijet background event, which are likely to be back-to-back, the expected signal jets must balance the significant \pT of the Higgs boson, leading to a smaller azimuthal separation.
As in the jet+\met analysis, extra jets are allowed in the SR to increase the acceptance and reduce the associated modelling systematic uncertainties. However, given the absence of colour connection between the two VBF quarks, the VBF process has less hadronic activity in the \emph{central} rapidity region between the two leading jets. Furthermore, the dijet invariant mass constructed from a jet radiated by a VBF quark and one of the two leading jets should be small relative to the invariant mass of the two leading jets. Consequently, up to two extra jets are allowed, but only if they are compatible with the VBF process, i.e.\ by requiring both their centrality as defined in Ref.~\cite{HIGG-2013-13} and their dijet invariant mass to be small.
Further suppression of the multijet background is achieved by requiring the \pT of the jet system (including jets tagged as pile-up jets by the JVT algorithm) to be large. The \met soft term is also required to be small
to remove \Wmunujets events in which the muon is not identified but is still seen as a high-\pT track in the inner detector. Finally, a veto on the presence of more than one \btagged jet is imposed to ensure orthogonality with the $\ttbar H$ search described above. This rejects very few events, as the VBF jets are mostly forward and hence outside the inner-detector acceptance which is used in $b$-tagging.

In order to increase the sensitivity, the SR is subdivided into 16 different regions, according to the \met, number of jets, dijet invariant mass, and azimuthal angle between the two leading jets.
As in the jet+$\met$ analysis, the main backgrounds are \Znunujets and \Wlnujets, and the same strategy is used to evaluate them: 1- and 2-lepton CRs are built, and dedicated NLO theoretical calculations~\cite{Lindert:2022ejn} performed in the relevant phase space are used to fix the $Z/W$ ratio, reducing the statistical uncertainties in the determination of the dominant \Znunujets background by using all leptonic CRs simultaneously to constrain it.
The smaller multijet background is determined in a fully data-driven way, using two independent methods because it can come either from jet mismeasurements or from a pile-up jet being wrongly identified as a VBF jet. Although a smaller component than the $V$+jets background, unlike in the ggF analysis, the multijet background is not found to be negligible in all SR bins, as can be seen in Figure~\ref{fig:ttHZMET}(c), representing from 0.4\% up to almost 14\% of the total background.

Since no significant excess is seen, an upper limit of 0.145 is set on the branching ratio of invisible Higgs boson decays, in agreement with the expected limit of $0.103^{+0.041}_{-0.028}$, where the main uncertainties are related to the data statistics, the multijet background estimate, lepton identification, and the jet energy resolution.

\textbf{VBF jets+{\met}+$\gamma$}\\
The VBF production mode is further exploited, by requiring the presence of an additional photon~\cite{EXOT-2021-17}, as depicted in Figure~\ref{fig:Hinv}(e) -- the photon veto in the VBF jets+\met  analysis makes these channels orthogonal.  Considerations similar to those described above are used to select events with a large \met, VBF jets and no leptons.
Specific requirements are then made on the photon. Since it is usually radiated from one of the scattering $W$ bosons, it is expected to be produced within the rapidity gap of the VBF jets, not be too energetic (to remove $\gamma$+jets background events), be well separated from the \met, and have a
trajectory that, when extrapolated to the beamline, is loosely compatible with the PV
(to remove non-collision background). Finally, a dense neural network is trained using the most significant kinematic features, and four bins in its output score form the SR.  The main $Z\gamma$+jets and $W\gamma$+jets backgrounds are estimated using CRs, built by reversing the photon centrality requirement or by requiring one lepton in the final state, respectively.

The observed limit set by this analysis on the branching ratio of invisible Higgs boson decays is 0.37, in line with the expected value of $0.34^{+0.15}_{-0.10}$, where the uncertainty is dominated by the data statistics.

\subsubsection{Combination of all channels}

A summary of the upper limits obtained in the channels described above is shown in Figure~\ref{fig:ttHZMET}(d), along with a combination of these results alone and with the previous \RunOne results~\cite{HIGG-2021-05}. The \RunTwo combination gives an observed (expected) upper value of 0.113 ($0.080^{+0.031}_{-0.022}$), an improvement of 22\% on the sensitivity of the most sensitive VBF jets+\met channel. In combination with the \RunOne results, the expected sensitivity is further improved by 4\%, with an upper limit at 0.107 ($0.077^{+0.030}_{-0.022}$).
In the \RunTwo combination, the leading systematic uncertainties are related to the $W/Z$+jets modelling uncertainties; without any systematic uncertainties, the upper limit would improve by 50\%.

\begin{figure}[htbp]
\begin{center}
\subfloat[]{\includegraphics[width=0.45\textwidth]{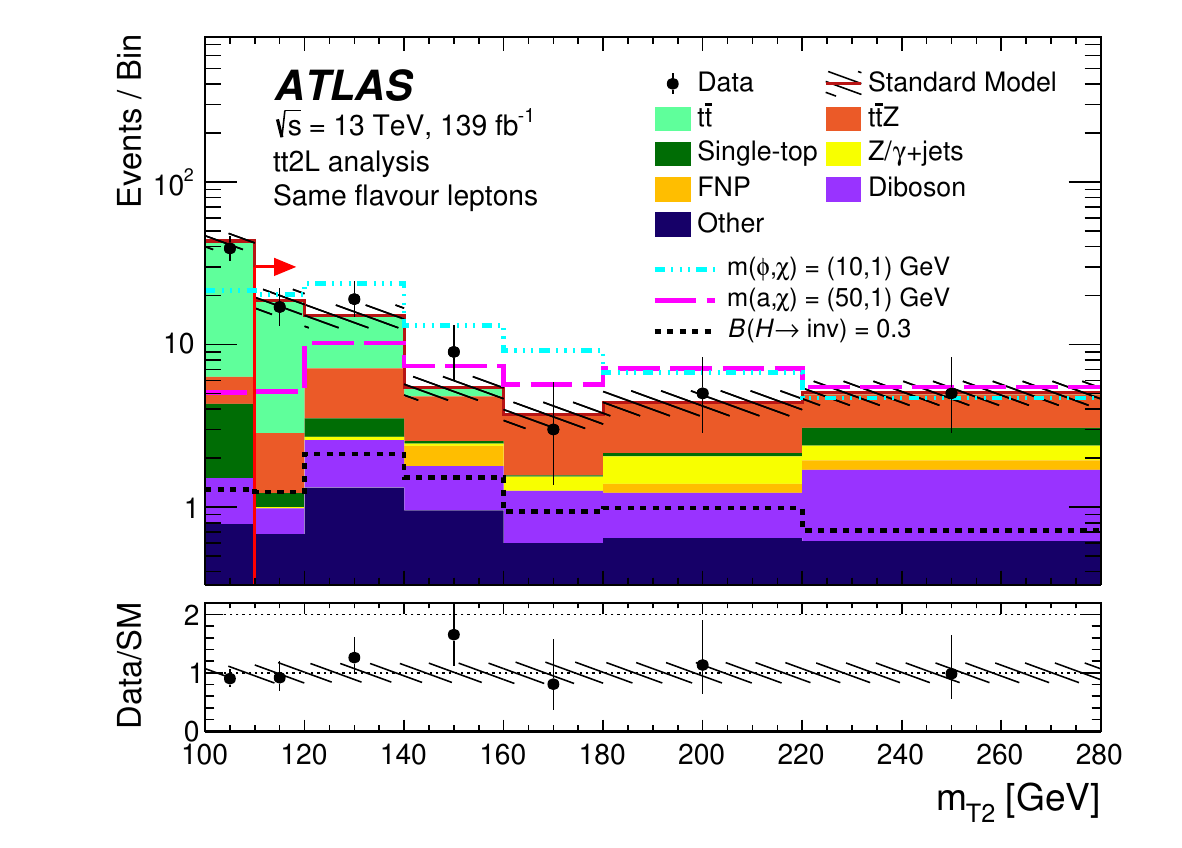}}
\qquad
\subfloat[]{\includegraphics[width=0.35\textwidth]{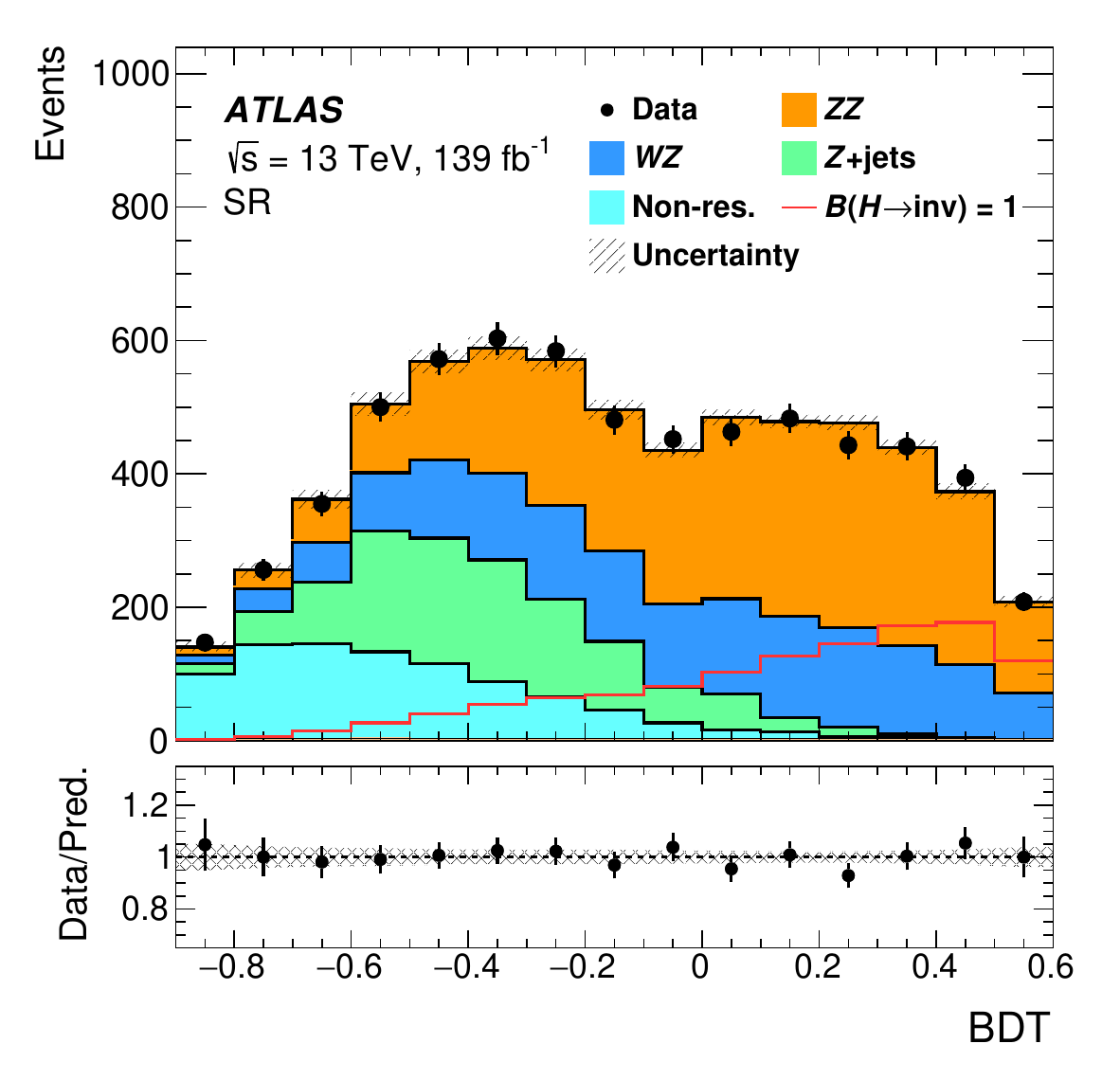}}
\qquad
\subfloat[]{\includegraphics[width=0.55\textwidth]{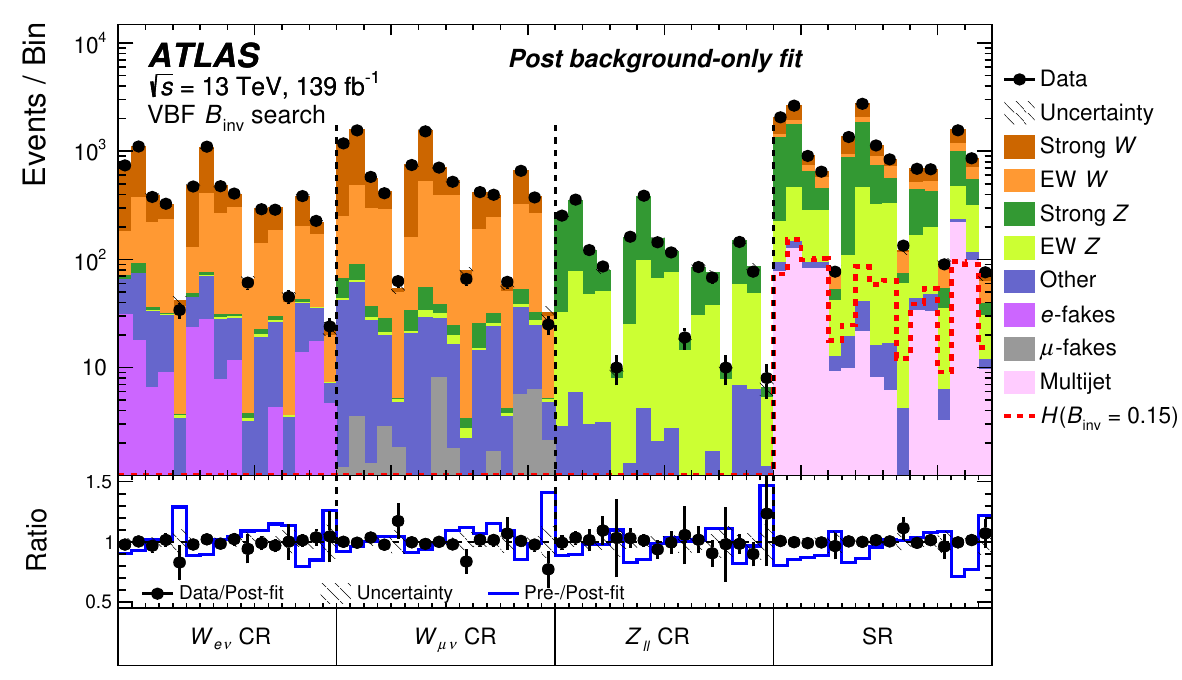}}
\qquad
\subfloat[]{\includegraphics[width=0.55\textwidth]{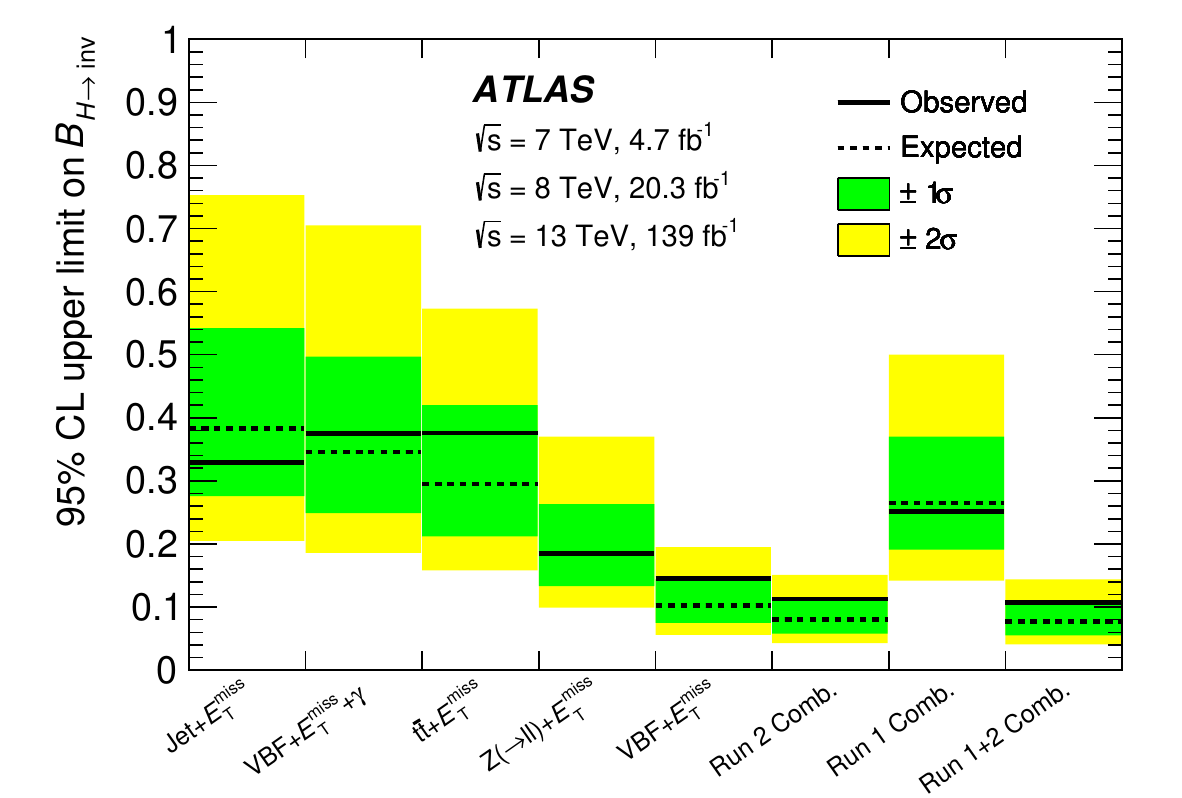}}
\end{center}
\caption{Measured distributions of
(a) $m_\mathrm{T2}$ in the 2-lepton $\ttbar H$ SR~\cite{SUSY-2018-08} where the SR requirement is shown by the red arrow, (b) the  BDT output distribution in the $Z$+\met SR~\cite{HIGG-2018-26} and (c) the yields in each of the CR and SR bins of the VBF jets+\met analysis~\cite{EXOT-2020-11}, compared to the estimated background and an invisible-Higgs-decay signal with a (a) 30\%, (b) 100\% or (c) 15\% branching ratio. The observed and expected 95\% CL upper limits (d) on the Higgs-to-invisible branching ratio for the individual \RunTwo analyses, along with the combination of these results alone and with the previous \RunOne results~\cite{HIGG-2021-05}. }
\label{fig:ttHZMET}
\end{figure}

\subsection{Pseudoscalar portal}
\label{sec:2hdmA}

Another possibility for a DM mediator would be the addition of a pseudoscalar portal. As $t$-channel interactions via a pseudoscalar would be suppressed in the non-relativistic limit, the sensitivity of direct-detection experiments in this case would be very low, so it is especially important to consider this portal at the LHC because it could offer a unique opportunity for detection.

The model considered here is the 2HDM+$a$ model suggested by the LHC DM Working Group~\cite{LHCDarkMatterWorkingGroup:2018ufk}, which is the simplest gauge-invariant and renormalizable ultraviolet completion of the simplified pseudoscalar model initially recommended by the LHC DM Forum~\cite{Abercrombie:2015wmb}, which only contained the DM candidate and the mediator. This model is a type-II two-Higgs-doublet (2HDM) model~\cite{Gunion:2002zf} to which an additional pseudoscalar $a$ and a fermionic DM candidate $\chi$ are added. After electroweak symmetry breaking, the 2HDM contains five Higgs bosons: a lighter CP-even boson, $h$, a heavier CP-even boson, $H$, a CP-odd boson, $A$, and two charged bosons, $H^{\pm}$. While the phenomenology of the model would be determined by 14 free parameters, some benchmark choices are made in order to match $h$ with the observed SM Higgs boson, to ensure the stability of the Higgs potential, or to evade  electroweak precision measurement constraints. In the end, the benchmarks are defined by five parameters: the mass of the heavy Higgs bosons, which are taken to be degenerate, $m_A = m_H = m_{H^\pm}$; the mass of the pseudoscalar mediator, $m_a$; the mass of the DM particle, $m_\chi$; the mixing angle $\theta$ between the two CP-odd states $a$ and $A$; and the ratio of the vacuum expectation values of the two Higgs doublets, $\tan{\beta}$.

This model leads to a large number of final states which can be probed. Some examples of their production modes are shown in Figure~\ref{fig:2hdma}. The following final states were explored in \RunTwo, and some are introduced in previous sections: the ubiquitous jet+\met (see Section~\ref{sec:jetmet}), $Z{+}\met$ (see Section~\ref{sec:Zmet}),
$h{+}\met$~\cite{EXOT-2018-46,HIGG-2019-02,HDBS-2018-50}, $Wt{+}\met$~\cite{EXOT-2018-43,EXOT-2021-01},
$tt{+}\met$ (see Section~\ref{sec:ttmet}), $tbtb$~\cite{HDBS-2018-51}, $bbbb$, $tttt$, $bb$ and $tt$ (see Section~\ref{sec:gauge}). A statistical combination of some of these has also been performed~\cite{EXOT-2018-64}. In this section, the searches not already discussed in the context of other models are introduced, and their complementarity in covering the model parameter space is discussed. Since the $tbH^{\pm}(tb)$ search~\cite{HDBS-2018-51} does not specifically target this model, but in general the production of a charged Higgs boson in association with a top quark and bottom quark, as shown in Figure~\ref{fig:2hdma}(g), it is reviewed in another report discussing extended Higgs sectors but not here.%

\begin{figure}[tb]
\begin{center}
\subfloat[]{\includegraphics[width=0.2\textwidth]{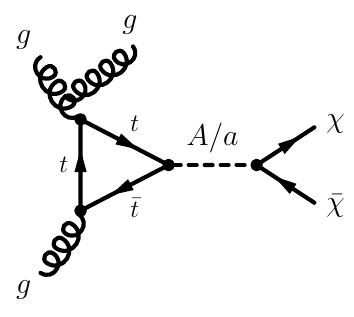}}
\qquad
\subfloat[]{\includegraphics[width=0.2\textwidth]{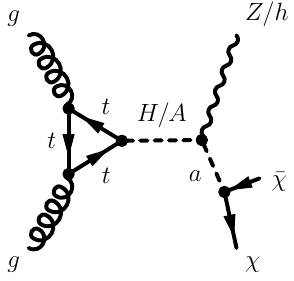}}
\qquad
\subfloat[]{\includegraphics[width=0.2\textwidth]{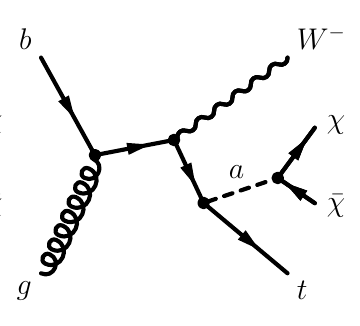}}
\qquad
\subfloat[]{\includegraphics[width=0.2\textwidth]{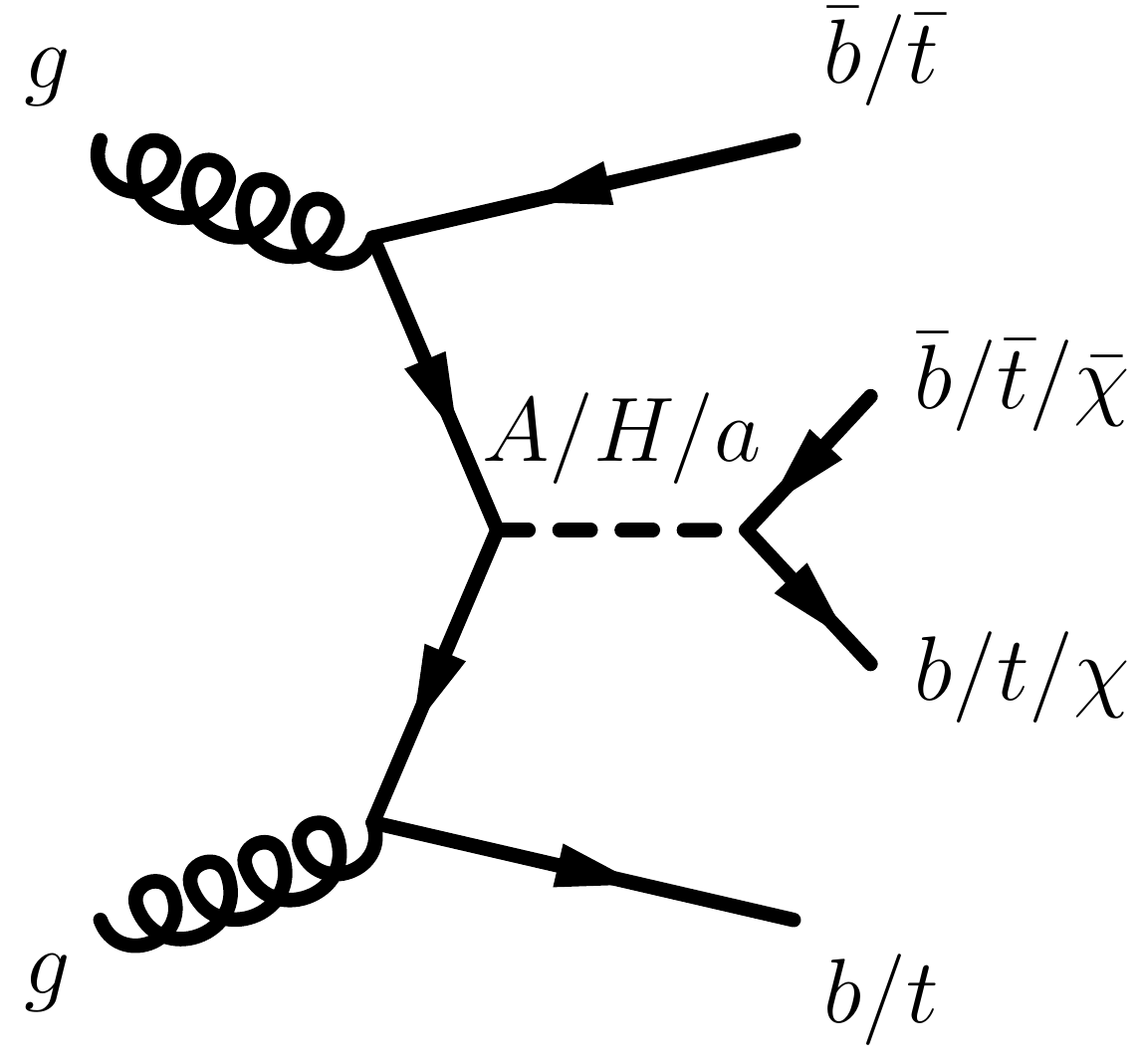}}
\qquad
\subfloat[]{\includegraphics[width=0.2\textwidth]{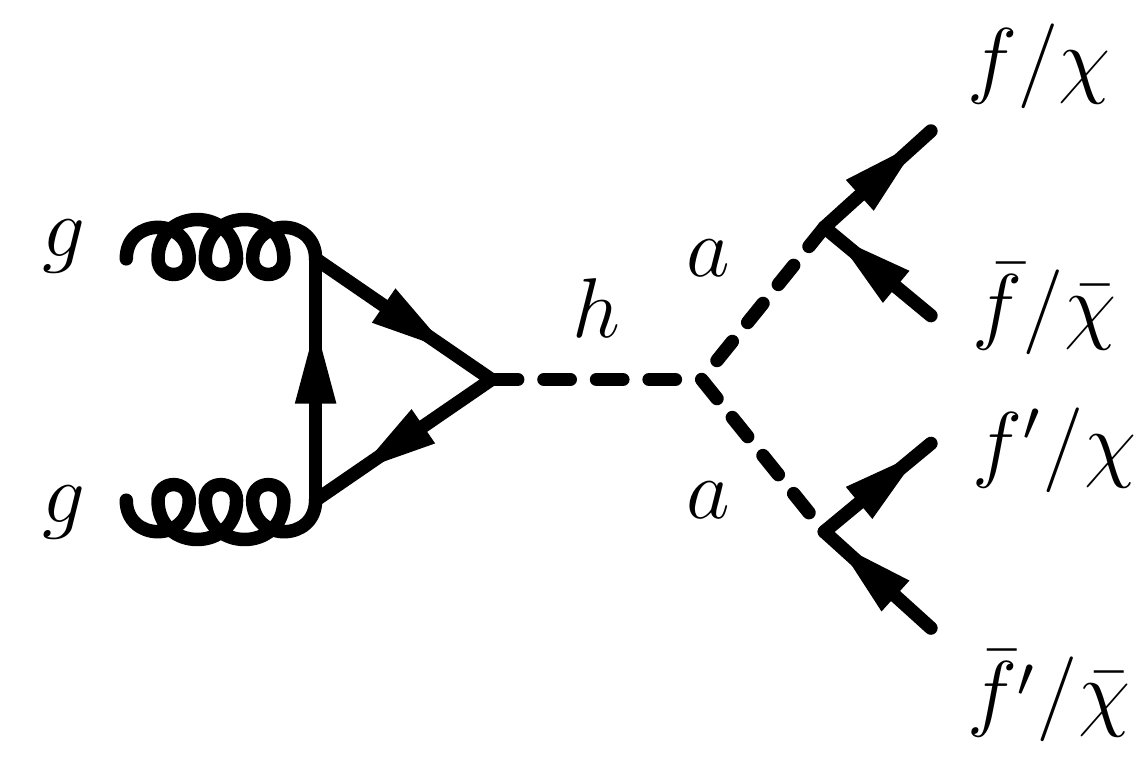}}
\qquad
\subfloat[]{\includegraphics[width=0.2\textwidth]{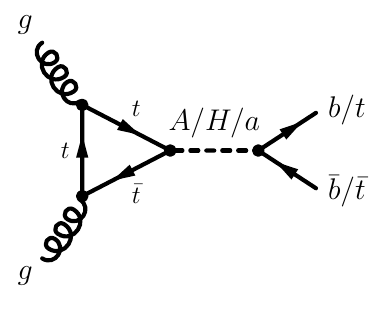}}
\qquad
\subfloat[]{\includegraphics[width=0.17\textwidth]{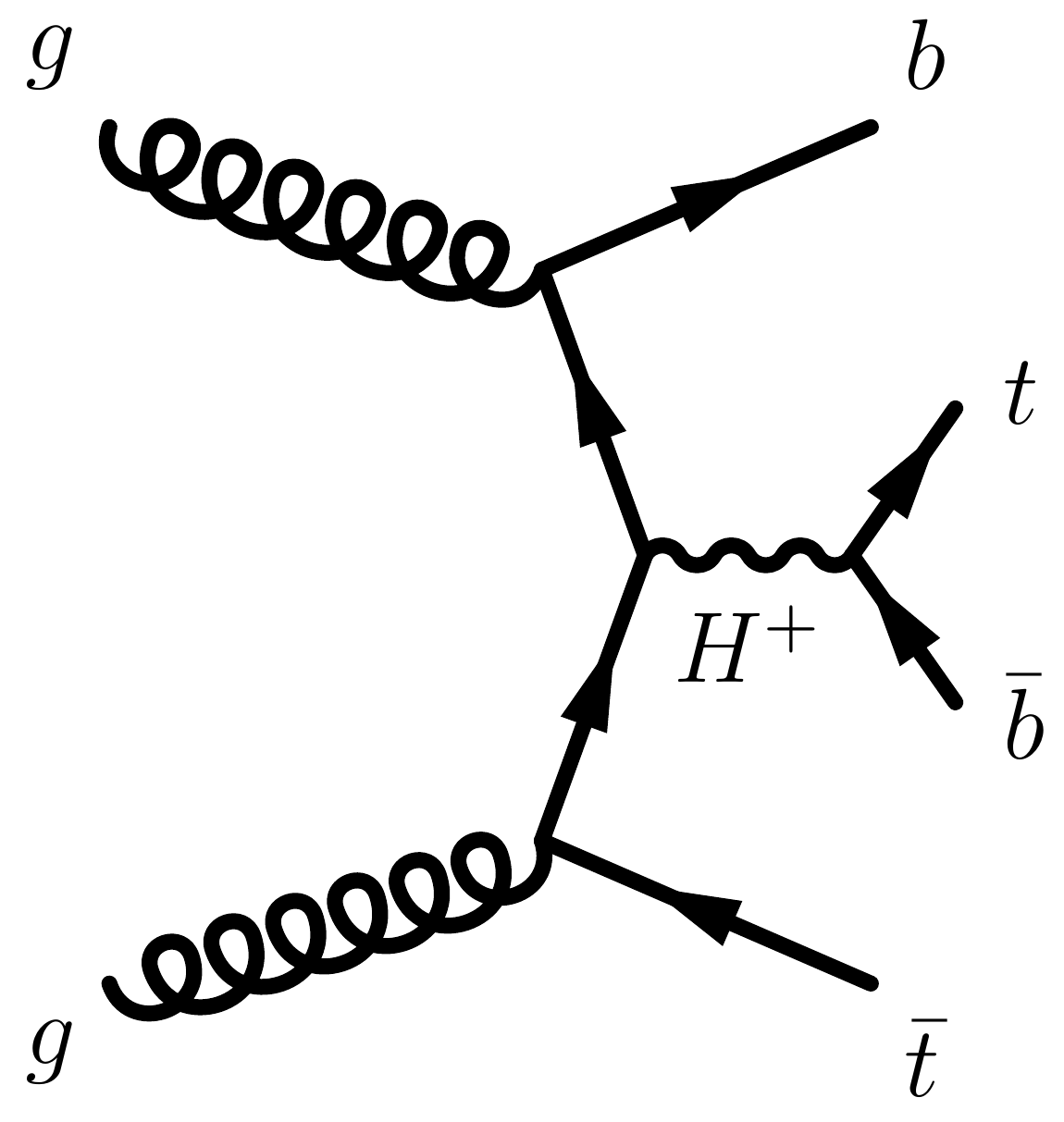}}
\end{center}
\caption{Examples of Feynman diagrams leading to various final states expected in the 2HDM+$a$ model: (a) jet+\met, (b) $Z$+\met and $h$+\met resonant production, (c) $Wt$+\met, (d) $tttt/ttbb/bbbb/tt{+}\met/bb{+}\met$ final states through associated $A/H/a$ production, (e) decay of a SM Higgs boson into a pair of $a$-bosons that subsequently decay into fermions or DM, (f) resonant $tt$ or $bb$ production, and (g) $tbH^{\pm}(tb)$ production. }
\label{fig:2hdma}
\end{figure}

\subsubsection{The $h{+}\met$ searches}
The 2HDM+$a$ model provides mechanisms in which the $h{+}\met$ signature could be produced resonantly (see Figure~\ref{fig:2hdma}(b)), making this final state a particularly sensitive probe for this model, unlike those in which the visible object comes from ISR. Three $h{+}\met$ channels with different Higgs boson decays have been explored, the most sensitive one being the $h(bb){+}\met$ channel~\cite{EXOT-2018-46}, followed by the $h(\gamma\gamma){+}\met$~\cite{HIGG-2019-02} and $h(\tau\tau){+}\met$~\cite{HDBS-2018-50} channels. The two most sensitive ones are discussed below.

\textbf{The $h(bb){+}\met$ channel}\\
Two regions are defined in this search, based on an \met trigger, depending on the expected boost of the Higgs boson: a resolved region, defined at lower \met values,
and a merged region, at higher values.
A lepton ($e$,$\mu$,$\tau$) veto is imposed and a minimum angular separation between the \met and the \pT of each of the three leading jets is required in both SRs.
The reconstructed Higgs boson mass $m_h$ is corrected for nearby muons to improve its accuracy when $b$-hadrons decay semileptonically. For better sensitivity, both SRs are binned in \met, in the number of \btagged jets (two or at least three), and in $m_h$.

In the resolved SR, the Higgs boson candidate is reconstructed from the two \btagged small-$R$ jets with the highest \pT. It must have a mass loosely compatible with the Higgs boson and a large \pT.
To remove the multijet background, the \met must be significant.
The dominant semileptonic \ttbar background is suppressed by using $m_\mathrm{T}(b,\met)$ built from each \btagged jet.
Because this background enters the SR when the lepton is missed, the leptonically decaying $W$ boson would be seen as \met, and this transverse mass would have an endpoint at the top-quark mass, while the signal could have higher values.
Furthermore, as the signal usually contains fewer jets than the background, up to four (five) jets are allowed when there are two (at least three) \btagged jets.  In the merged SR, at least one \largeR jet is required. In order to identify the \btagged subjets coming from the Higgs decay, even in highly boosted scenarios, variable-$R$ track-jets are used.~\cite{Krohn:2009zg}.
A Higgs boson candidate is identified as the large-$R$ jet if its two leading associated track-subjets are \btagged, and if its muon-corrected mass falls in a range loosely compatible with the Higgs boson.

After these selections, the main backgrounds, the \ttbar and $W/Z\,+$\,heavy-flavour (HF) processes, are estimated using 1-muon and 2-lepton CRs.
In Figure~\ref{fig:monohbb}(a), the resulting yields measured in data as a function of \met are compared with the background expectations in the 3-\btagged SRs. Good agreement is observed in all SRs within the uncertainties, which are dominated by statistical uncertainties in the merged regions and by systematic uncertainties in the resolved regions, the latter being mostly affected by the modelling of the \ttbar background, the uncertainties in the jet calibration, and the limited MC sample size.

\begin{figure}[tb]
\begin{center}
\subfloat[]{\includegraphics[width=0.45\textwidth]{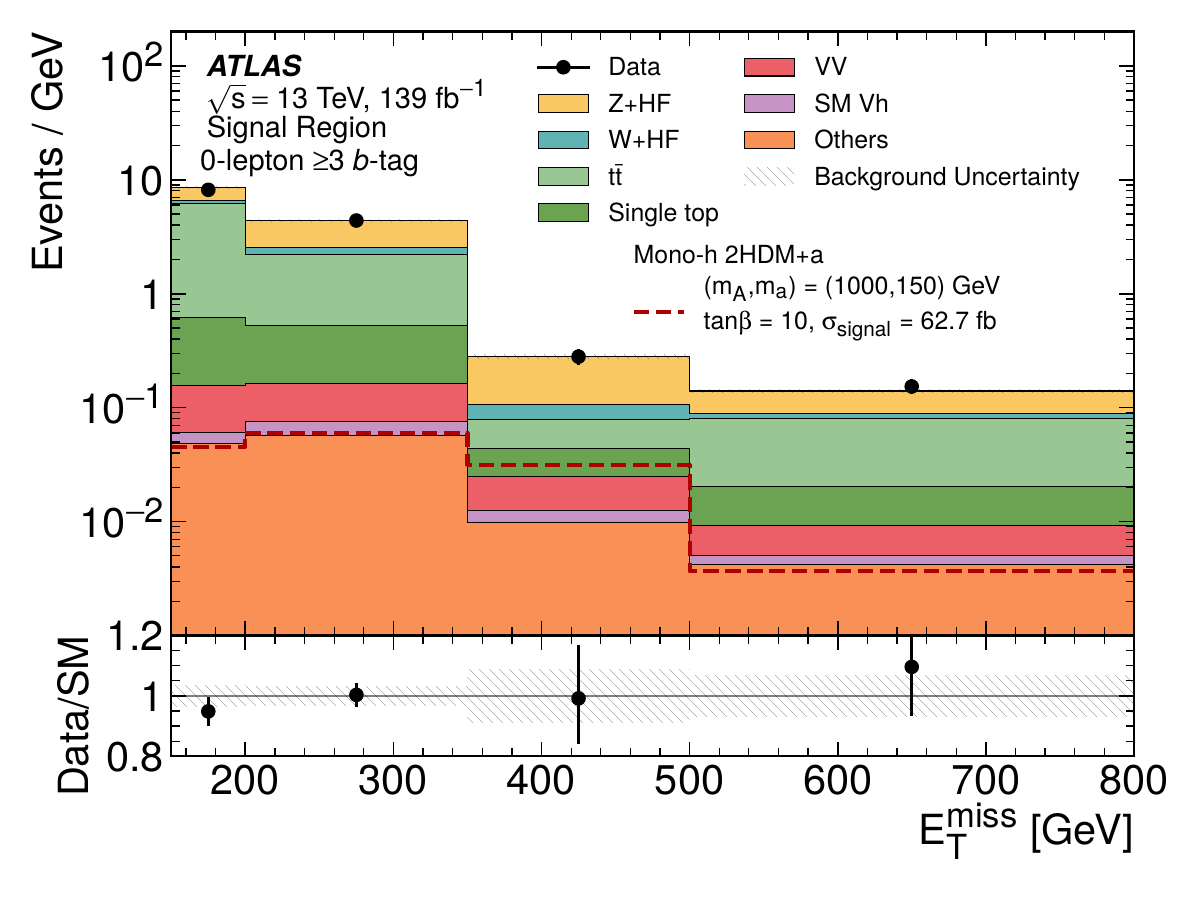}}
\qquad
\subfloat[]{\raisebox{0.05\height}{\includegraphics[width=0.45\textwidth]{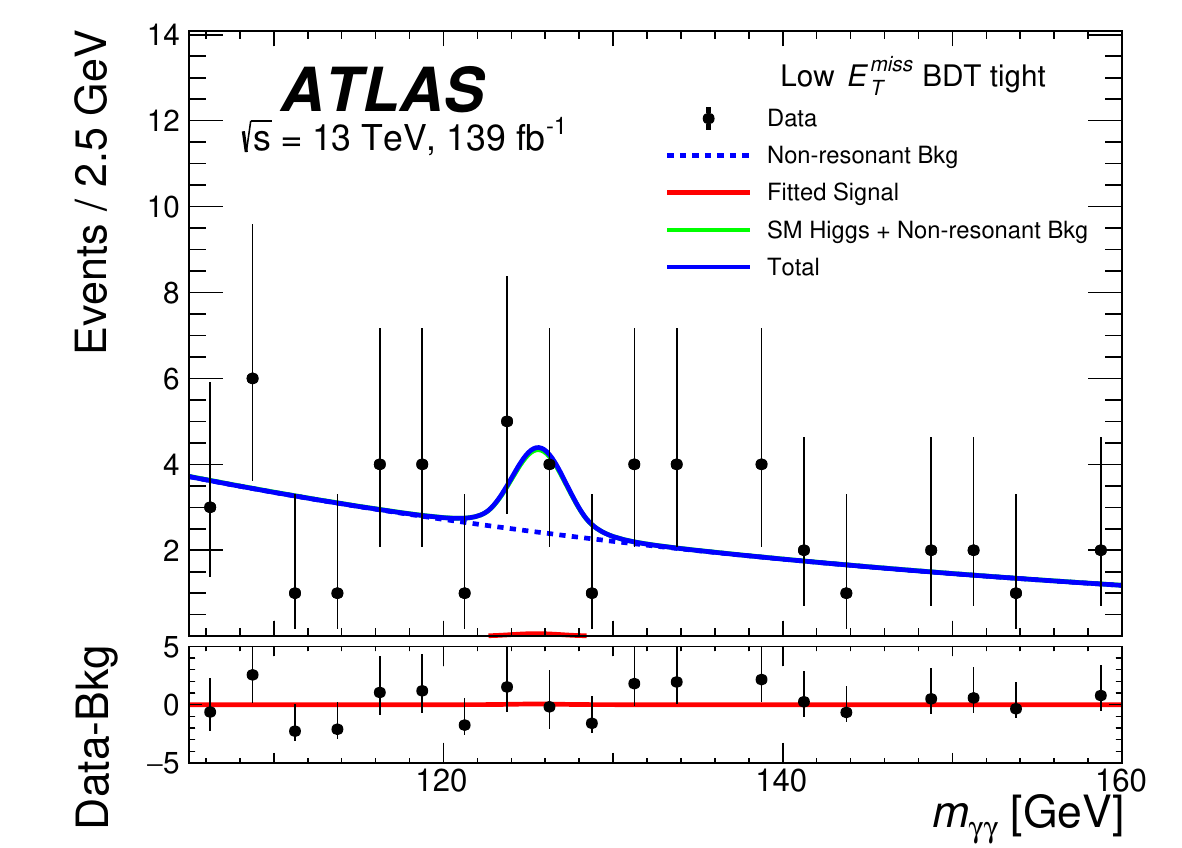}}}
\end{center}
\caption{(a) The measured \met distribution in the 3-\btagged resolved and merged SRs of the $h(bb){+}\met$ search~\cite{EXOT-2018-46} and (b) the measured $m_{\gamma\gamma}$ distribution in the BDT tight signal regions with low \met used in the $h(\gamma\gamma){+}\met$ search~\cite{HIGG-2019-02}, compared to expected background and an example 2HDM+$a$ signal.}
\label{fig:monohbb}
\end{figure}

\textbf{The $h(\gamma\gamma){+}\met$ channel}\\
In this channel~\cite{HIGG-2019-02}, the trigger is based on the photon pair. The \met requirement can thus be lower than in the $h(bb)$ channel, covering the parameter space in which a less-boosted Higgs boson is expected. The $m_{\gamma\gamma}$ value must be loosely compatible with the Higgs boson, and each photon must have a sizeable $\pT/m_{\gamma\gamma}$. A lepton ($e,\mu$) veto is applied. Given the low ID activity expected, the PV might not be well selected; to avoid these cases, selected events must not have an \met value which changes significantly if computed with an alternative NN-based PV algorithm~\cite{HIGG-2013-08}. Four SRs are built: at low and high \met, and with looser or tighter purity given by the score of a BDT (which uses $S_{\met}$ and the diphoton \pT as inputs). A functional form is fitted in each SR: for the signal and the SM Higgs boson background, a double-sided Crystal Ball function is used, while the non-resonant background is modelled by an exponential function of the type $\mathrm{e}^{am_{\gamma\gamma}}$, as shown in Figure~\ref{fig:monohbb}(b) in the low-\met, BDT tight SR. No significant excess is seen in any of the SRs for any of the signals considered.

\subsubsection{The $Wt{+}\met$ searches}

The rich phenomenology of the 2HDM+$a$ model also includes the production of a $Wt{+}\met$ final state, which can be produced when an initial $b$-quark radiates either a $W$ boson (as shown in Figure~\ref{fig:2hdma}(c)) or a charged Higgs boson which subsequently decays as $H^{-}\to a W^{-}$. Depending on the decay mode of the top quark and the $W$ boson, this can lead to final states containing zero, one or two leptons, all of which have been explored~\cite{EXOT-2018-43,EXOT-2021-01}.

In the 0-lepton channel~\cite{EXOT-2021-01}, at least four jets are required, exactly one of which must be $b$-tagged and another being a large-$R$ jet tagged as a $W$ boson. Since the boosted $W$ boson is assumed to come from the $H^{-}$ decay, the $W$-tagged jet and the $b$-tagged jet must not, when combined, be compatible with a top-quark decay in mass.
A large and significant value of \met is required,
which must be well separated from the selected jets. A large value of $\mT(b,\met)$
is also required
in order to remove the semileptonic \ttbar background in which the \met comes from a missed $W$ boson. The SR is binned in \met to improve the sensitivity. After these selections, the main backgrounds come from $V$+jets and semileptonic \ttbar events.

The 1-lepton channel~\cite{EXOT-2021-01} requires the presence of at least two jets, and the requirements placed on the $b$-tagged jet, $\met$, $S_{\met}$ and $\Delta\phi(\met,\mathrm{jet})$ are similar to those in the 0-lepton channel. The leptonically decaying $W$ boson can come from the decay of either the top quark or the $H^{-}$ in the signal, each case being the target of a different SR (SR$^{\mathrm{lep.top}}$ and SR$^{\mathrm{had.top}}$). In SR$^{\mathrm{lep.top}}$, the hadronically decaying $W$ boson can be boosted, so the presence of a $W$-tagged large-$R$ jet is required. In SR$^{\mathrm{had.top}}$, at least three jets are required. In both SRs, a large value of $\mT(\ell,\met)$ (compatible with the top-quark mass in SR$^{\mathrm{lep.top}}$ but not in SR$^{\mathrm{had.top}}$) is required, as well as an $am_\mathrm{T2}$ value above the top-quark mass.  Finally, the two SRs are kept orthogonal by requiring a lower value (SR$^{\mathrm{had.top}}$) or higher value (SR$^{\mathrm{lep.top}}$) of the invariant mass of the leading $b$-tagged jet and the highest-\pT jet which is not $b$-tagged. While the low number of events expected in SR$^{\mathrm{lep.top}}$ does not allow it to be subdivided further, five bins in \met are used in SR$^{\mathrm{had.top}}$. In these SRs, the main backgrounds are dileptonic \ttbar and either ${\ttbar}Z$ (in SR$^{\mathrm{lep.top}}$)  or $W$+jets (in SR$^{\mathrm{had.top}}$).

In the 2-lepton channel~\cite{EXOT-2018-43}, the events must contain OS leptons that are incompatible with the $Z$ boson mass (if they are of the same flavour), at least one jet (well separated from the \met), including at least one $b$-tagged jet. The dileptonic \ttbar and ${\ttbar}V$ backgrounds are reduced by three mass requirements. The first requires a large value of $m_\mathrm{T2}$.
The second imposes an upper bound on the minimum invariant mass found by combining the highest-\pT $b$-tagged jet with each of the leptons.
The last one is constructed by assembling the two leptons ($\ell_{1,2}$) with the two jets with the highest $b$-tagging score ($j_{1,2}$) in all lepton--jet combinations. Assuming they come from top-quark decays, requiring $m^{t}_{b\ell}=\min(\max(m_{\ell_1j_1},m_{\ell_2j_2}),\max(m_{\ell_1j_2},m_{\ell_2j_1}))$ to be above 150~\GeV suppresses these backgrounds. The main SM background contributions in this SR are \ttbar, ${\ttbar}Z$ and $tWZ$ processes, followed by diboson events.

In all these channels, dedicated CRs are used to constrain the main backgrounds. In the 0-lepton channel, 1- and 2-lepton CRs are defined to constrain the \ttbar and $W$+jets backgrounds.
In the 1-lepton channel, the \ttbar CR is built by reversing the requirement on $am_\mathrm{T2}$ and the veto on a second $b$-tagged jet. In the 2-lepton channel, it is built by reversing the $m_\mathrm{T2}$ and $m^t_{b\ell}$ requirements. In both the 1- and the 2-lepton channels, a 3-lepton 3-jet CR is used to constrain the ${\ttbar}Z$ background.
In the 2-lepton channel, a 3-lepton $WZ$ CR is also built, with a lower jet multiplicity.

The event yields in all SRs of the three analysis channels are shown in Figure~\ref{fig:wtmet01}. Reasonable agreement is found in all SRs, the discrepancy seen in the 2-lepton channel being lower than 2$\sigma$.

\begin{figure}[tb]
\begin{center}
\subfloat[]{\includegraphics[width=0.54\textwidth]{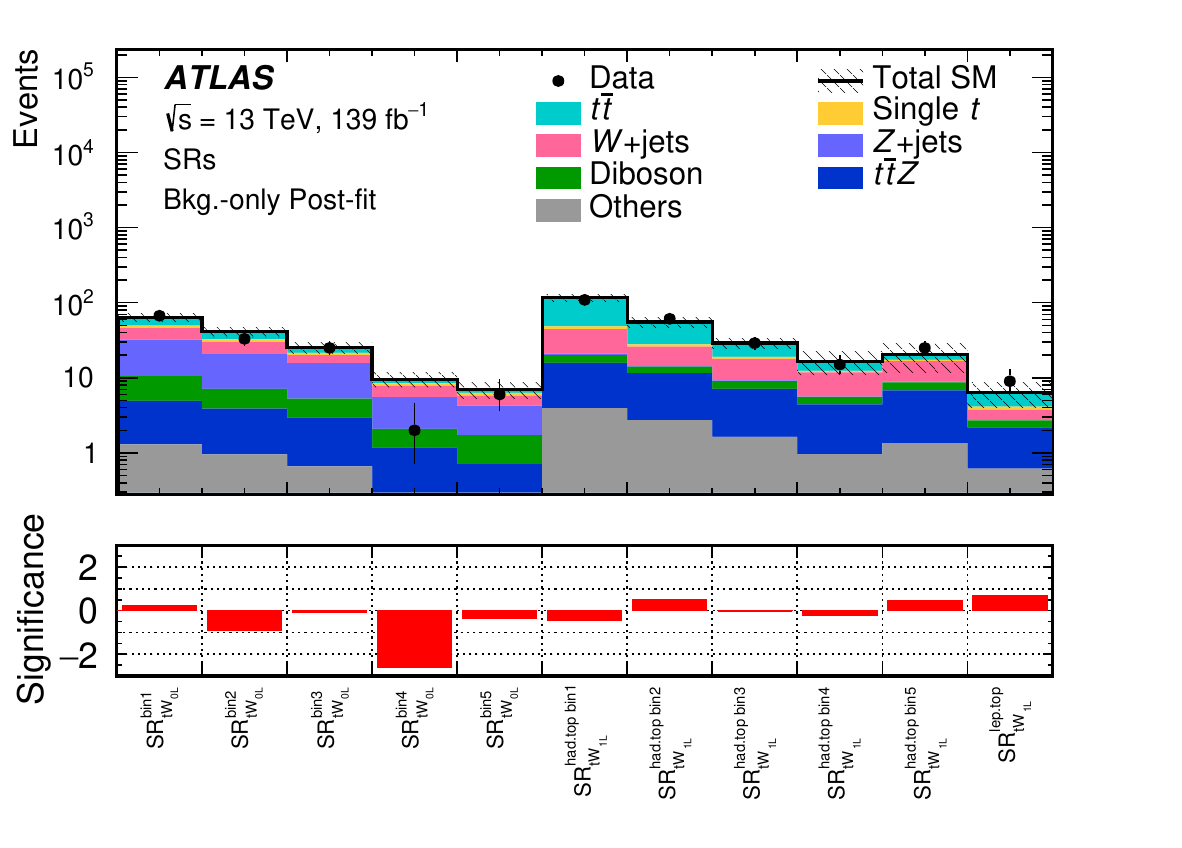}}
\qquad
\subfloat[]{\includegraphics[width=0.4\textwidth]{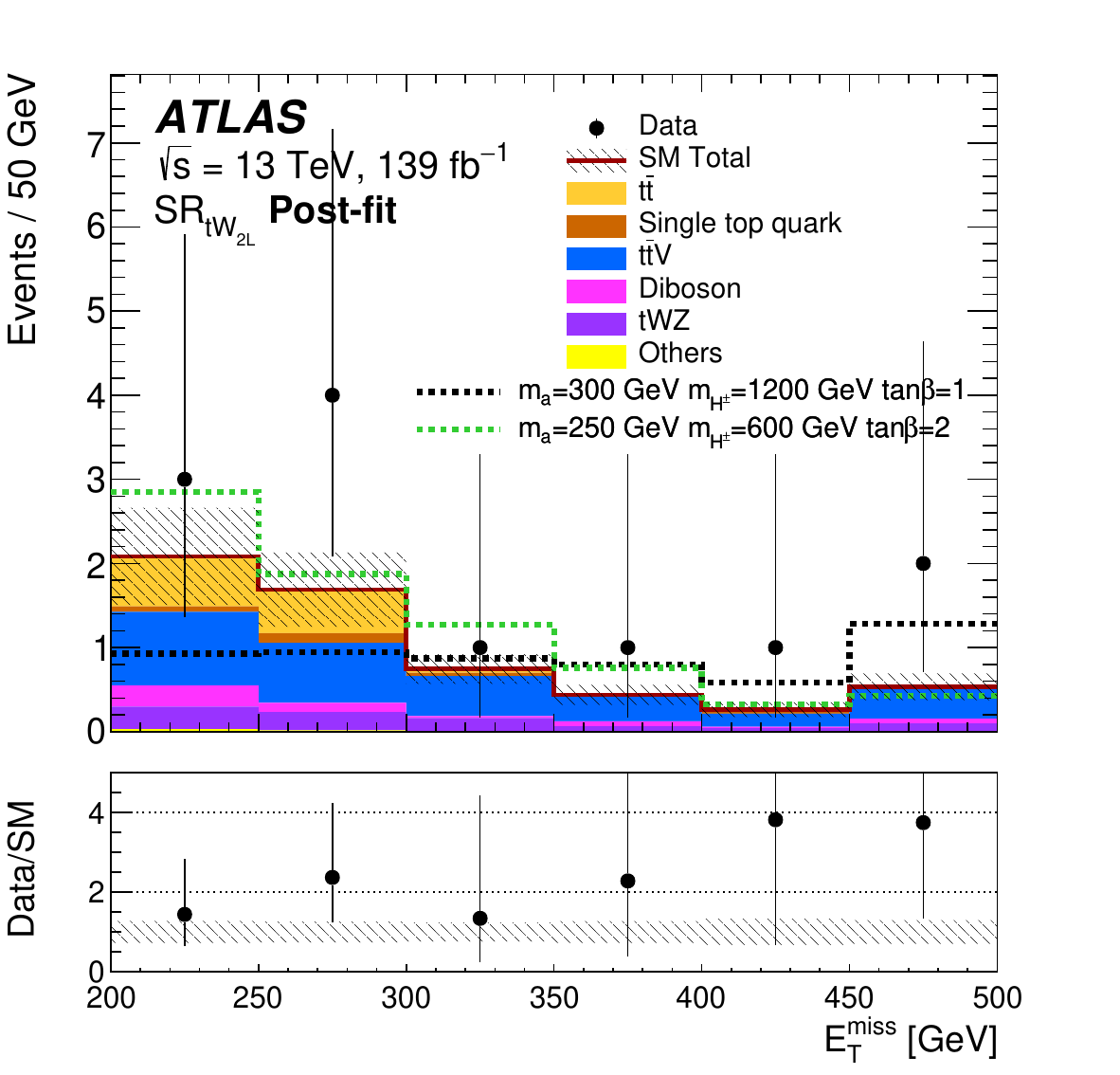}}
\end{center}
\caption{(a) Number of events found in each of the 0- and 1-lepton channel SRs~\cite{EXOT-2021-01} and (b) \met distribution in the 2-lepton SR~\cite{EXOT-2021-01} of the $Wt{+}\met$ search compared to the SM background expectation. In (b), examples of signal distributions are also shown.}
\label{fig:wtmet01}
\end{figure}

\subsubsection{Exclusions in the 2HDM+$a$ parameter space}

A statistical combination of the most sensitive channels, $h(bb){+}\met$, $Z(\to\ell\ell){+}\met$ and $tbH^{\pm}(tb)$, was performed in Ref.~\cite{EXOT-2018-64}, which provides an extensive set of exclusion limits. The constraints are evaluated for some representative benchmark scenarios, in which only one or two of the five free parameters introduced in Section~\ref{sec:2hdmA} are varied at a time, as shown for some examples in Figure~\ref{fig:2hdmaexcl}, thus showing some of the complementarity of the searches introduced above. The value of $m_\chi$, while having a strong effect on cosmological parameters such as the relic density, has only a limited impact on the collider searches for $m_\chi<m_a/2$ and is thus not varied in the figures shown here.

In the $(m_a,m_A)$ plane shown in Figure~\ref{fig:2hdmaexcl}(a), the pseudoscalar hierachy is explored, with $\tan\beta=1.0$ (favouring couplings to up-type quarks) and $\sin\theta=0.35$ (giving small $A{-}a$ mixing). The exclusion is dominated at lower $m_a$ by the $h(bb){+}\met$ and $Z(\to\ell\ell){+}\met$ channels, due to the resonant nature of the production mechanism. The $Z(\to\ell\ell){+}\met$ channel is able to probe lower $m_A$ values because a lower \met is expected there, and the $h(bb){+}\met$ analysis is based on an \met trigger. Conversely, the $h(bb){+}\met$ channel is more sensitive at higher values of $m_A$ because of a specific increase in the non-resonant $a^*\to ah$ production cross section there. At higher values of $m_a$, the $tbH^{\pm}(tb)$ channel dominates; as a direct search for charged Higgs production, it is relatively insensitive to the value of $m_a$.

In the $(\tan\beta, m_A)$ plane shown in Figure~\ref{fig:2hdmaexcl}(b), a value $m_a=250$~\GeV is chosen to suppress $a\to\ttbar$ and thus favour $a\to\chi\chi$.  The exclusion is mainly driven by the $Z(\to\ell\ell){+}\met$ analysis, except at larger values of $m_A$ as explained above, where the $h(bb){+}\met$ channel can dominate. The  sensitivity of the $tbH^{\pm}(tb)$ and $tttt$ channels is seen to increase at low $m_A$ , for which the heavy-Higgs production cross section is larger, and at low $\tan\beta$, as it favours third-generation couplings.

Finally, the exclusion as a function of $\sin{\theta}$ is shown in Figure~\ref{fig:2hdmaexcl}(c), which highlights the interplay between invisible and visible mediator decays due to the direct dependence of the couplings $g_{Aha}$, $g_{HZa}$ and $g_{a\ttbar}$ on $\sin\theta$: the higher values of $\sin{\theta}$ are covered by the $Z(\to\ell\ell){+}\met$ analysis, but its sensitivity at lower $\sin\theta$ falls quickly and the $tbH^{\pm}(tb)$ analysis becomes dominant. Although less sensitive than the $Z(\to\ell\ell){+}\met$ analysis, the $Wt{+}\met$ channel's sensitivity becomes comparable to that of $h(bb){+}\met$ at high $\sin{\theta}$ values, but the observed limit is weaker due to the small excess seen in the 2-lepton channel.

\begin{figure}[tb]
\begin{center}
\subfloat[]{\includegraphics[width=0.45\textwidth]{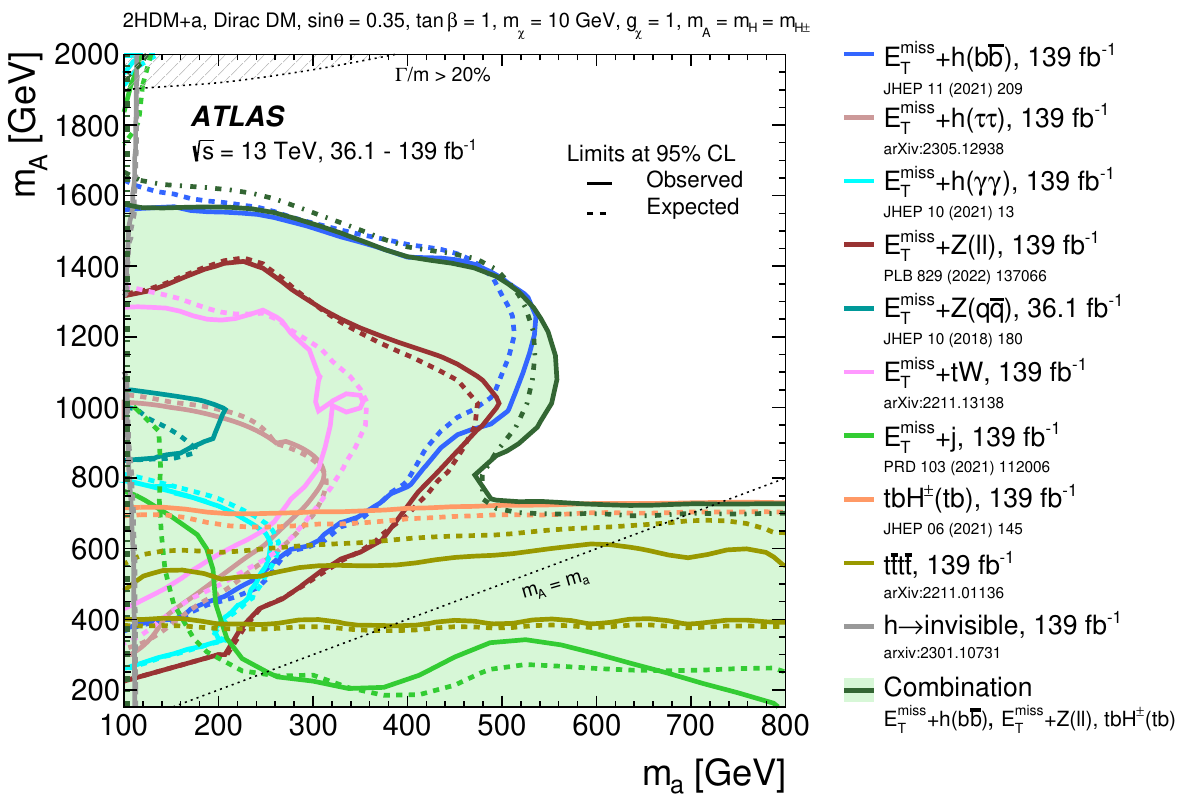}}
\qquad
\subfloat[]{\includegraphics[width=0.45\textwidth]{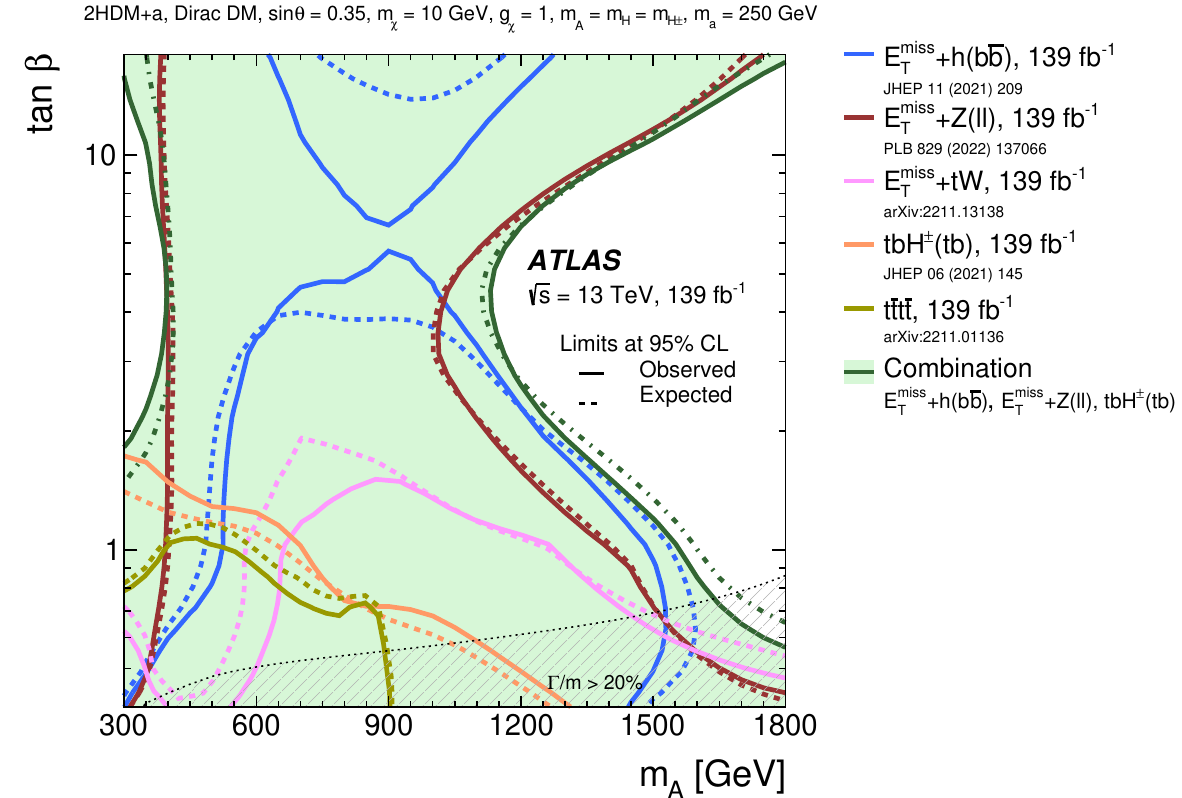}}
\qquad
\subfloat[]{\includegraphics[width=0.45\textwidth]{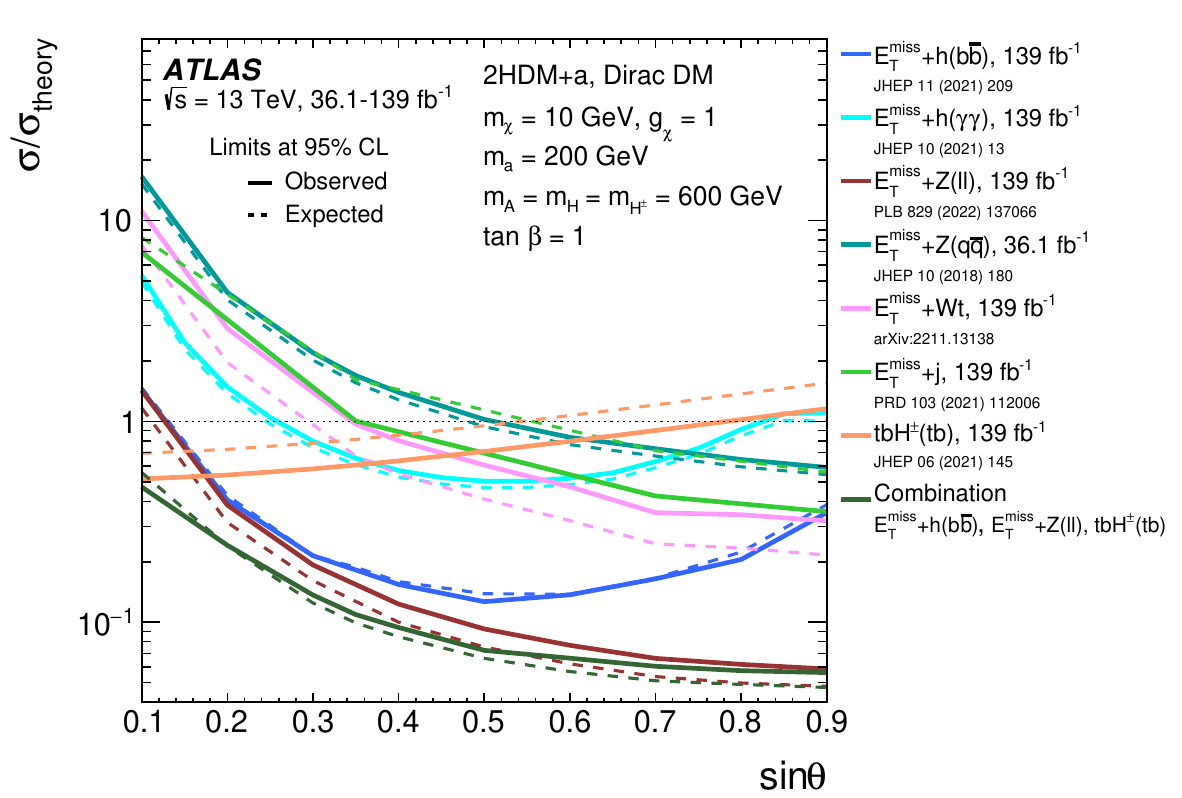}}
\end{center}
\caption{Observed (solid lines) and expected (dashed lines) exclusion limits~\cite{EXOT-2018-64} (a) in the $(m_a,m_A)$ plane with $\sin\theta=0.35$, $\tan\beta=1$, $m_\chi=10$~\GeV and $g_\chi=1$, (b) in the $(m_A,\tan\beta)$ plane for $\sin\theta=0.35$, $m_a=250$~\GeV, $m_\chi=10$~\GeV and $g_\chi=1$, and (c) as a function of $\sin{\theta}$ for  $\tan\beta=1$, $m_a=200$~\GeV, $m_\chi=10$~\GeV, $g_\chi=1$ and $m_A=600$~\GeV.}
\label{fig:2hdmaexcl}
\end{figure}

\subsection{Comparison with direct-detection experiments}
\label{sec:DMDD}

While the pseudoscalar portal discussed above would evade direct detection, the limits obtained in the V/A and Higgs portal models can, with some assumptions, be translated into limits on the WIMP--nucleon scattering cross section and thus compared with limits obtained by DM direct-detection experiments.
As described in Ref.~\cite{Boveia:2016mrp}, this translation can be made in the V/A simplified model once the couplings introduced in Section~\ref{sec:dmV} are fixed (to $(g_g,g_\chi,g_\ell)=(0.25,1,0)$ here).  A Higgs or vector mediator would lead to spin-independent scattering, while an axial-vector mediator would lead to spin-dependent scattering. For a DM Higgs portal, an effective-field theory framework is used~\cite{Djouadi:2012zc}, where the scale of new physics is well above the Higgs boson mass, and the DM particle is either a scalar or vector boson or a Majorana fermion. For a vector DM candidate, some ultraviolet-complete models in which it  is the gauge field of a new U(1)$'$ group are also considered~\cite{Zaazoua:2021xls,Arcadi:2020jqf,Baek:2014jga}. In these models a dark Higgs boson, with mass $m_2$ and mixing angle $\alpha$ with the SM Higgs boson, is introduced to generate the DM candidate's mass.

\begin{figure}[tb]
\begin{center}
\subfloat[]{\includegraphics[width=0.49\textwidth]{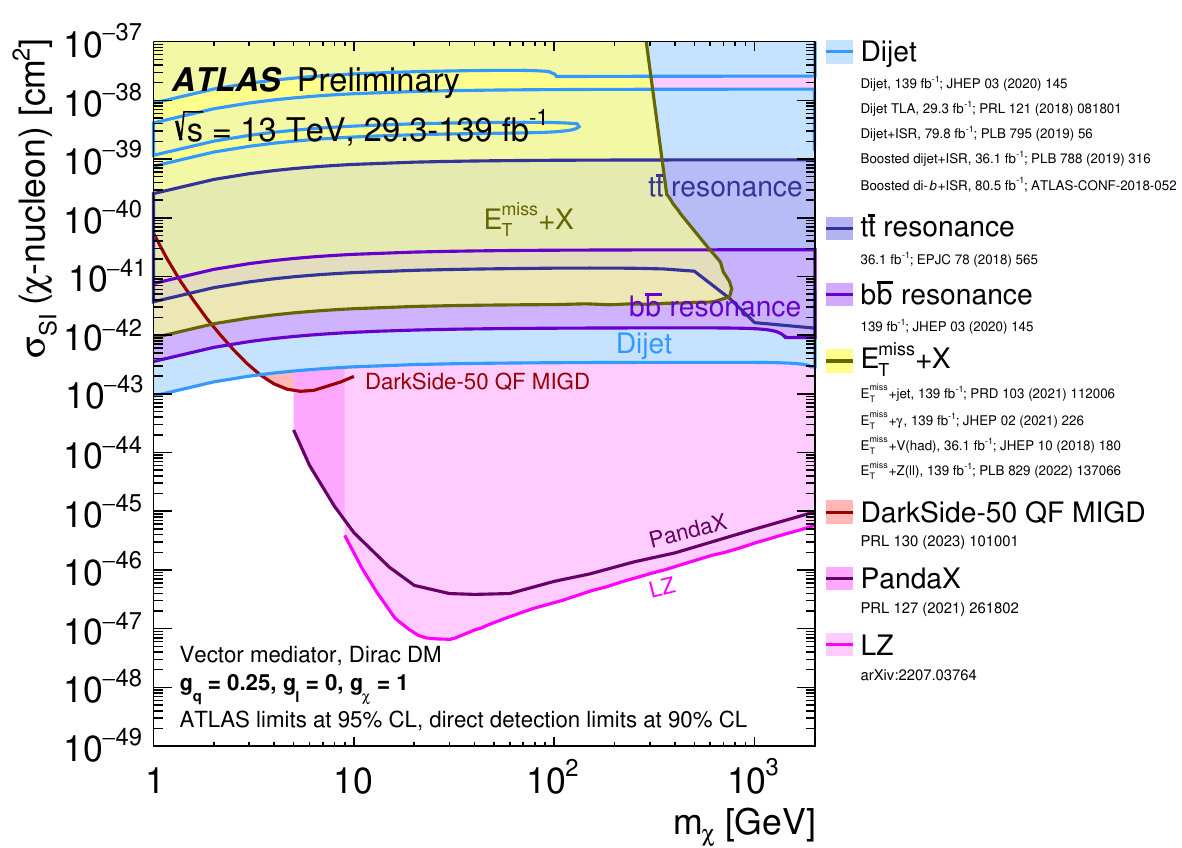}}
\subfloat[]{\includegraphics[width=0.49\textwidth]{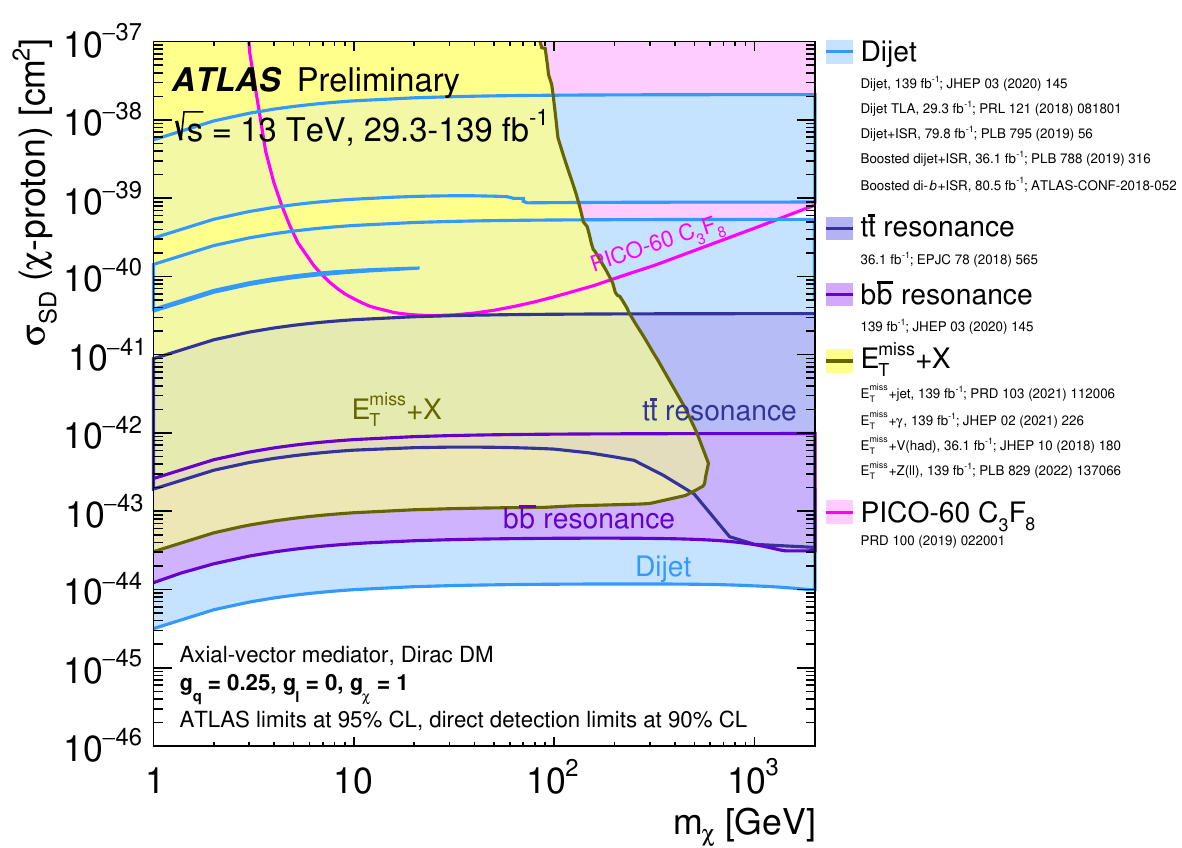}}
\qquad
\subfloat[]{\includegraphics[width=0.55\textwidth]{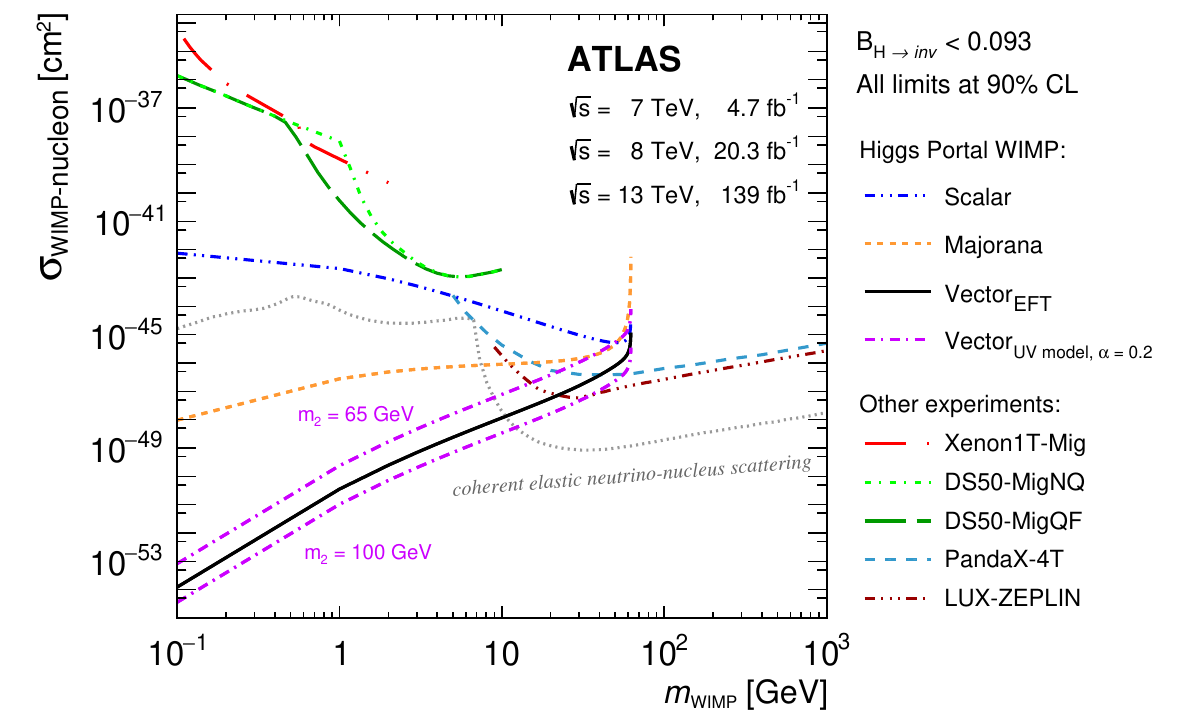}}
\end{center}
\caption{Comparison of the limits obtained (a,b) at 95\% CL by the resonant and $X$+\met searches in the V/A model or (c) at 90\% CL from the invisible-Higgs searches~\cite{HIGG-2021-05} with the limits obtained by direct-detection experiments in the plane of (a,c) the spin-independent $\chi$--nucleon scattering cross section versus $m_\chi$ or (b) the spin-dependent $\chi$--proton scattering cross section versus $m_\chi$ for different model assumptions. In (c), besides the direct-detection limits~\cite{PandaX-4T:2021bab,DarkSide:2018bpj,LZ:2022lsv,PhysRevLett.123.241803}, the coherent elastic neutrino--nucleus scattering \emph{neutrino fog} expected for a germanium target~\cite{Billard:2013qya,Ruppin:2014bra} is also shown.  }
\label{fig:DMDD}
\end{figure}

The limits obtained are shown in Figure~\ref{fig:DMDD}, where one can see the complementarity of the approaches. The collider searches presented here are particularly helpful in covering the low-mass region of the parameter space, which is kinematically easily accessible at the LHC but very difficult to probe in direct-detection experiments because of the very low induced recoil energies. The invisible-Higgs search can even probe below the \emph{neutrino fog} limit, the parameter space in which the neutrino background becomes dominant for direct-detection searches, but it becomes irrelevant for DM masses above $m_H/2$, where the Higgs boson would not decay invisibly. The spin-dependent interaction, which the direct-detection experiments constrain much less than the spin-independent one,
is probed very effectively by the collider searches.
\FloatBarrier

\subsection{Strongly coupled hidden sector}
\label{sec:svj}

If there is a strongly coupled dark sector, its matter fields (the \emph{dark quarks}) produced from SM particles through a mediator could hadronize into dark hadrons~\cite{Strassler:2006im,Strassler:2006ri,Han:2007ae,Schwaller:2015gea}. Depending on the parameters of the models, some dark mesons would be stable (and be DM candidates) while some would decay back into SM particles either promptly or with long lifetimes. Two searches for such dark sectors are presented here: one in which there is a bi-fundamental mediator $\Phi$ which couples SM quarks to dark quarks via the $t$-channel~\cite{EXOT-2022-37}, and one in which the mediator is a \Zprime boson, producing the dark quarks in the $s$-channel~\cite{HDBS-2018-45}.  In both these searches, the models considered imply that the unstable dark hadrons decay promptly.

In the $t$-channel production search~\cite{EXOT-2022-37}, two jets would be produced in a non-resonant way; since they are composed of SM particles and invisible dark hadrons, they are known as semi-visible jets~\cite{Cohen:2017pzm}. Some \met is thus expected and could even point in the direction of a jet, a possibility which is usually excluded by azimuthal separation requirements in other LHC DM searches.
The dark QCD parameters of the signal samples are fixed in accord with Refs.~\cite{Cohen:2017pzm,Albouy:2022cin}, with the $\Phi q q_d$ coupling, $\lambda$, set to 1. The free parameters considered are thus the branching ratio for decay of unstable dark mesons into stable dark mesons, $R_\mathrm{inv}$, and the mediator mass, $m_\Phi$. Since the analysis targets high mediator masses, and mismeasured jets from the multijet background need to be suppressed, the analysis selects events with $\met$ and $H_\mathrm{T} $ values above 600~\GeV. The events must contain at least two jets, with one being $\Delta\phi<2.0$ away from the \met direction, and no lepton. Taking $j_1$ and $j_2$ to be the closest and farthest jets in azimuth from the \met direction, their \pT balance, $\pT^\mathrm{bal}=|\vec{\pT}(j1)-\vec{\pT}(j2)|/ (\pT(j1)+\pT(j2))$, and their azimuthal separation are two largely uncorrelated discriminating variables, which are binned to form nine SRs. The main $W/Z$+jets and semileptonic \ttbar backgrounds are constrained using three muon CRs: one with two OS muons, and two with one muon and no or one \btagged jet.
The small multijet background is normalized in a CR at lower \met and high azimuthal separation.
The event yield in the SRs agrees well with the background expections,
so limits are drawn in the $(m_\Phi, R_\mathrm{inv})$ plane, as shown in Figure~\ref{fig:svj}(a), excluding the region up to $m_\Phi=2.7$~\TeV. At large values of $R_\mathrm{inv}$, for which a jet could simply disappear, the jet+\met analysis introduced in Section~\ref{sec:jetmet} complements this analysis, excluding higher $m_\Phi$ values.

In the $s$-channel production search~\cite{HDBS-2018-45}, the dark mesons are assumed to decay promptly into SM particles and the fraction of invisible components is assumed to be negligible, following the benchmark models from Ref.~\cite{Park:2017rfb} (for which the lightest dark baryons could still be DM candidates). Due to the double hadronization process (first in the dark sector and then in the SM), the dark jets considered are typically wider than the SM QCD jets and, for the dark-sector models considered, have a higher associated charged-particle multiplicity due to the multiplicity of dark mesons produced in the dark shower and their decays. The analysis thus selects events containing two high-\pT large-$R$ jets with high track multiplicity, and looks for a resonant excess in $m_\mathrm{JJ}$. The smooth $m_\mathrm{JJ}$ shape of the dominant SM QCD multijet background is determined in a data-driven way. Since no significant excess is seen, limits are placed on the cross section times branching ratio for the production of dark quarks through a \Zprime mediator as a function of the \Zprime mass, as shown for one benchmark in Figure~\ref{fig:svj}(b). In this figure, the limits are compared with theory predictions for given values of $g_q$ and $g_{q_d}$, the couplings of the \Zprime to SM quarks and to dark quarks, respectively. For the values chosen, $g_q=0.05$ and $g_{q_d}=0.2$, this analysis is able to exclude \Zprime masses up to 3.0~\TeV for this model, while the $\Zprime\to qq$ dijet resonance search constraints~\cite{EXOT-2019-03} are evaded due to the small $g_q$ value.

\begin{figure}[tb]
\begin{center}
\subfloat[]{\includegraphics[width=0.38\textwidth]{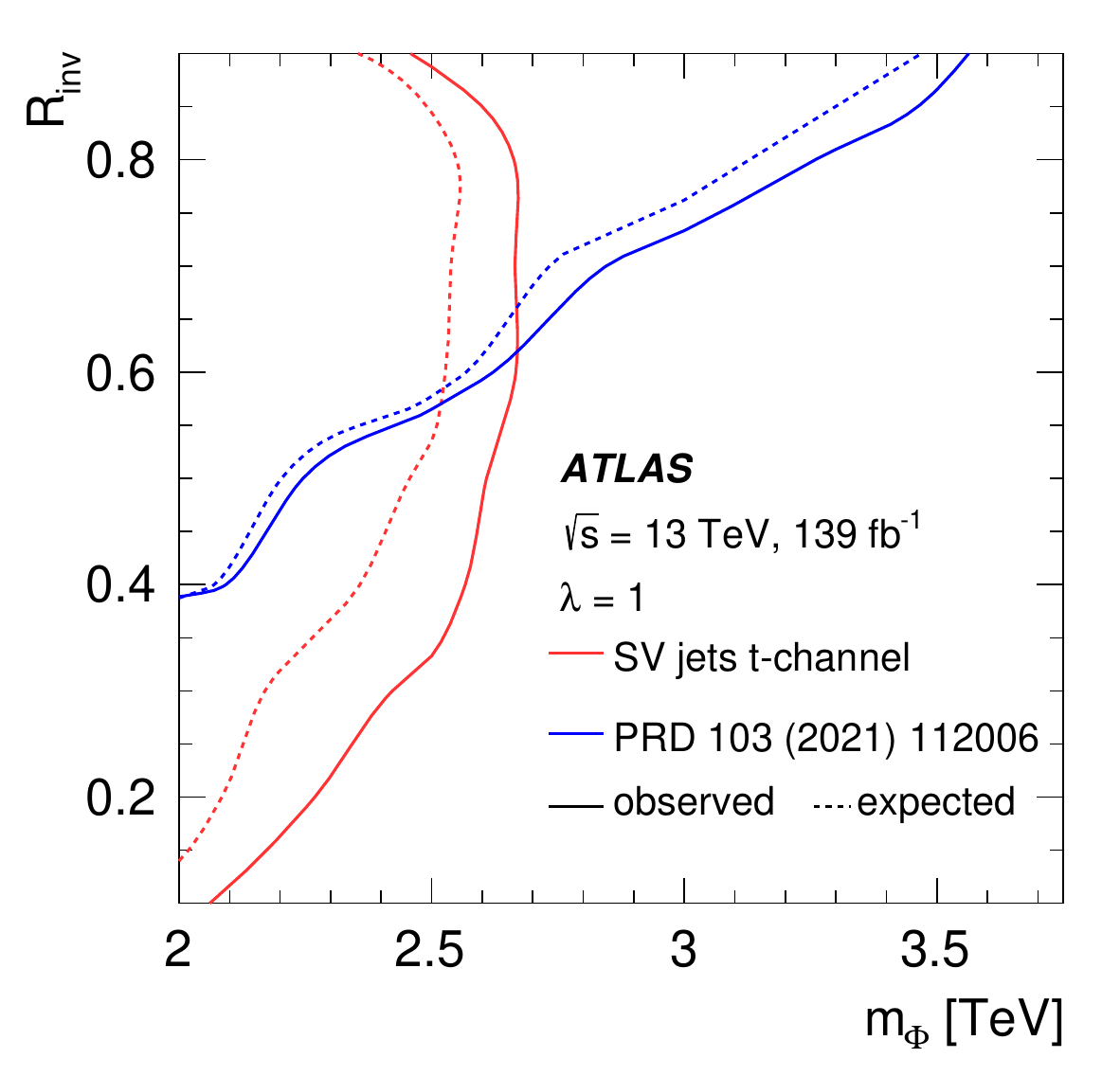}}
\qquad
\subfloat[]{\includegraphics[width=0.52\textwidth]{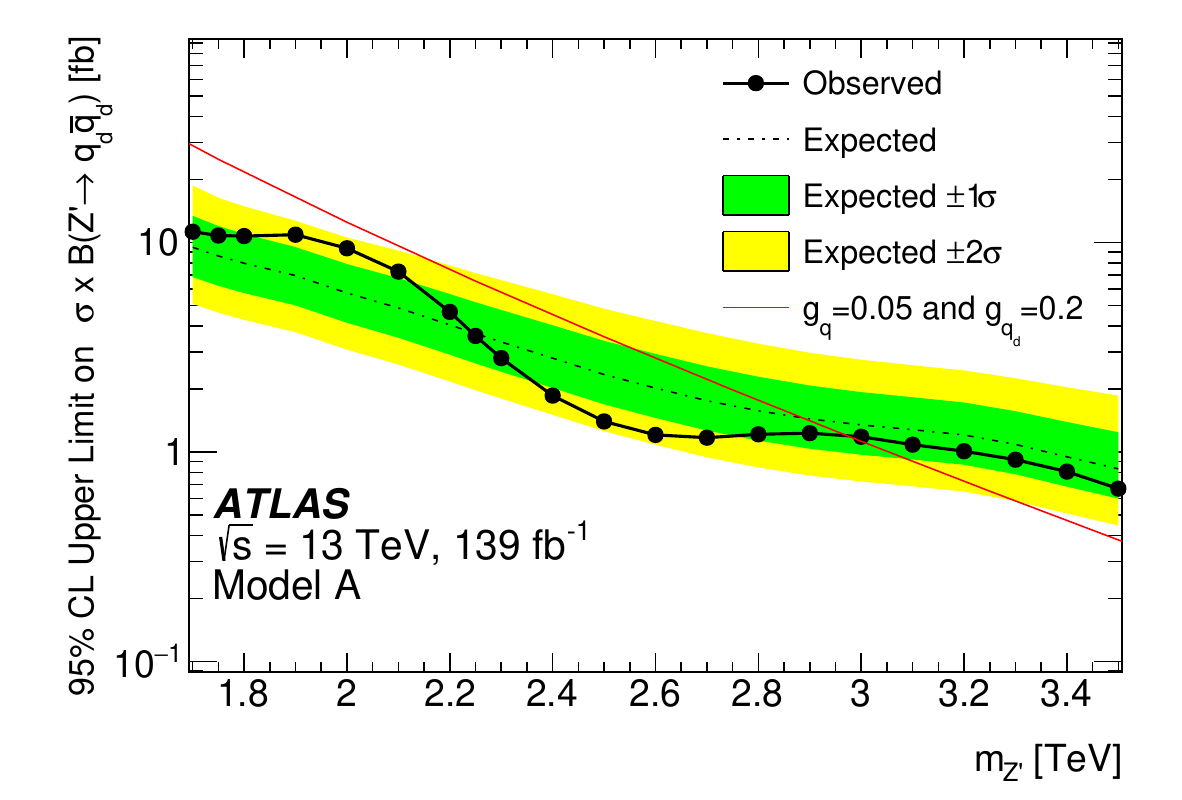}}
\end{center}
\caption{(a) Limits in the $(m_\Phi, R_\mathrm{inv})$ plane obtained in the $t$-channel search for semi-visible (SV) jets~\cite{EXOT-2022-37} and in the jet+\met search. (b) Limits set on the cross section times branching ratio for the production of dark quarks through a \Zprime mediator as a function of the \Zprime mass for one of the benchmark considered in the $s$-channel search~\cite{HDBS-2018-45}, compared to a theory prediction.}
\label{fig:svj}
\end{figure}


%
\section{Long-lived multi-charged or highly ionizing particles}
\label{sec:mcp}

Long-lived particles need not be neutral like the ones discussed in the last two sections. Indeed, various BSM theories predict the existence of particles with more than one unit of electric charge.

They can have relatively low charge multiplicity $|z|$, being known as multi-charged particles (MCPs), such as the two doubly charged fermions predicted by the almost-commutative model~\cite{Stephan:2005uj}, the stable multi-charged technibaryons predicted in the walking technicolour model~\cite{Sannino:2004qp}, or the long-lived doubly charged Higgs bosons predicted in the left--right symmetric model~\cite{Mohapatra:1974hk}. In this case, they would behave like higher-mass, more ionizing muons in the detector.

They can also have very high charge multiplicity, being known as highly ionizing particles (HIPs), such as strange matter~\cite{PhysRevD.30.2379} and Q-balls~\cite{Coleman:1985ki}. Dirac magnetic monopoles~\cite{PhysRev.74.817} are also classified as HIPs; they have a magnetic charge $Ne/2\alpha\approx68.5Ne$ (where $N$ is an integer, $e$ is the electron charge and $\alpha\approx1/137$ is the fine-structure constant), and an $N=1$ Dirac monopole would have an energy loss in the detectors comparable to that of an ion of electric charge $|z|=68.5$.

Searches for these two signatures were conducted as described below, with the MCP search targeting the range $2<|z|<7$~\cite{EXOT-2018-54} and the HIP search covering $20<|z|<100$~\cite{EXOT-2019-33}. In both cases, Drell--Yan (DY) and photon-fusion (PF) pair-production processes are considered,
following the model described in Ref.~\cite{Song:2021vpo}.

As mentioned above, the MCPs would look like heavy muons with a higher specific energy loss $\dif{E}/\dif{x}$ in the pixel, TRT and MDT subdetectors. Since the muon trigger is only  sensitive to particles with $\beta=v/c>0.65$, due to built-in timing restrictions, two other triggers are also used: the calorimeter-based \met trigger, which relies on the presence of ISR in the signal events, and a late-muon trigger which fires in events which have a jet in the current bunch-crossing and a muon in the next one. The events are then selected by requiring the presence of a central ID--MS combined muon, which must be well isolated from other tracks found in the silicon detectors. This isolation criterion is useful for removing events in which the large ionization loss would be due to multiple particle crossings instead of a unique MCP. Four variables are then used to build the signal regions: the $\dif{E}/\dif{x}$ significance,  $S(\dif{E}/\dif{x})$, in each of the three subdetectors, and $f^\mathrm{HT}$, the number of high-threshold (HT) hits\footnote{The HT is designed to discriminate between energy depositions from transition-radiation photons and the energy loss by minimum-ionizing particles.} divided by the number of low-threshold hits on the track as measured in the TRT. The $S(\dif{E}/\dif{x})$ variables are defined as the difference between the observed signal and the average value expected for a relativistic muon, divided by its root-mean-square width, where the expected values are measured in a $Z\to\mu\mu$ control region. A SR is defined for $z=2$, requiring $S(\mathrm{pixel~}\dif{E}/\dif{x})>13$, $S(\mathrm{TRT~}\dif{E}/\dif{x})>2$ and $S(\mathrm{MDT~}\dif{E}/\dif{x})>4$. Since the pixel $\dif{E}/\dif{x}$ measurement saturates for higher $z$ values, and the corresponding hits are not recorded, this variable is not used in the SR defined to target $z>2$, which instead requires $f^\mathrm{HT}>0.7$ and $S(\mathrm{MDT~}\dif{E}/\dif{x})>7$. In both regions, the background is estimated in a data-driven way using an ABCD method. In the $z=2$ SR, the ABCD regions are defined in the TRT versus MDT $\dif{E}/\dif{x}$ plane, while in the $z>2$ SR, $f^\mathrm{HT}$ and $S(\mathrm{MDT~}\dif{E}/\dif{x})$ are used instead. This results in an expected background of $1.6\pm0.4\,\text{(stat.)}\pm0.5$\,(syst.)  events in the $z=2$ SR, where 4 events are observed, and of $0.034\pm0.002\,\text{(stat.)}\pm0.004$\,(syst.) events in the $z>2$ SR, where no events are observed.

At higher electric charge multiplicity, a HIP passing through the TRT would not only produce a HT hit in a given straw but also produce other HT hits in neighbouring straws via the many $\delta$-rays it generates. Furthermore, because a HIP is too heavy to produce a shower in the EM calorimeter, its energy deposit in this detector would remain narrow even though it is likely to be stopped there. The event selection for such events starts with a custom HIP trigger based on these properties. The usual tracking algorithms are not used in this search since they can be confused by the multiple $\delta$-rays, and because magnetic monopoles, which are one of the signals targeted in this search, would bend in the $r{-}z$ (and not the usual $r{-}\phi$) plane. Instead, HIP reconstruction proceeds in three steps, illustrated in Figure~\ref{fig:mcphip}(a): it starts with an EM cluster which defines a $\phi$ direction near which TRT hits are counted; this direction is refined to align with the richest HT-hits region; and then this new $\phi$ direction is used as the centre of an 8-mm-wide rectangular road in the barrel (or a 12~mrad $r{-}\phi$ wedge in the endcap), which should capture the energy deposited while not including too many \pileup hits. As in the MCP case, the fraction of HT hits $f^\mathrm{HT}$ found in the road (or wedge) is used as a discriminating variable. Another powerful variable in this search is $w$, which is the average of the three fractions of EM cluster energy found in the $N$ most energetic cells of the presampler ($N=2$) and the first ($N=4$) and second ($N=5$) layers of the EM calorimeter, where the energy in each of these layers must exceed some minimum value. The SR is defined by requiring $f^\mathrm{HT}\ge0.77$ and $w\ge0.93$. These two variables also form the ABCD plane that is used to estimate the background. After these selections, $0.15\pm0.04\,\text{(stat.)}\pm0.05$\,(syst.) background events are expected in the SR, and no events are observed.

Both the MCP and HIP searches allow limits to be placed in the plane of the production cross section versus the electric charge, as shown for the Drell--Yan production mode in Figure~\ref{fig:mcphip}(b). Similar limits are obtained for the PF production mode. The HIP search is also interpreted in terms of magnetic monopoles: for example, for an $N=1$ spin-0 magnetic monopole, the search excludes masses up to 2.1~\TeV (3.4~\TeV) in the DY (PF) production mode.

\begin{figure}[tb]
\begin{center}
\subfloat[]{\raisebox{0.1\height}{\includegraphics[width=0.45\textwidth]{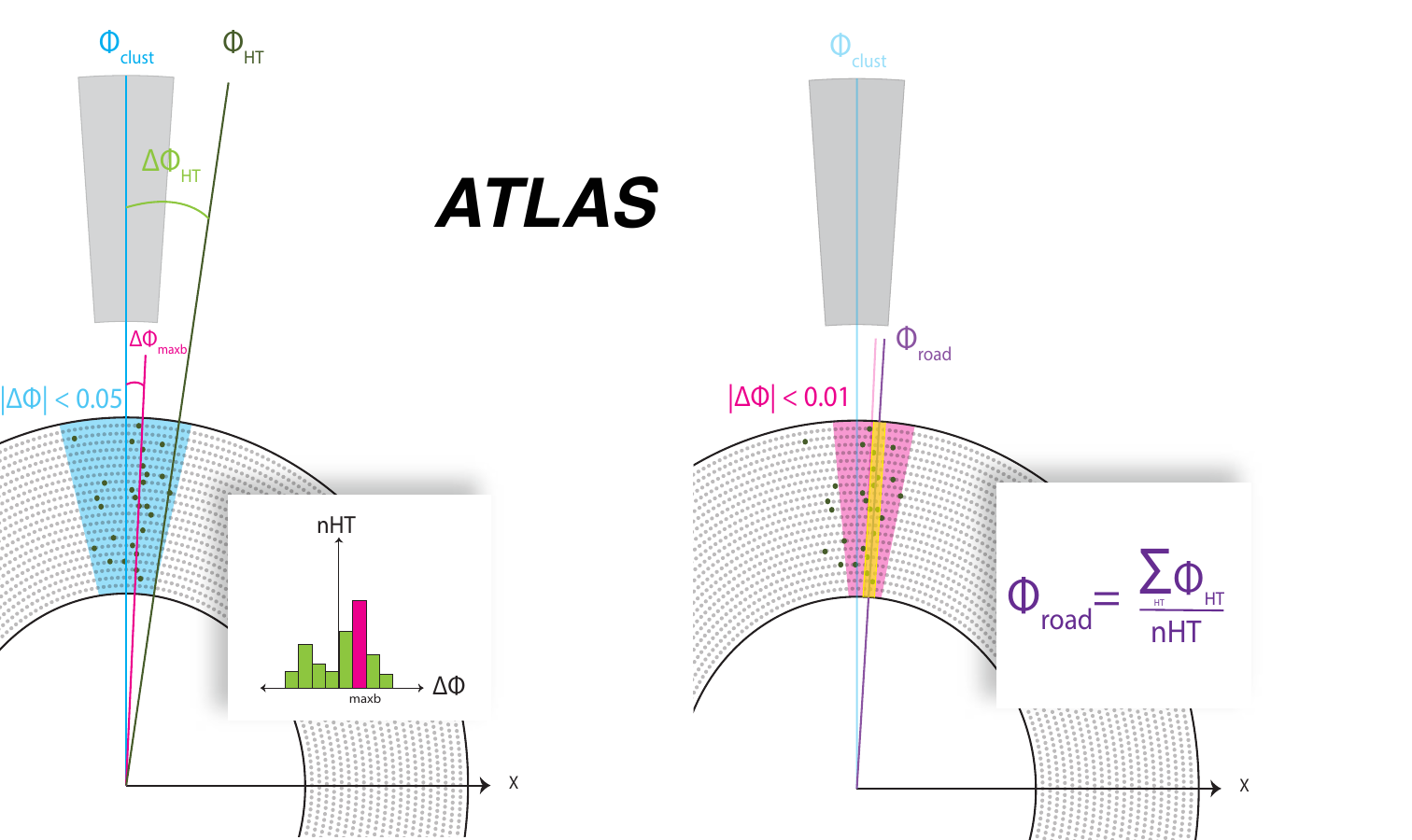}}}
\qquad
\subfloat[]{{\includegraphics[width=0.45\textwidth]{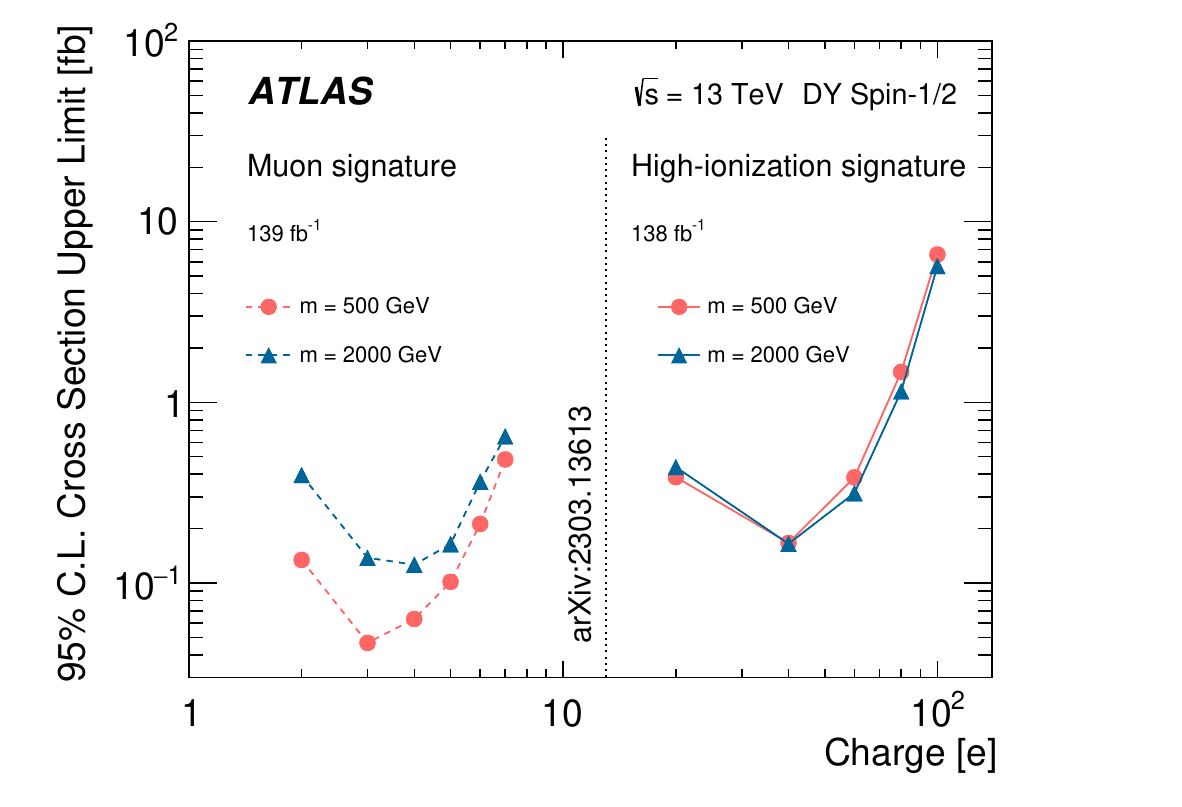}}}
\end{center}
\caption{(a) Schematic illustration of the TRT road-centering and hit-counting algorithms for HIP candidates: in the first stage (left), the wedge is defined around the azimuthal angle of the EM cluster (blue solid line) near which TRT hits are counted; this direction is refined to align with the richest HT-hits (nHT) region and this new direction (right) is used as the centre of a road (purple line) which should capture the energy deposited. (b) Observed exclusion limit on the cross section versus electric charge multiplicity as seen in the MCP (muon signature) and HIP (high-ionization signature) searches~\cite{EXOT-2018-54,EXOT-2019-33}, for DY production. The models with 500~\GeV masses are strongly excluded, given that a DY production cross section ($\sigma_\mathrm{DY}$) of at least a few fb is expected for the lowest charge considered and grows with the charge. At 2~\TeV, none of the MCP models are excluded because $\sigma_\mathrm{DY}\le O(0.01)$~fb is expected over the charge range considered, while for HIPs, the limits obtained between $40e$ and $80e$ are lower than the expected $\sigma_\mathrm{DY}$. }
\label{fig:mcphip}
\end{figure}



%
\section{Extra dimensions, gravitons and quantum black holes}
\label{sec:gravity}

The ATLAS experiment is also able to explore scenarios trying to bridge the gap between particle physics and gravity. Multiple BSM theories seek to explain the weakness of gravity relative to the other known fundamental interactions, sometimes using extra dimensions (EDs) to solve this puzzle~\cite{Arkani-Hamed:1998jmv,Randall:1999ee}. If our four dimensions exist on a sheet in a higher-dimensional space, and if gravitons, the hypothetical spin-2 mediators of gravity, are able to propagate across these EDs, then gravity could appear weaker than it actually is.

\subsection{Gravitons in the ADD and RS scenarios}
\label{sec:ADDRS}
The Arkani-Hamed, Dimopoulos and Dvali (ADD) model comprises $n$ large extra dimensions of size $R$~\cite{Arkani-Hamed:1998jmv} which transforms the fundamental scale $M_D$ of the 4+$n$-dimensional theory into the Planck scale via $M_\mathrm{Planck}^2\sim M_D^{2+n}R^n$. The compactification of the EDs results in a Kaluza--Klein (KK) tower of massive graviton modes with a mass splitting that is inversely proportional to $R$. Since large EDs are postulated, the observable graviton states appear as a continuum. If produced in the $pp$ collisions, the KK graviton (${G_\mathrm{KK}}$) escapes into the EDs, and the searches must rely, as in the DM searches discussed in Section~\ref{sec:dm}, on the production of a visible object, such as an ISR jet or photon. The jet+\met analysis (see Section~\ref{sec:jetmet}) has looked for such KK gravitons,
setting limits on $M_D$ at the multi-\TeV level for $n$ values from 2 to 6,\footnote{The $n=1$ case is already largely ruled out by gravitational measurements, as the corresponding $R$ to have $M_D$ at the \TeV scale would be larger than the Earth--Sun distance, while for $n=2$, it is already at the mm scale.} as can be seen in Figure~\ref{fig:ED}(a).  The photon+\met search for this signal was also pursued, with early \RunTwo data, but proved less sensitive~\cite{EXOT-2015-05}.

KK gravitons can also be obtained in a Randall--Sundrum (RS) model of EDs~\cite{Randall:1999ee}. In the original version of the RS model, here called RS1, a five-dimensional anti-de Sitter spacetime is postulated, in which SM particles and gravity are localized on two different branes. Since the fifth (extra) dimension is warped, the strength of gravity appears exponentially suppressed by the warp factor on the SM brane. In the bulk version of the RS model, only the Higgs boson is fixed on the TeV brane, while the other SM fields are allowed to propagate in the bulk of the ED. The SM particles are the zero-modes of these five-dimensional fields, and their mass hierarchies can be explained by localizing the heavy-fermion fields nearer the \TeV brane, and the lighter ones, nearer the Planck brane. Since the ${G_\mathrm{KK}}$ are localized near the \TeV brane, their couplings to light fermions are suppressed, contrary to the RS1 scenario. The signal samples in the search for RS KK gravitons, which can be seen as well-separated narrow resonances, are produced
for given choices of $k/M_\mathrm{Planck}$, where $k$ is the curvature parameter.

The RS KK graviton can decay into gluons or quarks~\cite{Allanach:2002gn} and the dijet resonance analysis (see Section~\ref{sec:qstar}) is therefore sensitive to part of the parameter space. The model considered is one of a RS1 KK graviton decaying into $b$-quarks with $k/M_\mathrm{Planck}=0.2$. The limit set on the lightest graviton's mass ($m_{G_\mathrm{KK}}$) in this model, obtained by using the $b$-tagged SR of the dijet analysis, is given in Table~\ref{tab:ED}. A search was also performed in the semileptonic \ttbar final state with a partial \RunTwo dataset~\cite{EXOT-2015-04}: its interpretation in terms of bulk RS gravitons gives weaker limits than those reported below, as shown in Table~\ref{tab:ED}. It was also interpreted in terms of a search for the first mode of the bulk RS KK gluon, excluding masses below 3.8~\TeV for a 15\% resonance width. The dilepton resonance search introduced in Section~\ref{sec:resqqll}, although not giving a mass limit for these specific models, provides auxiliary material for reinterpreting the obtained generic limits in terms of spin-2 resonances.

RS1 KK gravitons can also decay into photons, and a search for a high-mass $\gamma\gamma$ resonance~\cite{HIGG-2018-27} was performed for signals with $k/M_\mathrm{Planck}\le0.1$. In this analysis, two tight and isolated photons are required, with $m_{\gamma\gamma}>150$~\GeV, and with the leading (subleading) photon having $\pT>0.3\,(0.25)\,m_{\gamma\gamma}$. Templates of the $m_{\gamma\gamma}$ distributions are obtained by fitting a functional form to the distributions obtained either from MC simulation (for the non-resonant $\gamma\gamma$ background) or from a CR reversing some of the photon identification selection (for the $\gamma$+jet background). The relative amount of each background is determined using two-dimensional sidebands around the signal region for each photon. These sidebands are built by relaxing the photon isolation and identification criteria, and the $\gamma\gamma$ purity in the SR is found to be around 97\% above 400~\GeV.  Since no resonant excess is seen above this background in data, a limit is set on $m_{G_\mathrm{KK}}$, as shown in Table~\ref{tab:ED}.

A pair of Higgs bosons can also result from the bulk RS $G_\mathrm{KK}$ decay; such an interpretation was given in the resonant $HH\to bbbb$ search~\cite{HDBS-2018-41}, which uses a resolved SR and a boosted SR to probe lower or higher $m_{G_\mathrm{KK}}$.  In the resolved channel, four $b$-tagged jets are required and their pairing into two Higgs boson candidates $H_1$ and $H_2$ is decided by using a BDT. In the boosted SR, two high-\pT large-$R$ jets form the candidates, with two to four associated $b$-tagged track-jets. To reject the multijet background, $H_1$ and $H_2$ must not be too far apart in $\eta$, and in the resolved SR the \ttbar background is suppressed with a top-quark veto, which is based on attempts to reconstruct top quarks from the jets at hand. The SR in both regimes is then defined as an enclosed space in the $(m_{H_1},m_{H_2})$ plane around the expected reconstructed Higgs boson masses, with a annulus around it defining a CR which is used in the dominant multijet background's estimation. In the resolved SR, the background is estimated from data in a 2-$b$-tagged region, whose shape is corrected using event kinematic knowledge obtained from the annular CR with a neural network. In the boosted regime, events in regions with a lower number of $b$-tagged track-jets are used in conjunction with the annular CR in a simultaneous fit. The di-Higgs mass is the final discriminant and, in the resolved SR, it is corrected for better accuracy by rescaling the reconstructed $H_1$ and $H_2$ masses to 125~\GeV. Since no significant excess is seen, the two channels are combined to provide a limit on $m_{G_\mathrm{KK}}$, listed in Table~\ref{tab:ED}.

Finally, diboson resonance searches, which are discussed as \Wprime and \Zprime searches in Section~\ref{sec:gauge}, are also sensitive to scenarios with bulk RS KK gravitons, as summarized in Ref.~\cite{ATL-PHYS-PUB-2023-007}. A comparison of the limits on the production cross section in the three $ZZ$ channels (fully hadronic, semileptonic and fully leptonic) is shown in Figure~\ref{fig:ED}(b), while the limits on $m_{G_\mathrm{KK}}$ obtained by the various analyses for a value of $k/M_\mathrm{Planck}=1.0$ are summarized in Table~\ref{tab:ED}. While most of the analyses focus on ggF production, the semileptonic $WW/ZZ$ search also probes VBF production of the graviton, which is also interesting given the light-fermion suppression of the bulk RS ${G_\mathrm{KK}}$ couplings.

\begin{figure}[tb]
\begin{center}
\subfloat[]{\includegraphics[width=0.35\textwidth]{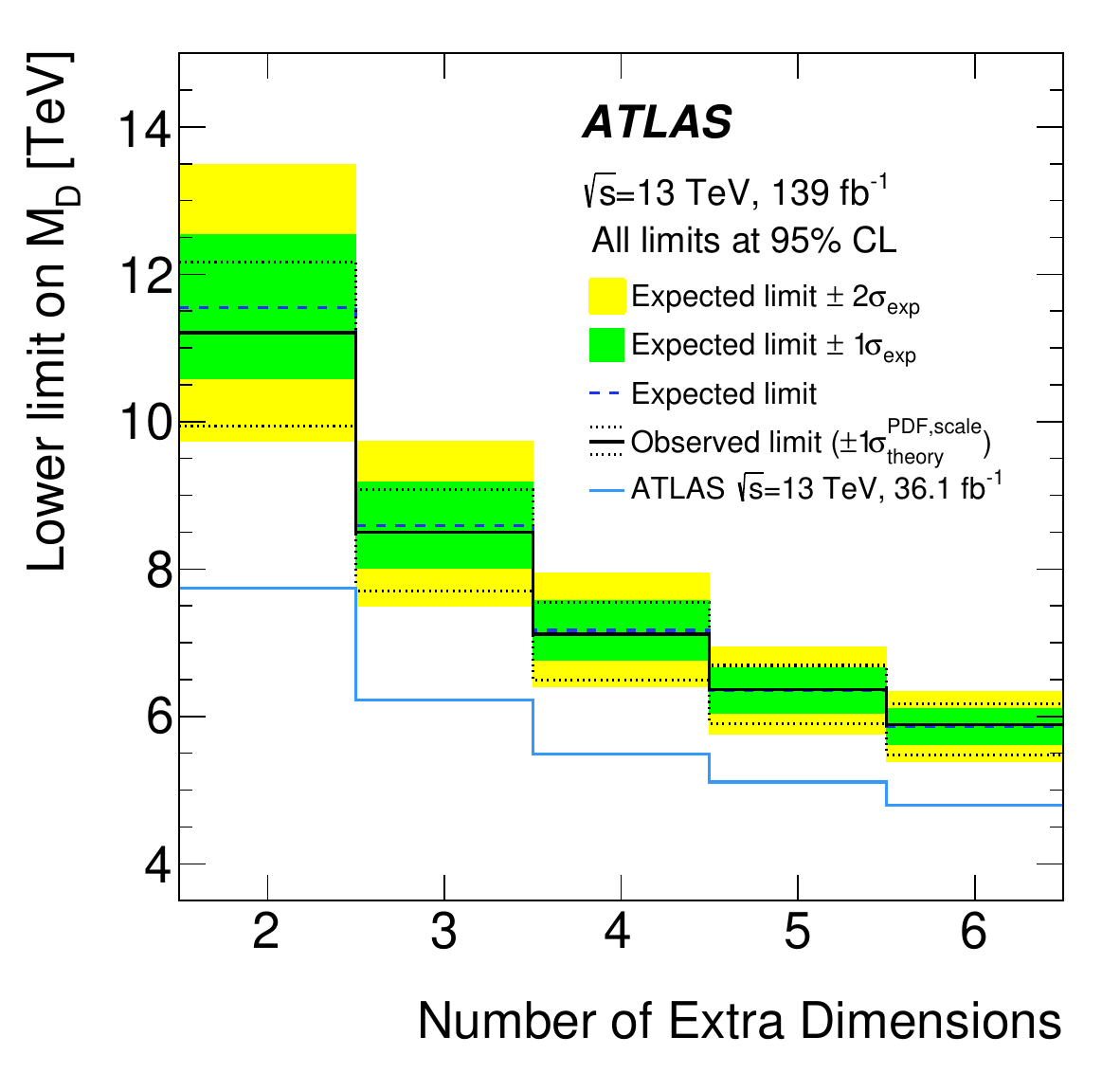}}
\qquad
\subfloat[]{\includegraphics[width=0.49\textwidth]{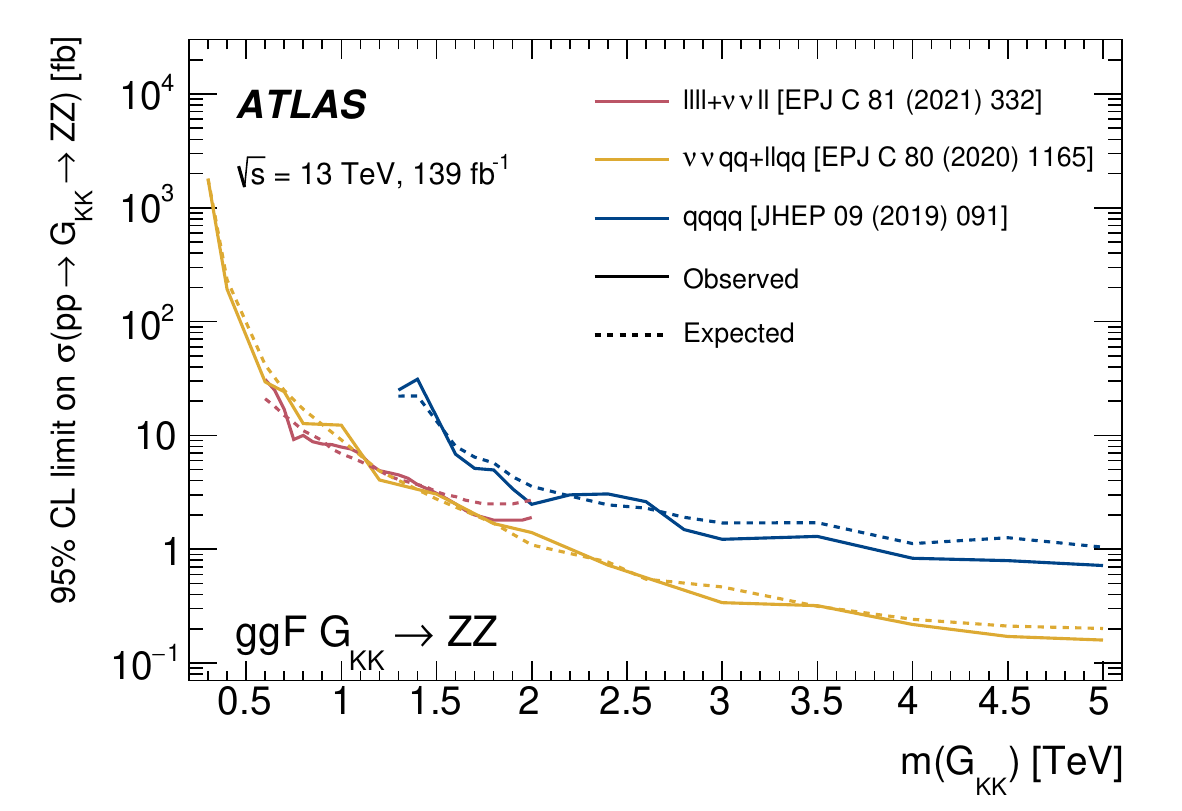}}
\end{center}
\caption{Observed exclusion limit on (a) the fundamental scale $M_D$ versus the number of extra dimensions in the ADD model, obtained by the jet+\met analysis~\cite{EXOT-2018-06}, and on (b) the cross section times branching ratio versus the mass of the graviton in a bulk RS model, obtained in the $ZZ$ resonance searches~\cite{ATL-PHYS-PUB-2023-007}. }
\label{fig:ED}
\end{figure}

\begin{table*}[tp]
\begin{center}
\caption{95\% CL lower limits on the lightest KK graviton mass obtained in various analyses for various RS scenarios. }
\begin{tabular}{l c c c}
\hline
\hline
Analysis final state &  Model & $k/M_\mathrm{Planck}$ & Excluded mass range for $m_{G_\mathrm{KK}}$ [$\text{T\electronvolt}$] \\\hline
$bb$ & RS1 & 0.2 & $<2.8$ ~\cite{EXOT-2019-03} \\\hline
$\gamma\gamma$ & RS1 & 0.1 & $<4.5$ ~\cite{HIGG-2018-27} \\\hline
Semileptonic \ttbar (36.1 \ifb) & bulk RS & 1.0 & 0.45--0.65  ~\cite{EXOT-2015-04} \\\hline
$HH\to bbbb$ & bulk RS & 1.0 & 0.298--1.46~~\cite{HDBS-2018-41}  \\\hline
$WW/ZZ\to qqqq$ & bulk RS & 1.0 & 1.3--1.8 ~\cite{HDBS-2018-31} \\\hline
$WW\to\ell\nu qq + ZZ\to\ell\ell q q$ & & & \\
ggF production & bulk RS & 1.0 & < 2.0 ~\cite{HDBS-2018-10} \\
VBF production & bulk RS & 1.0 & ~~< 0.76 ~\cite{HDBS-2018-10} \\\hline
$ZZ\to \ell\ell\ell\ell + ZZ\to\ell\ell\nu\nu$ & bulk RS & 1.0 & ~~< 1.83 ~\cite{HIGG-2018-09} \\
\hline
\hline
\end{tabular}
\label{tab:ED}
\end{center}
\end{table*}

\subsection{Gravitons in the clockwork gravity model}

The continuum clockwork gravity model~\cite{Giudice:2016yja,Giudice:2017fmj}, which has a five-dimensional spacetime metric, is also related to the gravity/weak-scale hierarchy problem. It predicts a narrowly-spaced spectrum of KK gravitons which can appear as a long-range semi-periodic structure in the invariant mass distribution: near the onset of the graviton spectrum, which is governed by a mass parameter $k_\mathrm{CW}$, the mass splittings are generally of the order of a few percent and can be resolved in a $\gamma\gamma$ or $ee$ search, while the splitting decreases and becomes unresolvable for higher graviton modes. The cross section is determined by $M_5$, the five-dimensional reduced Planck mass. A search for this periodic spectrum of resonances was performed in the $ee$ channel~\cite{EXOT-2019-40}, using the invariant mass spectrum from the $\Zprime\to\ell\ell$ search discussed in Section~\ref{sec:gauge_resonant}, and in the diphoton channel, using the spectrum from the search described in the previous section. In both cases, a functional form is fitted to the data to estimate the background. The signals are modelled using analytic invariant-mass templates including detector resolution effects, an example of which can be seen in Figure~\ref{fig:CW}(a). Continuous wavelet transforms (using a Morlet wavelet definition~\cite{10.1190/1.1441328}) are used to analyse the mass spectra in the frequency domain, transforming these spectra into images (called scalograms) displaying the wavelet amplitude in the frequency versus invariant mass plane. In these scalograms, a signal would appear as a region of high amplitude localized in frequency and mass above a more continuous background, as shown in Figure~\ref{fig:CW}(b). In order to analyse these images in terms of the clockwork model, a NN classifier is used. No significant excess is seen and limits are set, as shown in Figures~\ref{fig:CW}(c) for the $ee$ channel and~\ref{fig:CW}(d) for the diphoton channel. In the latter, the areas in the $k_\mathrm{CW}$ versus $M_5$ plane where the observed limits are stronger than the expected limits indicate that the data is effectively smoother than the expected fluctuations arising from the uncertainties.

\begin{figure}[tb]
\begin{center}
\subfloat[]{\includegraphics[width=0.35\textwidth]{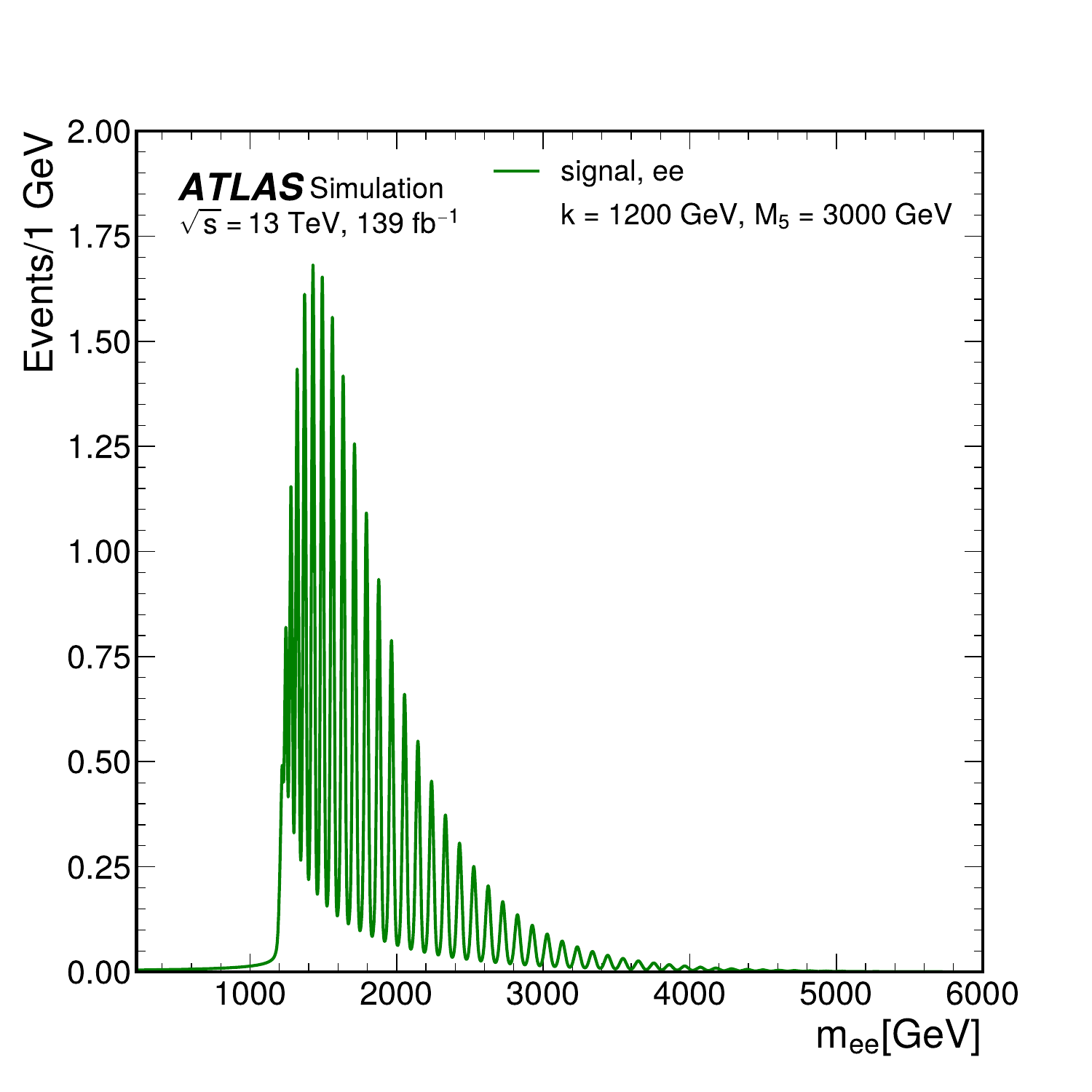}}
\qquad
\subfloat[]{\includegraphics[width=0.35\textwidth]{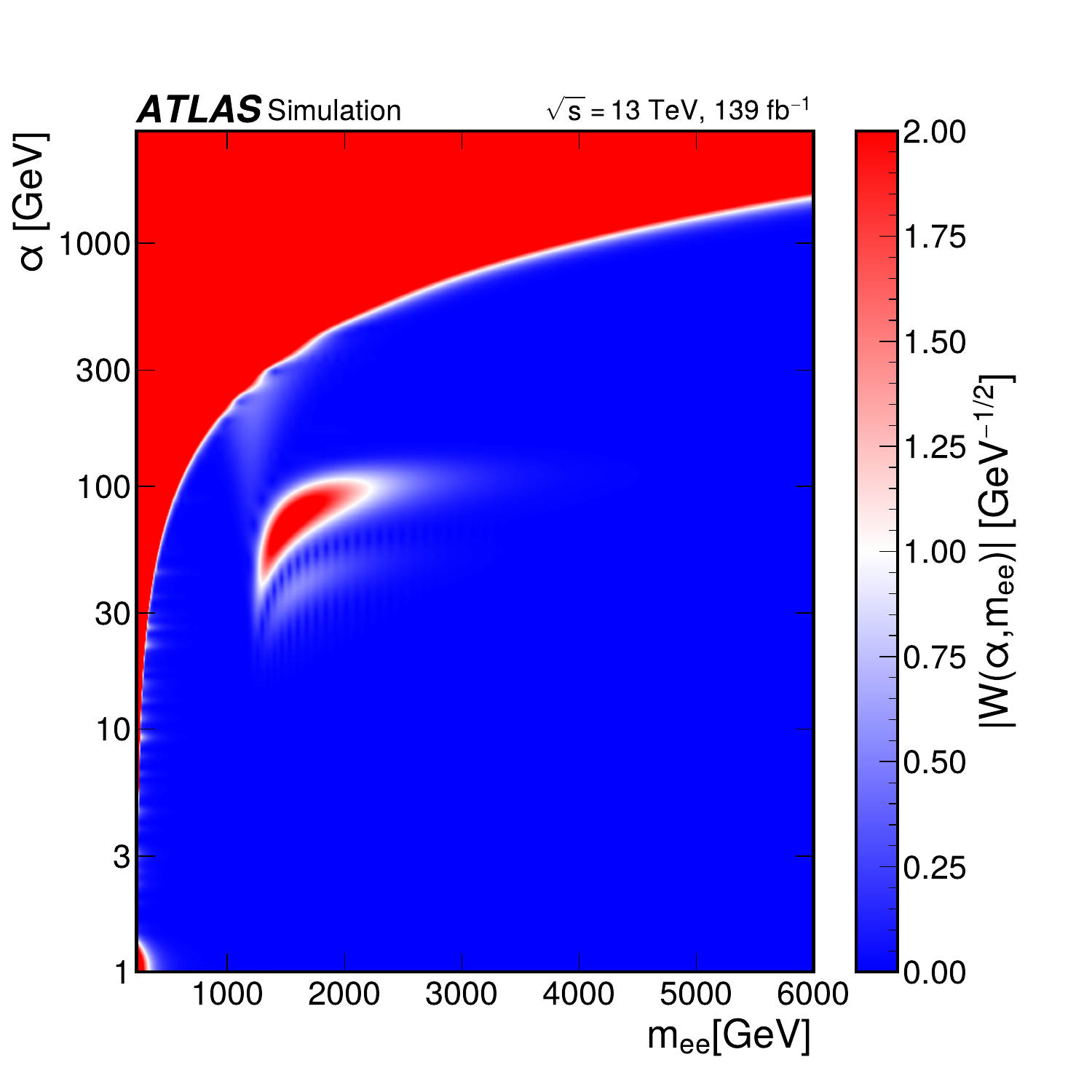}}
\qquad
\subfloat[]{\includegraphics[width=0.35\textwidth]{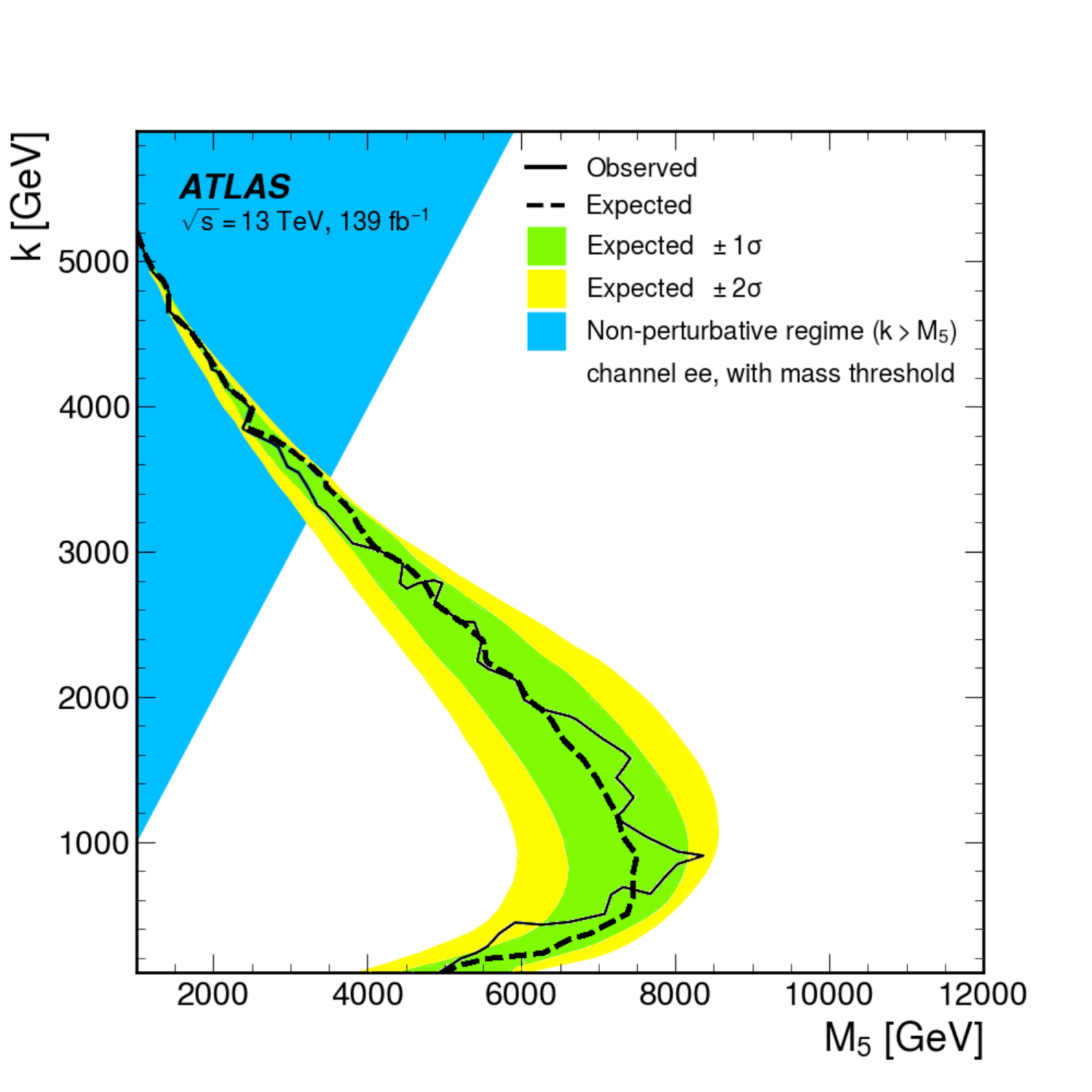}}
\qquad
\subfloat[]{\includegraphics[width=0.35\textwidth]{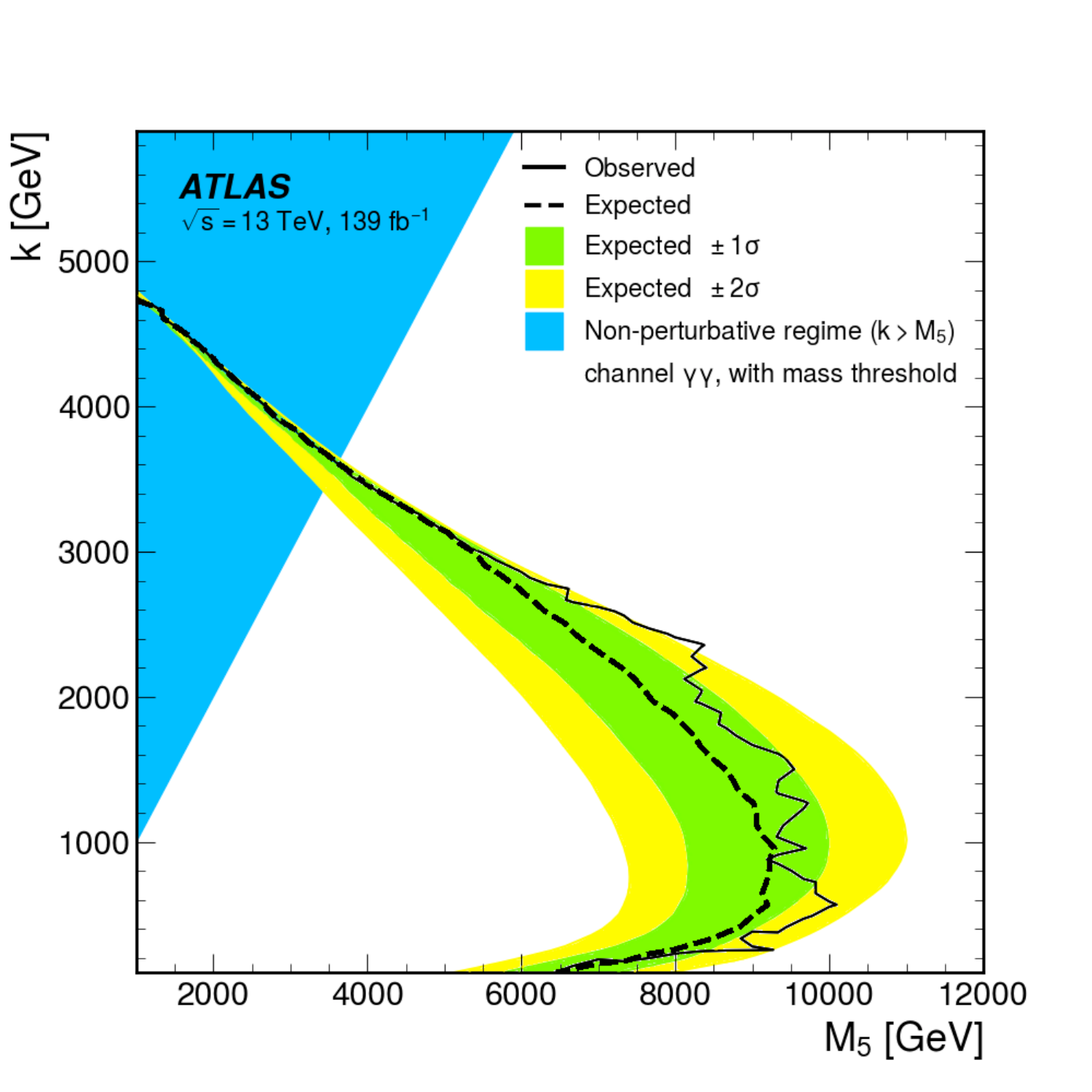}}
\end{center}
\caption{Example, in the $ee$ channel of the clockwork analysis~\cite{EXOT-2019-40} and for $k_\mathrm{CW}=1200$~\GeV (here called $k$) and $M_5=3000$~\TeV, of (a) a signal invariant mass distribution and (b) a simulated signal-plus-background scalogram in which $\alpha$ is inversely proportional to the frequency and $W$ is the amplitude coefficient. The limits in the $k_\mathrm{CW}$ versus $M_5$ plane is shown for (c) the $ee$ channel and (d) the $\gamma\gamma$ channel; in these plots the area excluded is to the left of the exclusion line.}
\label{fig:CW}
\end{figure}

\subsection{Quantum black holes}
In the ADD and the RS models, quantum black holes (QBH) could potentially be produced at the LHC~\cite{Gingrich:2009hj,Calmet:2008dg} when the energy is above the fundamental Planck scale $M_D$. Unlike semi-classical black holes, which decay due to Hawking radiation  into multiparticle final states, QBHs with masses near $M_D$ decay into two-particle final states~\cite{Anchordoqui:2001cg}. The dijet search can therefore also be used in this case. Signal events were generated for the ADD case with \textsc{BlackMax}~\cite{Dai:2007ki} for $n=6$ and various $M_D$ values, and the results of the dijet search in its inclusive signal region were reinterpreted to obtain limits on the cross section as a function of the QBH mass, $m_\mathrm{QBH}$.

The search for QBHs can also be performed in final states that violate SM global symmetries: this is done by looking for a resonance in the $eq$ and $\mu q$ final states~\cite{EXOT-2018-14}. In this search, \textsc{QBH}\,3.0~\cite{Gingrich:2009da} is used in conjunction with \PYTHIA[8] to generate events in the ADD (with $n=6$) and RS1 scenarios. The events are selected by requiring the presence of exactly one high-\pT jet and one high-\pT lepton with $m_{\ell j}>2$~\TeV.\footnote{Given that the signal is generated at LO, vetoing additional jet activity affects the signal acceptance in a way that can be corrected for by studying the veto effect in NLO $V$+jets events.} The main background comes from $W$+jets events, with some smaller contributions from $\ttbar$ and $Z$+jets events. These are estimated using CRs at lower $m_{\ell j}$, requiring or not requiring the presence of $b$-tagged jets, \met, or an extra lepton. In the electron channel, the fake-lepton background contribution is estimated via a matrix method. The main uncertainties in the background estimation come from uncertainties in the jet energy resolution and lepton modelling. The data agree with background-only expectations, and the limits from the electron and muon channels,  yielding very similar sensitivities, are combined.

For the ADD model, the dijet search places a lower limit on $m_\mathrm{QBH}$ at 9.4~\TeV, while the limit set by the $e/\mu$+jet search is 9.2~\TeV. The $e/\mu$+jet analysis is also interpreted in terms of an RS1 model, resulting in a limit of  $m_\mathrm{QBH}>6.8$~\TeV.


%
\section{Summary and conclusions}
\label{sec:conclusion}
The \RunTwo data have offered an unprecedented opportunity to search for answers to many of the fundamental questions still open today in high-energy physics. While no significant excess of events in data has been seen, the rich harvest of results has redefined the possible paths for physics beyond the Standard Model, putting more stringent constraints on the scales and couplings of new states as shown in this report and also summarized in Ref.~\cite{ATL-PHYS-PUB-2023-008}. A comparison of limits obtained by analyses presented in this report with those obtained in similar analyses performed with the \RunOne dataset is presented in Table~\ref{tab:Run1VsRun2}, showing the evolution of the constraints.

\begin{table*}[tb]
\begin{center}
\caption{Comparison between some 95\% CL lower limits obtained by \RunTwo analyses presented in the indicated sections of this report and the limits obtained in similar analyses with the \RunOne dataset.}
\begin{tabular}{l @{}c c c}
\hline
\hline
Model and final state & Section & \multicolumn{2}{c}{Excluded Range} \\
& & Run 1 & Run 2 \\\hline\hline
$q^*$ in a dijet resonance & \ref{sec:qstar} & $m<4.06$~\TeV ~\cite{EXOT-2013-11} & $m<6.7$~\TeV ~\cite{EXOT-2019-03} \\\hline
$\Zprime_\mathrm{SSM}$ in a dilepton resonance &  \ref{sec:resqqll} &  $m<2.90$~\TeV~\cite{EXOT-2012-23}  &  $m<5.1$~\TeV ~\cite{EXOT-2018-08} \\\hline
Type-III seesaw heavy leptons  & \ref{sec:typeiii} & $m<335$~\GeV~\cite{EXOT-2014-07} & $m<790$~\GeV ~\cite{EXOT-2018-33} \\
in $\ell\ell\nu\nu q q$  & & & \\\hline
VLQ $T$ (Singlet, $2\ell+3\ell$) & \ref{sec:vlq} & $m < 0.66$~\TeV~\cite{EXOT-2013-17}  & $m < 1.27$~\TeV ~\cite{EXOT-2018-58}\\\hline
Scalar $LQ^{u}_3$ ($LQLQ\to t\nu t\nu$)  & \ref{sec:lq} & $m < 640$~\GeV~\cite{EXOT-2014-03} & $m<1240$~\GeV ~\cite{SUSY-2018-12}\\\hline
LFV $Z\to e\mu$ &  \ref{sec:lfvz} & ${\cal B}>7.5\times10^{-7}$~\cite{EXOT-2013-02} & ${\cal B}>2.62\times10^{-7}$ ~\cite{EXOT-2018-35} \\\hline
FRVZ $\gamma_d$ in $H\to2\gamma_d+X$ & \ref{sec:dpj}  & $15 < c\tau < 260$ mm~\cite{EXOT-2013-22} & $0.42 < c\tau < 1001$ mm ~\cite{EXOT-2022-15} \\
with ${\cal B}(H\to2\gamma_d)=10\%$  & & & \\
and $m_{\gamma_d}=0.4~\GeV$ & & &  \\\hline
$H\to$ invisible combination & \ref{sec:hinv} & ${\cal B}>0.252$~\cite{HIGG-2015-03} & ${\cal B}>0.113$ ~\cite{HIGG-2021-05}\\\hline
Multi-charged particle & \ref{sec:mcp} &  $m<660$~\GeV~\cite{EXOT-2013-20} & $m<1060$~\GeV ~\cite{EXOT-2018-54} \\
with $|z|=2$  & & & \\\hline
ADD with $n=6$ in jet+\met & \ref{sec:ADDRS} & $M_D<3.06$~\TeV~\cite{EXOT-2013-13} & $M_D<5.9$~\TeV ~\cite{EXOT-2018-06} \\
\hline
\hline
\end{tabular}
\label{tab:Run1VsRun2}
\end{center}
\end{table*}

While the increases in integrated luminosity and centre-of-mass energy have both greatly contributed to the improvement in the limits, new or improved techniques for object identification, such as the large-radius tracking used in some long-lived-particle searches, new theoretical computations, such as the NLO $Z/W$+jets dedicated computations used in the jet+\met and vector-boson-fusion jets+\met analyses, and better analysis techniques, relying for example on machine learning, have all played their part in beating the simple scaling of the \RunOne results in many cases.
But the improvements have not stopped there either -- the \RunTwo analyses have also pushed back the frontiers by looking for new final states, such as semi-visible jets or clockwork gravitons, which could have passed through the net cast by the Collaboration in \RunOne.

With \RunThr now well underway, this strategy will be pursued further. While the increase in centre-of-mass energy from 13 to 13.6~\TeV might seem modest compared to the 8 to 13~\TeV jump from \RunOne to \RunTwo,  it translates to a production cross-section increase which is not negligible for high-mass states. This, coupled with the increased global data sample, upgrades to the detector (notably trigger-related ones) and the relentless performance and analysis improvement efforts, will further boost the sensitivity of the searches. \RunThr will be followed by the HL-LHC phase, which will increase the data sample tenfold with a further upgraded detector, allowing even more uncharted parameter space to be probed. The voyage of exploration is far from over!
\section*{Acknowledgements}
%

%
%

%
%

We thank CERN for the very successful operation of the LHC and its injectors, as well as the support staff at
CERN and at our institutions worldwide without whom ATLAS could not be operated efficiently.

The crucial computing support from all WLCG partners is acknowledged gratefully, in particular from CERN, the ATLAS Tier-1 facilities at TRIUMF/SFU (Canada), NDGF (Denmark, Norway, Sweden), CC-IN2P3 (France), KIT/GridKA (Germany), INFN-CNAF (Italy), NL-T1 (Netherlands), PIC (Spain), RAL (UK) and BNL (USA), the Tier-2 facilities worldwide and large non-WLCG resource providers. Major contributors of computing resources are listed in Ref.~\cite{ATL-SOFT-PUB-2023-001}.

We gratefully acknowledge the support of ANPCyT, Argentina; YerPhI, Armenia; ARC, Australia; BMWFW and FWF, Austria; ANAS, Azerbaijan; CNPq and FAPESP, Brazil; NSERC, NRC and CFI, Canada; CERN; ANID, Chile; CAS, MOST and NSFC, China; Minciencias, Colombia; MEYS CR, Czech Republic; DNRF and DNSRC, Denmark; IN2P3-CNRS and CEA-DRF/IRFU, France; SRNSFG, Georgia; BMBF, HGF and MPG, Germany; GSRI, Greece; RGC and Hong Kong SAR, China; ISF and Benoziyo Center, Israel; INFN, Italy; MEXT and JSPS, Japan; CNRST, Morocco; NWO, Netherlands; RCN, Norway; MNiSW, Poland; FCT, Portugal; MNE/IFA, Romania; MESTD, Serbia; MSSR, Slovakia; ARIS and MVZI, Slovenia; DSI/NRF, South Africa; MICIU/AEI, Spain; SRC and Wallenberg Foundation, Sweden; SERI, SNSF and Cantons of Bern and Geneva, Switzerland; NSTC, Taipei; TENMAK, T\"urkiye; STFC/UKRI, United Kingdom; DOE and NSF, United States of America.

Individual groups and members have received support from BCKDF, CANARIE, CRC and DRAC, Canada; CERN-CZ, PRIMUS 21/SCI/017 and UNCE SCI/013, Czech Republic; COST, ERC, ERDF, Horizon 2020, ICSC-NextGenerationEU and Marie Sk{\l}odowska-Curie Actions, European Union; Investissements d'Avenir Labex, Investissements d'Avenir Idex and ANR, France; DFG and AvH Foundation, Germany; Herakleitos, Thales and Aristeia programmes co-financed by EU-ESF and the Greek NSRF, Greece; BSF-NSF and MINERVA, Israel; Norwegian Financial Mechanism 2014-2021, Norway; NCN and NAWA, Poland; La Caixa Banking Foundation, CERCA Programme Generalitat de Catalunya and PROMETEO and GenT Programmes Generalitat Valenciana, Spain; G\"{o}ran Gustafssons Stiftelse, Sweden; The Royal Society and Leverhulme Trust, United Kingdom.

In addition, individual members wish to acknowledge support from CERN: European Organization for Nuclear Research (CERN PJAS); Chile: Agencia Nacional de Investigaci\'on y Desarrollo (FONDECYT 1190886, FONDECYT 1210400, FONDECYT 1230812, FONDECYT 1230987); China: Chinese Ministry of Science and Technology (MOST-2023YFA1605700), National Natural Science Foundation of China (NSFC - 12175119, NSFC 12275265, NSFC-12075060); Czech Republic: PRIMUS Research Programme (PRIMUS/21/SCI/017); EU: H2020 European Research Council (ERC - 101002463); European Union: European Research Council (ERC - 948254, ERC 101089007), Horizon 2020 Framework Programme (MUCCA - CHIST-ERA-19-XAI-00), European Union, Future Artificial Intelligence Research (FAIR-NextGenerationEU PE00000013), Italian Center for High Performance Computing, Big Data and Quantum Computing (ICSC, NextGenerationEU); France: Agence Nationale de la Recherche (ANR-20-CE31-0013, ANR-21-CE31-0013, ANR-21-CE31-0022, ANR-22-EDIR-0002), Investissements d'Avenir Labex (ANR-11-LABX-0012); Germany: Baden-Württemberg Stiftung (BW Stiftung-Postdoc Eliteprogramme), Deutsche Forschungsgemeinschaft (DFG - 469666862, DFG - CR 312/5-2); Italy: Istituto Nazionale di Fisica Nucleare (ICSC, NextGenerationEU), Ministero dell'Università e della Ricerca (PRIN - 20223N7F8K - PNRR M4.C2.1.1); Japan: Japan Society for the Promotion of Science (JSPS KAKENHI JP21H05085, JSPS KAKENHI JP22H01227, JSPS KAKENHI JP22H04944, JSPS KAKENHI JP22KK0227); Netherlands: Netherlands Organisation for Scientific Research (NWO Veni 2020 - VI.Veni.202.179); Norway: Research Council of Norway (RCN-314472); Poland: Polish National Agency for Academic Exchange (PPN/PPO/2020/1/00002/U/00001), Polish National Science Centre (NCN 2021/42/E/ST2/00350, NCN OPUS nr 2022/47/B/ST2/03059, NCN UMO-2019/34/E/ST2/00393, UMO-2020/37/B/ST2/01043, UMO-2021/40/C/ST2/00187, UMO-2022/47/O/ST2/00148); Slovenia: Slovenian Research Agency (ARIS grant J1-3010); Spain: BBVA Foundation (LEO22-1-603), Generalitat Valenciana (Artemisa, FEDER, IDIFEDER/2018/048), Ministry of Science and Innovation (MCIN \& NextGenEU PCI2022-135018-2, MICIN \& FEDER PID2021-125273NB, RYC2019-028510-I, RYC2020-030254-I, RYC2021-031273-I, RYC2022-038164-I), PROMETEO and GenT Programmes Generalitat Valenciana (CIDEGENT/2019/023, CIDEGENT/2019/027); Sweden: Swedish Research Council (VR 2018-00482, VR 2022-03845, VR 2022-04683, VR grant 2021-03651), Knut and Alice Wallenberg Foundation (KAW 2017.0100, KAW 2018.0157, KAW 2018.0458, KAW 2019.0447, KAW 2022.0358); Switzerland: Swiss National Science Foundation (SNSF - PCEFP2\_194658); United Kingdom: Leverhulme Trust (Leverhulme Trust RPG-2020-004), Royal Society (NIF-R1-231091); United States of America: U.S. Department of Energy (ECA DE-AC02-76SF00515), Neubauer Family Foundation.

%
%


%
%
%
%
%
%
%

%
%
%
%
%
%

%
%

%
%
%
\printbibliography
\clearpage
 
\begin{flushleft}
\hypersetup{urlcolor=black}
{\Large The ATLAS Collaboration}

\bigskip

\AtlasOrcid[0000-0002-6665-4934]{G.~Aad}$^\textrm{\scriptsize 103}$,
\AtlasOrcid[0000-0001-7616-1554]{E.~Aakvaag}$^\textrm{\scriptsize 16}$,
\AtlasOrcid[0000-0002-5888-2734]{B.~Abbott}$^\textrm{\scriptsize 121}$,
\AtlasOrcid[0000-0002-1002-1652]{K.~Abeling}$^\textrm{\scriptsize 55}$,
\AtlasOrcid[0000-0001-5763-2760]{N.J.~Abicht}$^\textrm{\scriptsize 49}$,
\AtlasOrcid[0000-0002-8496-9294]{S.H.~Abidi}$^\textrm{\scriptsize 29}$,
\AtlasOrcid[0009-0003-6578-220X]{M.~Aboelela}$^\textrm{\scriptsize 44}$,
\AtlasOrcid[0000-0002-9987-2292]{A.~Aboulhorma}$^\textrm{\scriptsize 35e}$,
\AtlasOrcid[0000-0001-5329-6640]{H.~Abramowicz}$^\textrm{\scriptsize 152}$,
\AtlasOrcid[0000-0002-1599-2896]{H.~Abreu}$^\textrm{\scriptsize 151}$
% ...
\bigskip
\\

% ...
$^{16}$Department for Physics and Technology, University of Bergen, Bergen; Norway.\\
% ...
$^{29}$Physics Department, Brookhaven National Laboratory, Upton NY; United States of America.\\
% ...
$^{35}$$^{(a)}$Facult\'e des Sciences Ain Chock, Universit\'e Hassan II de Casablanca;$^{(b)}$Facult\'{e} des Sciences, Universit\'{e} Ibn-Tofail, K\'{e}nitra;$^{(c)}$Facult\'e des Sciences Semlalia, Universit\'e Cadi Ayyad, LPHEA-Marrakech;$^{(d)}$LPMR, Facult\'e des Sciences, Universit\'e Mohamed Premier, Oujda;$^{(e)}$Facult\'e des sciences, Universit\'e Mohammed V, Rabat;$^{(f)}$Institute of Applied Physics, Mohammed VI Polytechnic University, Ben Guerir; Morocco.\\
% ...
$^{44}$Physics Department, Southern Methodist University, Dallas TX; United States of America.\\
% ...
$^{49}$Fakult\"{a}t Physik , Technische Universit{\"a}t Dortmund, Dortmund; Germany.\\
% ...
$^{55}$II. Physikalisches Institut, Georg-August-Universit\"{a}t G\"ottingen, G\"ottingen; Germany.\\
% ...
$^{103}$CPPM, Aix-Marseille Universit\'e, CNRS/IN2P3, Marseille; France.\\
% ...
$^{121}$Homer L. Dodge Department of Physics and Astronomy, University of Oklahoma, Norman OK; United States of America.\\
% ...
$^{151}$Department of Physics, Technion, Israel Institute of Technology, Haifa; Israel.\\
$^{152}$Raymond and Beverly Sackler School of Physics and Astronomy, Tel Aviv University, Tel Aviv; Israel.\\
% ...
$^{a}$ Also Affiliated with an institute covered by a cooperation agreement with CERN.\\
$^{b}$ Also at An-Najah National University, Nablus; Palestine.\\
$^{c}$ Also at Borough of Manhattan Community College, City University of New York, New York NY; United States of America.\\
$^{d}$ Also at Center for High Energy Physics, Peking University; China.\\
$^{e}$ Also at Center for Interdisciplinary Research and Innovation (CIRI-AUTH), Thessaloniki; Greece.\\
$^{f}$ Also at Centro Studi e Ricerche Enrico Fermi; Italy.\\
% ...

\end{flushleft}


%

%
%
%
%

%
%
%
%

\end{document}